\documentclass[11pt]{elsarticle}

\catcode`\@=11
\global\arraycolsep=2pt
\usepackage{ascmac}
\usepackage{amssymb}
\usepackage{amsmath}
\usepackage[hypertex]{hyperref}

\makeatletter
 
 \@addtoreset{equation}{section}
\makeatother

\title{Scale invariance vs conformal invariance}
\author{Yu Nakayama,}

\begin{document}
\maketitle

\vspace*{2.0cm}
\begin{center}
{\it California Institute of Technology,  \\ 
452-48, Pasadena, California 91125, USA}
\vspace{2.8cm}
\end{center}

\section*{Abstract}
In this review article, we discuss the distinction and possible equivalence between scale invariance and conformal invariance in relativistic quantum field theories. Under some technical assumptions, we can prove that scale invariant quantum field theories in $d=2$ dimension necessarily possess the enhanced conformal symmetry.  The use of the conformal symmetry is well appreciated in the literature, but the fact that all the scale invariant phenomena in $d=2$ dimension enjoy the conformal property relies on the deep structure of the renormalization group. The outstanding question is whether this feature is specific to $d=2$ dimension or it holds in higher dimensions, too. 
As of January 2014, our consensus is that there is no known example of scale invariant but non-conformal field theories in $d=4$ dimension under the assumptions of (1) unitarity,  (2) Poincar\'e invariance (causality), (3) discrete spectrum in scaling dimension, (4) existence of scale current and (5) unbroken scale invariance in the vacuum. 
We have a perturbative proof of the enhancement of conformal invariance from scale invariance based on the higher dimensional analogue of Zamolodchikov's $c$-theorem, but the non-perturbative proof is yet to come. 
As a reference we have tried to collect as many interesting examples of scale invariance in relativistic quantum field theories as possible in this article.
We give a complementary holographic argument based on the energy-condition of the gravitational system and the space-time diffeomorphism in order  to support the claim of the symmetry enhancement. 
 We believe that the possible enhancement of conformal invariance from scale invariance reveals the sublime nature of the renormalization group and space-time with holography.

This review is based on a lecture note on scale invariance vs conformal invariance, on which
 the author gave lectures at Taiwan Central University for the 5th Taiwan School on Strings and Fields.
%I appreciate your comments very much.\footnote{Reach me at nakayama@theory.caltech.edu .} 

\newpage
\tableofcontents
\newpage
\section{Introduction}

In late 1970s, there was a legendary international conference at Dubna, a city in the former Soviet Union. The theme of the workshop was scale invariance in physics. The organizer was N.~N.~ Bogoliubov, who is one of the founders of the renormalization group and who first introduced the concept of scale invariance in quantum field theory. His influence on renormalization group, among other numerous contributions to mathematical physics, was enormous.
For example, K.~Wilson later recalls at the Nobel lecture that it was one mysterious chapter in his famous textbook \cite{Bogo} that mesmerized him when he was a PhD student and eventually led him to the later study of the renormalization group.

At one point of the conference after the plenary session for scale invariance, one western physicist asked a question, which KGB might have called ``provocative" in those days: What is the difference between scale symmetry and conformal symmetry? The speaker got stuck and hesitated to answer. The chairman of the session, Bogoliubov, however, immediately took the microphone and spoke with dignity
``There is no mathematical difference, but when some young people want to use a fancy word they call it Conformal Symmetry". One young eastern physicist could not stand this answer, stood up and yelled ``15 parameters!". The yell echoed but did not reach. The adjournment of the meeting was quickly announced.

The name of the eastern young physicist is A.A~Migdal, who is one of the earlier advocates of the use of conformal invariance in theoretical physics. This anecdote can be found in his biography ``Ancient history of conformal field theory" \cite{Migdal}. The aim of this review article is first to understand the mathematical as well physical idea  behind this 40 year long debate about the relation between scale invariance and conformal invariance. As we go along, we will see the amusing twist and turn: {\it mathematically} what Migdal yelled was correct, but {\it physically} what Bogoliubov answered may be true!
Under some mild assumptions, scale invariant quantum field theories always seem to be conformal invariant! To appreciate the statement better, we have to start with the distinction between scale invariance and conformal invariance.

In elementary school, we learn a rectangle is not a square. In graduate school, we, however, learn (?) scale invariance is conformal invariance. In the era of AdS/CFT, everybody touts conformal invariance, so without much reflection we are somehow accustomed to the ``belief" that scale invariant quantum field theories show conformal invariance. Let us, however, pose here and ask ourselves the questions: has it been proved? Are we really sure our beloved $\mathcal{N}=4$ supersymmetric Yang-Mills theory is conformal invariant? Are we just some young people who want to use a fancy word? 

Of course scale invariance does not imply conformal invariance at least at the level of the superficial mathematical definition. Otherwise, we do not need two different names for the identical concept. However,
our nature may be more beautiful than we naively expect. This hidden beauty of the nature is the essential reason why we are interested in the question about the relation between scale invariance and conformal invariance.
Our ultimate goal of the review is to understand the mysterious symmetry enhancement in quantum field theories and gravitational systems.

Indeed, such a symmetry enhancement may happen. In general relativity, for instance, there is a famous theorem by Israel \cite{Israel} that states  all axisymmetric static black holes must be spherically symmetric in $d=4$ dimension (i.e. Schwartzshild black hole solution)  as vacuum solutions of Einstein's equation. This is a symmetry enhancement from the axisymmetry to the spherical symmetry due to the dynamics of the classical gravitational system. Presumably it has a deep quantum gravitational origin. The black hole has no hair, and any classical probe cannot distinguish the microscopic degrees of freedom. We expect that the symmetry enhancement does not occur with no good reason.

Scale invariance is ubiquitous in our nature. We can easily find them in our daily lives. The coastline, snowflakes, lightening, and stock charts, all show scale invariance or fractal structures. Even at a supermarket we can find a vegetable called roman cauliflower (a.k.a broccoli romanesque), which shows a beautiful discrete scale invariant structure.\footnote{It is good with pasta, in particular with oil-based source.}  The (discrete) scale invariance here is realized as a self-similarity: if we look at the same system closer or further away, it looks similar. Does our society show  a self-similar organization structure?
The repetitive structure begs the question: is there any fundamental component in such a self-similar or scale invariant object? Or, is the self-similar structure itself the fundamental organizing principle?

The most notable application of the scale invariance in theoretical physics is the renormalization group flow. One of the central dogmas in the 20th century physics was Wilson's renormalization group (see e.g. \cite{Wilson:1973jj}). In a plain word, his idea of the renormalization  group is a successive application of  scale transformation and coarse graining. If we are interested in the long range universal physics, we can then integrate out the  short-range degrees of freedom that might show non-universal dynamics. Afterwards we can talk about the  effective theory of the long range degrees of freedom by keeping only relevant degrees of freedom of the theory and focusing on the relevant parameters. All the detailed short  range information is judiciously encoded in the process of the renormalization of the relevant parameters.

Schematically, renormalization group transformation is realized as the path integral form:
\begin{align}
e^{-\mathcal{S}_{\mathrm{eff}}[\tilde{\Phi};g(\Lambda)]} = \int_{\Lambda}^{\Lambda_0} \mathcal{D}\Phi  e^{-\mathcal{S}_0[\Phi; g(\Lambda_0)]}    
\end{align}
where $\Lambda$ is the renormalization scale. Such path integrals appear either in quantum field theories or in the statistical mechanical ensembles. In either situations, we integrate out the ``high energy degrees of freedom" from the scale between $\Lambda_0$ and $\Lambda$ contained in a field $\Phi$ by keeping only the low energy mode below $\Lambda$.
We expect that when we take $\Lambda$ sufficiently small, there are only a few universal parameters that will characterize the system. The success of the idea of the renormalization group transformation explains why we can understand our world with sufficient accuracy even without knowing the most detailed elementary microscopic physics (e.g. string theory?). Without exaggeration, this is how our standard model works in elementary particle physics, how Einstein's general relativity works in gravitational physics, how the hydrodynamics works in such a way that our airplane flies, and  how the idea of universality of the statistical mechanics work in condensed matter physics.

From Wilson's renormalization group viewpoint, it may be natural that we would expect a scale invariant fixed point in the infrared (IR) limit. The intuition is that if we encounter any non-trivial energy scale, we can simply integrate out the corresponding degrees of freedom. Eventually, there should be no scale at all!
 As a consequence, the universality class of the long range behavior in quantum field theories or many body systems is characterized by the fixed point of the renormalization group flow. This idea has a great success in particular in $d=2$ dimension, dimension where the universality combined with conformal invariance made it possible to classify the critical phenomena.

\begin{figure}[tbh]
\begin{center}
\includegraphics[width=\linewidth]{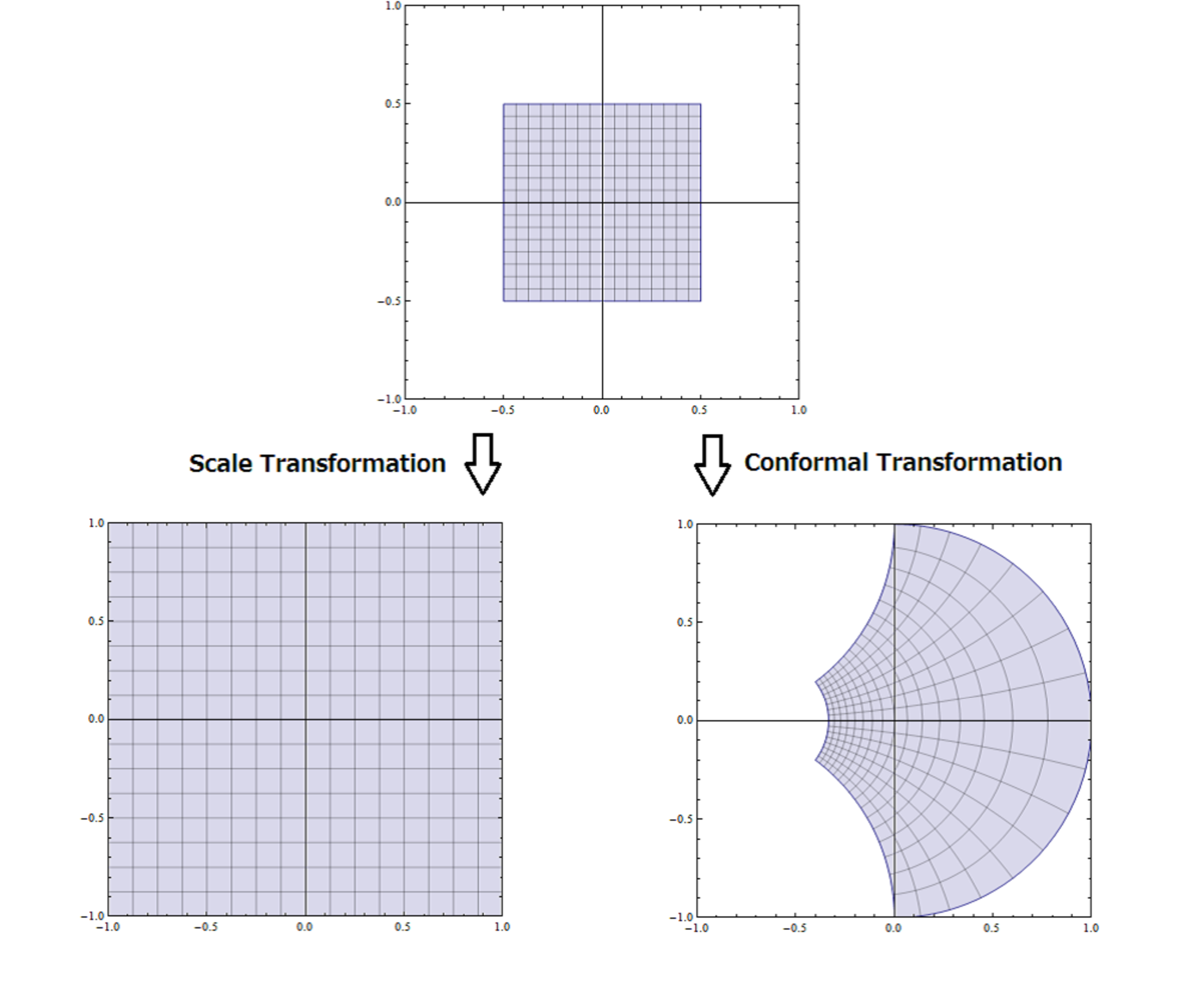}
\end{center}
\caption{We see a graphical distinction between scale invariance and conformal invariance in $d=2$ dimension. Our perception is approximately invariant under scale transformation but not invariant under conformal transformation. Do you think  conformal transformation keeps the ``same shape"?}
\label{fig1}
\end{figure}

Conformal invariance is the other leading actor of this review article.
What is conformal invariance? While we will describe the mathematical definition of the conformal invariance momentarily, an intuitive idea can be guessed from the root: ``con-formal" comes from a Latin word {\it conformalis} ``having the same shape".  It is the transformation that leaves the size of the angle between corresponding curves unchanged. Such transformations are more general than the scale transformation. See fig \ref{fig1}. In $d=4$ dimension, the parameter of the conformal transformation is 15 rather than 11 of Poincar\'e + scale symmetry.

Therefore there is no a priori reason why a given scale invariant system must show conformal invariance. Nevertheless, as we mentioned at the beginning, there should be a reason why theoretical physicists in our generation have not made a clear  distinction in our everyday research. We believe it is because we have empirical knowledge that almost all scale invariant quantum field theories that we know show conformal invariance so there is no point in emphasizing it in our textbooks. Do we have to talk about unicorns or dragons in zoology lectures or in textbooks (even though they might exist in principle due to our limited knowledge)? 

But it is still a mystery: why does scale invariance have to accompany conformal invariance?
The aim of this review article is to uncover the puzzle behind the enhancement of conformal invariance from scale invariance. As we will see, the underlying reason must be  related to a deep property of the renormalization group flow. In particular, the notion of irreversibility of the renormalization group flow and counting degrees of freedom will play a crucial role in our discussions. 

\begin{figure}[tbh]
\begin{center}
\includegraphics[width=\linewidth]{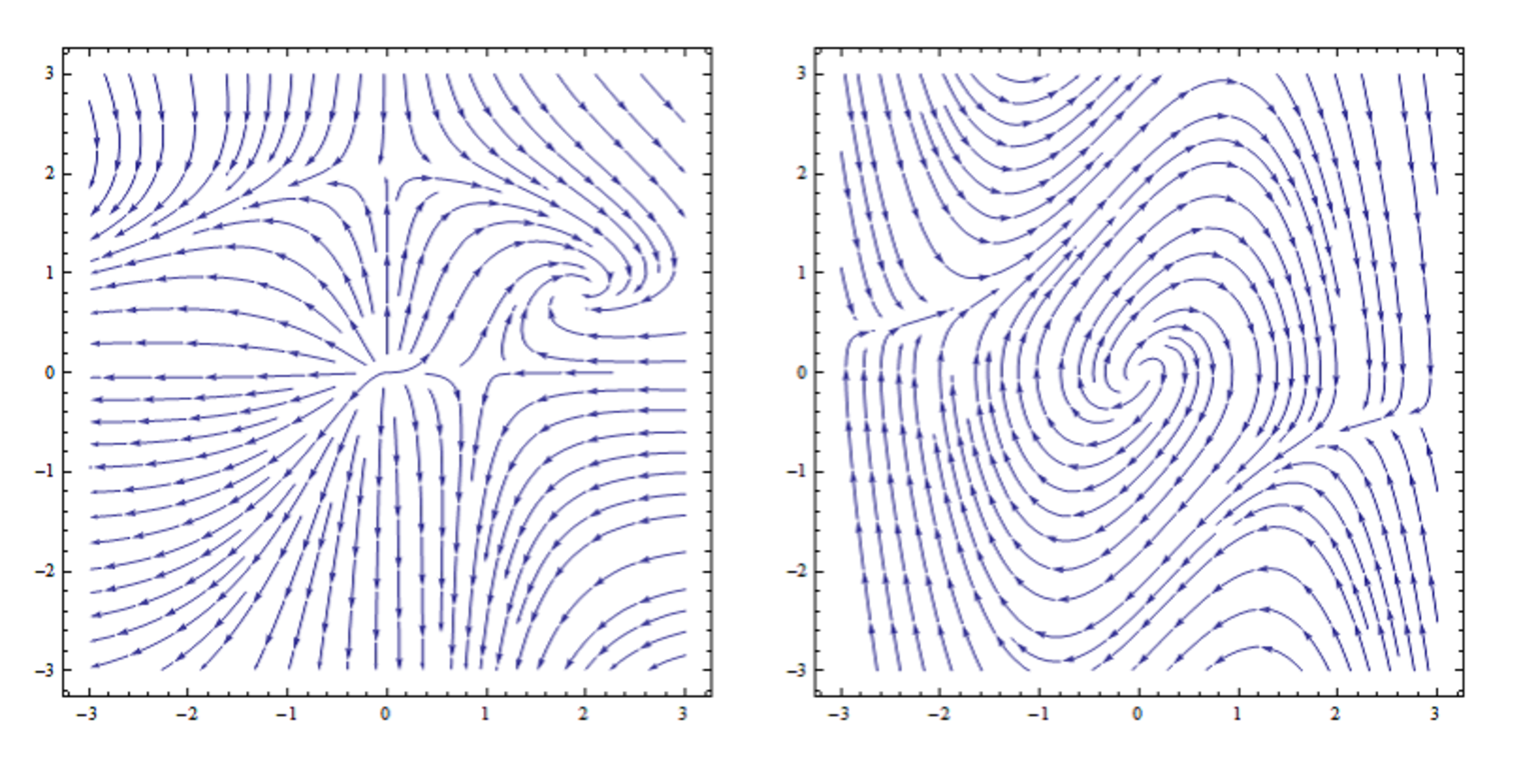}
\end{center}
\caption{We show artificially generated examples of (possible?) renormalization group flow. The left hand side contains UV fixed point as well as IR fixed point. The right hand side shows a cyclic behavior with UV fixed point.}
\label{cyclic}
\end{figure}

The idea of irreversibility of the renormalization group flow can be understood in an intuitive way: as mentioned, the renormalization group flow is accompanied with a kind of coarse graining. We lose information along the flow. It is very counterintuitive if the renormalization group flow shows a cyclic or chaotic behavior (although it was envisaged by the pioneers \cite{Wilson:1970ag}\cite{Bogo}). See fig \ref{cyclic} for illustration. The field theory understanding of this coarse graining is supported by the so-called ``$c$-theorem"  that dictates there exists a function (called ``$c$"-function) that monotonically decreases along the renormalization group flow. Roughly speaking, this function counts the degrees of freedom at a given energy scale.
If such a function exists, the cyclic or chaotic behavior in renormalization group flow cannot occur. In relativistic quantum field theories in $d=2$ dimension, there is a proof that such a function indeed exists and the renormalization group flow is irreversible \cite{Zamolodchikov:1986gt}.

As we will see, scale invariant but non-conformal field theories are intimately connected with  the possibility of such a cyclic or chaotic behavior in the renormalization group flow. At least this is the case within perturbation theory in $d=4$ dimension as first emphasized in \cite{Fortin:2011sz}. There is a clear tension between them. The above mentioned proof in $d=2$ dimension of the irreversibility of the renormalization group flow implies that scale invariance must be enhanced to conformal invariance in $d=2$ dimension. In this review article, we would like to report the current status of the situations in higher dimensions. Unfortunately, as of January 2014, there is no compelling non-perturbative proof that scale invariance is enhanced to conformal invariance in higher dimensions with no counterexamples reported so far under some reasonable assumptions. However, our accumulated knowledge of the subject in recent years in line with the development of the higher dimensional analogue of the $c$-theorem may suggest that the complete proof will be close by.

To look for a non-perturbative evidence, we will study the holographic argument. This is the second aim of this review article, which was mainly developed by the present author.
Holographic principle is by far the most profound but conjectural principle that connects non-gravitational quantum physics and the corresponding (quantum) gravity. It has a beautiful concrete realization known as AdS/CFT correspondence, and we have culminating evidence that it is true. With the holographic principle in mind, our idea is to explore the hidden side of the quantum field theories from the analysis in gravitational backgrounds. According to the   AdS/CFT correspondence, the classical gravity will describe a certain strongly coupled limit of the dual quantum field theories, and it is expected that  it provides non-perturbative understanding of them.

As we have already mentioned in the black hole example,  gravitational systems show their own symmetry enhancement mechanism. We conjecture that the consistency of the quantum gravity is encoded in the consistency of the renormalization group flow through the holographic equivalence, and vice versa. The $c$-function associated with the renormalization group flow can be viewed as ``entropy" of the gravitational system, which should be monotonically decreasing  along the evolution. Along the same line of reasoning, we will argue that it is  reasonable that scale invariant holographic configurations show the enhanced enhanced symmetry corresponding to conformal invariance. Indeed, we will see that we can provide a holographic proof of the statement based on a certain energy-condition.

At the same time, we would like to ask some pertinent questions in quantum gravity. What would be the fundamental mechanism to exclude seemingly pathological geometries from quantum gravity such as superluminal propagation of information, closed time-like curves, and so on? We believe that the consistency of the renormalization group flow e.g. absence of the limit cycle or chaotic behavior would give a hint to understand the fundamental aspects of quantum geometry and quantum gravity through the holography. Our earlier attempt to discuss the issue from the holographic approach is summarized in a review paper \cite{Nakayama:2010zz}.

There was a debate whether scale invariance without conformal invariance is possible or not in $d=4$ dimension over the last couple of years, but we are happy to announce that we converge to the point our holographic argument predicts \cite{Nakayama:2012nd}\cite{Luty:2012ww}\cite{Fortin:2012hn}. However, the complete proof is yet to come.
We wish the complete field theory proof would appear soon to make this review article partly obsolete.

\subsection{Physical applications we have in mind}

Suppose we could show that scale invariant quantum field theories necessarily possess the enhanced symmetry of conformal invariance. What could we learn? What would be the practical benefit of conformal invariance compared with the mere scale invariance? Are there any real applications of conformal invariance relevant for our understanding of physical phenomena in nature? Or, alternatively, is there any physics which relies on  scale invariance without conformal invariance? Before going into the technical analysis of the relation between scale invariance and conformal invariance from section \ref{section2} on, we would first like to show how the distinction and possible equivalence between scale invariance and conformal invariance plays an important role in various examples in condensed matter physics, high energy physics, gravitational physics, as well as other arenas of physics and beyond. The list is far from complete, so we would like to encourage interested readers to look for possible further applications. 

\begin{itemize}

\item  Critical phenomena:

Boiling water may be the first example of phase transition we learn in the science classroom during our elementary education. The phase transition under the normal pressure we typically experience on earth is first order. However, it is known that if we tune the pressure, the thermal transition of water can be second order (at $ T = 273.16 \mathrm{K}$ and $P = 0.61 \mathrm{kPa}$). The second order phase transition there is known as an example of critical phenomena (see e.g. \cite{Stanley} for a review).
A surprising fact is that the second order phase transition of the water shows the universality. It means that various critical exponents take the universal values:
the jump of the density near the critical temperature $T_c$ for example shows
\begin{align}
\delta \rho (T) \sim (T-T_c)^\beta \label{cr1}
\end{align}
where $\beta \sim 0.325$, and the specific heat shows 
\begin{align}
C \sim (T-T_c)^{-\alpha} \label{cr2}
\end{align}
where $\alpha \sim 0.11$.
These numbers are precisely  same as that of the $d=3$ dimensional Ising model at the critical point, where for example the density $\rho$ in the above water example is replaced by the spontaneous magnetization, up to experimental/numerical errors. One of the goal of the theoretical physics is to understand these universal properties associated with the critical phenomena. How do they occur emergently and why do they show the universality?

The crucial observation is that at the critical point, the so-called ``scaling hypothesis" applies \cite{Widom}, which means that the thermodynamic quantities show the scaling behavior. As already mentioned, the best way to understand the critical behavior with the scale invariance would be the renormalization group. The second order phase transition and the associated critical phenomena are characterized by the scale invariant fixed point of the effective Hamiltonian (or free energy).

The scaling hypothesis alone cannot determine the critical exponent while it may give some relations among them. Here is the power of conformal invariance. Suppose that the critical phenomena are classified by the conformal invariant fixed point. Then the classification of the critical phenomena is equivalent to the classification of conformal field theories. For instance, in $d=2$ dimension, a complete classification (for small central charge, relevant for condensed matter applications) of the conformal field theories is possible \cite{Belavin:1984vu} , and the critical exponent is solely determined by the constraint from the conformal invariance. 

It is very likely that the critical exponents of the higher dimensional critical phenomena are controlled by the dynamics of conformal field theories as in $d=2$ dimension. Indeed, the recent development (which we will briefly mention in section \ref{cci}) shows that the critical exponents such as \eqref{cr1} \eqref{cr2} are completely consistent with the conformal hypothesis. It is this ``conformal hypothesis" than the mere scaling hypothesis that enables us to compute the critical exponents by using the constraint from the conformal symmetry.

It is therefore of theoretical as well as practical importance to understand under which conditions we expect the conformal invariance in critical phenomena. Our study of the relation between scale invariance and conformal invariance is essential because the definition of the critical phenomena and the simple renormalization group argument by themselves  do not imply the emergence of the conformal invariance (at least without further thoughts). Nevertheless, we are led to the belief that the conformal hypothesis is more or less valid through our experience. One of the final goals of this study is to give a firm basis of this belief.

\item Particle physics:

Quantum field theories are foundation of particle physics. The notion of approximate or asymptotic scale invariance in quantum field theories describing the elementary particles played a significant role in understanding the high energy properties of QCD, which is asymptotic to the trivial Gaussian scale invariant fixed point (massless free field theories) in the ultraviolet (UV) limit \cite{Gross:1973id}\cite{Politzer:1973fx}. The free Gaussian fixed point there turns out to be also conformal invariant. A slight violation of the scale invariance is known as the Bjorken scaling in QCD phenomenology \cite{Bjorken:1968dy}. 
Historically speaking, the conformal symmetry itself was discovered in the study of Maxwell theory of electromagnetism, which is probably the best known example of conformal field theories realized in nature. 

The scale invariance is tightly related to massless properties of the quantized field. It is obvious that the mass ``scale" will break the scale invariance. However, it is slightly non-trivial that classical scale invariance will be typically broken in quantum field theories due to the renormalization group effect. Thus, the emergence of scale invariance in particle physics is quite non-trivial.
We should again emphasize that the conformal invariance is stronger than the mere scale invariance at this point. Although related, the massless field theories (even if we ignore the interaction) do not necessarily possess conformal invariance. A good example is the Einstein gravity with zero cosmological constant. At the linearized level, it describes the massless scale invariant graviton, but the theory is not conformal invariant off-shell (while the dispersion of the on-shell graviton is conformal invariant).

In some beyond-the-standard-model scenarios, the existence of  non-trivial scale invariant fixed points plays some important roles (see e.g. \cite{Luty:2004ye} for a review). Our standard model has the so-called hierarchy problem. The weak-energy scale is more than fifteen digit smaller than the Planck scale, and the naive power-counting suppressed by the Planck scale (or any other high energy scale) gives various constraints on the model building without allowing fine-tuning. One typical idea to use the scale invariant non-trivial fixed point as a part of the model building tools is to rely on the ``anomalous dimensions" of various operators to deviate from the naive engineering dimensional counting.

However, the extra constraint from the conformal invariance may give further restrictions on the possible amount of anomalous dimensions \cite{Rattazzi:2008pe}. The structure of the renormalization group such as the irreversibility may further restrict the possibility. On the other hand, if the scale invariance without conformal invariance were allowed, the constraint could be relaxed with a slight loss of predictive powers. It relies on our philosophical viewpoint which is better (i.e. generalization vs predictive power), but the discussion would be moot for the latter possibility of using scale invariant but non-conformal field theories if it turns out that such theories do not exist at all.

There is also a speculative idea that our standard model may be embedded in the scale invariant UV fixed point. A partial implementation of the idea led to the prediction of 126 GeV Higgs mass \cite{Shaposhnikov:2009pv} (see also \cite{Tavares:2013dga}). The idea is tightly related to the asymptotic safety scenario, which we would like to turn next in the context of gravity.

\item Cosmology and gravitational physics:

There are two major directions to apply scale invariance in cosmology and gravitational physics. The first one is the possibility of asymptotic safety in quantum gravity \cite{Weinberg:1980gg} (see e.g. \cite{Nagy:2012ef} for a review). The conjecture is that  although Einstein gravity is non-renormalizable in the power-counting sense and it is strongly coupled in the ultraviolet, it may have a non-trivial UV fixed point and the lack of the predictive power in non-renormalizable theory may be circumvented by assuming that we are on the flow emanating from this hypothetical UV fixed point of quantum gravity. Obviously, the UV fixed point must be scale invariant. It is less obvious if such a fixed point (if any) would show the conformal invariance. As we will discuss in this review, the gravitational theory (either scale invariant or conformal invariant) is special and does not satisfy the usual assumptions we would make when we talk about the enhancement of conformal invariance from scale invariance.

Another possible application of scale invariance in cosmology is the primordial fluctuations in cosmic microwave background \cite{Komatsu:2010fb}. The fluctuation of the cosmic microwave background is known to be nearly scale invariant from the observational data, and one explanation is from the approximate de-Sitter invariance during the inflation. The crucial point is the de-Sitter invariance could predict the enhanced (nearly) conformal invariant spectrum rather than the mere scale invariant spectrum \cite{Maldacena:2002vr}. So far, observationally we have no evidence for or against the enhanced conformal invariance in the cosmic microwave background. It is very interesting to see if this hypothetical conformal invariance could be observed in the cosmic microwave background in the study of the higher point functions in the tensor mode. At the same time, it would be very interesting if we could construct a cosmological model that predicts only the scale invariance without the enhanced conformal invariance.
Based on the holography, it may be related to the possibility to have scale invariant but non-conformal field theories in $d=3$ dimension. 

\item String theory:

The foundation of the worldsheet formulation of the first quantized string theory is a conformal field theory (see e.g. \cite{Polchinski:1998rq}\cite{Polchinski:1998rr} for a review). In order to get rid of the space-time ghost and construct the consistent loop amplitudes, it is not enough to have scale invariance: the worldsheet theory must possess the conformal invariance. The distinction between scale invariance and conformal invariance is crucial, and there can be a target space which is not consistent as a string background due to the lack of the worldsheet conformal invariance although it is scale invariant. Note that the situation is a slightly peculiar because the general theorem of enhancement of conformal invariance from scale invariance in $d=2$ dimension does not apply here because the worldsheet theory violates some of the assumptions made. 

As we have already mentioned in the introduction the holography is a novel idea to understand the nature of strongly coupled quantum field theories from gravitational analysis. It was originally developed in the string theory by connecting the symmetry of the AdS space-time and the symmetry of the conformal field theories. Here, the conformal invariance is taken for granted. 
In this review, we would like to ask the question why and how the holography relies on the underlying conformal invariance of the dual field theories for the consistency. At the same time, we would like to reveal the nature of the space-time from the possible constraint on the dual field theories.

\item Biology, economy, human behavior etc:

There are many other arenas of science in which we would encounter scale invariance without conformal invariance. As we have already mentioned in the introduction, in our daily lives there is no shortage of scale invariant phenomena. For another amusing example, Zipf's law \cite{Zipf} predicts that the frequency of the appearance of words in this review (say, the most common word ``the" gets a rank $r$ of 1 and "broccoli" among others would get the lowest rank) shows the scaling law $Q(r) = Fr^{-1/\alpha}$. Here $\alpha$ measures the richness of one's vocabulary. As far as the author knows, there is no conformal extension of this observation. 

There are two different categories here. In the first case, there is no obvious group theoretic way to extend the scaling symmetry to conformal symmetry. In this case, we can easily explain the non-enhancement of symmetry from scale invariance: simply we cannot do it. Presumably, the stock market economy, which shows power-law scaling behavior falls in this category while there is a possibility of conformal invariance in bond market because it may be described by a free string action (see e.g. \cite{Dash} and reference therein).
The second case, which is more non-trivial and is similar to the quantum field theory examples so far, is that although the algebraic enhancement of the conformal symmetry from scale invariance is possible, in reality it does not show it. Let us only mention one example: human perception of visionary. Our visual perception is approximately Euclidean invariant as well as scale invariant (so there is a group theoretic chance of enhancement to conformal invariance), but actually we know it is not conformal invariant. If it were conformal invariant, we would not perceive the right panel of Figure \ref{fig1} as somewhat distorted compared with the left one! Indeed, as we will mention in this review article, there may be a field theoretic understanding of human perception based on scale invariant but non-conformal field theories \cite{Bialek:1986it}\cite{Bialek:1987qc}\cite{Nakayama:2010ye} as we mentioned in the introduction.

\end{itemize}

\subsection{Organization of the review and conventions}

This review is based on a lecture note on scale invariance vs conformal invariance, on which
 the author gave lectures at Taiwan Central University for the 5th Taiwan School on Strings and Fields. Originally, we had three lectures, and the plan was as follows.
In Lecture 1, we began with the definition of scale invariance and conformal invariance, and gave criteria to distinguish between the two. Then we showed various examples of scale invariant systems with or without conformal invariance. In Lecture 2, we tried to give a field theoretic argument why scale invariant phenomena typically show enhanced conformal invariance under several assumptions. In Lecture 3, we gave holographic discussions on the distinction and showed possible enhancement from scale invariance to conformal invariance from the properties of the space-time dynamics.

In this review article, we have reorganized the lectures into 11 sections for a more coherent and self-contained presentation. Lecture 1 corresponds to section 2,3, and 4. Lecture 2 corresponds to section 5,6,7,8, and 9. Lecture 3 corresponds to section 10. Of course, we did not cover everything presented here in the lectures, so the review is significantly expanded compared with what the author discussed in the school. We have also added more recent developments after the school to catch up with the rapid pace of the researches in this field.

Before going into the main discussions, we would like to summarize the basic conventions of the review for future reference. We also provide some small tips for the readers how to read the review.
\begin{itemize}
\item As broader  audience  in mind, let us fix our notation for the space-time dimensionality. When we say $d$-dimension, it means either $d$-dimensional Euclidean (statistical) system, or relativistic $d$-dimensional space-time which is given by $d-1$ dimensional space and $1$ dimensional time.  $\mu=0,1,2 \cdots d-1$ refers to Lorentz index with the Minkowski metric $\eta_{\mu\nu}  = (-++\cdots)$. We also work in the Euclidean signature field theories with no particular mentioning of Wick rotation\footnote{With this regards, we will not discuss some subtle aspects of the global conformal transformation in Minkowski space-time. See section \ref{contro}.} $x_d = ix_0$.  In the Euclidean signature, we use $\mu = 1,2 \cdots d$ with the Euclidean metric $\delta_{\mu\nu} = (+++ \cdots)$. The antisymmetrization of the tensor indices is represented by $[IJ]$, and the symmetrization is represented by $(IJ)$.
Only when we discuss field theories in $d=2$ dimension, we will use the complex coordinate which will be explained in section \ref{cfunc}.

\item In section \ref{section10}, we add the extra holographic coordinate and use $M= 0,1,  \cdots d-1, r$ or $M=0,1, \cdots d-1 ,z$  referring to  $d+1$-dimensional space-time coordinate. We also work in the Euclidean signature as above.

\item We use the natural unit: $\hbar = c = 1$. In section \ref{section10}, we further assume the Planck constant $\kappa_d = 1$.

\item Otherwise stated, the summation convention of Einstein is used (e.g. $A_\mu B^\mu \equiv \sum_{\mu} A_\mu B^\mu$). The coordinate indices are raised and lowered with an appropriate metric $g_{\mu\nu}$. Such a ``metric" may or may not be naturally available when tensor indices take the value in more abstract spaces such as ``coupling constant spaces", and we will be explicit about raising and lowering indices then. $\Box = D^\mu D_\mu$ denotes the Laplacian (or D'Alembertian), and in flat space-time it is given by $\partial^\mu \partial_\mu$. The use of $\Box$ is not restricted to $d=4$ in this review article.  In some obvious cases without confusion, we even omit the summation conventions: e.g. $x^2 = x^\mu x_\mu$.

\item Spacetime argument $x$ in field variables such as a scalar field $\Phi(x)$
is sometimes omitted as $\Phi$ to avoid the clumsy expressions when the space-time dependence is not important.

\item Our metric and curvature convention is same as that of Wald's textbook \cite{Wald:1984rg}. See \ref{convention} for more about our conventions.

\item In most of the sections dealing with interacting  quantum field theories, we assume fields appearing in various formulae are all finitely renormalized, and we typically put the suffix $0$ otherwise. Although we do give some basic explanations of the renormalization procedure in the review article, it is beyond our scope to perform the explicit renormalization program, and we refer to textbooks (e.g. \cite{Collins:1984xc},\cite{ZinnJustin:2002ru},\cite{Weinberg:1995mt}\cite{Weinberg:1996kr}). Unfortunately, at a certain point, we have to go beyond the textbook treatment because the distinction between scale invariance and conformal invariance is so subtle. We hope interested readers will find reference provided throughout the review article useful and fill the gap if necessary.\\

\item At many places, we state that the renormalization group flow is a quantum effect. In the (classical) statistical systems, It should be understood that the effect is induced by the statistical fluctuation. They are equivalent up on Wick rotation.

\item When we talk about quantum gauge theories, the gauge fixing procedure will be always implicit. After the gauge fixing, we have to add various terms both in the action and the energy-momentum tensor. However, all these terms (that could violate scale invariance or conformal invariance) are BRST trivial, so the physical discussions will not be affected.

\item We do not cover various techniques in conformal field theories developed in particular in $d=2$ dimension. We refer to \cite{Ginsparg:1988ui}\cite{DiFrancesco:1997nk} and references therein. We also refer to the lecture note \cite{Slava} for a complementary approach in higher dimensional conformal field theories.

\item We try to avoid spinors and supersymmetry as much as possible. In some advanced sections, we assume some basic knowledge (e.g. $\gamma_\mu$ denotes the Dirac gamma matrix). For supersymmetry, we refer to textbooks  \cite{Wess:1992cp}\cite{Weinberg:2000cr} and a lecture note \cite{Argyres}. %\footnote{It is my pleasure to recall that I first encountered the mentioning of distinction between scale invariance and conformal invariance in the lecture note \cite{Argyres} when I was a grad student, which refers to the original paper \cite{Polchinski:1987dy}.} (But in reality, I'm more afraid that most of the high-energy-theory-oriented readers know what the $\mathcal{R}$-current is, while they don't know what the virial current is.)
In various symbolic formulae such as a field variation $\frac{\delta \mathcal{S}}{\delta \phi}$, we implicitly pretend as if $\phi$ were bosonic and suitable modifications are necessary for anti-commuting fermionic fields.

\item The discussion on the holographic approach in section \ref{section10} is relatively independent. The other sections are self-contained within field theory arguments.  Those who are not interested in holography can skip section \ref{section10} entirely.
 On the other hand, although understanding of section \ref{section10} requires some basic facts presented in the earlier sections, one may directly start with section \ref{section10} to grasp the holographic approach. In the latter case, we also recommend  \cite{Nakayama:2010zz} for reference.

%\item A few exercises are scattered throughout the lecture note for some random reasons. They are boxed such as

%\

%\begin{shadebox}
%(Exercise) Find as many typos and sign errors in the lecture note as possible.\footnote{Don't forget to report them!}
%\end{shadebox}

%\

%Most of them are easy, and they are more or less irrelevant for the main discussions of the lectures. However, I think some of them are fun problems to solve.

%\item If you find the reference list incomplete, please  let me know. In particular, I'm more on the high energy physics side, so if you find anything missing from the statistical mechanics side, I welcome a lot.

\item Minor revisions will be updated at https://sites.google.com/site/scalevsconformal/

\end{itemize}

\newpage

\section{Statement of the problem}\label{section2}
The aim of this section is to formulate the statement of the problem on the relation between scale invariance and conformal invariance. We begin with the mathematical distinction between scale invariance and conformal invariance, and then we proceed to discuss the criteria to distinguish them in quantum field theories. As we will argue in section \ref{local}, in local quantum field theories, the criteria are stated as a property of the energy-momentum tensor. Our discussions in this section are restricted to the case in flat Minkowski or Euclidean space-time, but in section \ref{emr}, we will learn that the problem is better understood in the curved background with more general background sources.

%We then show various examples of scale invariant field theories that often (but not always) show conformal invariance. In the examples we will discuss, I tried to collect all examples of scale invariant but non-conformal relativistic field theories as far as I have recognized. 

\subsection{Conformal algebra as maximal bosonic space-time symmetry}\label{symmetrya}
In special relativity, we postulate the Poincar\'e  algebra as the fundamental symmetry of our space-time:
\begin{align}
i[J^{\mu\nu},J^{\rho\sigma}] &=
\eta^{\nu\rho}J^{\mu\sigma}-\eta^{\mu\rho}J^{\nu\sigma} -
\eta^{\sigma\mu}J^{\rho\nu} + \eta^{\sigma\nu}J^{\rho\mu} \cr
i[P^\mu,J^{\rho\sigma}] &= \eta^{\mu\rho}P^{\sigma} -
\eta^{\mu\sigma} P^\rho \cr[P^\mu,P^\nu] &= 0 \ .
\end{align}
$P^\mu$ is the translation generator and $J^{\mu\nu}$ is the Lorentz transformation generator.
In quantum mechanics, they are realized by Hermitian operators acting on a given Hilbert space.
The representation of Poincar\'e algebra in terms of particles will naturally lead to the formalism of the quantum field theory \cite{Weinberg:1995mt}.

For a massless scale invariant theory, one can augment this Poincar\'e
algebra by adding the dilatation generator $D$ as
\begin{align}
[P^\mu,D] &= i P^\mu \cr [J^{\mu\nu},D] &= 0  \ . \label{dilatationc}
\end{align}
We use the terminology ``dilatation" and ``scale transformation" interchangeably throughout the review.
The representation theory of the Poincar\'e algebra augmented with the dilatation naturally leads to the notion of unparticles \cite{Georgi:2007ek}\cite{Georgi:2007si}. The theory of unparticles sometimes relies on a delicate difference between scale invariance and conformal invariance (see e.g. \cite{Nakayama:2007qu}). 

The generalization of the Coleman-Mandula theorem \cite{Coleman:1967ad}\cite{Haag:1974qh}
asserts (for $d\ge 3$) that the maximally enhanced bosonic symmetry
of the space-time for massless particles is obtained\footnote{With this assertion, we have to be careful about the 
assumption in the Haag-Lopuszanski-Sohnius theorem. In particular, the analyticity assumption
of S-matrix can be violated with an IR divergence in interacting
conformal field theories. For example, as noted in Weinberg's textbook, the validity of the Haag-Lopuszanski-Sohnius theorem (even the Coleman-Mandula theorem) for the Banks-Zaks fixed point has not been proved \cite{Weinberg:2000cr}. The more recent discussions on the allowed space-time symmetry {\it with the assumption of conformal invariance} can be found in \cite{Maldacena:2011jn}.} by adding 
 the special conformal transformation $K^\mu$:
\begin{align}
[K^\mu,D] &= -iK^\mu \cr [P^\mu,K^\nu] &= 2i\eta^{\mu\nu}D+
2iJ^{\mu\nu} \cr [K^\mu,K^\nu] &= 0 \cr [J^{\rho\sigma},K^\mu] &=
i\eta^{\mu\rho} K^\sigma - i\eta^{\mu\sigma} K^\rho \ . \label{groupsp}
\end{align}

As is clear from the group theory structure above, the conformal
 invariance demands scale invariance from the closure of the algebra in (\ref{groupsp}) but the converse is not
necessarily true: scale invariance does not always imply conformal
invariance. However, in many field theory examples as we will show in section \ref{examples}, we typically observe the emergence of the full
conformal invariance rather than the scale invariance alone.
The goal of this review article is to uncover deep dynamical 
reasons behind the enhancement of the symmetry from  scale invariance to conformal invariance.

\

%\begin{shadebox}
%(Exercise) Count the number of generators for the conformal group in $d=4$ dimension to show it is $15$. 
%\end{shadebox}

\

\subsection{Local field theory realization}
As extensively discussed in \cite{Weinberg:1995mt}, the realization of the Poincar\'e algebra in quantum mechanics with a particle interpretation naturally leads to quantum field theories. In this review article, we discuss relativistic quantum field theories with scale invariance or with conformal invariance. The former is referred as scale invariant field theory (SFT) and the latter is referred as conformal field theory (CFT) in the literature. We should note that the physical realization of the conformal symmetry may not be restricted to the quantum field theories based on the particle picture. It may be possible to formulate a scale invariant or conformal invariant string field theory, and more generally higher membrane theories, but we will not discuss these rather exotic possibilities in this review article.  

We may realize the Poincar\'e algebra together with dilatation and conformal transformation on space-time functions as differential operators
\begin{align}
\mathcal{P}_\mu &= i\partial_\mu \cr
\mathcal{J}_{\mu\nu} &= i(x_\mu \partial_\nu - x_\nu \partial_\mu) \cr
\mathcal{D} & = i(x^\mu \partial_\mu) \cr
\mathcal{K}_\mu &= 2ix_\mu x^\rho\partial_\rho - ix^2 \partial_\mu \ .
\end{align}
In quantum field theories, we should realize these symmetries as operators acting on Hilbert space (Schr\"odiner picture) or equivalently on local operators (Heisenberg picture). 

One basic assumption in the local quantum field theory in the Heisenberg picture is that the Poincar\'e generators act on local fields (or more generally local operators) as
\begin{align}
[P_\mu, O(x)] &= -i\partial_\mu O(x) \cr
[J_{\mu\nu},O(x)] &= (\Sigma_{\mu\nu} - i (x_\mu \partial_\nu -x_\nu \partial_\mu)) O(x) \ ,
\end{align}
where $\Sigma_{\mu\nu}$ is the $SO(d-1,1)$ spin matrix acting on generically non-scalar operator $O(x)$. 

The action of the dilatation on local fields must be consistent with the algebra \eqref{dilatationc} introduced in the last subsection:
\begin{align}
[D, O(x) ] = - i(\Delta + x^\mu \partial_\mu) O(x) \ ,
\end{align}
where $\Delta$ is the scaling dimension matrix acting on the collection of the local operators $O(x)$ (with fixed spin), which may not be digagonalizable in general. Actually, at this point, we can add any conserved charge $Q$, which commutes with the Poincar\'e generators, to $D$ without changing the commutation relations \eqref{dilatationc}. In general, $Q$ acts as anti-symmetric rotation on $O(x)$ rather than the engineering scaling transformation, which typically acts as ``symmetric part". In conformal field theories, the closure of the conformal algebra will uniquely determine the action of $D$ in most cases. From the scaling algebra alone, we cannot say that the anti-symmetric part of $\Delta$ is not forbidden.

Finally, let us consider the action of the conformal transformation on the local operators. We here only consider the unitary highest weight representation. 
In this case, we know that the dilatation generator is diagonalizable and its spectrum is positive definite (see section \ref{cci} for further constraint from unitarity). The highest weight representation is also known as (quasi-)primary field or (quasi-)primary operator, and it is annihilated by $K_\mu$ at the origin of the space-time (i.e. $x^\mu = 0$).
For primary operators $O(x)$, the action of the conformal generator is given by 
\begin{align}
[K_\mu, O(x)] = (-2i x_\mu \Delta - 2x^\lambda \Sigma_{\lambda \mu} - 2i x_\mu x^\rho \partial_\rho + ix^2 \partial_\mu) O(x) \ .
\end{align} 
The action on the non-primary operators may be obtained by acting $P_\mu$ further.

As in general quantum mechanics in Heisenberg picture, the assumption of the invariance under Poincar\'e, dilatation and conformal transformation is dictated by the requirement 
\begin{align}
P_\mu |0\rangle = J_{\mu\nu} |0 \rangle = D  | 0 \rangle  = K_\mu | 0 \rangle = 0 \ , \label{vass}
\end{align}
where $| 0\rangle $ is the vacuum state of the quantum field theories under consideration. 
This vacuum properties can be translated into the Ward-Takahashi identities for correlation functions in quantum field theories under the Noether assumption that we will discuss in section \ref{local}. We should emphasize that in most part of our discussions, we do assume that dilatation and conformal symmetry are not spontaneously broken as is clear from our vacuum assumption \eqref{vass}. We should note however that this does not mean that the dilatation or conformal symmetry are not broken spontaneously. For example, $\mathcal{N}=4$ super Yang-Mills theory spontaneously breaks the conformal invariance in the coulomb branch where the scalar field obtains a non-zero vacuum expectation value.

\subsection{Space-time symmetry and energy-momentum tensor}\label{local}
In section \ref{symmetrya}, we introduced the symmetry of a given quantum system as an algebra of conserved charges
that act on the Hilbert space (or S-matrix in the Haag-Lopuszanski-Sohnius   theorem\footnote{The reason why they discussed the symmetry of the S-matrix rather than the symmetry of the Hilbert space is based on the hypothesis that the Hilbert space may not be good observables in relativistically interacting systems. According to the purists at the time, only S-matrix was observable. We will not be so pedantic about it.}). In quantum field theories,
we usually postulate that these symmetries are realized by local conserved currents. Obviously, if a current $j_\mu$ is conserved: $\partial^\mu j_\mu =0$, one can construct the conserved charge
\begin{align}
Q = \int d^{d-1} x j_0 \ .
\end{align}
Strictly  speaking, the existence of the current rather than the charge is not necessary for the existence of the symmetry, but this assumption (Noether assumption)
 covers most of interesting examples we will discuss in this review article. The assumption implies that
 we will exclusively consider local quantum field theories. 
 Although it is an interesting possibility as we mentioned, we will not discuss,
 for instance, conformal or scale invariant string field theories or more generally membrane field theories if any. 

With the Noether assumption, the  translational invariance means that the theory possesses a conserved energy-momentum  tensor:
\begin{align}
\partial^\mu T_{\mu\nu} = 0 \label{conserve}
\end{align}
The Lorentz invariance further demands that the energy momentum tensor can be chosen to be symmetric (known as the Belinfante  prescription):
\begin{align}
T_{\mu\nu} = T_{\nu\mu}
\end{align}
so that the Lorentz current $J_\rho^{J_{\mu\nu}}= x^{[\mu}T^{\nu]}_{\rho}$ is conserved.

The scale invariance $(x^\mu \to \lambda x^\mu)$ requires that
\begin{align}
T^\mu_{ \ \mu} =\partial^\mu J_\mu
\end{align}
so that $D_\mu= x^\rho T_{\mu\rho} -J_\mu$ is the conserved scale current (dilatation current).\footnote{Actually, we do not have to assume that the energy-momentum tensor is symmetric: We do not 
need Lorentz invariance for scale invariance.} 
As we will see, roughly speaking, the first term generates the space-time dilatation while the second term generates the additional scaling of fields to preserve the total scale invariance of the theory. The current $J_{\mu}$ is known as the virial current \cite{Coleman:1970je}. The word ``virial" is derived from Latin ``vis" meaning ``power" or ``energy". %According to Wikipedia, 
It was Clausius in 19th century who first used the name with the definition $\sum x^i p_i$, which reminds us of the virial current for a free scalar field theory we will describe in section \ref{frees}. Probably it refers to the ``internal" degrees of freedom responsible for the scale transformation like those in ``molecules".

The special conformal invariance is a symmetry under
\begin{align}
x^\mu \to \frac{x^\mu + v^\mu x^2}{1+2v^\mu x_\mu + v^2 x^2} \ . \label{specialconfi}
\end{align}
It requires that the energy-momentum tensor is traceless:
\begin{align}
T^{\mu}_{\ \mu} = 0
\end{align}
so that we can construct the special conformal current $K_{\mu}^{(\rho)} = \left[\rho_\nu x^2 - 2x_\nu(\rho_\sigma x^\sigma)\right] T^{\nu}_{\ \mu}$.\footnote{Note that unlike scale invariance, we have to assume that the energy-momentum tensor is symmetric here.
We need the Lorentz invariance for the  special conformal invariance to exist as can be seen from the algebra (\ref{groupsp}).}

In the literature, it is often claimed that the inversion and the translation generate the full conformal transformation.
This is true because $K_\mu = I P_\mu I$ with $I$ generating space-time inversion $ x^\mu \to \frac{{x}^\mu}{{x}^2}$, but the converse may not hold. Invariance under the conformal algebra does not imply invariance
under the inversion (see e.g. \cite{Hortacsu:2001bp}\cite{Weinberg:2010fx} for a related comment). The point is that inversion is a disconnected component of the conformal group and it is only an outer automorphism. We can see it
explicitly if we recall that the action of inversion on the cylinder $\mathbb{S}^{d-1} \times \mathbb{R}^1$ is given by the time reversal (on $\mathbb{R}^1$) in the radial quantization of conformal field theories. We refer e.g. to \cite{Ginsparg:1988ui}\cite{DiFrancesco:1997nk} for the radial quantization in $d=2$ dimension. The similar construction is possible in any $d\ge 2$ \cite{Fubini:1972mf}. The time-reversal
may or may not be a symmetry of the theory on $\mathbb{S}^{d-1} \times \mathbb{R}^1$ as is the case in flat Minkowski space-time.

Energy-momentum tensor is not unique: one can improve it without spoiling the conservation law (\ref{conserve}):
\begin{align}
T_{\mu\nu} \to T_{\mu\nu} + \partial^\rho B_{\mu\nu\rho} \ . \ \ B_{\mu\nu\rho} = -B_{\rho \nu \mu} \ . 
\end{align}
Belinfante showed \cite{Belinfante} that by using the ambiguity, one can always make it  symmetric when the theory is Poincar\'e invariant (see also \cite{Dumitrescu:2011iu}) by explicitly constructing $B_{\mu\nu\rho}$ from the spin current 
\begin{align}
B_{\mu\nu\rho} = \frac{1}{2}(s_{\nu\rho\mu} + s_{\mu\nu\rho} + s_{\mu\rho\nu})
\end{align} 
where the Lorentz current is given by $J^{J_{\mu\nu}}_{\rho} = x^{[\mu}\hat{T}^{\nu]}_{\rho} + s^{\mu\nu}_{\ \ \ \rho}$ with possibly non-symmetric energy-momentum tensor $\hat{T}_{\mu\nu}$.

The non-uniqueness of the energy-momentum tensor has an important consequence in conformal invariance. 
Suppose the energy-momentum tensor is given by
\begin{align}
T^\mu_{\ \mu} &= \partial^\mu \partial^\nu L_{\mu\nu} \ \ (d \ge 3) \cr
T^\mu_{\ \mu} &= \partial^\mu \partial_\mu L \ \ \ (d = 2) \label{suppem}
\end{align}
with certain {\it local} operators $L_{\mu\nu}$ and $L$,
then by using this ambiguity, one can define the improved energy-momentum tensor (see e.g. \cite{Coleman:1970je}\cite{Callan:1970ze}\cite{Polchinski:1987dy})
\begin{align}
\Theta_{\mu\nu} =& T_{\mu\nu} + \frac{1}{d-2}\left(\partial_\mu \partial_\alpha L^\alpha_{\ \nu} + \partial_\nu \partial_\alpha L^\alpha_{\ \mu} - \Box   L_{\mu \nu} - \eta_{\mu\nu} \partial_\alpha \partial_\beta L^{\alpha \beta} \right) \cr
&+\frac{1}{(d-2)(d-1)}\left(\eta_{\mu\nu}\Box L^{\alpha}_{\ \alpha} - \partial_{\mu}\partial_\nu L^{\alpha}_{\ \alpha}\right) 
\end{align}
for $d\ge 3$, and 
\begin{align}
\Theta_{\mu\nu} = T_{\mu\nu} + \left(\eta_{\mu\nu} \Box  L^{\alpha}_{\ \alpha} - \partial_{\mu}\partial_\nu L^{\alpha}_{\ \alpha}\right) 
\end{align}
for $d=2$. The improved energy-momentum tensor $\Theta_{\mu\nu}$ is traceless (as well as symmetric and conserved). Thus the precise condition for the conformal invariance is (\ref{suppem}).
The traceless energy-momentum tensor may not be unique because we can still add $\partial^\rho \partial^\sigma \Upsilon_{\mu\rho\nu\sigma}$ with $\Upsilon_{\mu\rho\nu\sigma}$ possessing the symmetry of Weyl tensor (symmetry properties of Riemann tensor plus traceless condition. See \ref{convention}). When there is such a possibility, a different choice will give a different Weyl invariant theory in the curved background as we will describe more about the Weyl transformation in section \ref{emr}.\footnote{Fortunately, the unitarity of the operator dimension restrict the possibilities of improvement, and we will not encounter such inequivalent Weyl invariant theories in $d\ge3$ with the assumption of unitarity.} If we allow more than two derivative modifications of the energy-momentum tensor, it is possible to introduce further higher derivative improvement terms in $d>4$, but we will not discuss them in this review article. As we will see in  section \ref{conss}, unitarity demands that the only allowed improvement term in unitary quantum field theories in $d\ge 3$ is from $L_{\mu\nu} = \eta_{\mu\nu} L$ with a dimension $d-2$ scalar operator $L$ if we demand the energy-momentum tensor has the canonical scaling dimension $d$.

\subsection{Unitarity, causality and discreteness of spectrum}\label{uniReeh}
As we discussed in section \ref{symmetrya}, the scale transformation is merely a subgroup of the conformal algebra, and there is no apparent reason that scale invariance must imply conformal invariance. However, we are not interested in general representations of the symmetry algebra. We are only interested in unitary representations realized by local quantum field theories. We have discussed in section \ref{local} how the condition can be stated in terms of the properties of the energy-momentum tensor in local quantum field theories with the Noether assumption.
In this subsection, we clarify the other tacitly implied assumptions when we talk about unitary relativistic quantum field theories.

The first important notion is unitarity. We have already assumed that the symmetry algebra is represented by unitary operators acting on the Hilbert space. 
In quantum field theories, we can derive various important properties from the unitarity assumption.
First of all, the unitary representation of the Poincar\'e algebra leads to the spectral decomposition of the two-point function. Let us examine the Feynmann $T^*$ two-point function for the scalar operator $O(x)$ for simplicity.\footnote{In the following, without further mentioning, the correlation functions must be understood as the Feynmann $T^*$ function in the Lorentzian field theories and the Schwinger function in the Euclidean field theories. In most of the discussions, we will neglect the contact terms.} It satisfies the spectral decomposition
\begin{align}
\langle O(x) O^\dagger (y) \rangle &= \sum_n e^{-ik_n(x-y)} |\langle 0|O(0)|n\rangle |^2 \cr
&=
i \int_0^{\infty} d\sigma^2 \rho(\sigma^2) \int \frac{d^d{k}}{(2\pi)^d} \frac{e^{-ik(x-y)}}{{\sigma^2 + k^2 -i\epsilon}} \ 
\end{align}
with the positive definite spectral function 
\begin{align}
\rho(\sigma^2) \ge 0 \ 
\end{align}
by assuming that the physical Hilbert space has a positive definite norm.
If we further assume scale invariance, the spectral function for the operator $O(x)$ with a definite scaling dimension $\Delta$ has the scaling form
\begin{align}
\rho(\sigma^2) \propto (\sigma^2)^{\Delta-d/2} \ .
\end{align}
When $\Delta -d/2$ is an integer, the momentum space two-point function can have scaling anomaly as we will see some examples later (e.g. in section \ref{sad})\footnote{Otherwise, the two-point function is ultralocal and has only support at the coincident point.}. For this particular correlation functions, there is no further constraint from conformal invariance.

There are other consequences of unitarity. In particular, the unitarity of the S-matrix 
\begin{align}
S^\dagger S = 1 
\end{align}
yields various important properties such as the optical theorem that we will use later in understanding the relation between scale invariance and conformal invariance in section \ref{section8}, and these are briefly reviewed and summarized in \ref{smatrix}. We should note that strictly speaking S-matrix does not exist for scale invariant or conformal invariant theories due to the IR divergence, so we have to think of it as in a regularized sense.

After Wick rotation, the statement of the unitarity in the correlation functions is equivalent to the reflection positivity. Let us define the Euclidean adjoint\begin{align}
\Theta[O(x_1, \cdots x_d)] = O^\dagger(x_1,\cdots -x_d) \ ,
\end{align}
where $x_d = ix_0$ is from the  Wick rotation.
Then the reflection positivity states
\begin{align}
\langle O(x) \Theta[O(x)] \rangle \ge 0 \ .
\end{align}
Consequently, the scalar Euclidean two-point function has the spectral decomposition
\begin{align}
\langle O(x) O^\dagger (y) \rangle = \int_0^{\infty} d\sigma^2 \rho(\sigma^2) \int \frac{d^d{k}}{(2\pi)^d} \frac{e^{-ik(x-y)}}{{\sigma^2 + k^2 }} \ 
\end{align}
with the same spectral density $\rho(\sigma^2)$ when the Wick rotation is well-defined. When we use the reflection positivity for tensor operators, it is important to flip the sign whenever we have the $x^d$ tensorial indices.

While statistical models realized as Euclidean field theories do not necessarily satisfy the reflection positivity (see e.g. the last example in section \ref{frees}), in most of the review, we do assume the reflection positivity. Indeed, it is known that scale invariant but non-conformal field theories exist if we abandon the reflection positivity, and we argue that the reflection positivity plays a crucial role in understanding the enhancement of scale invariance to conformal invariance in Euclidean field theories. In particular, the proof in $d=2$ dimension relies on it as we will see in section \ref{proof2}.

The invariance under Poincar\'e symmetry naturally introduces the notion of causality. 
We assume that no information can propagate faster than the speed of light. This by itself is not a consequence of the Poincar\'e invariance (because the representation theory allows tachyon), but it is natural to assume this property.
In relativistic quantum field theories, there are several different notions of causality proposed in the literature, but in this review paper we assume one of the stronger version called microscopic causality. The microscopic causality demands that every local operators $O_i(x)$ (anti-)commute 
\begin{align}
[O_i(x), O_j(y)] = 0 
\end{align}
when the separation is space-like $(x-y)^2 >0$. 
If we start with a renormalizable Lagrangian, the microscopic causality is satisfied at the formal level of argument based on the covariant canonical quantization by demanding the spectrum does not contain any tachyonic states.

One important application of the unitarity and the causality is the so-called Reeh-Schlieder theorem, which we use heavily in the following discussions. 
One consequence of the Reeh-Schlieder theorem is $O(x)|0\rangle =  0 \iff O(x) = 0$ in general quantum field theories. Obviously this is not true for non-local operators such as  charge or momentum.

The proof is not elementary \cite{RS}, so we only give a sketch of the physical idea with the emphasis on the role of the microscopic causality.
What we would like to show boils down to the claim that in any correlation functions, the insertion of $O(x)$ is zero except for contact terms when $O(x) |0\rangle = 0$: 
\begin{align}
\langle 0 | O_1(x_1) \cdots {O}(x) \cdots  O_n(x_n) |0 \rangle = 0 \ .
\end{align}
To see this, we suppose that the insertion point $x$ is space-like separated with all the other $x_i$. Then 
the microscopic causality demands $[O_i(x_i), {O}(x) ] = 0$, so 
\begin{align}
\langle 0 | O_1(x_1) \cdots {O}(x) \cdots  O_n(x_n) |0 \rangle =  \langle 0 | O_1(x_1) \cdots  \cdots  O_n(x_n) {O}(x)|0 \rangle 
\end{align}
vanishes by acting ${O}(x)$ on the vacuum from the assumption $O(x) |0\rangle = 0$. The correlation functions (more precisely Feynmann $T^*$ function) in the other causal domains are related by analytic continuation, so they must vanish identically in any causal domain.

The argument crucially relies on causality, so if the theory does not have a notion of causality, the proof fails. 
A good example is the Schr\"odinger field theory in which the annihilation operator $\Psi(x)$, which is local, annihilates the vacuum, but it is obviously not zero identically. We also note that for non-local operators like charge or momentum, the above argument does not apply. Again, as reflection positivity, there is no fundamental reason to believe (let alone prove) the Reeh-Schlieder theorem in Euclidean statistical mechanics, but in most part of  the  review, we assume this property.

The Reeh-Schlieder theorem is essentially the baby version of the state operator correspondence in conformal field theories. It enable us to access any quantum state from local operators acting on the vacuum. At this point, we should emphasize that our field theory discussions only concern the local observables and  we do not make any statement about the global aspects of the quantum field theories. 

Finally, we would like to address the unitarity in conformal field theories. In conformal field theories, there is another related but slightly different notion of unitarity because (Euclidean) conformal algebra admits a different inner product than the conventional Dirac conjugation assumed above where all the conformal generators are Hermitian e.g. $P_\mu^\dagger = P_\mu$, $K_\mu^\dagger = K_\mu$. We can check that the Euclidean conformal algebra has the involutive anti-automorphism given by
\begin{align}
w(P_\mu) &= K_\mu \cr
w(K_\mu) &= P_\mu \cr
w(J_{\mu\nu}) &= J_{\mu\nu} \cr
w(D) &= D \ \label{auto}
\end{align} 
This anti-automorphism is known as the Belavin-Polyakov-Zamolodchikov (BPZ) conjugation and it is suitable for discussing the radial quantization with the state operator correspondence. Indeed, this is how we assign Hermitian conjugation in quantum field theories defined on $\mathbb{S}^{d-1} \times \mathbb{R}$ after conformal mapping from the flat Euclidean space.

Note that the dilatation may be regarded as the radial Hamiltonian $\mathbb{S}^{d-1} \times \mathbb{R}$  in this viewpoint. 
In stating the condition for the enhancement of conformal invariance from scale invariance, we introduce the notion of ``discreteness of the spectrum", and the nomenclature comes from the radial quantization.
When we say that a conformal field theory has a discrete spectrum, we really mean that the spectrum under this radial time evolution is discrete and from the viewpoint of the original theory on $\mathbb{R}^{d-1,1}$ the dilatation eigenvalues are discrete. 

One simple consequence of the existence of this BPZ conjugation and positivity of the norm with respect to this inner product is the simpler derivation of the above-mentioned Reeh-Schlieder theorem in conformal field theories. The unitarity under the BPZ conjugation \eqref{auto} demands that the radial inner product is positive definite, and it follows that for any Hermitian operator $O(x)$ (i.e. under the BPZ conjugation so that $w(O(x)) = O(x)$)
\begin{align}
\langle O(x) O(y) \rangle = \langle O | O \rangle = 0 
\end{align}
if and only if $O(x) = 0$. The second equality is nothing but the state operator correspondence in conformal field theories.

\subsubsection{Consequence of scale invariance alone}\label{conss}

We now list some basic consequences of the scale invariance in relativistic quantum field theories.

\begin{itemize}
\item The scale invariance constrains the form of correlation functions, but more weakly than in the conformal invariant case we will discuss in section \ref{cci}.
All correlation functions must scale accordingly to the scaling dimension matrix due to the Ward-Takahashi identity for the dilatation. In contrast to the conformal invariant situations that will be discussed in section \ref{cci}, the two-point functions may not be diagonal with respect to 
scaling dimensions: for instance, scalar two-point functions are given by
\begin{align}
\langle O_1(x_1) O_2(x_2) \rangle  = \frac{c_{12}}{(x_1-x_2)^{\Delta_1+\Delta_2}} \ . \label{twop}
\end{align}
Note that we do not have to demand $\Delta_1 = \Delta_2$ here.

Three-point functions take the less restrictive form (see e.g. \cite{Fortin:2011sz}):
\begin{align}
\langle O_1(x_1)O_2(x_2)O_3(x_3) \rangle = \sum_{\delta_{1},\delta_{2}} \frac{c^{\delta_1, \delta_2}_{123}}{(x_1-x_2)^{\delta_1} (x_2-x_3)^{\delta_{2}}(x_3-x_1)^{\Delta_{123}-\delta_1-\delta_2}} \label{threep}
\end{align}
with $\Delta_{123} = \Delta_1 + \Delta_2 + \Delta_3$.  In \eqref{twop} and \eqref{threep}, we have assumed that the dilatation operator can be diagonalized with the eigenvalues $\Delta_i$.

\item
The representation theory does not tell that the scaling dimensions in scale invariant field theories are diagonalizable. In most generality, we may still assume that the scaling dimension matrix takes the Jordan normal form by an appropriate change of basis. In such bases, the position space two-point function may contain $\log$ term. The higher point functions will become more complicated accordingly. We, however, do not know any unitary examples in which this is the case. A non-unitary example of non-diagonalizability of operator spectrum in conformal field theory is known as logarithmic conformal field theory.

\item 
Assuming the diagonalizability (as well as reality) of the  scaling dimension, one can show the unitarity bound on scaling dimensions \cite{Grinstein:2008qk}: For instance in $d=4$ dimension, the scaling dimension $\Delta$ of operators with $(j_1,j_2)$ Lorentz spin must satisfy
\begin{align}
\Delta \ge j_1 + j_2 + 1 \ ,
\end{align}
which is weaker than the unitarity bound in conformal field theories discussed in section \ref{cci}.

\end{itemize}

\subsubsection{Consequence of conformal invariance}\label{cci}
If we further assume conformal invariance, we naturally impose more constraints  on various observables. We list some of them appearing in correlation functions. 
\begin{itemize}
\item 
The conformal Ward-Takahashi identity constrains the forms of correlation functions. For a review,  we refer to \cite{Ginsparg:1988ui}\cite{DiFrancesco:1997nk} for $d=2$. In higher dimensions, the discussion in \cite{Osborn:1993cr} would be the most comprehensive one except that they assume invariance under inversion and CP transformation (see also \cite{Bzowski:2013sza} for the momentum space approach).
Two-point functions of primary operators are diagonal with respect to their conformal dimensions. For instance, scalar two-point functions are given by
\begin{align}
\langle O_1(x_1) O_2(x_2) \rangle = \frac{c_{12} \delta_{\Delta_1 \Delta_2}}{(x_1-x_2)^{2\Delta_1}} \ . 
\end{align}

Three-point functions of scalar primary operators are uniquely fixed \cite{Polyakov:1970xd}\cite{Migdal:1971xh}\cite{Migdal:1972tk}
\begin{align}
\langle O_1(x_1) O_2(x_2) O_3(x_3) \rangle =  \frac{c_{123}}{(x_1 - x_2)^{-\Delta_3 + \Delta_1 + \Delta_2} (x_2 - x_3)^{-\Delta_1 + \Delta_2 + \Delta_3} (x_1 - x_3)^{-\Delta_2 + \Delta_1 + \Delta_3}} \ .
\end{align}

\item
Due to the unitarity constraint, the conformal dimension of primary operators are bounded \cite{Mack:1975je} (see \cite{Minwalla:1997ka} for a pedagogical review). For instance in $d=4$ dimension, the conformal dimension of primary operators with $(j_1,j_2)$ Lorentz spin must satisfy
\begin{align}
\Delta \ge j_1 + j_2 + 2 - \delta_{j_1j_2,0} \ .
\end{align}

\item The four-point functions of conformal field theories satisfy the beautiful bootstrap equations.\footnote{Historically, there is another ``bootstrap equation" that is the Schwinger-Dyson-like self-consistent equations with anomalous dimensions (see e.g. \cite{Migdal:1972tk} for its demonstration in conformal field theories). It is also known as the skeleton expansion. We would like to thank S.~Rychkov and A.~Migdal for the reference.}
 Let us consider the simplest one ($x_{ij} = x_i - x_j$ for a shorter notation) with four identical scalar operators with conformal dimension $\Delta$:
\begin{align}
\langle O(x_1)O(x_2)O(x_3)O(x_4) \rangle  = \frac{g(u,v)}{x^{2\Delta}_{12} x_{34}^{2\Delta}} \ ,
\end{align}
with $u = \frac{x_{12}^2 x_{34}^2}{x_{13}^2x_{24}^2}$, $v= \frac{x_{14}^2x_{23}^2}{x_{13}^2x_{24}^2}$.
From the operator product expansion (OPE) \cite{Wilson} (see also \cite{Migdal:1971xh} for the earlier application of the OPE in conformal field theories)
\begin{align}
O(x)O(0) = \sum_{_i} C_{OOi} \mathcal{C}(x_\mu,\partial_\mu) O^i(x) \ ,
\end{align}
where $\mathcal{C}(x_\mu,\partial_\mu)$ gives the sub-leading non-primary operator contributions fixed by conformal invariance, we see that the four-point function is determined\begin{align}
 g(u,v) =  \sum_{i}(C_{OOi})^2 g^{(l)}_{\Delta}(u,v) \ ,
\end{align}
where $g^{(l)}_{\Delta} (u,v)$ are explicitly known conformal blocks with spin $(l)$ and conformal dimension $\Delta$ \cite{Dolan:2000ut}\cite{Dolan:2003hv}. On the other hand, the four-point function must satisfy the crossing symmetry
\begin{align}
g(u,v) = (u/v)^{\Delta} g(v,u) \ .
\end{align}
With the unitarity constraint, it gives interesting constraints on the spectrum $\Delta$ and the OPE coefficients $C_{OOi}$ of the theory. 
We should note that the OPE in conformal field theories is proved to be convergent \cite{Pappadopulo:2012jk}
rather than just asymptotic as in non-conformal field theories.
In $d=2$ dimension, the program of conformal bootstrap was carried out in the seminal work \cite{Belavin:1984vu} thanks to the infinite dimensional extra constraint from the Virasoro symmetry.

\item

 In contrast, in scale invariant field theories, the use of OPE is less effective. First of all, the convergence of the OPE is not proved, and there are more complications because (1) the two-point functions of higher spin operators are not uniquely fixed by the symmetry, (2) the two-point functions may not be diagonalized with respect to the scaling dimension and (3) therefore OPE can be much more complicated. Thus, at this point, the usage of the bootstrap technique was not fully appreciated in the literature.

\end{itemize}

\subsection{Formulation of the problem}
Let us summarize the discussions in this section and finally formulate the problem of scale invariance vs conformal invariance in relativistic quantum field theories. The question is under which conditions a given scale invariant field theory has the enhanced conformal symmetry. 

Under the Noether assumption, a relativistic quantum field theory is scale invariant if and only if the trace of the energy-momentum tensor is given by the divergence of the local virial current
\begin{align}
T^{\mu}_{\ \mu} = \partial^\mu J_\mu \ .
\end{align}
When the energy-momentum tensor can be improved to be traceless the theory is conformal invariant. Assuming the unitarity, the condition of the possibility of the improvement is that the Virial current is a divergence of a certain local scalar operator:
\begin{align}
J_\mu = \partial_\mu L \ .
\end{align}

Therefore the strategy to show the enhancement of the conformal invariance from scale invariance will be as follows. Given a quantum field theory, we compute the energy-momentum tensor, and see if the trace can be written as the divergence of the virial current. When this is the case, we try to remove the virial current by available improvement of the energy-momentum tensor in the theory under consideration. The claimed enhancement from scale invariance to conformal invariance means that this is always possible under certain assumptions.

With the use of the Reeh-Schlieder theorem, one obvious goal is to show
\begin{align} 
\langle T^{\mu}_{\ \mu}(x) T^{\mu}_{\ \mu}(0) \rangle = 0 \ \label{rtt}
\end{align}
in a given theory, which immediately implies $T^{\mu}_{\ \mu}(x) = 0$ as an operator identity. With various consistency conditions of local quantum field theories, it may be able to derive this equality (see section \ref{proof2} for the discussions in $d=2$ dimension).
Of course, when the improvement is possible, we have to give the prescription to choose the ``correct" $T^\mu_{\ \mu}(x)$ here. Otherwise, the left hand side of \eqref{rtt} may not vanish even though the theory is improved to be conformal.

At this point, we would like to emphasize that the energy-momentum tensor in interacting quantum field theories is a delicate object because it must incorporate the effects of the renormalization group. Indeed, one can convince ourselves that the trace of the energy-momentum tensor is nothing but the response of the scale transformation and the non-trivial effects (invisible at the classical level) induced by the change of the renormalization scale are precisely regarded as a quantum contribution to the scaling transformations. Thus, the energy-momentum tensor in quantum field theories should know the effects of the renormalization group.
In addition, the above mentioned improvement ambiguity must be taken into account together with the renormalization group effects.
Like any other symmetries, it is possible that the renormalization group may introduce the apparent violation of the symmetries we would like to retain, which is the conformal symmetry in our case. We are forced to look for the particular good class of renormalization scheme (if any) that shows the conformal invariance in a manifest manner.

Therefore, it is crucial to understand how to compute the trace of the energy-momentum tensor and study its properties in general interacting quantum field theories. Preferably it is better not to commit ourselves to one particular renormalization scheme, but rather develop the general argument applicable in any reasonable renormalization scheme.
In the next section, we present the general formalism to compute the energy-momentum tensor and study its relation to the renormalization group.

\newpage

\section{Energy-momentum tensor and renormalization group}\label{emr}
In section \ref{section2}, we have showed that in order to understand the relation between scale invariance and conformal invariance, we need to understand the quantum properties of the trace of the energy-momentum tensor. In particular, we have emphasized its relation to the renormalization group transformation. The goal of this section is to develop the general discussions on the renormalization group and its relation to the properties of the energy-momentum tensor and its anomaly.

\subsection{Schwinger functional}\label{Schwingerfunction}
\subsubsection{Promoting coupling constants to background fields}
The energy-momentum tensor is a composite operator in general Lagrangian quantum field theories, and we have to understand the composite operator renormalization to address its properties under renormalization group. Of course, one may be able to renormalize the composite operators such as the energy-momentum tensor by using the perturbation theory with Ward-Takahashi identity satisfied at each order in perturbation. The procedure is well-described in the textbook on the renormalization, and we will not dwell on the technical details in this review article. We primarily use the fact that the composite operator renormalization can be done in a consistent manner (aside from possible anomalies we will mention later) and we only rely on the generic properties of the renormalization group. The most of the formal argument in this section can be understood without going into the details of the renormalization. We will give some examples in section \ref{examples}. 

Once we admit the renormalizability, it is useful to consider the generating functional for the correlation functions of these renormalized composite operators. This is formally summarized by the so-called (renormalized) Schwinger functional \cite{Schwinger:1951xk}. The formal path integral expression of the Schwinger functional would be 
\begin{align}
e^{-W[g^I(x)]} = \int \mathcal{D}X e^{-S_0[X] - \int d^d x \lambda^I(x) \mathcal{O}_I(x) } \ , 
\end{align}
where $\lambda^I(x)$ are collections of source terms for each operator $\mathcal{O}_I(x)$ in the theory specified by $S_0[X]$, which are not necessarily scalars. In particular, we are promoting every coupling constants into background fields. 
In principle, the Schwinger functional contains all the local information of the given quantum field theory. 

We do not necessarily require the existence of the path integral expression over the ``fundamental fields" $X$ to discuss the Schwinger functional.  One important defining property of the Schwinger functional is the Schwinger action principle:
\begin{align}
\langle \mathcal{O}_I (x) \rangle = \frac{\delta W}{\delta \mathbf{\lambda}^I(x)} \  \label{Sact}
\end{align}
as well as higher point functions defined in a similar way by repeated derivatives.
We assume that the Schwinger functional is renormalized in the sense that the correlation functions so obtained as in \eqref{Sact} is finitely renormalized. Again this can be formally satisfied by the composite operator renormalizations of $\mathcal{O}_I$ if the quantum field theory under consideration is renormalizable. We can impose the renormalization condition of the Schwinger functional by formal power series of the renormalized correlation functions.

This definition of the renormalized  Schwinger functional is useful to understand the renormalization group properties of the quantum field theories, but it
possesses various subtle issues. First of all, the definition is ambiguous due to contact terms, renormalization scheme dependence, local counterterms and so on. Also, the parametrization of the theory space by $\lambda^I(x)$ can be redundant due to various operator relations among $\mathcal{O}_I$ (see section \ref{ambl}). Extra assumptions such as symmetry consideration  (in particular supersymmetry)  may ameliorate these problems to some extent. We will come back to them when necessary in the following argument.

\subsubsection{Curved background}\label{cuvedb}
One of the most important composite operators in our discussion is the energy-momentum tensor, and the source of the energy-momentum tensor in the Schwinger functional is the background metric tensor. This inevitably forces us to consider quantum field theories in the curved background.
So far, we have discussed the space-time symmetry of the flat Minkowski (or Euclidean) space-time.
The properties of the energy-momentum tensor studied in section \ref{local} are more succinctly derived in the curved background. With this motivation, let us discuss the energy-momentum tensor in the curved background and its relation to the problem of scale invariance vs conformal invariance.

One of the most important facts in general relativity is that the energy-momentum tensor is a source of gravity through the coupling to the metric $g_{\mu\nu}$.
Suppose we can couple our Poincar\'e invariant quantum field theory to a general covariant gravitational theory (not necessarily in a unique fashion).
Let us conventionally define the energy-momentum tensor for matter fields $\Phi$ as\footnote{Another common definition of the energy-momentum tensor is the Noether energy-momentum tensor:  $T_{\mu\nu} = \frac{\partial \mathcal{L}}{\partial (\partial^\mu \Phi)} \partial_\nu \Phi - \eta_{\mu\nu} \mathcal{L}$ for the classical two-derivative action  $\mathcal{S}= \int d^d{x} \mathcal{L}(\Phi, \partial_\mu \Phi)$. It is related to the energy-momentum tensor in general relativity  (\ref{emg}) in the flat space-time limit $g_{\mu\nu} = \eta_{\mu\nu}$ by improvement we discussed in section \ref{local}.}
\begin{align}
T_{\mu\nu} = \frac{2}{\sqrt{|g|}}\frac{\delta \mathcal{S}}{\delta g^{\mu\nu}} \ , \ \ T^{\mu\nu} = -\frac{2}{\sqrt{|g|}}\frac{\delta \mathcal{S}}{\delta g_{\mu\nu}} \ ,   \label{emg}
\end{align}
where $ \mathcal{S} = \int d^d{x}\sqrt{|g|} \mathcal{L}(\Phi,g_{\mu\nu})$ is the matter action with the action density $\mathcal{L}(\Phi,g_{\mu\nu})$.\footnote{Due to our convention, the action density is minus of the Lagrangian density in  the Lorentzian signature.}
The diffeomorphism invariance of the action $\mathcal{S}$ automatically gives the conservation
$D^\mu T_{\mu\nu} = 0$ with the covariant derivative $D^{\mu}$ up to the equations of motion, and the symmetry $T_{\mu\nu}= T_{\nu\mu}$. Of course, when the theory contains a spinor, we have to consider the spin connection with vielbein $e_{\mu}^a$ as fundamental gravitational degrees of freedom (rather than the metric $g_{\mu\nu} = e_{\mu}^a e_{\nu a}$ itself). The symmetric energy momentum tensor is then defined by
\begin{align}
T^{\mu\nu} &= \frac{1}{2} (T^{\mu}_{\ a} e^{a\nu} + T^{\nu}_{\ a} e^{a\mu}) \cr
T^{\mu}_{ \ a} &= \frac{\delta \mathcal{S}}{\delta e^a_{\ \mu}} \ .
\end{align}
We will not talk about the vielbein formalism explicitly but the generalization  of the following argument should be obvious (except for the gravitational anomaly whose absence will be always assumed). 

If the action is scale invariant or constant Weyl invariant, i.e. $g_{\mu\nu} \to e^{2\bar{\sigma}} g_{\mu\nu}$, where $\bar{\sigma}$ is a space-time independent constant, then
the action density is scale invariant up to a total derivative term $\delta \mathcal{L} = -\bar{\sigma} D^\mu J_\mu$, so
\begin{align}
T^\mu_{\ \mu} = \frac{2}{\sqrt{|g|}}g^{\mu\nu}\frac{\delta \mathcal{S}}{\delta g^{\mu\nu}} = D^\mu J_\mu \ .
\end{align}
This is the origin of the virial current from the viewpoint of the curved background.

On the other hand, if the action is Weyl invariant, i.e. $g_{\mu\nu} \to e^{2\sigma(x)} g_{\mu\nu}$, where $\sigma(x)$ is space-time dependent arbitrary scalar function, then
the action density itself must be invariant under the Weyl rescaling, so the energy-momentum tensor is traceless.
\begin{align}
T^\mu_{\ \mu} = \frac{2}{\sqrt{|g|}}g^{\mu\nu}\frac{\delta \mathcal{S}}{\delta g^{\mu\nu}} = 0 \ .
\end{align}

We recall that conformal killing vectors of the $d$-dimensional $(d>2)$ Minkowski space-time are generated by the conformal algebra $so(2,d)$ (see e.g. \cite{Ginsparg:1988ui}\cite{DiFrancesco:1997nk}).\footnote{The conformal killing vectors in $d=2$ dimension gives the infinite dimensional Virasoro algebra.} They are precisely translation, Lorentz rotation, dilatation, and special conformal transformation.
The conformal killing vector induces the diffeomorphism
\begin{align}
ds^2 = \Omega(\tilde{x}) \eta_{\mu\nu}d\tilde{x}^\mu d\tilde{x}^\nu = \eta_{\mu\nu} dx^\mu dx^\nu 
\end{align}
that makes the metric invariant up to overall Weyl factor $\Omega$. When the theory is Weyl invariant, if we restrict ourselves to the Minkowski space-time, the above argument implies that the theory under study is automatically conformal invariant \cite{Zumino:1970}. Since the energy-momentum tensor may not be unique for a given conformal field theory, the converse may not be true: we might be able to couple a conformal field theory to gravity in a non-Weyl invariant way. We will see a free scalar example in section \ref{frees}.

To close this subsection, let us briefly discuss the interpretation of the improvement of the energy-momentum tensor mentioned in section \ref{local} from the curved space-time viewpoint. 
If we add the curved space action
\begin{align}
\mathcal{S}_{\mathrm{imp}}  = \int d^d x \sqrt{|g|} \left( R L + R_{\mu\nu} L^{\mu\nu} + R_{\mu\nu\rho\sigma} L^{\mu\nu\rho\sigma} \right) \ , \label{improvecurved}
\end{align}
the energy-momentum tensor defined in (\ref{emg}) obtains extra contributions such as $(\partial_\mu \partial_\nu - \eta_{\mu\nu} \Box) L$ in the flat space-time limit $g_{\mu\nu} = \eta_{\mu\nu}$. This is the origin of the improvement terms and ambiguities of the energy-momentum tensor from the non-uniqueness of the curved space-time action given a flat space theory. In a certain situation, by choosing appropriate terms in (\ref{improvecurved}), one may be able to construct the traceless energy-momentum tensor, and then the theory is Weyl invariant.

\subsection{Weyl Anomaly}\label{Weyla}
In section \ref{local}, we learned that local conformal field theories require $T^{\mu}_{\ \mu} = 0$ in the flat space-time. 
In curved background with various background sources, however, the conformal field theory breaks Weyl invariance due to the so-called 
Weyl anomaly \cite{Capper:1974ic}\cite{Deser:1976yx} (see e.g. \cite{Duff:1993wm}\cite{Deser:1996na} for historical reviews).\footnote{We use the terminology Weyl anomaly and trace anomaly interchangeably. We also assume that the Weyl symmetry is not spontaneously broken. The trace anomaly in the spontaneously broken phase has been studied e.g. in \cite{Schwimmer:2010za}\cite{Armillis:2013wya}\cite{Gretsch:2013ooa}.}
Weyl anomaly plays a central role in discussing scale invariance vs conformal invariance from the viewpoint of the Schwinger functional, and in this section, we consider the situation in which only the metric is a non-trivial position dependent source. We will continue our discussions on the Weyl anomaly in the Schwinger functional  with more general sources in section \ref{localr}.

Let us consider the one-point functions of the energy-momentum tensor from the Schwinger functional 
\begin{align}
\langle T_{\mu\nu} \rangle = \frac{2}{\sqrt{|g|}} \frac{\delta W[g_{\mu\nu]}}{\delta g^{\mu\nu}} \ .
\end{align}
Its trace is suppose to be zero in conformal field theories, but in general it is not in the curved background due to the Weyl anomaly.
In $d=2$ dimension, the Weyl anomaly is proportional to the scalar curvature $R$ as\footnote{The factor  $-1/2\pi$ is due to our different normalization than the most common convention in the string theory literature. We conventionally normalize $c=1$ for a free scalar in $d=2$ dimension. There seems no standard convention for the Weyl anomaly in higher dimensions.}
\begin{align}
\langle T^{\mu}_{\ \mu} \rangle = +\frac{1}{2\pi} \frac{c}{12} R \ .
\end{align}
Despite the anomaly, it still makes sense to say the theory is conformal invariant because it vanishes in the flat space-time and the flat-space dilatation as well as conformal current conservation is not affected.
The constant number $c$ is known as the central charge and its value depends on the conformal field theory under  consideration. It is because in $d=2$ dimension, the conformal symmetry is enhanced to the infinite dimensional symmetry known as Virasoro symmetry, and $c$ coincides with the center of the Virasoro algebra. As we will see in section \ref{proof2}, the central charge $c$ has a physical interpretation of counting massless degrees of freedom at the conformal fixed point. For instance, a free massless scalar has $c=1$ while a free massless Majorana fermion has $c= 1/2$.

In $d=4$ dimension, the most generic possibility of the Weyl anomaly from naive dimensional analysis (without any other background sources) is
\begin{align}
\langle T^{\mu}_{\ \mu} \rangle =c (\mathrm{Weyl})^2 - a \mathrm{Euler} + bR^2 +  \tilde{b} \Box R + d \epsilon^{\mu\nu\rho\sigma} R_{\mu\nu}^{\ \ \alpha\beta}R_{\alpha\beta \rho \sigma} \ ,
\end{align}
where $\mathrm{Euler} =R_{\mu\nu\rho\sigma}^2 -4R_{\mu\nu}^2+ R^2 $C$\mathrm{Weyl}^2 = R_{\mu\nu\rho\sigma}^2 -2R_{\mu\nu}^2 +\frac{1}{3}R^2$, $\Box R = D^{\mu}D_{\mu} R$ and Levi-Civita symbol $\epsilon^{\mu\nu\rho\sigma}$ is an antisymmetric tensor (rather than tensor density). 
The term $\tilde{b}\Box R$ can be removed by adding a local counterterm proportional to $\int d^4x \sqrt{|g|} \tilde{b}R^2$ to the effective action, so it is not an anomaly in a conventional sense.
In addition, for conformal field theories, we can show $b=0$ due to the Wess-Zumino consistency condition as we will discuss momentarily.
The Pontryagin term $\epsilon^{\mu\nu\rho\sigma} R_{\mu\nu}^{\ \ \alpha\beta}R_{\alpha\beta \rho \sigma}$ is consistent, but it breaks the CP transformation \cite{Nakayama:2012gu}. There is no known unitary field theory model that gives the Pontryagin term as Weyl anomaly (the self-dual two-form gauge field theory in the  Euclidean signature may be an exception \cite{Duff:1980qv}). A further study of this CP odd anomaly in relation to supergravity can be found in \cite{Bonora:2013rta}.

All anomalies must satisfy the Wess-Zumino consistency condition. Or more precisely, we can add  local non-gauge invariant counterterms so that the effective action satisfies it \cite{Bardeen:1984pm} (see also \cite{Bonora:1985cq}\cite{Boulanger:2007st} for the discussions of such counterterms in the context of Weyl anomaly: the upshot is we can use them so that the Weyl anomaly is given by diffeomorphism invariant terms).
The Weyl transformation is Abelian, and the Wess-Zumino consistency condition for the Weyl variation  $\delta_{\sigma(x)} = -2 \int d^4x \sqrt{|g|} \sigma(x) g^{\mu\nu}(x)\frac{\delta}{\delta g^{\mu\nu}(x)}$ is simply (see e.g. \cite{Bonora:1983ff}\cite{Bonora:1985cq}\cite{Cappelli:1988vw}\cite{Osborn:1991gm})
\begin{align}
[\delta_{\sigma(x)}, \delta_{\tilde{\sigma}(x')} ] W[g_{\mu\nu}] = 0 \ ,
\end{align} 
and the first order variation is given by the Weyl anomaly $\delta_{\sigma(x)} W[g_{\mu\nu}] = -\int d^4x \sqrt{|g|} \sigma(x) \sqrt{|g|} T^\mu_{\ \mu}(x)$ if the theory is conformal invariant in the flat space-time.
It is trivial to see that $\mathrm{Weyl}^2$ and Pontryagin term in the Weyl anomaly satisfy this condition because they are Weyl invariant by themselves. The Euler term non-trivially satisfies the condition after partial integration. However, we can check $R^2$ term does not satisfy the condition due to the term proportional to $\sigma \partial_\mu \tilde{\sigma} - \tilde{\sigma} \partial_\mu \sigma$, so $b$ must vanish for Weyl invariant field theories.

For free field theories, the Weyl anomaly can be read from the Schwinger-De~Wit computation of the one-loop determinant (see e.g. \cite{Vassilevich:2003xt} for a review). For demonstration, let us consider the Laplace-type differential operator $\Delta$ (e.g. $\Delta = -\Box + \xi R$ acting on a scalar) with eigenvalues $\Delta \phi_n = \lambda_n \phi_n$. We introduce the Schwinger-De~Wit heat kernel
\begin{align}
F(x,y, \rho) = \sum_n e^{-\rho \lambda_n} \phi_n(x) \phi_n(y) \ .
\end{align}
When we expand the diagonal heat kernel 
\begin{align}
F(x,x,\rho) = \sum_m \rho^{m-\frac{d}{2}} \int d^d x \sqrt{g} b_m(x) \ ,
\end{align}
we can obtain the regularized one-loop determinant $W[g_{\mu\nu}] = \log\mathrm{Det} \Delta^{-1/2}$ for the (bare) Schwinger functional. 
For a non-minimally coupled scalar, the explicit computation gives
\begin{align}
b_2(x) = \frac{1}{16\pi^2} \left( \frac{1}{6}\left(\frac{1}{5}-\xi \right) \Box R + \frac{1}{2}\left(\frac{1}{6}-\xi\right)^2 R^2 - \frac{1}{180} R_{\mu\nu}^2 + \frac{1}{180} R_{\mu\nu\rho\sigma}^2 \right) \ . \label{sd}
\end{align}
Since the Schwinger-De Wit heat kernel computes the one-loop logarithmic divergence that gives renormalization of the corresponding terms in gravity in even dimension (gravitational beta function), we have the formula\footnote{If the original theory is not conformal invariant in the flat space-time, we should regard the Weyl anomaly here as the additional violation due to the curved background \cite{Duff:1993wm}.}  that relates it with the Weyl anomaly:
\begin{align}
\langle T^{\mu}_{\ \mu} \rangle = b_{d/2}(x) \ .
\end{align}
For example, we obtain the Weyl anomaly for a non-minimally coupled scalar in $d=4$ dimension from  (\ref{sd}).
Note that with this definition of the Weyl anomaly, $b R^2$ term is non-zero if we consider the non-conformal scalar with $\xi \neq \frac{1}{6}$. In (\ref{sd}), $\zeta$-function regularization is assumed and a different regularization gives a different coefficient in the $\Box R$ term (see e.g. \cite{Asorey:2003uf}).

Alternatively, as is the case with all the other anomalies, we may regard the Weyl anomaly as the non-invariance of the path integral measure under the Weyl transformation. After carefully choosing the path integral variables to preserve the diffeomorphism invariance, the free field computation of the anomalous Jacobian gives the same result. See e.g. \cite{Fujikawa:2004cx} for a review of the path integral approach.

For reference, we give the free field values for the Weyl anomaly in $d=4$ dimension. The Euler term $a$ for a real scalar, a Dirac fermion, and a real vector is given by $\frac{1}{90 (8\pi)^2}, \frac{11}{90 (8\pi)^2}$ and $\frac{62}{90 (8\pi)^2}$ with our normalization. The $\mathrm{Weyl}^2$ term $c$ for a real scalar, a Dirac fermion, and a real vector is  $\frac{1}{30 (8\pi)^2}, \frac{6}{30 (8\pi)^2}$ and $\frac{12}{30 (8\pi)^2}$ with our normalization.
It is not immediately obvious which combination of $a$ and $c$ will count the degrees of freedom in $d=4$ dimension compared with the situation in $d=2$ dimension, where there is no other choice but to use $c$. Note that the number $a$ and $c$ are quite different from the naive ``number of helicities" that usually appear in  thermal properties of non-interacting massless particles.
We will revisit the problem in section \ref{Cardyc}.

Although we will not discuss it further in this review article, the Weyl anomaly can be generalized to any even dimensions. See \cite{Deser:1993yx} for the complete classification. In odd dimensions, we can convince ourselves that there is no Weyl anomaly from the naive dimensional counting (without any other sources than the background metric). However, there is a more subtle anomaly in contact terms that may be inconsistent with the conformal invariance \cite{Closset:2012vp}. We may also have CP violating trace anomaly that may come from the other space-time dependent sources than the gravitational background even in odd dimensions as we will see in section \ref{localr}.

\subsection{Trace of energy-momentum tensor in perturbatively renormalizable theory}\label{temt}

\subsubsection{Interacting theories and renormalization group}\label{Interar}
The computation of the trace of the energy-momentum tensor in interacting quantum field theories is intimately related to the properties of the renormalization group of the quantum field theories under consideration. In this subsection, we develop the formal argument to relate the trace of the energy-momentum tensor to the renormalization group flow. While most of the discussions are (approximately) valid in Wilsonian renormalization group, we focus on the conventional power-counting renormalization scheme with perturbative computations in mind. We will make a comment on the generalization at the end of this subsection.

The renormalization group invariance of the quantum field theories tells that when we change the renormalization scale (or cut-off in the Wilsonian sense), the physical quantities does not change if we simultaneously change the coupling constant and redefine operators. This procedure is known as the renormalization group transformation. The renormalization group transformation can be regarded as the other side of the coin of the scale transformation we want to discuss. Indeed, from the renormalization group invariance, we can convince ourselves that the study of the renormalization group is equivalent to the study of the response of the theory under the scale transformation.
More concretely, the renormalization group invariance can be stated as the Callan-Symanzik equation \cite{Callan:1970yg}\cite{Symanzik:1970rt}\cite{Symanzik:1971vw} with respect to the renormalization scale $\Lambda$:\footnote{Since we are not considering the explicit mass terms in the most of our discussions, probably it is more appropriate to call it Gell-Mann Low renormalization group equation \cite{GL}.}
\begin{align}
\left(\frac{\partial}{\partial\log\Lambda} + \beta^I\frac{\partial}{\partial g^I} \right) \left \langle \phi_{i_1}  \cdots \phi_{i_n} \right\rangle = \gamma_{i_1}^{\ j_1}  \left \langle \phi_{j_1}  \cdots \phi_{i_n} \right\rangle + \cdots + \gamma_{i_n}^{\ j_n}  \left \langle \phi_{i_1}  \cdots \phi_{j_n} \right\rangle \ , \label{globalCS}
\end{align}
where $\beta^I$ is the renormalization group beta function 
\begin{align}
\beta^I(g) = \left. \frac{\partial g^I}{\partial \log\Lambda}\right|_{g_0^I} 
\end{align}
that encodes the information of how the coupling constant changes along the renormalization group flow,
and $\gamma_{i}^{\ j}$ is the anomalous dimension matrix that encodes how the fields are renormalized along the renormalization group flow.
Here, we have treated $\phi_i$ as if it were the ``fundamental" (scalar) field, but any composite operators satisfy essentially the same set of equations once they are properly renormalized as we will describe in the following.

The renormalization group equation suggests that in quantum field theories, the scaling transformation is affected by the presence of the renormalization group beta functions (as well as the anomalous dimensions). The effects of the latter may be absorbed by assigning the renormalized scaling dimensions to the operators, but the former should be regarded as a quantum violation of the scaling symmetry of the quantum field theories.

How does this affect the dilatation generator and the energy-momentum tensor? More importantly, how does this violation is related to the conformal invariance?
To address these issues, in particular, in relation to the conformal invariance rather than the mere scale invariance, it is convenient to consider the local renormalization group equation for the Schwinger functional.

We recall that the Schwinger functional \cite{Schwinger:1951xk}, which we have introduced in section \ref{Schwingerfunction}, is obtained by promoting coupling constants $g^I$ to space-time dependent background fields $g^{I}(x)$.
\begin{align}
e^{-W[g^I(x)]} = \int \mathcal{D} X e^{-S_0[X] - \int d^dx \sqrt{|g|} (g^I(x) O_I(x) + a^a_\mu(x) J_a^\mu(x) + \mathcal{O}(g^2)) } \ ,
\end{align}
where $O_I(x)$ are scalar operators under consideration and $J_a^\mu(x)$ are vector operators. As discussed in section \ref{cuvedb}, we have introduced the source for the energy-momentum tensor by considering the (weakly) curved space-time with the metric $g_{\mu\nu}(x)$. We could have introduced higher tensor operators, but they are not important in the following discussions. Higher order terms $\mathcal{O}(g^2)$ can be ambiguous due to the ambiguities in the renormalization group procedure and we will further discuss them in section \ref{ambl}.

The crucial assumption in the following is that the Schwinger functional is finitely renormalized (referred as ``renormalizability assumption"). Theoretically this assumption is a great advantage because varying the renormalized Schwinger functional automatically takes into account the renormalization of the composite operators. The renormalization group equation for the renormalized Schwinger functional is called the local renormalization group equation  \cite{Osborn:1991gm} because we perform the space-time dependent change of coupling constants as well as the renormalization scale. This has a huge advantage in discussing the conformal invariance (rather than merely scale invariance) because it directly provides the response to the non-constant Weyl transformation as we discussed in section \ref{cuvedb}. In contrast, the conventional Callan-Symanzik equation only knows the response to the constant scale transformation.

Throughout this section, we concentrate on the so-called massless renormalization group flow in which we have no dimensionful coupling constants.
Without any dimensionful coupling constant at hand, the local renormalization group operator can be expressed as
\begin{align}
\Delta_{\sigma} &= \int d^dx \sqrt{|g|} \left( 2\sigma g_{\mu\nu} \frac{\delta}{\delta g_{\mu\nu}} + \sigma \beta^I \frac{\delta}{\delta g^I}  \right. + \left. \left( \sigma \rho_I^a D_{\mu} g^I - (\partial_\mu \sigma) v^a \right) \frac{\delta}{\delta a^a_{\mu}} \right) \  
\end{align}
under the assumption of power-counting renormalization scheme. Note that the change of the scale $\sigma(x)$ is a space-time dependent scalar function here. The meaning of the covariant derivative $D_\mu$ will be explained in section \ref{ambl} below in details.
The assumption of the local renormalizability is equivalent to the claim that the Schwinger functional is annihilated by $\Delta_{\sigma}$ up to the Weyl anomaly that is a local functional of the renormalized sources:
\begin{align}
\Delta_{\sigma} W [g_{\mu\nu}, g^I, a^a_\mu] = A_{\sigma}[g_{\mu\nu},g^I,a^a_\mu] \ .
\end{align} 
This equation is known as the local renormalization group equation or local Callan-Symanzik equation. The precise relation to the global Callan-Symanzik equation \eqref{globalCS} will be studied in section \ref{contact}.

The each term in $\Delta_{\sigma}$ has a simple interpretation. The first term $2\sigma g_{\mu\nu} \frac{\delta}{\delta g_{\mu\nu}}$  generates nothing but the Weyl rescaling of the metric by the Weyl factor $\sigma(x)$: $\delta_{\sigma}g_{\mu\nu}(x) = 2\sigma(x) g_{\mu\nu}(x)$. The renormalization of the coupling constants introduce running of the coupling constants under the change of the local scale transformation:
$\beta^I$ is the scalar beta function for the corresponding operator $O_I$ which is necessary to cancel the divergence appearing in the coupling constant renormalization for $g^I$.   Less familiar terms $\rho_I^a$ and $v^a$ are related to the renormalization group running for the vector background source $a^a_{\mu}$, and called the vector beta functions. We emphasize that once the coupling constant $g^I(x)$ is space-time dependent, we have an extra divergence in relation to vector operators that must be cancelled by renormalizing the background vector fields $a_{\mu}$. Even in the flat space-time limit, such effects are actually visible as the renormalization of the composite vector operators as we will see in section \ref{contact}.

The invariance of the Schwinger functional under the local renormalization group (up to anomaly) corresponds to the trace identity
\begin{align}
T^{\mu}_{\ \mu} = \beta^I O_I + (\rho_I^a D_\mu g^I) J_a^\mu + D_\mu (v^aJ_a^\mu) + A_\mathrm{anomaly} \ \label{traceiden}
\end{align}
once we use the Schwinger action principle \eqref{Sact}. 
This trace identity plays the central role in the following argument because we know that the study of the scale invariance and conformal invariance boils down to the properties of the trace of the energy-momentum tensor and
 this identity tells us how we can compute the trace of the energy-momentum tensor from the local renormalization group flow.

In most of our analysis on the local renormalizaiton group, we typically assume the power-counting renormalization scheme, so for instance the beta functions do not contain higher derivative terms of the coupling constants or the local scale transformation $\sigma(x)$. In Wilsonian framework, there is no reason that these terms are not generated. However, with the usual argument of irrelevance of the non-renormalizable operators, these will not affect most of the perturbative renormalization group flow.\footnote{Non-perturbatively, there is a possibility that seemingly irrelevant deformations in the ultraviolet may be relevant in the infrared due to large anomalous dimensions during the renormalization group flow. In such cases, the deformation is called ``dangerously irrelevant".} With the same reason, if we follow the power-counting renormalization scheme, the renormalizability of the Schwinger functional is equivalent to the usual renormalizability of the perturbative quantum field theories if we treat the renormalized Schwinger functional as a formal power series of renormalized correlation functions. At the technical level. the mass independent renormalization scheme with the dimensional regularization is frequently used to compute the various renormalization group functions.

\subsubsection{Ambiguities in local renormalization group}\label{ambl}

The $\mathcal{O}(g^2)$ higher order terms in the definition of the renormalized Schwinger functional contain some arbitrariness related to contact terms and scheme dependence. At this point we should mention that there are two types of important background fields whose structure of the contact terms may be constrained by requiring the relevant Ward-Takahashi identities.
The first one is the background metric $g_{\mu\nu}(x) = \eta_{\mu\nu} + h_{\mu\nu}(x) + \cdots$ (here $\eta_{\mu\nu}$ is the flat space-time metric) that naturally couples with the energy-momentum tensor as $h_{\mu\nu} T^{\mu\nu} + O(h^2)$. The arbitrariness for the coupling to the background metric is reduced by requiring that the Schwinger functional $W[g_{\mu\nu}(x), g^I(x)]$ is diffeomorphism invariant with respect to the background metric $ds^2 = g_{\mu\nu}(x) dx^{\mu} dx^{\nu}$. Still, it does not fix the arbitrariness entirely because there are higher curvature corrections such as the $\xi R \phi^2$ term in scalar field theories as we mentioned in section \ref{cuvedb}. We could also add the local counterterms constructed out of metric which is diffeomorphism invariant.

The second important example is the background vector fields $a^a_{\mu}(x)$ that couple to not-necessarily-conserved vector operators $J_a^\mu(x)$.  Generically, the vector operators $J^\mu_a$ are not conserved due to the source terms $g^I(x)O_I(x)$ in the interaction. In order to systematically implement the broken Ward-Takahashi identities for the vector operators $J_a^\mu$, it is convenient to introduce the compensated gauge transformations for the source of the violation such as $g^I(x)$ so that the Schwinger functional $W[g_{\mu\nu}(x), g^I(x), a^a_{\mu}(x)]$ is invariant under the compensated gauge transformation:
\begin{align}
\delta a_\mu(x) &= D_{\mu} w(x) \cr
\delta g^I(x) &= -(wg)^I(x) \  \label{compensg}
\end{align}
with the gauge parameter $w$.
Here we assume that the ``free part" of the action $S_0[X]$ has the symmetry $\mathcal{G}$ and the background gauge fields $a_{\mu}(x)$ lies in the corresponding Lie algebra $\mathfrak{g}$. The coupling constants $g^I(x)$ form a certain representation under $\mathcal{G}$. 
We will denote the covariant derivative $D_{\mu} = \partial_\mu + a_{\mu}$ and the field strength $f_{\mu\nu} = \partial_\mu a_\nu - \partial_\nu a_{\mu} + [a_{\mu},a_{\nu}]$ as usual in the matrix notation. More explicitly, we have  $(wg)^I = h_{ab}w^a T^{bI}_{\ J} g^J$ and $D_\mu g^I = \partial_\mu g^I + h_{ab}a_\mu^a T^{bI}_{\ J}g^J$ with the representation matrix $T^{aI}_{\ J}$. 
When non-zero, the bi-linear form $h_{ab}$ can be taken to be unity by rescaling of fields.
When the covariant derivative acts on tensors, they must contain the additional space-time connection.
This compensated gauge invariance plays a significant role in understanding the  importance of operator identities in the local renormalization group analysis \cite{Osborn:1991gm}\cite{JO}.

Due to this ambiguity, the Schwinger functional must be invariant under the compensated gauge transformation \eqref{compensg}: 
\begin{align}
\Delta_{w} W[g_{\mu\nu},g^I, a_{\mu}]  = \int d^dx \sqrt{|g|} \left(D_\mu w \cdot \frac{\delta}{\delta a_{\mu}} - (wg)^I \frac{\delta}{\delta g^I} \right) W[g_{\mu\nu},g^I, a_{\mu}] = 0 \ \label{gaugetrans}
\end{align}
for any Lie algebra element $w \in \mathfrak{g}$ that generates the compensated symmetry $\mathcal{G}$. Hereafter $\cdot$ denotes the invariant scalar product on $\mathfrak{g}$ (proportional to $h_{ab}$) and we often suppress $a$ indices for a shorter notation. Thus, the local renormalization group operator can be equivalently rewritten as
\begin{align}
\Delta_{\sigma} &= \int d^dx \sqrt{|g|} \left( 2\sigma g_{\mu\nu} \frac{\delta}{\delta g_{\mu\nu}} + \sigma \mathcal{B}^I \frac{\delta}{\delta g^I}  \right. + \left. \left( \sigma \hat{\rho}_I D_{\mu} g^I \right)\cdot \frac{\delta}{\delta a_{\mu}} \right) \ , \label{lrgrv} 
\end{align}
when we act on the gauge invariant $W[g_{\mu\nu},g^I, a_{\mu}]$, where
\begin{align}
\mathcal{B}^I &= \beta^I - (vg)^I \cr
\hat{\rho}_I &= \rho_I + \partial_I v \ . \label{gaugeinvb}
\end{align}
In the language of the trace identity, rewriting here corresponds to the use of the operator identity or the equations of motion\footnote{This equation may seem to assume implicitly that the tree level equations motion are the same as the renormalized ones. Depending on the renormalization scheme, it may not be the case and it is possible to have corrections such that $(wg)^I$ is effectively replaced by $(Xwg)^I$, where $X = 1+ O(g^I)$ now contains the higher order corrections. Such a possibility is unavoidable in $d=3$ dimension due to possible gauge anomaly in the right hand side of \eqref{gaugetrans}. We do not expect the gauge anomaly in $d=4$ dimension, but we may have (fractional) Chern-Simons counterterms we will discuss later. In any case, after rewriting it as in \eqref{lrgrv} with whatever renormalized operator identity we have in the theory, there will be no significant difference in the following.}
\begin{align}
v \cdot D_\mu J^\mu = -(vg)^I O_I \ 
\end{align}
so that we have the equivalent expression \cite{Osborn:1991gm}
\begin{align}
T^{\mu}_{\ \mu} = \mathcal{B}^I O_I + (\hat{\rho}_I D_\mu g^I) \cdot J^\mu + A_\mathrm{anomaly} \ . \label{trirv}
\end{align}

Although the physics does not change with the gauge (for the background fields) which we choose, we will mostly stick to the conventional choice \eqref{lrgrv} and \eqref{trirv} in the following. This choice has a great advantage in the flat space-time limit because $\mathcal{B}^I = 0$ directly implies the conformal invariance (i.e. $T^{\mu}_{\ \mu}|_{g_{\mu\nu} = \eta_{\mu\nu}} = 0$). If we used the other choice, we would have to keep track of both $\beta^I$ and $v$ to compute $\mathcal{B}^I = \beta^I - (vg)^I$ in order to discuss the conformal invariance. For this reason, it is most convenient \cite{JO}\cite{Nakayama:2012nd} to define the renormalization group equation for the running background source fields by
\begin{align}
\frac{dg^I}{d\sigma} &= \mathcal{B}^I \cr
\frac{da_{\mu}}{d\sigma} &= \hat{\rho}_I D_{\mu} g^I \ . 
\end{align}
Again, we could evolve the coupling constants in whatever gauge we like (i.e. $\frac{dg^I}{d\sigma} = \beta^I$), and the physics does not change. However, the conformal invariance at the fixed point would be disguised.

\subsubsection{Anomalous dimensions and global Callan-Symanzik equation}\label{contact}

Let us try to read physical information on correlation functions from the local Callan-Symanzik equation. In particular, we would like to derive the formula for the anomalous dimensions from various beta functions. In order to obtain the global Callan-Symanzik equations for correlation functions in flat space-time, we derive the Schwinger functional with respect to $g^I(x)$ and $a^a_\mu(x)$. After setting $D_\mu g^I =0$ with $a_\mu = 0$ and integrating over the space-time once to get rid of one delta function, it gives 
\begin{align}
&\left(\frac{\partial}{\partial \log\Lambda} + \mathcal{B}^I \frac{\partial}{\partial g^I}  \right) \langle O_{I_1}(x_1) O_{I_2}(x_2) \cdots J_{a_1}^\mu(y_1)J_{a_2}^\mu(y_2) \cdots \rangle \cr
 &=   \gamma^{J_1}_{I_1} \langle O_{J_1}(x_1) O_{I_2}(x_2) \cdots J_{a_1}^\mu(y_1) J_{a_2}^\mu(y_2)\cdots  \rangle +  \gamma^{J_2}_{I_2} \langle O_{I_1}(x_1) O_{J_2}(x_2) J_{a_1}^\mu(y_1) J_{a_2}^\mu(y_2) \cdots \rangle + \cdots \cr
& +  \gamma_{a_1}^{b_1} \langle O_{I_1}(x_1) O_{I_2}(x_2) \cdots J_{b_1}^\mu(y_1) J_{a_2}^\mu(y_2) \cdots  \rangle  +  \gamma_{a_2}^{b_2} \langle O_{I_1}(x_1) O_{I_2}(x_2) \cdots J_{a_1}^\mu(y_1) J_{b_2}^\mu(y_2) \cdots  \rangle + \cdots \cr
& + \text{contact terms} \ . 
\end{align}
up to contact terms with extra delta functions.  Here, the anomalous dimension matrix for the scalar operator is given by
\begin{align}
-\gamma^I_J = \partial_J \mathcal{B}^I + (h_{ab} \hat{\rho}_J^a  T^{bI}_{\ \ K} g^K) \ , \label{anomalouss}
\end{align}
whose origin can be seen by applying $\frac{\delta}{\delta g^I}$ to the local renormalization group operator. In particular, we note that the second term comes from the vector beta functions.
Similarly the anomalous dimension matrix for the vector operator is given by 
\begin{align}
-\gamma^{a}_{b} = \hat{\rho}_I^a h_{bc} T^{cI}_{ \ \ J} g^J \ , \label{anovect}
\end{align}
whose origin can be seen by applying $\frac{\delta}{\delta a_\mu^a}$ to the local renormalization group operator. 

Note that while we have computed the anomalous dimension in a particular gauge,  the physical consequence from the  results \eqref{anomalouss} \eqref{anovect} do not change in a different gauge. For this gauge invariance to hold, it is crucial that we have additional contributions from the vector beta functions to the anomalous dimension for the scalar operator in \eqref{anomalouss}.

We have argued that the gauge invariance of the Schwinger functional is a consequence of the operator identity. In classical field theories, the operator identities are typically derived from the use of the equations of motion.
In classical field theories, there is nothing subtle about using equations of motion to simplify the trace of the energy-momentum tensor. Indeed, we see that the trace of the improved energy-momentum tensor for a free scalar vanishes up on the usage of the equations of motion. In quantum mechanics, the use of the equations of motion introduces contact terms in correlation functions, and they play an important role in deriving the Ward-Takahashi identities and the scaling properties of the correlation functions.

To see this, let us consider the path integral of a scalar field theory with action $\mathcal{S}[\phi]$ as an example. By using the field redefinition
\begin{align}
\phi (x) \to (1+\alpha(x)) \phi(x)
\end{align}
with the invariance of the path integral measure\footnote{This is non-trivial because we are interested in the composite operator insertion. Indeed, Konishi anomaly \cite{Konishi:1983hf} of the supersymmetric gauge theories does suggest the violation here. In our discussion of the operator identity, we have to use the correct quantum operator identity.} within the path integral expression
\begin{align}
\left \langle \phi(x_1) \cdots \phi(x_n) \right\rangle = \int \mathcal{D} \phi \phi(x_1) \cdots \phi(x_n) \exp(-\mathcal{S}[\phi]) \ ,
\end{align}
 and taking $\alpha(x) \to \delta (x)$ limit, we can formally obtain
\begin{align}
\left\langle \phi(x) \frac{\delta \mathcal{S}[\phi]}{\delta \phi(x)} \phi(x_1) \cdots \phi(x_n) \right \rangle = \sum_i \left\langle \delta(x_i -x) \phi(x_i) \prod_{i\neq j}\phi(x_j)\right\rangle \ . \label{peom}
\end{align}
This means that the equations of motion  $\frac{\delta \mathcal{S}[\phi]}{\delta \phi(x)} = 0$ is valid up to contact terms (if the anomaly were present, the use of the equations of motion within a composite operator would be modified). In addition, if we integrate (\ref{peom}) over $x$, the insertion of the equations motion can be used to rescale the bare fields in the path integral computation of the correlation functions.
For example, in a free scalar field theory, we will find in section \ref{frees} that the trace of the energy-momentum tensor is a total derivative (or zero in the improved case) up to the equations of motion $\frac{2-d}{2}\phi\Box \phi$. This explains the canonical scaling dimension of scalar under scale transformation after inserting the trace of the energy-momentum tensor for scale transformation in the path integral.

The contact terms associated with the equations of motion play an important role in understanding the renormalization group flow. To see this,
let us now consider the effects of different choices of gauge in the beta functions on the correlation functions on fundamental fields that appear in the global Callan-Symanzik equations \eqref{globalCS}. The choice of the gauge corresponded to the use of the operator identities in the trace identities and the contact terms there  will affect the anomalous dimensions of fundamental fields. 

For this purpose, it is again convenient to introduce the source fields for the ``fundamental fields" $\phi_i$ as $\delta S = \int d^dx \sqrt{|g|} J^i \phi_i$, and compute the Schwinger functional. The local Callan-Symanzik equation contains the extra variation
\begin{align}
\delta \Delta_{\sigma} = \int d^d x\sqrt{|g|} \hat{\gamma}^{i}_{\ j}(g^I) J^j\frac{\delta}{\delta J^i} \ ,
\end{align}
where $\hat{\gamma}^i_{\ j}$ is the scaling dimension of the field $\phi_i$. 
Accordingly, the trace of the energy-momentum tensor contains the extra contribution
\begin{align}
\delta  T^{\mu}_{\ \mu} = \hat{\gamma}^{i}_{\ j} (g^I) J^j \phi_i \ 
\end{align}
The global Callan-Syamnzik equation \eqref{globalCS} is simply obtained by differentiating the Schwinger functional with respect to $J^i$, and integrating over the space-time once to get rid of one delta function.

At this point, we realize that the gauge transformation of the background vector fields $a_\mu^a$ affect also the anomalous dimensions of $\phi^i$ because the current (non-)conservation law is modified by the existence of the source field $J^i$ once it is charged under $\mathcal{G}$:
\begin{align}
D^\mu J_\mu^a = g^I T^{aI}_{\ \ J} O^J + J^j T^{ai}_{ \ \ j} \phi_i \ .
\end{align}
Therefore, the gauge choice associated with $v^a$ in \eqref{gaugeinvb} gives the extra (antisymmetric) contribution to the anomalous dimensions of fields $\phi^i$ 
\begin{align}
\hat{\gamma}^{i}_{\ j} \to \hat{\gamma}^i_{j} + h_{ab}v^a T^{b i}_{ \ \ j} \ ,
\end{align}
where $T^{a i}_{ \ \ j}$ is the representation matrix of fields $\phi_i$ under $\mathcal{G}$.

In the Lagrangian field theories, this ambiguity is precisely those coming from the use of the equations of motion when we evaluate the trace of the energy-momentum tensor. If we keep track of the equations of motion, the trace of the energy-momentum tensor is
\begin{align}
T^{\mu}_{\ \mu} = \beta^I O_I +v^a (\partial_\mu J^\mu_a) +  (d_0+\gamma)\int \phi \frac{\delta \mathcal{S}}{\delta \phi} \ 
\end{align}
in the flat space-time limit with all Lorentz non-invariant sources turned off,
where the matrix structure of $\gamma$ acting on $\phi$ is suppressed. Here $d_0$ is the ``canonical dimension" of the field $\phi$. The inclusion of $d_0$ here is rather conventional in the Callan-Symanzik equation with reference to ``free" field theories. The sum $\hat{\gamma} = d_0 + \gamma$ gives the total scaling dimension of the field $\phi$, which has an intrinsic meaning without referring to reference free field theories.

Whenever $\beta^I O_I$ can be transformed into the virial current, the Callan-Symanzik equation can be further transformed as
\begin{align}
\left(\frac{\partial}{\partial \log\Lambda} + \tilde{\beta}^I\frac{\partial}{\partial g^I} \right) \left \langle \phi_{i_1}  \cdots \phi_{i_n} \right\rangle  = (\gamma_{i_1}^{\ j_1} + S_{i_n}^{\ j_n})  \left \langle \phi_{j_1}  \cdots \phi_{i_n} \right\rangle + \cdots + (\gamma_{i_n}^{\ j_n} + S_{i_n}^{\ j_n})  \left \langle \phi_{i_1}  \cdots \phi_{j_n} \right\rangle \ \label{css}
\end{align}
by introducing the ``flavor"\footnote{Throughout the review article, the ``flavor" symmetry with quotation mark refers to the spurious broken symmetry acting on the interaction terms. In perturbation theory, it is the symmetry of the kinetic terms (e.g. $O(N_b)$ for scalars and $U(n_f)$ for fermions), but is broken by the interaction terms such as Yukawa interactions or scalar self-interactions.} rotation matrix $S^{i}_{\ j}$ with $\beta^I O_I = \tilde{\beta}^I O_I + \partial^\mu J_\mu$ up to equations of motion because the change of the coupling constant in the virial current direction can be absorbed by the rotations of fields (or more abstractly operators). 

Correspondingly, the trace of the energy-momentum tensor is rewritten as
\begin{align}
T^{\mu}_{\ \mu} = \tilde{\beta}^I O_I +\tilde{v}^a (\partial_\mu J^\mu_a) + (d_0+\gamma + S)\int \phi \frac{\delta \mathcal{S}}{\delta \phi} \ \label{emrw}
\end{align}
by using the equations of motion (operator identity). The use of the equations of motion is manifest in the last term of (\ref{emrw}) so that it gives the extra wavefunction renormalization factor $S$ in the Callan-Symanzik equation (\ref{css}). When $\tilde{\beta}^I$ vanishes by choosing a wavefunction renormalization factor $S$, the theory is indeed scale invariant. If in addition, all the vector beta functions $\tilde{v}^a$ vanish in this choice of $S$, then the theory is conformal invariant. 
Although the  wavefunction renormalization factor $S$ introduces non-standard antisymmetric part (rather than symmetric part) \cite{Fortin:2011sz}, we may diagonalize the dilatation operator if it is diagonalizable. Without conformal invariance, the diagonalization may not be possible but at least we could simplify it in the Jordan normal form.

Unfortunately, the global Callan-Symanzik equation says nothing about the distinction between scale invariance and conformal invariance. We have to study the unintegrated trace of the energy-momentum tensor to see the distinction. Here the local version of the renormalization group has advantage because we can understand the total derivative contributions to the trace of the energy-momentum tensor. We will further discuss the method of the local renormalization group in relation to the distinction between scale invariance and conformal invariance in section \ref{localr}.

\subsection{Computation of trace of energy-momentum tensor}\label{red}
So far, our discussions on the renormalization group have been abstract. Let us now present a concrete way of finding scale invariant but non-conformal field theories within power-counting renormalization scheme of perturbatively renormalizable quantum field theories in $d=4$ dimension. 
We also would like to clarify the origin of the $v^a (\partial_\mu J^\mu_a)$ term in the trace of the energy-momentum tensor in dimensional regularization. 
In this section, the coupling constants are taken to be position independent. This will give a recipe to connect the local renormalization group computation to  a more conventional global renormalization group computation. However, the price to pay is that we should carefully analyze the composite operator renormalization which we will explain in the following.

Our discussion here is based on the dimensional regularization with minimal subtraction, but since our final expression is renormalization group scheme covariant, any other regularization should work in principle. The scheme covariance of the cyclic renormalization group flow was discussed in \cite{Fortin:2012ic} under the change of the coordinate transformation in the coupling constant space $g^I \to \tilde{g}^I(g)$. We will further discuss the more non-trivial scheme associated with the ``gauge transformation" on the coupling constant space in the following. The discussion of this section is based on \cite{Jack:1990eb}\cite{Osborn:1991gm} (c.f. \cite{Fortin:2012hn} for a concise summary). Some concrete examples of the renormalization procedure will be presented in section \ref{examples}. 

First of all, we recall that all classically scale invariant power-counting renormalizable quantum field theories have the classical energy-momentum tensor whose trace is zero up on improvement (classical Weyl invariance) in $d=4$ dimension \cite{Callan:1970ze}.  
To regularize the divergence in quantum field theory within perturbation theory, we use the dimensional regularization and evaluate the trace of the energy-momentum tensor in $d=4-\epsilon$ dimension. The trace is proportional to the total action density\footnote{To assure this, we have to include suitable improvement terms for scalars.} up to the terms that vanish with equations of motion
\begin{align}
T^{\mu}_{\ \mu} = \epsilon \mathcal{L} + \int \phi \frac{\delta \mathcal{S}}{\delta \phi} \ .
\end{align}
We will renormalize the action density operator $\mathcal{L}$ so that it satisfies the renormalization group equation in $d=4-\epsilon$ dimension
\begin{align}
\left(\hat{\beta^I}\frac{\partial}{\partial g^I} + \hat{\gamma} \phi \frac{\partial}{\partial \phi} - \epsilon\right) \mathcal{L} = 0 \ , \label{rengrp}
\end{align}
where $\hat{\beta}^I = \epsilon (k g)^I + \beta^I(g)$ are beta functions in $d = 4-\epsilon$ dimension ($k$ is a constant that depends on the power of the coupling constants $g^I$ appearing in the action) and  $\hat{\gamma} = \epsilon + \gamma$ are anomalous dimension in $d = 4-\epsilon$. In massless QCD, for instance, there is no complication at this point, and we can simply take $\epsilon \to 0$ and rederive (\ref{ta}). The formal justification of the renormalization group equation \eqref{rengrp} can be found in the appendix of \cite{Jack:2013sha}.

In a more complicated situation, this naive limit must be modified in a subtle way. The point is that although $\hat{\beta}^I\frac{\partial}{\partial g^I}\mathcal{L}$ is a finite operator, $\frac{\partial}{\partial g^I}\mathcal{L}$ might not be. We have to expand $\frac{\partial }{\partial g^I}\mathcal{L} = [O_I] + N_{I}^a \partial_\mu[J^\mu_a] + M_{Ik} \Box[O^{(2)}_k]$, where all $[O]$ are finite operators\footnote{In this section, we make a careful distinction between unrenormalized operators $O$ and the finitely renormalized composite operators $[O]$. We should remember that most of the other part of the review article, the composite operators $O$ are finitely renormalized implicitly and they could have been written as $[O]$ as in this section.} while $N_{I}^a$ and $M_{Ik}$ can contain $\epsilon^{-1}$ and higher poles. Note that $\int d^dx \frac{\partial}{\partial g^I}\mathcal{L}$ must be finite so the divergence appears only in derivatives. Thus, if we express the trace of energy-momentum tensor in terms of finite operators, we should obtain
\begin{align}
T^{\mu}_{\ \mu} = \beta^I [O_I]  + \partial^\mu [J_\mu] + \Box [O^{(2)}] + (d_0+\gamma)\int \phi  \frac{\delta \mathcal{S}}{\delta \phi} \ , \label{finiteop}
\end{align}
where we have taken $\epsilon \to 0$ limit safely because all the operators are finite now. 

One important point to notice is that for $[J_\mu]$ to be finite, we have to cancel the poles in $N_{I}^a$ and linear $\epsilon$ terms in $\hat{\beta}^I$.\footnote{For a technical reason, it is important that we use minimal subtraction here because $\epsilon$ only appears in the first term in 
$\hat{\beta}^I (g,\epsilon) = \epsilon (k g)^I + \beta(g) $ 
and the higher $\epsilon$ terms does not appear in the beta function.} 
This means that at the leading order, we obtain $[J^\mu] = v^a [J_a^\mu] = g^I N_{I}^{a (1)} J_a^{\mu}$ with $N_{I}^{a(1)}$ is the $\epsilon^{-1}$ term in $N_{I}^a$. The higher terms are also constrained because of the delicate cancellation between $N_{I}^a$ and $\epsilon (kg)^I$. The coefficient $v^a$ is interpreted as the beta function for the divergence of a vector current $\partial^\mu J_a^\mu$. 
A similar argument applies for the dimension two operators $[O^{(2)}_k]$, but it is of little relevance for our perturbative discussions. In the following, we assume $\Box[O^{(2)}_k]$ term is removed by improvement of the renormalized energy-momentum tensor (see e.g. \cite{Brown:1980qq}\cite{Hathrell:1981zb} for reference).

However, this is not the end of the story because there is an operator identity
 (equations of motion) to relate $\partial_\mu [J_a^\mu]$ to sum of $[O^I]$s. Therefore, the separation between $\beta^I [O_I]$ and $v^a \partial_\mu [J_a^\mu]$ is actually arbitrary. After all, the possibility of the equality 
\begin{align}
T^{\mu}_{\ \mu} = \beta^I [O_I] + v^a \partial_\mu [J^\mu_a] = \partial^\mu [J_\mu] \ 
\end{align}
up to equations of motion, which we are looking for the scale invariant field theories,
 assumes the operator identity such as $\beta_I[O^I] = \partial^\mu [K_\mu]$ for a certain current operator $[K_\mu]$.

With this operator identity, the trace of the energy-momentum tensor is invariant under  
\begin{align}
\beta^I &\to \beta^I + (w\cdot g)^I \cr
v^a &\to v^a + w^a   \ , \label{gaugebeta}
\end{align}
where $w$ acts on coupling constant as an element of the ``flavor" symmetry generator (i.e. $(w\cdot g)^I = h_{ab}w^a T^{b I}_{\ \ J} g^J$ with a representation matrix $T^{a I}_{\ \ J}$ for the symmetry as before).\footnote{For instance in $\phi^4$ theory, the coupling constants $\lambda^{ijkl}$ transforms as forth rank symmetric tensor under the $O(N_b)$ rotation induced by the wavefunction renormalization $S^{ij}$ on $\phi_i$.} Thus, the beta functions are ambiguous in the dimensional regularization computation. This is precisely what we have discussed in terms of the renormalized Schwinger functional in section \ref{ambl}. Since we have not introduced the position dependence of $g^I$, the derivative part of the vector beta function $\rho_I D_\mu g^I$ remains zero.

To cancel the ambiguity, it is customary to introduce the $\mathcal{B}$ function \cite{Jack:1990eb}\cite{Osborn:1991gm}, which is defined by the full trace of the energy momentum tensor,
\begin{align}
T = \mathcal{B}^I [O_I] + (d_0+\gamma + v) \int \phi \frac{\delta \mathcal{S}}{\delta \phi} = \beta^I [O_I] + v^a \partial_\mu [J^\mu_a] + (d_0+\gamma)\int \phi \frac{\delta \mathcal{S}}{\delta \phi} \ .
\end{align}
We can see that $\mathcal{B}$ function is  invariant under the gauge transformation (\ref{gaugebeta}).
Clearly, the conformal invariance requires vanishing of the $\mathcal{B}$ functions rather than the  vanishing of beta functions.

We note the appearance of the additional equations of motion operator with $v$ if we  use $\mathcal{B}$ function as a renormalization group flow of the coupling constant $g^I$: $\frac{dg^I}{d\log\mu} = \mathcal{B}^I$. This changes the wavefunction renormalization factor compared with the ``standard" one $\frac{dg^I}{d\log\mu} = \mathcal{\beta}^I$ which we started with. Actually, this could have been asked at (\ref{rengrp}) because the renormalization group equation itself was ambiguous as discussed around (\ref{css}). If we had renormalized the action density operator $\mathcal{L}$ with the usage of the additional wavefunction factor $v$ and the corresponding $\mathcal{B}$ function, we would not have to introduce the divergence part of the vector beta functions $v^a$ when we rewrite the bare operator into the finite ones because the same renormalization prescription removes $\epsilon^{-1}$ poles in $N_I^a$. In this way, the renormalization group flow has various ambiguities if we allow the appearance of virial current operators, but they all cancel out in the final expression for the trace of the energy-momentum tensor, and the question over scale invariance vs conformal invariance is a physically well-posed one.

So far, we have not discussed how to compute  the divergence part of the vector beta function $v^a$ in practice. In general, the renormalization of the composite operator discussed above is complicated. Conceptually, it is easier to consider the space-time dependent coupling constant $g^I(x)$, and introduce the additional counterterms $ \int d^d x N^a(x) \partial_\mu J_a^\mu$ in the action. As we have mentioned $v^a$ can be regarded as the beta function for $N^a(x)$. 

More generically, we can consider the counterterm $\int d^d x N^a_I(g) \partial_\mu g^I J^\mu_a$, part of which generates $ \int d^d x N^a(x) \partial_\mu J_a^\mu$ after partial integration (i.e. ``symmetric part").\footnote{The term that cannot be written as $ \int d^d x N^a(x) \partial_\mu J_a^\mu$  (i.e. ``antisymmetric part") is related to the extra term in the trace of the energy-momentum tensor $ \rho^I_a (D_\mu g_I) J^a$ that appears when the coupling constant is position dependent (see section \ref{localr} for more details). \label{fonot}}
 In the dimensional reguralization, we may identify $N^a_I(g)$ here with the operator renormalization factor $N^a_I$ used in the computation of the  current contribution $\partial^\mu J_\mu$ to the trace of the energy-momentum tensor in (\ref{finiteop}) because the functional derivative in the local Callan-Symanzik operator $\hat{\beta}^I \frac{\delta}{\delta g^I(x)}$ will act on the renormalized action to give the finite operator relation $\frac{\delta}{\delta g^I(x)} \mathcal{S}|_{D_\mu g^I = 0} = [O_I] + N_I^a \partial_\mu [J^\mu_a] + M_{Ik} \Box[O^{(2)}_k]$. 

In this way, in the dimensional regularization with minimal subtraction, the computation of the vector beta function from the counterterm $N^a$ gives the vector beta function through $N^a_I$.  For an explicit computation of the counterterm $N^a_I$, we can study the antisymmetric wavefunction renormalization with additional momentum flow to accommodate the position dependence of $g^I$.
We refer to the literature  \cite{Jack:1990eb}\cite{Fortin:2012hn}  how to  compute the diverging part of $N^a$ and consequently $v^a$. If we use the dimensional regularization with the prescription that the anomalous dimension matrix $\gamma$ is symmetric, $g^IN_I^{a(1)}\partial_\mu J_a^\mu$ vanishes up to two loops. At three loops, there is a non-trivial contribution in this prescription and we will quote the result in section \ref{Intera}, where non-zero term $g^IN_I^{a(1)}\partial_\mu J_a^\mu$ in the trace of the energy-momentum tensor played a crucial role in confirming conformal invariance of the fixed point at three-loop order.

\subsection{(Redundant) conformal perturbation theory}\label{reddd}
As a complementary but concrete approach to the discussions in the previous subsections, we will try to understand the role of the redundant operators and the computation of the beta functions in conformal perturbation theory in this final section of section \ref{local}. After all, the ambiguities we have encountered due to the equations of motion are nothing but due to the redundancy of our description of the quantum field theory under consideration.
It will also  give some general perspectives on the perturbative searches for scale invariant but non-conformal field theories. 

First of all, we should recall that quantum field theories have intrinsic ambiguities due to the field redefinition. In high energy-physics, this is manifested in the invariance of the S-matrix under the field redefinition \cite{Chis}\cite{Kame}, and in statistical physics, it is know as the invariance of the partition function under the change of the integration variables \cite{Wegner}. Correspondingly, the deformation of the effective action that is related to total derivative terms up on using the equations of motion is the so-called redundant perturbation because it does not affect any physics. Clearly, it is of importance to tame the redundant perturbation to discuss the perturbative scale invariance without conformal invariance.

The conformal perturbation theory \cite{Zamolodchikov:1987ti}\cite{Cappelli:1989yu} is defined by perturbing the reference conformal field theory by adding relevant or marginal perturbations $\delta S = \int d^d x g^I O_I(x)$. From the unitarity, $O_I(x)$ must be conformal primary operators of the reference conformal field theory. For technical simplicity, we focus on the situation when all $O_I(x)$ have conformal dimension $d$, but the generalization of the following argument for  including slightly relevant deformations is possible.  

We assume that $O_I(x)$ have the canonical normalization in the reference conformal field theory:
\begin{align}
\langle O_I(x) O_J(y) \rangle_0 = \frac{\delta_{IJ}}{(x-y)^{2d}} \ . 
\end{align} 
The conformal invariance demands that the three-point functions among $O_I(x)$ must be given by
\begin{align}
\langle O_I(x)  O_J(y) O_K(z) \rangle_0 = \frac{C_{IJK}}{(x-y)^{d}(y-z)^{d}(z-x)^{d}} \   . \label{threepoinconf}
\end{align} 
In these expressions, the subscript $0$ means the expectation value in the reference conformal field theory.
So far, it is a standard conformal perturbation theory setup. In order to allow the non-trivial existence of the virial current, we allow the appearance of the  conserved current $J_a^\mu$ in the reference conformal field theory in the OPE
(see e.g. \cite{Friedan:2012hi}cite{Behr:2013vta} for a similar argument in $d=2$ dimension)
\begin{align}
O_I(x) O_J(y) = \frac{C_{IJK}}{(x-y)^{d}} O_K(y) + \frac{C^{a}_{IJ} (x-y)_{\mu}}{(x-y)^{d+2}} J_a^\mu(y) + \cdots \ . \label{opecc}
\end{align}
Here $C_{IJK}$ is totally symmetric while $C^{a}_{IJ} = -C^{a}_{JI}$ is a certain representation matrix of the ``flavor symmetry" generated by $J_a^{\mu}$.

Before going on, let us discuss the current contribution in the OPE (\ref{opecc}). From the unitarity, we require $D^\mu J_\mu^a = 0$ in the reference conformal field theory, so the possible addition of $D^\mu J_a^\mu$ in the action is a redundant perturbation in a double sense because (A) it is a total derivative, and (B) it vanishes by conservation. However, the OPE (\ref{opecc}) means that some operators $O^I$ are charged under the ``flavor symmetry" because we can derive the Ward-Takahashi identity
\begin{align}
\langle D^\mu J^a_\mu(x) O^I(x_1) \cdots \rangle_0 = \delta(x-x_1)C_{IJ}^a \langle O^J(x_1) \cdots \rangle_0 \ 
\end{align}
from the OPE. It follows that in the perturbed conformal field theory, we have the violation of the symmetry as
\begin{align}
D^\mu J_\mu^a = g^I C_{IJ}^a O^J \ . \label{broken}
\end{align}
The equation will get renormalized at the higher order, but since it is outside of our scope to develop a systematic higher order conformal perturbation theory, it will not be important.

The conformal perturbation theory begins with the formal definition
\begin{align}
\langle \cdots \rangle = \langle e^{-\int d^dx g^I(x) O_I(x)} \cdots \rangle_0 \ 
\end{align}
for the correlation functions of the perturbed theory as a perturbative series in $g^I$. The right hand side is typically divergent and we need a suitable renormalization. To discuss the vector beta functions, we have promoted the coupling constant $g^I$ to be space-time dependent as mentioned in section \ref{red}. Accordingly, we need more counterterms, which is suppressed here (see also section \ref{localr}).

Let us compute the beta function in a conventional way.\footnote{One cautious remark is that we do not pay attention to the ``Lagrangian density operator" of the reference theory, which gives an additional redundant deformation. In usual quantum field theories, we do renormalize the wavefunction to reduce the number of independent running coupling constants, but this has not been attempted here. Anyway, it will be higher order corrections than we study here.} 
At the second order in perturbation theory, we encounter the divergence in the ``vacuum diagram" by colliding $\int d^d x g^I(x) O_I(x)$ and $\int d^d y g^J(y) O_J(y)$ near $x\sim y$. The divergence
from the scalar three-point function
\begin{align}
\int d^dx d^d y g^I(x) O_I(x) g^J(y) O_J(y) \sim \log \mu \int d^d z C_{IJK} g^I(z) g^J(z) O_K(z)
\end{align}
can be removed by renormalizing the coupling constant with the beta function\footnote{In the following discussions of the conformal perturbation theory, $d$-dependent numerical factors that appear in the integration over the space-time are omitted. One may always absorb them in the normalization of $g^I$.}
\begin{align}
{\beta}^I = \frac{d g^{I}}{d\log\mu} =  C_{IKL} g^K g^L + \mathcal{O}(g^3) \ .
\end{align}

As we mentioned in section \ref{red}, the vector beta functions could have been obtained from the divergence in  $\int d^dx g^I \partial^\mu g^J N_{IJ}^a J_\mu^a$ (with symmetric $N_{IJ}^a$). At the second order in conformal perturbation theory with the above conventional prescription, however, the would-be divergent term is only
\begin{align}
\log \mu \int d^dz  g^I(z)\partial_\mu g^J(z) C_{IJ}^a J^\mu_a \label{convenv}
\end{align}
which does not affect the vector beta function because $C_{IJ}^a$ is antisymmetric. This term itself is renormalized by $\rho_I$ term in the space-time dependent coupling constant term in the trace of the energy-momentum tensor (\ref{traceiden}) that was mentioned in footnote \ref{fonot}
\begin{align}
\rho_I^a = C_{IJ}^a g^J 
\end{align}
and we will discuss more in section \ref{localr}, but it has nothing to do with the discussion relevant for the computation of $\mathcal{B}^I$ function here.
The symmetric part does not appear due to the conservation $D^\mu J_\mu^a =0$ in the reference theory.
 Thus the divergence part of the vector beta functions $v^a$ are zero in this prescription at this order. We therefore conclude
\begin{align}
\mathcal{B}^I =  C_{IKL} g^K g^L + \mathcal{O}(g^3) \ , \label{Bfunction}
\end{align}
and we observe it is given by the gradient flow with the potential
\begin{align}
\tilde{c} = \frac{1}{3}C_{IJK} g^I g^J g^K  + \mathcal{O}(g^4) \ .
\end{align}
so that $\partial_I \tilde{c} = C_{IJK} g^J g^K = \mathcal{B}^I$. We have more to say about the gradient formula in section \ref{proof2}. For later reference, we note that the Zamolodchikov metric, which we will discuss in section \ref{proof2}, is $\chi_{IJ} = \delta_{IJ}$, and the antisymmetric part vanishes.

The potential $\tilde{c}$  is invariant under the ``flavor" symmetry transformation $\delta^a g^I = C^a_{IL} g^L$.
As a consequence, we obtain
\begin{align}
\delta^a g^I \cdot \mathcal{B}^I = 0 \ ,
\end{align}
which means at the leading order in conformal perturbation theory, the renormalization group flow is orthogonal to the ``flavor symmetry" transformation and the virial current must vanish.

Let us briefly discuss the ambiguities of the beta functions with this setup.
The point is that we could subtract more in the scalar operator beta functions as long as we add more to the vector beta functions. We consider the  counterterm
\begin{align}
\log \mu \int d^d z g^I w_I^a \partial_\mu J^{\mu}_a \label{vectorcounter}
\end{align}
with $w_I^a$ of $\mathcal{O}(1)$, which is arbitrary. In contrast to the conventional counterterm (\ref{convenv}), it is non-zero at the second order in perturbation theory because of the broken conservation law (\ref{broken}). Or if we stick to the reference conformal field theory, one can perturb it once more by $g^IO_I$. It gives a contribution to the divergence part of the vector beta function
\begin{align}
\tilde{v}^a = g^I w_I^a  \ . \label{modifiedv}
\end{align}

Clearly the added term (\ref{vectorcounter}) by itself is divergent and we have to cancel it. This is done by further adding the scalar operator counterterm
\begin{align}
\log \mu \int d^d z g^I w_I^a g^K C_{KL}^a O_L \ ,
\end{align}
which precisely cancels with (\ref{vectorcounter}) after using the equations of motion. It gives the scalar operator beta function at the second order in addition to the original one that was needed to cancel the OPE singularity:
\begin{align}
\tilde{\beta}^I =  C_{IKL} g^K g^L + (g^J w_J^a C_{KI}^a) g^K \ . \label{modifiedbeta}
\end{align}
Of course, such artificial adding and subtracting the same term up to the equations of motion does not change the physics, and this is what we called the ambiguities in the beta functions discussed in section \ref{red}. Although we may think the conventional computation seems more natural at this order, at higher orders in perturbation theory it becomes more non-trivial. 
In anyway, the most important object is the the invariant $\mathcal{B}$ function (\ref{Bfunction}) that appears in the total trace of the energy-momentum tensor.
We can confirm that it does not change under the ambiguity since the contribution from \eqref{modifiedv} is cancelled against the second term in \eqref{modifiedbeta}.

\newpage

\section{Examples}\label{examples}
In this section, we will present several examples of scale invariant field theories that may or may not show conformal invariance in various dimensions.
\subsection{Free theories} \label{frees}
A free massless scalar theory in $d$ dimension has the action minimally coupled  with gravity:
\begin{align}
 \mathcal{S} = \frac{1}{2}\int d^d x\sqrt{|g|} (\partial^\mu \phi \partial_\mu \phi) \ . \label{minimalscalar}
\end{align}
The (canonical) energy-momentum tensor $T_{\mu\nu} = \frac{2}{\sqrt{|g|}}\frac{\delta \mathcal{S}}{\delta g^{\mu\nu}}|_{g_{\mu\nu}=\eta_{\mu\nu}} $ from the action \eqref{minimalscalar} is
\begin{align}
 T_{\mu\nu} = \partial_\mu \phi \partial_\nu \phi - \frac{\eta_{\mu\nu}}{2} (\partial_\rho \phi)^2 \ .
\end{align}
The trace can be computed as
\begin{align}
 T^\mu_{\ \mu} = \frac{2-d}{2}(\partial_\mu \phi)^2 = \frac{2-d}{4}(\Box \phi^2) \ .
\end{align}
In the last line, we have used the equations of motion (EOM). In classical field theories, there is nothing wrong with the usage of equations of motion in deriving conserved currents.\footnote{One exceptional subtlety may be that it is possible that the symmetry algebra may only close up to the equations of motion (on-shell symmetry rather than off-shell symmetry). Correspondingly, we have a so-called zilch symmetry whose variation is proportional to the equations of motion, which does not have the corresponding Noether current. They are related to field redefinition ambiguities.} Even in quantum mechanics, the equations of motion hold as an operator identity (as long as there is no anomaly) in a suitably renormalized sense.

The free massless scalar  is obviously scale invariant. The virial current is given by
\begin{align}
J_\mu = \frac{2-d}{2} \phi \partial_\mu \phi \ .
\end{align}
Moreover, it is conformal invariant in any dimension because $T^{\mu}_{\ \mu} = \partial^{\mu} \partial^\nu L_{\mu\nu}$ with 
\begin{align}
L_{\mu\nu}= \frac{2-d}{4} \eta_{\mu\nu}\phi^2 \ .
\end{align}
Indeed, one can improve the curved space action by adding $\frac{1}{2} \int d^dx \sqrt{|g|}  \frac{d-2}{12}R\phi^2$ so that
the theory is manifestly Weyl invariant, and the energy-momentum tensor is traceless.
The improved action (e.g. $\frac{1}{2}\int d^4x \sqrt{|g|} (\partial^\mu \phi \partial_\mu \phi  + \frac{R}{6}\phi^2)$ in $d=4$ dimension) is known as conformal scalar action.

Although we can improve the energy-momentum tensor as we wish, there can be a conflict with other symmetries. For instance, a free massless scalar theory can possess the shift symmetry $\phi \to \phi +c$. A physically relevant situation is when  the massless scalar is given by a Nambu-Goldstone boson. In such a case, it is unnatural to improve the energy-momentum tensor because $\frac{1}{2}  \int d^dx \sqrt{|g|} \frac{d-2}{12} R\phi^2$ term will be incompatible with the shift symmetry. Indeed, the shift symmetry does not commute with the scale transformation or special conformal transformation when $d\neq 2$. 

%\

%\begin{shadebox}
%(Exercise) Show free massless Dirac fermion is conformal invariant in any space-time dimension.
%\end{shadebox}

%\

For a free massless Dirac fermion, the energy-momentum tensor can be computed as
\begin{align}
T_{\mu\nu} = i\frac{1}{2}\bar{\psi}(\gamma_\mu \partial_\nu + \gamma_\nu \partial_\mu) \psi -  i\eta_{\mu\nu} \bar{\psi} \gamma^\rho \partial_\rho \psi \ ,
\end{align}
and by using the Dirac equation, it is shown to be traceless in any dimension. Thus the massless free fermion is conformal invariant in any dimension. We remark that in $d=2$ dimension we do not have to use the Dirac equation to show that the trace of the energy-momentum tensor vanishes because the canonical scaling dimension agrees with the geometric dimension of the spinor.

Another interesting example is free $U(1)$ Maxwell theory in $d$ dimension \cite{Jackiw:2011vz}\cite{ElShowk:2011gz}.
\begin{align}
\mathcal{S} = \int d^d x\sqrt{|g|} \frac{1}{4}F^{\mu\nu}F_{\mu\nu} \ .
\end{align}
Canonical gauge invariant energy-momentum tensor  $T_{\mu\nu} = \frac{2}{\sqrt{|g|}}\frac{\delta \mathcal{S}}{\delta g^{\mu\nu}}|_{g_{\mu\nu}=\eta_{\mu\nu}}$ can be computed as 
\begin{align}
T_{\mu\nu}= F_{\mu\rho}F^{\rho}_{\ \nu}- \frac{\eta_{\mu\nu}}{4}(F_{\rho\sigma})^2 
\end{align}
Its trace does not vanish when $d\neq 4$:
\begin{align}
T^\mu_{\ \mu} =\frac{4-d}{4}(F_{\rho\sigma})^2 = \frac{4-d}{8} \partial_\mu (A_\rho F^{\mu\rho}) \ ,
\end{align}
but it is a divergence of a current by using the free Maxwell equation. 
Therefore, the massless free vector field is scale invariant with the virial current
\begin{align}
J_\mu = \frac{4-d}{8} A^\nu F_{\mu\nu} \ . \label{Maxvir}
\end{align}
When $d=4$, it is well-known that Maxwell theory is conformal invariant. 
However in the other dimensions, we cannot improve the energy-momentum tensor so that it is traceless.
Therefore Maxwell theory in $d\neq4$ is scale invariant, but not conformal invariant.

We can alternatively study the correlation functions in $d$-dimensional Maxwell theory and see that it does not satisfy the conformal Ward-Takahashi identity.
For instance, the direct computation shows that
\begin{equation}
\left\langle F_{\mu\nu}(x)F_{\lambda\sigma}(0)\right\rangle =\frac{2d-4}{(x^{2})^{d/2}}\left[\left(\eta_{\mu\lambda}-\frac{d}{2}\frac{x_{\mu}x_{\lambda}}{x^{2}}\right)\left(\eta_{\nu\sigma}-\frac{d}{2}\frac{x_{\nu}x_{\sigma}}{x^{2}}\right)-\mu\leftrightarrow\nu\right],\label{eq:FF-1}
\end{equation}
cannot be conformal two-point function of primary two-form fields, and the three-point functions of the scalar primary scalar operator $\Phi = (F_{\mu\nu})^2$ is obtained by
 \begin{align}
\left\langle \Phi(x_{1})\Phi(x_{2})\Phi(x_{3})\right\rangle  & =\frac{-8(d-2)^{3}(d-4)d}{(x_{12}^{2})^{d/2}(x_{13}^{2})^{d/2}(x_{23}^{2})^{d/2}}\times\nonumber \\
 & \left[2+d^{2}\frac{(x_{12}.x_{13})(x_{12}.x_{23})(x_{13}.x_{23})}{x_{12}^{2}x_{13}^{2}x_{23}^{2}}-d\frac{(x_{12}.x_{23})x_{13}^{2}+2\text{ perms}}{x_{12}^{2}x_{13}^{2}x_{23}^{2}}\right]\ \label{F2-1} \ ,
\end{align}
where $x_{ij}\equiv x_{i}-x_{j}$ and it is scale invariant but not conformal invariant (except in $d=4$ dimension) by comparing it with \eqref{threepoinconf} predicted from the conformal invariance.

One peculiar feature of the scale invariance of the free Maxwell theory in $d\neq 4$ is that  the scale current $D_\mu = x^\nu T_{\mu\nu} - J_\mu$ is not gauge invariant due to the gauge non-invariance of the virial current (\ref{Maxvir}). This is related to the fact that the scale dimension of the vector potential is different from the geometric dimension of 1-form. 
Because of this fact, strictly speaking, the Noether assumption is violated. Nevertheless the scale charge $ D = \int d^{d-1} x D_0$ is obviously gauge invariant (after partial integration of the gauge parameter), and all the correlation functions scale as they should.

One may note that in $d=3$ dimension, something special happens. A free massless vector is dual to a free massless scalar $\phi$ in $d=3$ dimension by dualizing $F_{\mu\nu}=\epsilon_{\mu\nu\rho}\partial \phi$ with the dual action $\int d^3x \partial^\mu \phi \partial_\mu \phi$, so we may reformulate it with the scalar field, and we can see that the virial current is then given by $J_\mu \sim \partial_\mu (\phi^2)$. Note that the dual scalar must accompany the gauged shift symmetry, so the theory cannot be Weyl invariant (because would-be improvement term is not gauge invariant). It is still embedded in a conformal field theory \cite{Jackiw:2011vz}\cite{ElShowk:2011gz}.

In the above discussions, we have been careless about the gauge fixing, but the  conclusion does not change by the gauge fixing procedure. In $d=4$ dimension, the gauge fixing term in the Maxwell theory violates the conformal invariance, but the violation is BRST trivial.
It is interesting to note, however, the introduction of the BRST charge together with the hidden ``conformal generator" will generate infinite dimensional graded algebra \cite{ElShowk:2011gz} in $d\neq 4$. In this case, the ``conformal symmetry" is not the symmetry of the physical spectrum because it does not commute with the BRST charge. Throughout the review article, we concentrate on the symmetries that commute with the BRST charge when we talk about the gauge theories.

Generic massless vector field theories without gauge invariance (thus without unitarity, or reflection positivity) are scale invariant but not conformal invariant in any dimension as emphasized by Riva and Cardy \cite{Riva:2005gd}
\begin{align}
 \mathcal{S}= \int d^dx \left(\frac{1}{4}(\partial_\mu v_\nu - \partial_\nu v_\mu)^2 - \frac{\alpha}{2} (\partial^\mu v_\mu)^2 \right)\ .
\end{align}
with 
\begin{align}
T^{\mu}_{\ \mu} = \left(2-\frac{d}{2}\right)(\partial_\mu v_\nu \partial^\mu v^\nu - \partial_\mu v_\nu \partial^\nu v^\mu) -\alpha\left((2-d)v_\mu \partial^\nu \partial^\mu v_\nu - \frac{d}{2}(\partial^\mu v_\mu)^2 \right) \ .
\end{align}
This can be improved to be traceless only when $\alpha = \frac{d-4}{d}$ (see e.g. \cite{ElShowk:2011gz}). 
 In the Euclidean signature, this model is regarded as a theory of elasticity \cite{Elas}, where $v_\mu$ is the displacement vector. The model can be also regarded as a free field theory describing the theory of perception \cite{Bialek:1986it}\cite{Bialek:1987qc}\cite{Nakayama:2010ye}.

\subsection{Interacting theories}\label{Intera}
In $d=4$ dimension, we can argue that classically all unitary (power-counting) renormalizable scale invariant actions are conformal invariant \cite{Callan:1970ze}. The statement can be verified by explicitly writing down all the possible renormalizable interaction terms.
The quantum corrections due to the renormalization and necessity of cut-off hiddenly introduces the scale, so the scale invariance can be broken. As already argued in section \ref{emr}, the effect of the renormalization  plays a crucial role in our discussions on scale invariance and conformal invariance. In particular, the trace of the energy-momentum  tensor obtains a quantum correction
\begin{align}
T^{\mu}_{\ \mu} =  \beta^I O_I +v^a (\partial_\mu J^\mu_a) \ , \label{eomq}
\end{align}
where $\beta^I =\frac{dg^I}{d\log\mu}$ is the beta function for the coupling constant $g^I$ and tells how much our coupling constant runs along the renormalization group flow. We have already discussed possibly unfamiliar second term $v^a$ (the so-called divergence part of the vector beta function) in section \ref{red}. Roughly speaking, it is the beta function that removes the divergence of the background space-time dependent current coupling $\int d^4 x ( N^a) \partial_\mu J^\mu_a$ with a space-time dependent background field $N^a$. The general expression (\ref{eomq}) for the violation of the trace of the energy-momentum tensor by quantum corrections is known as the trace identity and we had more detailed discussions in section \ref{contact} and \ref{red}.

As discussed in \ref{local}, the energy-momentum tensor is not unique. In (\ref{eomq}), we have assumed that we can use the ambiguities to remove $\Box O$ term with a scalar operator $O$ (preferably dimension $2$ but not necessary). In interacting field theories, the term $\Box O$ can be renormalized and it may mix with the other terms, so this assumption is more non-trivial than we naively think, but we leave the problem set a side for now (see e.g. \cite{Brown:1980qq}\cite{Hathrell:1981zb} for reference), and come back to the point when necessary.
We also note that the expression for the trace of the energy-momentum tensor (\ref{eomq}) is only up to the usage of the equations of motion, and we have discussed the consequence of the equations of motion in quantum field theories in section \ref{contact} in relation to the renormalization group equation.

As an example, let us consider massless many flavor QCD \cite{Caswell:1974gg} (a.k.a Banks-Zaks model \cite{Banks:1981nn}) of $SU(N_c)$ gauge group with $N_f$  pairs of Dirac fermions in fundamental representation.
The two-loop beta functions are given by
\begin{align}
\beta(g) &= -\frac{g^3}{48\pi^2}(11N_c-2N_f) - \frac{g^5}{(16\pi^2)^2}\left(\frac{34}{3}N_c^2 - \frac{1}{2}N_f\left(\frac{16}{3}+\frac{20}{3}N_c\right)\right) \cr
v^a & = 0 \ .
\end{align}
The absence of the vector beta functions in this theory is essentially because there is no parity-even non-conserved gauge invariant vector operator with dimension $3$. 
Note that when $J^{\mu}_a$ is conserved, it does not give any contribution in (\ref{eomq}). We also do not have any good candidate for $\Box O$ term  due to the lack of classical dimension $2$ operators.

We recall that the scale invariance demands that the trace of the energy-momentum tensor is given by the divergence of a virial current: $T^{\mu}_{\ \mu} = -\frac{\beta(g)}{2g^3}\mathrm{Tr}F^{\mu\nu}F_{\mu\nu} = \partial^\mu J_\mu$,
but there is no candidate for the virial current $J_\mu$ in perturbation theory with the same reasons for the absence of $v^a \partial_\mu J_a^\mu$. As a consequence, the requirement of scale invariance reduces to $\beta(g) =0$ 
 and it automatically implies conformal invariance. Indeed  for $N_f \sim \frac{11 N_c}{2}$, we can find a perturbative 
conformal fixed point \cite{Banks:1981nn}.

It is not our main scope to discuss the details of renormalization and compute beta functions (in particular at higher loops), but let us present how the trace of the energy-momentum tensor can appear and why it is related to the beta functions in dimensional regularization at one-loop level. As can be inferred from section \ref{frees} with the $U(1)$ Maxwell field theory example, in $d=4-\epsilon$ dimension, the trace of the energy-momentum tensor in massless QCD is given by
\begin{align}
T^{\mu}_{\ \mu} = \frac{\epsilon}{4g^2_0}\mathrm{Tr} (F^{(0) }_{\rho\sigma})^2 \  \label{baret}
\end{align}
up to terms that vanish with the equations of motion.
If we naively take $\epsilon \to 0$, this vanishes, but we have to renormalized the bare field strength operator $\mathrm{Tr}(F^{(0)}_{\mu\nu})^2$ and bare gauge coupling constant $g_0$. Indeed, both of them contain $\epsilon^{-1}$ pole in dimensional regularization, and it can result in  the cancellation in (\ref{baret}). Since the renormalization necessary here is the same as that for the bare Lagrangian density, we conclude that in $\epsilon \to 0$ limit
\begin{align}
T^{\mu}_{\ \mu} = -\frac{\beta(g)}{2g^3} \mathrm{Tr} [(F_{\rho\sigma})^2]  \ , \label{ta}
\end{align}
where $\mathrm{Tr} [(F_{\rho\sigma})^2] $ is the renormalized finite operator.
Note that the $\epsilon^{-1}$ poles are related to the beta function in the standard dimensional regularization.
Although the heuristic argument here is essentially correct for massless QCD, we have to be more careful about the renormalization of the composite operator in the derivation of the right hand side of (\ref{ta}), which is related to the appearance of $v^a (\partial_\mu J^\mu_a)$ in more complicated examples. See section \ref{red} for further details.

At one-loop level, we do not need to be careful about the composite operator renormalization and the simple application of the  background field method (see e.g. \cite{polyakov}) by decomposing $A^{(0)}_\mu = \bar{A}_\mu + \delta A_\mu$ and integrating out the fluctuation $\delta A_\mu$ from the one-loop determinant gives the $\epsilon^{-1}$ pole in
\begin{align}
\langle \mathrm{Tr}(F^{0}_{\rho\sigma}))^2 \rangle = g^2_0 \mathrm{Tr}(\bar{F}_{\rho\sigma})^2 \frac{b_0}{(4\pi)^2} \epsilon^{-1} + \cdots
\end{align}
so substituting it in (\ref{baret}) gives the one-loop formula for (\ref{ta}) with $\beta(g) = -\frac{b_0 g^3_0}{(4\pi)^2}$ with $b_0 = \frac{1}{3}(11N_c-2N_f)$. As in chiral anomaly, one-loop background field  computation gives the bare trace anomaly formula.\footnote{Alternatively, in the one-loop approximation of the background field method, one may be able to compute the anomalous Jacobian in the path integral measure under the Weyl transformation to obtain the same result as reviewed in \cite{Fujikawa:2004cx}.} We refer to \cite{Hortacsu:1972bw}\cite{Collins:1976yq}\cite{Nielsen:1977sy} for further details on the operator renormalization needed beyond one-loop. See also \cite{Hathrell:1981gz} for the detailed structure of the renormalized energy-momentum tensor and trace anomaly at higher loops in QED.

The above discussion is based on the perturbative power-counting renormalization, and we do not know whether in non-perturbative regime, there can be other (possibly emergent) operators that  appear in the trace of the energy-momentum tensor in massless QCD. This is a very difficult problem for many flavor massless QCD, and we do not have any good theoretical tool to investigate it while such a possibility is often neglected. Presumably lattice computer simulations will shed some lights on it. See \cite{review} for the current status of lattice simulations of the conformal windows of massless QCD. As far as we know, what they have computed so far could not distinguish scale invariance and conformal invariance.  Eventually, we hope to compute the three-point functions to see whether the conformal invariance is realized (see also \cite{DelDebbio:2013qta} for the distinction between scale invariance and conformal invariance on the lattice). We also would like to refer to our collaboration \cite{Iwasaki:2012kv}\cite{Ishikawa:2013wf}\cite{Ishikawa:2013tua} with this respect. 

We emphasize that it is not absurd to imagine such a possibility. For instance, when a chiral symmetry is broken, the Nambu-Goldstone boson appears and it does have a non-zero (but a kind of trivial) virial current as mentioned in section \ref{frees}, which is indeed emergent. Similarly, in the magnetic free phase of Seiberg-duality \cite{Seiberg:1994pq}, we have emergent conformal dimension two operator (due to the emergence of magnetic infrared free fields) that may appear in the trace of the energy momentum tensor as an improvement term. 

We now consider more non-trivial situations in which the symmetry does not forbid the non-trivial existence of the perturbative virial current.
The most general power-counting renormalizable classically scale invariant field theories in $d=4$ dimension have interactions with gauge couplings, Yukawa couplings $y^{abi}$, and $\phi^4$ scalar  self-interactions $\lambda^{ijkl}$. Each interactions may have non-trivial beta functions, so the trace of the energy-momentum tensor is schematically given by
\begin{align}
T^\mu_{\ \mu} = -\frac{\beta(g)}{2g^3} \mathrm{Tr} F^{\mu\nu} F_{\mu\nu} + \left(\beta_{y^{abi}} (\psi_a\psi_b)\phi_i + c.c.\right) +\beta_{\lambda^{ijkl}} \phi_i\phi_j\phi_k\phi_l + v^a \partial_\mu J_a^\mu \ . \label{ssssc}
\end{align}
We have assumed that $\theta$ angles are not renormalized.\footnote{More precisely, one may introduce the beta function for the $\theta$ angle as long as the contribution to the trace of the energy-momentum tensor is cancelled by the 
anomalous conservations of $J^\mu$ (so that $\mathcal{B}$ function is zero). This corresponds to introducing field rotations on chiral fermions as a  part of our renormalization scheme, whose anomaly is cancelled by the flow of the $\theta$ angle. Physically this is a redundant flow.} We also assume that we fine-tune mass terms  for $\phi$ and the cosmological constant to make them vanish during the renormalization. As a further technical assumption, we assume that the energy-momentum tensor stays in the improved form  (i.e. absence of $\Box O$ term) during the renormalization \cite{Callan:1970ze}\cite{Collins:1976vm}.

In these theories, we have several candidates for the non-trivial virial current
\begin{align}
J_\mu = q^{ij}(\phi_i \partial_\mu \phi_j) +  p^{ab}(\bar{\psi}_a \gamma_\mu\psi_b) \  \label{currentans}
\end{align}
that corresponds to $O(N_b)$ rotations for scalars ($q^{ij}$ is antisymmetric), and $U(n_f)$ rotations for fermions ($p^{ab}$ is anti-Hermitian). Depending on the details of the interactions, some of them are conserved and do not contribute to the virial current. The naive application of the equations of motion schematically gives
\begin{align}
\partial^\mu J_\mu = q^{im}\lambda^{mjkl}\phi_i\phi_j\phi_k\phi_l +  (q^{im}y^{abm} + p^{ac}y^{cbi}) \phi_i \psi_a \psi_b + c.c 
\end{align}
so it may be possible to rewrite some of the terms in (\ref{ssssc}) as a divergence of the virial current.
We note  that the number of possible candidate currents for the non-trivial virial current is same as that of the redundant perturbations (see section \ref{reddd} for more details about redundant perturbations).

A priori, if we look for a conformal invariant fixed point with $T^{\mu}_{\ \mu} = 0$, we have to solve the equations $\beta^I = 0$ (if $v^a = 0$) whose number is the same as that of the coupling constants $g^I$, and we expect that we typically find a fixed point. If we relax the condition so that $T^{\mu}_{\ \mu} = \partial^\mu J_\mu$, we naively expect more solutions because we have more free parameters available and it seems much easier to find a scale invariant but non-conformal fixed point. Does this expectation work in reality? Apparently, this seems in contradiction with what we empirically know about the difficulty to construct scale invariant but non-conformal field theories.
Actually it turns out that the naive expectation does not work for the physical reason we will argue in the following sections.

It was known that up to  two-loops within the minimal subtraction scheme with assuming that the wavefunction renormalization matrix is symmetric, $v^a =0$  and there is no non-trivial solutions of $J_\mu$ that would give scale invariant but non-conformal invariant fixed point (see also \cite{Polchinski:1987dy}\cite{Dorigoni:2009ra}\cite{Fortin:2011ks} for an attempt in $d=4-\epsilon$ dimensions). The significantly more complicated three-loop (four-loop for gauge interaction) computation was done in \cite{Fortin:2012ic} for diagrams that are relevant for our discussions, and they found that within  the minimal subtraction scheme, there exists a non-trivial solution to the equation
\begin{align}
\beta^I O_I = \partial^\mu K_\mu \ , \label{solq}
\end{align}
where $K_\mu$ has the same ansatz as (\ref{currentans}).

By construction, the eigenvalue of $\beta^I$ flow is pure imaginary when it is given by the divergence of a current because the current generates $O(N_b)$ or $U(n_f)$ rotations in perturbation theory. Thus if we define the renormalization group flow by $\frac{dg^I(\mu)}{d\log\mu} = \beta^I(g(\mu))$, the non-trivial solution of (\ref{solq}) gives the cyclic renormalization group flow. It was quite surprising, and it was interpreted that  it gives the first non-trivially interacting counterexample of scale but non-conformal field theories in $d=4$ dimension.

However, in order to understand the conformal invariance, we had to compute the  additional terms in  the trace of the energy-momentum tensor $v^a \partial_\mu J_a^\mu$ independently to argue if the {\it total} trace of the energy-momentum tensor vanishes.  We anticipated that this must cancel against  the beta functions because it looks inconsistent with the general argument from the strong version of the $a$-theorem as we will discuss in section \ref{section8}. Soon after,  the additional terms $v^a \partial_\mu J_{a}^\mu$ have been computed \cite{Fortin:2012hn}, and they exactly cancel against the  $\partial^\mu K_\mu$ term computed within the same regularization scheme (see appendix of \cite{Nakayama:2012nd} for the first observation and the physical explanation).

We will revisit how and why the naive expectation that it is much easier to solve the scale invariant but non-conformal condition than to solve the conformal invariant condition is not true in section \ref{section8}. As for the counting goes, we realize that whenever there is a candidate for the virial current, there is a corresponding symmetry in the coupling constant space, and the beta functions are no-longer independent. As a consequence, the number of free parameters does not increase as naively expected in the above considerations.

So far there is no known scale invariant but not conformal invariant unitary quantum field theory in $d=4$ dimension. Although there is no non-perturbative proof, the enhancement of conformal invariance from scale invariance is presumably true under some assumptions such as 
\begin{itemize}
\item unitarity
\item Poincar\'e invariance (causality)
\item discrete spectrum in scaling dimension
\item existence of scale current
\item unbroken scale invariance
\end{itemize}
We will see in section \ref{proof2}, these assumptions are sufficient to prove the conformal invariance in $d=2$ dimension. We also note that thanks to the fourth assumption, we can exclude the counterexample (free Maxwell theory in $d\neq 4$) in section \ref{frees}. As far as we know, there is no known counterexample of scale invariance without conformal invariance in other dimensions than $d=2,4$, either, with the above assumptions.

\subsection{More examples}\label{more}
We have more examples of possible scale invariance without conformal invariance discussed in various literatures. We list some of them in this subsection.

\begin{itemize}

\item Non-linear sigma model for a quasi-Ricci flat manifold by Hull and Townsend \cite{Hull:1985rc} 

Probably this is the first physical example in which the distinction between scale invariance and conformal invariance was emphasized (see also \cite{Tseytlin:1986tt}\cite{Shore:1986hk}\cite{Callan:1986jb}). 
They observed that in order to achieve the scale invariance of the non-linear sigma models in $d= 2$, whose action is $\mathcal{S}= \int d^2x \mathcal{G}_{IJ} (X) \partial_\mu X^I \partial^\mu X^J$ with $\mathcal{G}_{IJ}(X)$ being the metric for the target space $\mathcal{M}$ and $X^I$ being the map from two-dimensional space-time to $\mathcal{M}$,
we require that the target space metric must be quasi-Ricci flat \cite{Friedan:1980jm} in one-loop approximation:
\begin{align}
R_{IJ}(\mathcal{G}(X)) = D_I V_J(X) + D_I V_J(X) \ ,
\end{align}
where $R_{IJ}(\mathcal{G}(X))$ is the target space Ricci-tensor and $V_I(X)$ is a certain (non-zero) vector field in the target space.
On the other hand, conformal invariance demands that the target space metric is Ricci flat (up to possible dilaton improvement terms).
\begin{align}
R_{IJ}(\mathcal{G}(X)) = D_I D_J \Phi(X) \ . \label{riccid}
\end{align}
Here $\Phi(X)$ is a certain scalar in the target space, which can be removed by improvement.
Note that the pull-back of the vector filed $V_I$ (i.e. $J_\mu = \partial_\mu X^I V_I(X)$) will be identified with the virial current. 
The condition for scale invariance is weaker than conformal invariance.

As we will see in section \ref{proof2}, however, under certain assumptions, scale invariance must imply conformal invariance in two-dimensional non-linear sigma models. Indeed, it is a mathematical fact that any quasi-Ricci flat manifold $\mathcal{M}$ must be Ricci flat when $\mathcal{M}$ is compact, and therefore there is no scale invariant but non-conformal field theories realized by a compact non-linear sigma model. We are delighted to know that the field theory theorem is in perfect agreement with the mathematical theorem on Riemannian geometry \cite{Polchinski:1987dy}. 

The two-dimensional black hole (Euclidean cigar geometry) is an example of conformal field theory whose target space is quasi-Ricci flat \cite{Witten:1991yr}: 
\begin{align}
ds^2 = k(dr^2 + \tanh^2r d\theta^2) \ .
\end{align}
The target space is non-compact, and it solves (\ref{riccid}). It has an exact conformal field theory description by $SL(2,\mathbb{R})/U(1)$ coset model at level $k$. This type of quasi-Ricci flat space-time and its higher dimensional generalization was studied in \cite{Bonneau:1986if} (but they are all conformal invariant).

A scale invariant but non-conformal quasi-Ricci flat space-time may be obtained in the linearized order around the Minkowski space-time as a vector gravitational wave. Let $\mathcal{G}_{IJ} = \eta_{IJ} + h_{IJ} e^{ikX}$ with the small fluctuation $h_{IJ}$. The scale invariance requires (see e.g. section 3.6 of the textbook \cite{Polchinski:1998rq} with a slight generalization mentioned in exercise 15.12 \cite{Polchinski:1998rr})
\begin{align}
-k^2 h_{IJ} + k_I k^L h_{JL} + k_J k^L h_{LI} -k_Ik_J (h^L_{\ L}+\phi) = ik_I V_J + ik_J V_I
\end{align}
 with a constant vector $V_I$ and a constant scalar $\phi$. By introducing a particular little group and the vector $n_I$ such that $n^2 = 0$, $n_I k^I = 1$, we can assume $n^I h_{IJ} = 0$. The scale invariant condition becomes $k^2 = 0$ and 
\begin{align}
k^L h_{IL} = -ik_I(V_L n^L) + i V_I \ .
\end{align}
Clearly, when $V_L=0$, we have transverse traceless tensor mode as well as dilaton scalar mode for a conformal invariant solution. However, we have additional $d-2$ vector polarization for a scale invariant but non-conformal solution specified by $V_I$ (up to gauge transformation). This is only possible because we violate the unitarity and the discreteness of the spectrum. The example here clearly illustrates that in string theory scale invariance is not sufficient but we have to demand the full conformal invariance for the consistency.

\item
Wilson-Fisher fixed point \cite{Wilson:1971dc}: Is 3d Ising model conformal invariant?

It is an extremely interesting and important problem to show if the critical phenomena of 3d Ising model shows conformal invariance. In $d=2$ dimension, it is long known that the critical phenomenon is described by a free Majorana fermion, which is conformal invariant. The direct proof of conformal invariance from the statistical model is, however, mathematically very hard \cite{Smirnov:2007pm}. 

In $d=3$ dimension, the success of the bootstrap approach \cite{Rychkov:2011et}\cite{ElShowk:2012ht} suggests that it must show conformal invariance. How much do we know about it? If we assume that the critical phenomenon of the 3d Ising model has the same universality class as the Landau-Ginzburg model with $\lambda\Phi^4$ potential in $4-\epsilon$ dimension analytically continued to $\epsilon = 1$, the trace of the energy-momentum tensor is given by
\begin{align}
T^{\mu}_{\ \mu} = [-\epsilon \lambda + \beta(\lambda,\epsilon)]\Phi^4
\end{align}
within perturbation theory (after fine-tuning $m^2 \Phi^2$ term). If we employ minimal subtraction scheme, $\beta(\lambda,\epsilon)$ does not depend on $\epsilon$ and we can recycle the four-dimensional computation in the minimal subtraction scheme (see e.g. \cite{ZinnJustin:2002ru} for explanations). As in the Banks-Zaks theory, the significant feature is there is no candidate for the virial current in perturbation theory for one-component Landau-Ginzburg model. Therefore, perturbative fixed point (Wilson-Fisher fixed point) is necessarily conformal invariant.\footnote{We define conformal invariance in $d=4-\epsilon$ dimension as vanishing of the trace of the energy-momentum tensor.} 

Unfortunately, $\epsilon$ expansion is asymptotic, and it is not obvious if there is any non-perturbative emergence of virial current (but we will have little to say about scale invariance and conformal invariance in $d=3$ dimension in this review article except in section \ref{section9}). Since we know that at $\epsilon =2$ the theory is conformal because it is described by a free fermion, we anticipate that there is no such a possibility in between, but it is extremely important to give more rigorous non-perturbative argument for it. 

\item
The fermionic version of the Landau-Ginzburg theory is known as the G\"ursoy model \cite{Gursoy} (in $d=4$ dimension).\footnote{I would like to thank K.~Akama for the reference.} The action is $S = \int d^4x\left( i\bar{\psi}\gamma^\mu \partial_\mu \psi + \lambda (\bar{\psi} \gamma_H \psi)^{4/3}\right)$. It is classically scale invariant as well as conformal invariant.  In two-dimension it is known as the massless Thirring model, where interaction term is $\lambda (\bar{\psi} \gamma_\mu \psi)^{2}$ and it is conformal invariant quantum mechanically. In these models, there is a potential candidate for the virial current $\bar{\psi}\gamma_\mu \psi$, but in $d=2$ dimension, it is a total derivative after bosonization, so it can be improved away in any way. We do not know much about the situations in the other dimensions.

\item
Scalar Riegert Theory $\mathcal{S} = \int d^4x (\Box \phi)^2$ \cite{Riegert:1984kt}\cite{Fradkin:1983tg}.

In Polchinski's classic paper \cite{Polchinski:1987dy} (see the textbook \cite{ZinnJustin:2002ru} for the same remark), it was mentioned that a fourth derivative scalar theory is scale invariant but not conformal invariant, but we can find the Weyl invariant extension of the fourth derivative operator, which is commonly known as scalar Riegert operator\footnote{The operator was also found by Paneitz at the same time \cite{PA}. As far as we are aware, the same operator appeared earlier in the work by Fradkin and Tseytlin \cite{Fradkin:1982xc}\cite{Fradkin:1981jc} in the context of conformal supergravity.} in $d=4$ dimension: 
\begin{align}
\Delta_4 =\Box^2 + 2G_{\mu\nu}D^\mu D^\nu + \frac{1}{3}\Delta^\mu R \Delta_\mu \  \label{Riegert}
\end{align}
The corresponding action
\begin{align}
\mathcal{S} = \int d^4x \sqrt{|g|} \phi \Delta_4 \phi \ 
\end{align}
is Weyl invariant (with zero Weyl weight for the scalar).
Indeed, we could construct the improved energy-momentum tensor such that the theory
is conformal invariant (although it is not unitary).

This is not so surprising, but probably a more surprising thing is that there had been a claim that no supersymmetric Riegert operator at the non-linear level in the old minimal supergravity \cite{Buchbinder:1988yu}\cite{Grosse:2007au}.\footnote{The author would like to thank S.~Kuzenko for the reference and related discussions. After the lecture, we had learned that the supersymmetric Riegert operator was explicitly constructed in \cite{Butter:2013ura}. The existence of super Weyl action, which differs from the supersymmetric Riegert operator in \cite{Butter:2013ura} by superspace total derivative, was suggested in \cite{Private}\cite{Baumann:2011nm} and the super Weyl invariance was checked at the linearized order upon superspace integration by part.} The supersymmetric Riegert operator also exists in the new minimal supergravity \cite{Manvelyan:1995hz}.

In the other dimensions, when $d$ is odd, we can construct higher derivative Weyl invariant free scalar actions of order  $\Box^{n}$ for any $n$ \cite{FG}\cite{GG}\cite{G}. In even dimensions, the Weyl invariant higher derivative free scalar actions exist when $n \le d/2$. As an example, there  does not exist Weyl invariant fourth (or higher) order derivative actions in $d=2$ dimension. We can directly check that the  energy-momentum  tensor  for the action $\mathcal{S}= \int d^dx (\Box \phi)^2$ is given by
\begin{align}
T_{\mu\nu} = (\partial_\mu \phi \partial_\nu \Box \phi + \partial_\nu \phi \partial_\mu \Box \phi) - \eta_{\mu\nu}(\partial^\rho \phi\partial_\rho \Box \phi + (\Box\phi)^2/2) \ 
\end{align}
and see that it cannot be improved to be traceless in $d=2$ dimension (in contrast to the case $d\ge 3$, where it is possible).

\item Topological twist.

There is an interesting class of (non-unitary) scale invariant but non-conformal field theories obtained by the so-called topological twist. The idea of the topological twist is to start with the Euclidean field theory with extra internal global symmetry and declare that the new Euclidean rotation is obtained 
by mixing the original Euclidean rotation and the internal global symmetry. The resulting theory is typically non-unitary (and may violate spin-statistics). 
Moreover, even if we started with a unitary conformal field theory, the resulting theory could be only scale invariant with respect to the new Euclidean symmetry.

Let us consider the simple example of $O(4)$ symmetric free bosons in $d=4$ dimension.
\begin{align}
\mathcal{S} = \int d^4x \partial^\mu \phi^i \partial_\mu \phi^j \ .
\end{align}
The assumed $O(4)$ rotation acts on $i$ index.
The original theory is conformal invariant. Now, let us declare that the new Euclidean rotation is given by the diagonal sum of the original Euclidean rotation and the $O(4)$ rotation. Then we regard $\phi^i$ as the Euclidean vector under the newly defined ``topologically twisted" Euclidean rotation. The resulting theory is 
\begin{align}
\mathcal{S} = \int d^4x \partial^\mu \phi^\nu \partial_\mu \phi_\nu \ ,
\end{align}
which is a particular example of Riva-Cardy theory mentioned at the end of section \ref{frees}, and it is only scale invariant but not conformal invariant. Note that the theory still admits the twisted ``conformal invariance" whose algebra is different from the conventional conformal algebra discussed in section \ref{section2} because after all, all we did was just the renaming of the symmetry algebra. This is not inconsistent with the theorem discussed in later sections (in particular in $d=2$ dimension in section \ref{proof2}) because after the twist, the theory is non-unitary with respect to the new topologically twisted Hamiltonian.\footnote{This example is not topological at all, so it is a misnomer. When the energy-momentum tensor is exact with respect to the topologically twisted scalar supersymmetric charge, the theory becomes topological.}

\item
Spontaneous broken case with the non-linear action $\mathcal{S}= \int d^4x \phi^4 f\left(\frac{\partial^\mu \phi \partial_\mu \phi}{\phi^4} \right)$ \cite{Jackiw} \cite{Jackiw:2011vz}. 

This scale invariant action is not conformal invariant classically unless $f(x) = c_0 + c_1 x$, in which case, we have just $\phi^4$ self interaction with the conventional kinetic term. The scale invariant vacua at $\phi = 0$ is singular with higher terms. When $\phi \ne 0$, scale invariance is spontaneously broken.

Of course, the spontaneous breaking of scale invariance does not necessarily exclude conformal invariance (which is again spontaneously broken). One example is 
\begin{align}
 \mathcal{S} = \int d^4x \phi^{4} f\left(\frac{(\Box \phi)}{\phi^3}\right) \ .
\end{align}
Here the theory can be made manifestly Weyl invariant in the curved background by replacing $\Box$ with $\Box - \frac{1}{6}R$.

\item 
Liouville theory coupled with matter

The analogue of the above example in $d=2$ dimension is the Liouville action. 
\begin{align}
 \mathcal{S} = \frac{1}{4\pi}\int d^2 x\sqrt{|g|} \left(\partial^\mu \varphi \partial_\mu \varphi + R(b+b^{-1}) \varphi + \lambda e^{-2b \varphi} \right) \ .
\end{align}
It is Weyl invariant and importantly, the conformal invariance is not spontaneously broken, which is a special feature of $d=2$. The quantization of the Liouville theory can be performed without breaking the conformal invariance (see e.g. \cite{Nakayama:2004vk} and references therein). 

Once we couple the Liouville theory to a non-linear sigma model with the specific interaction
\begin{align}
 \mathcal{S} = \frac{1}{4\pi}\int d^2x \sqrt{|g|} \left(\mathcal{G}_{IJ}(X) \partial_\mu X^I \partial^\mu X^J + h(X) \partial_\mu \varphi \partial^\mu \varphi  + R(b+b^{-1}) \varphi + \lambda e^{-2b\varphi} \right) \ ,
\end{align} 
the theory is scale invariant but not conformal invariant classically \cite{Iorio:1996ad}. In  \cite{Ho:2008nr}, we have shown that by considering the light-like wave form for $h(X)$, we may be able to preserve the scale invariance without conformal invariance after quantization. The model breaks the assumption of discreteness of spectrum as well as unitarity (in the light-like wave case) to avoid Zamolodchikov-Polchinski theorem that claims scale invariance implies conformal invariance in $d=2$ dimension.

\item The above mentioned Liouville theory and Riegert theory were studied in the context of generating classical effective action (so-called Wess-Zumino action) for the Weyl anomaly. If we restrict ourselves to the A-type Weyl anomaly (Euler density term), the Wess-Zumino action is invariant under the constant Weyl transformation due to the space-time integration, but it is not invariant under the non-constant Weyl transformation.\footnote{The author would like to thank I.~Shapiro  for the comment and discussion.} This Wess-Zumino action plays a significant role in the next section. To avoid a confusion, in flat space-time, one can always make the Riegert and Liouville action conformal invariant. This is because the Weyl non-invariance is proportional to the curvature.

\item 

As a simple generalization of the free Maxwell theory example discussed in section \ref{frees}, in $d$-dimensional space-time, among various Abelian free form fields, only zero-form field (scalar), $d/2$-form field and $d-1$ form field (dual to scalar) are conformal invariant (see e.g. \cite{Pons:2009nb}). In the first quantized approach, this was discussed in \cite{Siegel:1988gd}.
Note that except for the conformal case, the scale current (rather than charge) in these examples is not well-defined.

\item 

Within the Lagrangian formulation in $d=4$ dimension, it was mentioned in  \cite{Mack:1969rr}\cite{Gross:1970tb} that the non-gauge invariant interaction terms such as $\phi \partial_\mu \phi A_\mu$ is scale invariant but not conformal invariant. It violates unitarity.

\end{itemize}

\subsection{Theory without action}\label{without}
As mentioned in section \ref{local}, it is possible that theories are scale or conformal invariant without having local currents. In particular, this is the case when the action principle does not exist.

In $d=4$ dimension, conformally covariant massless wave-equations (known as  Bargmann-Wigner equation) must take the form \cite{Bracken:1982ny}
\begin{align}
\partial^\mu \sigma^{\alpha \dot{\beta}}_\mu \Psi_{\alpha \beta \gamma \cdots} = 0 \ , \label{wave}
\end{align}
where $\Psi_{\alpha \beta \gamma \cdots}$ has completely symmetric spinor indices and $\sigma_{\mu}^{\alpha \dot{\beta}}$ is the Pauli matrix. The CPT conjugate equation is obtained with dotted spinors. We note that except for helicity $0,1/2,1$, there is no simple Lagrangian formulation whose equations of motion are equivalent to (\ref{wave}) with the local energy-momentum tensor. After the second quantization, the theory is conformal invariant in the sense that all the correlation functions transform in a conformally covariant manner as the helicity $h$ field $\Psi_{\alpha \beta \gamma \cdots}$  being conformal primaries with conformal dimension $h+1$. 

An interesting observation is that the massless rank-four spinor $\Psi_{\alpha\beta\gamma \delta}$, which has helicity two, can be regarded as the linearized Weyl tensor around the Minkowski vacuum. Indeed, the linearized Einstein equation makes  the other components of linearized curvature tensor vanish, and the Bianchi identity is nothing but the conformal wave equation (\ref{wave}). Therefore, the linearized Einstein gravity {\it is} on-shell conformal invariant. However, there is no conserved energy-momentum tensor, nor conformal current, so the Noether assumption is clearly violated \cite{Dorigoni:2009ra}.\footnote{The vacuum Einstein equation $R_{\mu\nu} = 0$ is not Weyl invariant, so the Minkowski vacuum after Weyl transformation would not solve the vacuum Einstein equation. The conformal transformation here only acts on the linearized variation from the Minkowski vacuum. We also note that $\Psi_{\alpha \beta \gamma \delta}$ has the conformal dimension $3$, which may be unexpected from the non-linear Einstein gravity.} Relatedly, the off-shell action does not have conformal invariance, and the scale current is not gauge invariant.

The discussion also applies to helicity 3/2 massless rank-three spinor $\Psi_{\alpha\beta\gamma}$ with the conformal wave-equation (\ref{wave}). The theory is equivalent to the on-shell Rarita-Schwinger theory of a massless spin 3/2 particle. Combining it with the above helicity 2 wave-equation, we can conclude that the linearized supergravity around the Minkowski vacuum is on-shell superconformal invariant. Again, we do not have a conserved supercurrent supermultiplet nor a superconformal current.\footnote{To some extent, this is just a peculiar coincidence in $d=4$ dimension. In higher dimensions, not all massless equations (like Maxwell theory or linearized gravity) are conformal invariant \cite{Siegel:1988gd}.}

Without assuming the existence of the action, or more precisely the existence of the energy-momentum tensor, it seems possible to construct unitary scale invariant field theories without conformal invariance by considering so-called generalized free field theories \cite{Dorigoni:2009ra}.
 For instance, we start with  the  two-point functions of the vector operator
\begin{align}
\langle A_{\mu}(x) A_\nu(0) \rangle = \frac{1}{x^{2\Delta}}\left(\alpha \eta_{\mu\nu} - 2\frac{x_\mu x_\nu}{x^2} \right)  \ \label{freeaaa}
\end{align}
and demand that the higher point  functions are defined by the Wick contraction. One could write down the  formal (possibly non-local) free action from the inverse of the two-point  function (\ref{freeaaa}). 

Unless we take a very particular value for $\alpha$ (e.g. $\alpha = 1$ or $\alpha = 1/\Delta$), the theory is not conformal invariant. Note that for a sufficiently large value of $\Delta$, the unitarity is preserved irrespective of the conformal invariance (see section \ref{conss}). Here we should note that the unitarity only refers to the positivity constraint on the two-point functions mentioned in section \ref{conss}. We may have other pathologies in relation to unitarity. Since the theory does not seem to possess the energy-momentum tensor, we do not  know how to couple it to gravity, and this peculiar property will enable us to  evade various assumptions in the proof of the enhancement from scale invariance to conformal field theories. Indeed, we will see that the existence of the energy-momentum tensor plays a crucial role in our arugment for the enhancement.

\subsection{Controversial examples}\label{contro}

In this subsection, we have tried to collect the list of all the known scale invariant but non-conformal field theories. Here, we will list some more controversial ones.
\begin{itemize}
\item In this review, we do not discuss a subtle aspect of global conformal invariance in Minkowski space-time. The problem is that the global conformal transformation (\ref{specialconfi}) can affect the causal structure by making the space-like separation into the time-like one and vice versa.
We are satisfied with the infinitesimal conformal symmetry, and do not discuss the global aspects (with possible breakdowns). See \cite{Hortacsu:1972bw}\cite{Schroer:1974ay}\cite{Schroer:1974de}\cite{Luscher:1974ez}\cite{Nikolov:2001iz} for reference and possible resolutions.\footnote{The author would like to thank Prof. Hortacsu for pointing out the reference.}

\item
In \cite{Nakayama:2012nd}, a non-unitary example of supersymmetric scale invariant but non conformal field theories was constructed. The structure of the renormalization group flow is Jordan block type, which we never expect in unitary conformal field theories. 

\item
An interesting but confusing example of possible scale invariant but not conformal field theories in $d=2$ dimension is the so-called time-like Liouville theory obtained by a certain ``analytic continuation" of the conventional Liouville theory. It was observed in \cite{Strominger:2003fn}\cite{Zamolodchikov:2005fy}\cite{McElgin:2007ak}, the two-point functions are not diagonal with respect to the operator dimension, suggesting a possible violation of conformal invariance. An alternative interpretation was presented in \cite{Harlow:2011ny}, but the situation is not conclusive.

\item 
In \cite{LeClair:2003hj}\cite{LeClair:2004ps}, it was pointed out that a certain hermitian deformation of the unitary minimal model shows the periodic structure in its S-matrix, suggesting that the  renormalization group flow is cyclic. A similar idea appeared in Zamolodchikov's work \cite{Zamolodchikov:1979ba}. The result seems to be inconsistent with the $c$-theorem we will discuss in section \ref{proof2}, and the author would be happy to know the resolution of the puzzle.\footnote{The current understanding of the author is as follows: in the perturbative regime, their renormalization group flow satisfies the gradient formula, and Zamolodchikov's $c$-theorem applies. The cyclic structure necessarily brings us to the non-perturbative regime, in which something wrong (e.g. confinement) could occur.} 

\item
As we will discuss in section \ref{section6}, the analogue of $c$-theorem in $d=4$ dimension is known as the $a$-theorem and it has a deep connection with the problem of the enhancement of conformal invariance from scale invariance.
From time to time, the violations of the weak version of the $a$-theorem were reported. In most cases, it turned out that either such hypothetical theories did not exist or computations of the central charge were erroneous in a subtle way. 
One example of the former is given by the series of AdS/CFT dual pairs with a certain Sasaki-Einstein manifold induced from  the cone $z_1^k + z_2^2 + z_3^2 + z_4^2 = 0$ \cite{Gauntlett:2006vf}\cite{Nakayama:2007sb}\cite{Shapere:2008un} whose dual field theory construction can be found in \cite{Cachazo:2001sg}.
The naive conclusion of the AdS/CFT correspondence is $a$ is increasing as $k$ decreases when $k\ge 4$, which is supposed to be a relevant flow. What is happening here is that such a Sasaki-Einstein manifold does not exist due to a geometrical obstruction \cite{Gauntlett:2006vf}. From the dual field theory perspective, the assumption of the existence of the non-trivial fixed point in $a$-maximization was incorrect.  Another example would be certain $\mathcal{N}=1$ gauge theories obtained from the strongly coupled cousins of $\mathcal{N}=2$ gauge theories induced by M5-branes wrapping around Riemann surfaces. Those theories do not possess the Lagrangian description, but some of the flow among them seem to violate the $a$-theorem. It was interpreted in \cite{Bah:2011je} that such theories should not exist, and it would be interesting to see what was wrong with them.

An example of the latter was reported in \cite{Shapere:2008un}, which was later rebutted in \cite{Gaiotto:2010jf}. The problem was due to a subtlety in taking the IR limit and the assumption that the IR fixed point is described by a single superconformal field theory (rather than many superconformal field theories weakly coupled with each other). It is important to note that after the physical proof of the $a$-theorem which we will review in section \ref{section8}, our gears are shifted so that we use the $a$-theorem to exclude these hypothetical possibilities by seeking the flaw in the arguments rather than claiming them as counterexamples (see e.g. \cite{Amariti:2012wc}\cite{Buican:2013ica} for such attempts in the recent literature). 

\end{itemize}

\

\

\begin{shadebox}
Birth of conformal symmetry and Weyl symmetry

\

The conformal invariance in relativistic systems was (as far as the author knows) first discussed in the context of the symmetry of the Maxwell theory (with massless matters) at the very beginning of the 20th century \cite{Cun}\cite{Bate}. We have learned that this discovery is essentially due to the fact that the dimensionality of our space-time is precisely $d=4$ dimension. A legend says that Poincar\'e should have known the special relativity before Einstein because he knew that the Maxwell equation is not invariant under the Galilei group, but is invariant under the Lorentz group. The author wonders what people had imagined for the discovery of the further symmetry there.\footnote{The Maxwell theory has yet another mysterious symmetry ``electric-magnetic duality", which has its own theoretical impact afterwards.}

Weyl, on the other hand, studied the space-time dilatation, and he introduced the concept of Weyl transformation \cite{Weyl}. His motivation was to explain the electromagnetism from geometry. He was the first one who introduced the concept of gauge invariance, and he believed that the Weyl transformation $g_{\mu\nu} \to e^{2\sigma} g_{\mu\nu}$ is related to the electro-magnetic gauge invariance $A_\mu \to A_\mu + \partial_\mu \sigma$. Actually, he further noticed that this idea crucially relies on the fact that our universe is $d=4$ dimension. Otherwise, the Maxwell theory is not invariant under the Weyl transformation. We refer to \cite{Kastrup:2008jn} for more details on the historical development.

By themselves, the Weyl invariance is given by the transformation of the metric, and the conformal invariance is given by the transformation of the field in flat space-time and they are different symmetries. These two are, however, related by the diffeomorphism invariance of the underlying quantum field theory coupled with the gravity. As far as the author knows, the clear understanding of this connection was presented by Zumino in \cite{Zumino:1970}.

Apart from the historical origin, The author cannot help but to think that the structure of the space-time is deeply related to the (non-existence) of the scale invariant quantum field theories without enhanced conformal symmetry. As a related observation, non-trivial existence of interacting quantum field theories in general crucially depends on the space-time dimensionality. The power-counting in relavitistic Lagrangian field theory (with unitarity) demands that all the interations become irrelevant in dimension greater than $d=6$ dimension in perturbation theory.
We refer to \cite{Seiberg:1997ax} for an interesting application of this idea to supersymmetric theories. 
Of course, these questions are ultimately related to the renormalization group flow through our attempts to classify all the quantum field theories. A possible constraint on the classification will be the main subject of the following sections.
\end{shadebox}

\newpage

\section{Proof in $d=2$ dimension} \label{proof2}

Starting from this section, we would like to discuss a possible proof of enhancement from scale invariance to conformal invariance. In this section, we begin with the well-established situation in $d=2$ dimension, and expand our surveys in higher dimensions in later sections, in which we have partial but promising results. Various attempts are reviewed in relation to the higher dimensional analogue of Zamolodchikov's $c$-theorem. 

\subsection{Zamolodchikov-Polchinski theorem}\label{cfunc}

In $d=2$ dimension, we can give a rigorous argument that scale invariance is enhanced to conformal invariance under the following assumptions \cite{Zamolodchikov:1986gt}\cite{Polchinski:1987dy} (see also \cite{Mack1})
\begin{itemize}
\item unitarity
\item Poincar\'e invariance (causality)
\item discrete spectrum in scaling dimension
\item existence of scale current
\item unbroken scale invariance
\end{itemize}
Most interesting classes of two-dimensional quantum field theories satisfy these assumptions, but we should remember that if we violate one of them, we may construct counterexamples (see various examples in section \ref{more}). One important class of exceptions is the string world-sheet theory, in which the assumption of unitarity and the discreteness of the spectrum are both violated. Thus in string perturbation theory, it is not enough to check the scale invariance, but we have to show the conformal invariance for its consistency.

Let us present the proof. In $d=2$ dimension, it is convenient to use the complex coordinate notation $z=\sigma + i\tau$. Accordingly, the energy-momentum tensor is denoted as $T = T_{zz}$ and $\Theta =T^{\mu}_{\ \mu}$.
The conservation of the energy-momentum tensor gives
\begin{align}
\bar{\partial}T + 4\partial \Theta = 0 \ , \label{consh}
\end{align}
and similarly for its ``complex conjugate" $\bar{T} = T_{\bar{z}\bar{z}}$. 
Following Zamolodchikov \cite{Zamolodchikov:1986gt}, we introduce
\begin{align}
F(|z|^2) &=  (2\pi)^2 z^4 \langle T(z,\bar{z}) {T}(0) \rangle  \cr
G(|z|^2) &=  (2\pi)^2z^3\bar{z} \langle \Theta(z,\bar{z}) {T}(0) \rangle  \cr
H(|z|^2) &=  (2\pi)^2 z^2\bar{z}^2 \langle \Theta(z,\bar{z}) \Theta(0) \rangle  \ ,
\end{align}
which only depend on $|z|$ (from Euclidean invariance) by the combination $\log|z|\Lambda$, where $\Lambda$ is the renormalization scale.\footnote{We use the convention that all the coupling constants are dimensionless.}
Let us  define the $c$-function
\begin{align}
C = 2\left(F-\frac{1}{2}G - \frac{3}{16}H \right) \  . \label{Zamoc}
\end{align}
The response to the renormalization group flow is fixed by the conservation (\ref{consh}) as
\begin{align}
\frac{dC}{d\log |z|^2} = -\frac{3}{4}  H  \le 0 
\end{align}
from the positivity of the two-point function for $\Theta$.
This is celebrated Zamolodchikov's $c$-theorem: the $c$-function decreases along the renormalization group flow, and it agrees with the central charge at the fixed point (since $H=G=0$ at the fixed point as we will see).

At the scale invariant fixed point, one can assume that  the energy-momentum  tensor shows 
canonical scaling behavior (see section \ref{techni} for a further discussion), so $T_{\mu\nu}$ has a canonical scaling dimension of $2$, and hence $C$ is a constant. Then 
\begin{align}
\langle \Theta(z,\bar{z}) \Theta(0) \rangle  = 0
\end{align}
which means from unitarity and causality (according to Reeh-Schlieder theorem \cite{RS}), $\Theta(z,\bar{z}) = 0$ as an operator identity.
 Since $\Theta$ is the trace of the energy-momentum tensor, the scale invariance implies conformal invariance in $d=2$ dimension.

For later purposes, let us expand $\Theta$ with respect to the operators in the theory $\Theta = \mathcal{B}^I O_I$, where $\mathcal{B}^I$ can be interpreted as the same $\mathcal{B}$ function introduced in section \ref{temt}. The $c$-theorem can be expressed as
\begin{align}
\frac{dc}{d\log\mu} &= \mathcal{B}^I \chi_{IJ} \mathcal{B}^J  \ge 0 \cr
\chi_{IJ} &=  \frac{3}{2} (2\pi)^2 |z|^4\langle O_I(z,\bar{z}) O_J (0) \rangle|_{|z|=\mu^{-1}}  \ . \label{zamocc}
\end{align}
At this point, we identify $C$ defined in (\ref{Zamoc}) with $c(g(\mu))$ as an interpolating function between the central charges $c$ at conformal fixed points. The manifestly positive definite metric $\chi_{IJ}$ is known as Zamolodchikov's metric. Since it is positive definite, the $c$-function stays constant along the renormalization group flow if and only if $\mathcal{B}^I$ vanishes and  the theory is conformal invariant. 

There is a physical meaning in $c$ as counting degrees of freedom. 
If we quantize the conformal field theory on a cylinder with the radial quantization, the scaling dimension of the operator in $\mathbb{R}^{1,1}$ is identified with the energy spectrum on the cylinder. The modular invariance of the partition function dictates that the asymptotic density of states with a given radial energy $E$ is 
\begin{align}
\rho(E) \sim \exp\left(4\pi\sqrt{\frac{cE}{6}}\right) \ .
\end{align}
This is known as Cardy formula \cite{Cardy:1986ie}, and it tells that the central charge dictates the effective degrees of freedom of the conformal field theory. 
It is therefore reasonable that the central charge decreases along the renormalization group flow from our intuition that the renormalization group flow gives a coarse graining and the effective reduction of the degrees of freedom.

We have one comment on Zamolodchikov's $c$-theorem. A priori, we know that the $c$-function (\ref{Zamoc}) at $|z|=\mu^{-1}$ is a function of the energy scale $\mu$, but it is not immediately obvious if it is a function of the running coupling constants alone (i.e. $c(\mu) = c(g^I(\mu))$ evaluated at the energy scale $\mu$, and does not depend on the energy scale $\mu$ explicitly. An intuitive reason why the dependence is only through the running coupling constants $g^I(\mu)$ is the renormalizability.
Since the renormalized two-point functions do not depend on the renormalization scale $\Lambda$, we obtain the Callan-Symanzik equation for the two-point functions, or more general correlation functions of the energy-momentum tensor (by assuming that there is no anomalous dimension for $T_{\mu\nu}$):
\begin{align}
\frac{d }{d \log\Lambda} \langle T_{\mu\nu} \cdots \rangle = \left(\frac{\partial }{\partial \log\Lambda} + \beta^I \frac{\partial }{\partial g^I}\right) \langle T_{\mu\nu} \cdots \rangle = 0 \ .
\end{align}
In particular it applies to the above $c$-function constructed out of energy-momentum tensor two-point functions.
On the other hand, since $c$ at $|z|=\mu^{-1}$ is a dimensionless quantity, we have the Euler identity
\begin{align}
\left(\frac{\partial}{\partial \log\mu} + \frac{\partial}{\partial \log \Lambda}\right) c = 0 \ . 
\end{align}
This explains the simple chain rule
\begin{align}
\frac{d}{d\log\mu} c = \beta^I \frac{\partial }{\partial g^I} c  \left(= \mathcal{B} ^I \frac{\partial }{\partial g^I} c\right)
\end{align}
with the running coupling constants $g^I(\mu)$.
Note that when $T_{\mu\nu}$ is a singlet under the ``flavor" rotation, there is no distinction between beta function and $\mathcal{B}$ function here. 
We also know that at the fixed point, $c(\mu)$ is a function of the running coupling constants and does not depend on the trajectory of the renormalization group flow since it is specified by the Weyl anomaly and therefore it is intrinsic to the conformal fixed point.

Indeed, the local renormalization group analysis (as we will review in section \ref{localr}) tells that the $c$-function is actually a function of the running coupling constants alone and does not depend on the trajectory of the renormalization group flow.
In particular, within the power-counting renormalization scheme, one can show the ``gradient formula" \cite{Osborn:1991gm}:
\begin{align}
8 \partial_I \tilde{c} = (\chi^g_{IJ} + w_{[IJ]}) \mathcal{B}^J + (\hat{\rho}_I  g)^J w_J \ , \label{gradc}
\end{align}
where $w_{[IJ]} = \partial_I w_J - \partial_J w_I$, 
and $\hat{\rho}_I$ is the vector beta function we introduced in section \ref{local} that gives an extra ``flavor" rotation. These renormalization group functions will be further explained in more detail later in section \ref{localr}.  
By multiplying it with $\mathcal{B}^I$, we obtain (we use $\hat{\rho}_I \mathcal{B}^I = 0$ which we will prove later in section \ref{localr})
\begin{align}
\frac{d \tilde{c}}{d\log\mu} = \mathcal{B}^I \partial_I \tilde{c} = \mathcal{B}^I \chi^{g}_{IJ} \mathcal{B}^J \ 
\end{align}
with the fact that $\tilde{c}$ only depends on $\mu$ through the running coupling constants $g^I(\mu)$.
In addition, we can use a certain freedom in local renormalization group flow in order to make $\chi^g_{IJ}$ coincide with the Zamolodchikov metric $\chi_{IJ}$ in (\ref{zamocc}). Therefore, Zamolodchikov's $c$-function coincides with the $\tilde{c}$-function that appeared in the local renormalization group analysis, and the $c$-function is really a function of the running coupling constants.

The flow equation (\ref{gradc}) is known as the ``gradient formula". It would have been a true gradient formula if there would be no $w_J$. See also 
\cite{Friedan:2009ik} for further details on the validity of the gradient formula in general quantum field theories in $d=2$ dimension.

\subsection{Canonical scaling of $T_{\mu\nu}$}\label{techni}

In Zamolodchikov's argument, we tacitly assumed that the energy-momentum tensor has a canonical scaling dimension.
This assertion can be proved in $d=2$ dimension with the assumption of the discreteness of  scaling dimensions of operators in the theory (in particular, there is no dimension zero operator other than the identity operator) \cite{Polchinski:1987dy}. The canonical scaling of the energy-momentum tensor is violated when $T_{\mu\nu}$ is not an eigenoperator under dilatation:
\begin{align}
i[D,T_{\mu\nu}] =  x^\rho \partial_\rho  T_{\mu\nu} +dT_{\mu\nu}+ y_a\partial^\sigma \partial^\rho\hat{Y}^a_{\mu\sigma\nu\rho} \ ,
\end{align}
where $\hat{Y}^b_{\mu\sigma\nu\rho}$ is the complete set of tensor operators (excluding the trivial contribution from the identity operator) that have the symmetry of Riemann tensor and the scaling properties
\begin{align}
i[D,\hat{Y}^a_{\mu\sigma\nu\rho}] =  x^\lambda \partial_\lambda  \hat{Y}^a_{\mu\sigma\nu\rho} + \hat{\gamma}^{a}_{\ b} \hat{Y}^b_{\mu\sigma\nu\rho} \ .
\end{align}
This is still consistent with the scale invariance because the algebra of $D$ and $P_{\mu}$ is not affected from this mixing thanks to the space integration.

Polchinski argued that  in $d=2$ dimension, one can always improve the energy-momentum  tensor so that it has a canonical scaling dimension as long as there is no dimension zero operators than the identity operator. 
He introduced the improved energy-momentum  tensor  by
\begin{align}
\Theta_{\mu\nu}' = T_{\mu\nu}  + y^a(d-2-\hat{\gamma})^{-1}_{ab} \partial^\sigma \partial^\rho \hat{Y}^b_{\mu\sigma\nu\rho} \ .
\end{align}
The discreteness of the scaling dimensions, and the absence of  dimension zero operators  allows us to  invert the matrix in $d=2$ dimension, so the energy-momentum  tensor with canonical scaling dimension always exists.

In the other dimensions, this argument is subtle because if the scaling dimension matrix $\hat{\gamma}$ has an eigenvalue $d-2$, %(I think Polchinski's paper has a typo)
then one cannot invert the matrix. 
Within power counting renormalization scheme in $d=4$ dimension, in many non-trivial examples \cite{Brown:1980qq}\cite{Hathrell:1981zb}, one can explicitly construct the energy-momentum tensor that is not renormalized and finite, and therefore it has the canonical scaling dimension. On the other hand, the new improved energy-momentum tensor \cite{Callan:1970ze} is defined so that its trace vanishes at the conformal fixed point and can be renormalized. Indeed, at the conformal fixed point, the Wess-Zumino consistency condition of the local renormalization group flow (see discussions in section \ref{localr}) constrains the mixing of operators in a non-trivial way, and it makes 
it possible to choose a basis in which energy-momentum tensor has a canonical scaling dimension (see \cite{Collins:1976vm}\cite{Yonekura:2012kb}), where the new improved energy-momentum tensor has the canonical scaling dimension. 
However, at the merely scale invariant fixed point without conformal invariance, there is no a priori guarantee that we can choose the renormalization scheme so that the energy-momentum tensor has the canonical scaling dimension.

\subsection{Alternative approach}

\subsubsection{Simple alternative derivation}\label{sad}
Without referring to the $c$-theorem, there is a more direct way to derive the enhancement from scale invariance to conformal invariance in $d=2$ dimension.\footnote{The argument here is close to the historically original one presented in \cite{Mack1}.} For this purpose, we study two-point functions of the energy-momentum tensor in momentum space. The point is that the conservation and the canonical scaling of the energy-momentum tensor gives the unique structure of the two-point functions so that the trace must vanish.

The assumption of the canonical scaling dimension in position space of the energy-momentum tensor leads to the requirement that the momentum space energy-momentum tensor two-point function must show (we use the complex momentum $k = k_\sigma + ik_\tau$ and $\bar{k} = k_\sigma - ik_\tau$)
\begin{align}
\langle T(k) T(p) \rangle &= c\frac{k^3}{\bar{k}} \delta(k+p) \cr
\langle T(k) \Theta(p) \rangle &= ek^2 \log|k|^2 \delta(k+p) \cr
\langle \Theta(k) \Theta(p) \rangle &= h|k|^2 \log|k|^2\delta(k+p) \cr
\langle T(k) \bar{T}(p) \rangle &= w|k|^2 \log |k|^2 \delta(k+p) \  , \label{assumpt}
\end{align}
where we have neglected the contact terms that are polynomial in $k$ and $\bar{k}$.  

The appearance of $\log|k|^2$ is not in contradiction with the assumed scale invariance because the scale transformation only gives the extra ultra local contact terms. These are related to the anomalous contribution in the local Callan-Symanzik equations.

The conservation of the energy-momentum tensor (again up to the contact terms) requires 
\begin{align}
e = 0 \ , \ \ h = 0 \ , \ \ w = 0 \  .
\end{align}
From the Reeh-Schlieder theorem by going back to the position space, we conclude that $\Theta(x) = 0$ as an operator identity. Thus, the scale invariance implies conformal invariance. If we had kept track of the contact terms, we could see $\langle \Theta(x) \Theta(0) \rangle$ contains the contact term proportional to $c$, which can be related to the Weyl anomaly from the second variation of the effective action with the metric.
%\footnote{The assumption of the canonical scaling dimension is not necessarily needed in the above discussion (irrespective of Polchinski's argument that it does not happen in $d=2$ as reviewed in section \eqref{techni}) because the ansatz of the two-point function \eqref{assumpt} is still consistent with the Jordan form of the scaling dimension matrix. 
%The only change needed is the relation between the two-point function and the trace anomaly coefficient, but it does not affect the argument.} 

We may attempt a similar derivation in higher dimensions \cite{Polchinski:1987dy}\cite{Dorigoni:2009ra}.  However, we can immediately realize that the number of  independent two-point functions of the energy-momentum tensor is larger than the constraint from conservation and unitarity even if we assumed the canonical scaling of the energy-momentum tensor, so we cannot derive the similar result in this way. We will explicitly see what happens in section \ref{scale2}.
In retrospect, we have a good reason for this: two-point functions of the energy-momentum tensor do not seem to be a good barometer to show the higher dimensional analogue of Zamolodchikov's $c$-theorem as we will see. 

Let us briefly discuss what would happen if we relaxed the condition of the canonical scaling of the energy-momentum tensor. For the simplest example, suppose the energy-momentum tensor renormalization has the Jordan block form (see section \ref{mixingsi} for more general situations)
\begin{align}
\frac{d}{d\log \mu}T_{\mu \nu} & = \eta(\partial_\mu \partial_\nu - \eta_{\mu\nu}\Box) O \cr
\frac{d}{d\log\mu} O &= 0 
\end{align}
where $O$ has the scaling dimension zero so that the scaling dimension matrix is not diagonalizable for non-zero $\eta$. We have relaxed the assumptions in Polchinski's argument (see section \ref{techni}) to allow dimension zero operator. This may happen in bosonic string theory with tachyon operators in the non-compact target space-time.

After imposing the conservation and the dilatation Ward-identity (with local anomalous sources), the momentum space two-point functions are given by
\begin{align}
\langle O(k) O(p) \rangle &= \frac{c_{OO}}{|k|^2} \delta(k+p) \cr
\langle O(k) T(p) \rangle & = \left(c_{OT} - 4\eta c_{OO} \log|k|^2 \right) \frac{k}{\bar{k}} \delta(k+p) \cr
\langle O(k) \Theta(p) \rangle & = \eta c_{OO} \log |k|^2 \delta(k+p) \cr
\langle T(k) T(p) \rangle &= \left(c- 8\eta c_{OT} \log|k|^2 + 16\eta^2 c_{OO} (\log|k|^2)^2 \right)\frac{k^3}{\bar{k}} \delta(k+p) \cr
\langle T(k) \Theta(p) \rangle &= (2\eta c_{OT} \log|k|^2 - 4\eta^2 c_{OO} (\log |k|^2)^2) k^2\delta(k+p) \cr
\langle \Theta(k) \Theta(p) \rangle &= \left(-\frac{1}{2}\eta c_{OT} \log|k|^2 +  \eta^2 c_{OO}  (\log|k|^2)^2 \right)|k|^2\delta(k+p) \label{logtp}
\end{align}
up to contact terms. We may improve the energy-momentum tensor so that $c_{OT}$ vanishes, but $\eta$ remains non-zero. 
 Irrespective of the improvement, we see that the conservation of the energy-momentum tensor alone does not lead to the conclusion $\Theta = 0$ as an operator identity.\footnote{However, we may suspect that the appearance of $(\log |k|^2)^2$ is inconsistent with the locality of the correlation functions. If this additional locality constraint is imposed, either $(c_{OO}, c_{OT})$ or $\eta$  must vanish at the scale invariant fixed point. In the former case, the unitarity is violated as in log CFT, and in the latter case, the scaling dimension matrix must be diagonalizable and reduced to the original argument. The author would like to thank A.~Bzowski and K.~Skenderis for the related discussions.} Note that Zamolodchikov's argument still holds, but the (effective) central charge can decrease with an arbitrary amount suggesting the violation of the unitary or the discreteness of the spectrum.\footnote{Physically, this is not unexpected because with a dimension zero operator at hand, one can make the effective central charge arbitrarily large by improvement (e.g. in Liouville theory), and what $\eta$ term does is the renormalization of the improvement term, so the effective central charge should change with an arbitrary amount during the renormalization group.}

We have one technical comment on the Weyl anomaly. If there were no mixing due to $\eta$, the logarithmic term in the third equation in \eqref{logtp} would suggest that the Weyl anomaly contains the ``tachyon coupling" $\int d^2x \sqrt{|g|} \Phi$, where $\Phi$ is the source for the dimension zero operator $O$ (see section \ref{mixingsi} and \cite{Osborn:1991gm}). However, the violation of the naive scaling symmetry in log here is cancelled by relating it to the other correlation functions from the non-diagonal scaling dimension matrix (i.e. wavefunction renormalization) rather than contributing to the Weyl anomaly  $\int d^2x \sqrt{|g|} \Phi$. The similar comments apply to the other terms in \eqref{logtp} so that the Weyl anomaly is actually local.

\subsubsection{Averaged c-theorem}\label{moron}

Zamolodchikov's argument can be presented in a slightly different way \cite{Friedan:2009ik}.
Define
\begin{align}
c_{(2)}^M &=  -\int d^2x G_{(\mu)} \langle \Theta(x) \Theta(0) \rangle \cr
\Theta &= \mathcal{B}^I O_I \cr
\chi_{IJ} &= -\frac{d}{d\log\mu} \int d^2 x G_{(\mu)}(x) \langle O_I(x) O_J(0) \rangle \label{averagedcc} 
\end{align}
where
\begin{align}
G_{(\mu)}(x) = 3\pi x^2 \theta(1-\mu |x|) \ ,
\end{align}
with $\theta(x)$ being the step function so that $G_{(\mu)}(x)$ has only support  $|x| \ge \mu^{-1}$.
The metric $\chi_{IJ}$ is positive definite from unitarity (no dangerous contact term will contribute because $\frac{dG_{(\mu)}}{d\log\mu}$ has a support only when $O^I$ are separated).
It can be easily shown that
\begin{align}
\frac{dc_{(2)}^M}{d\log\mu} = \mathcal{B}^I\chi_{IJ} \mathcal{B}^J \ge 0 \ .
\end{align}
We note that this $c^{M}_{(2)}$-function is equivalent to the one in section \ref{cfunc}.

One can now repeat the same analysis in $d \ge 2$. 
We define
\begin{align}
c_{(d)}^M &=  -\int d^dx G^{(d)}_{(\mu)} \langle \Theta(x) \Theta(0) \rangle \cr
\Theta &= T^{\mu}_{\ \mu} = \mathcal{B}^I O_I \cr
\chi_{IJ} &= -\frac{d}{d\log\mu} \int d^d x G^{(d)}_{(\mu)} \langle O_I(x) O_J(0) \rangle 
\end{align} \label{moronh}
where
\begin{align}
G^{(d)}_{(\mu)}(x) = 3\pi x^d \theta(1-\mu |x|) \ .
\end{align}
The metric is again positive definite from unitarity.
It can be easily shown that
\begin{align}
\frac{dc_{(d)}^M}{d\log\mu} = \mathcal{B}^I\chi_{IJ} \mathcal{B}^J \ge 0 \ . \label{moroncccc}
\end{align}
Can we declare the proof of $c$-theorem in any dimension?
Does it mean scale invariance implies conformal invariance in any dimension?

A related idea was explored in \cite{Cappelli:1991ke}. The integrated $c_{(d)}^M$ is known as the averaged $c$-function. In the later works \cite{Anselmi:1999xk}\cite{Anselmi:2001yp}\cite{Anselmi:2002fk}, it was argued that the integral of the two-point functions of the trace of the energy-momentum tensor (\ref{moronh}) is directly related to the difference of $\tilde{b}$ coefficient in the Weyl anomaly in $d=4$ dimension.\footnote{More precisely, we had to fine-tune local counterterms (see section \ref{localr}) to achieve this claim.}
 It was also argued that although $\tilde{b}$ itself is scheme dependent, the ambiguity cancels in the difference of the UV fixed point and IR fixed point, and the integral only depends on the trajectory of the renormalization group flow. Since it depends on the trajectory, the quantity has a very different nature than Zamolodchikov's $c$-function in $d=2$ dimension. We have no direct way to connect the averaged $c$-function to the local correlation function so we cannot use the Callan-Symanzik equation to trade the $\mu$ dependence with beta functions. 

What is special in $d=2$ dimension is the identity  \cite{Friedan:2009ik}
\begin{align} 
\partial^\mu [(2x^\nu x^{\rho}x^\sigma - 2x^2 x^\nu \eta^{\rho\sigma} -x^2 x^\sigma \eta^{\nu \rho} ) \langle T_{\mu\nu}(x) T_{\rho\sigma}(0) \rangle] = -3 x^2 \langle \Theta(x) \Theta(0) \rangle \ .
\end{align}
It enable us to integrate $c_{(2)}^M$ by part to rewrite the averaged quantity (\ref{averagedcc}) into the local form
\begin{align}
c_{(2)}^M(\mu) = 2\pi^2(2x^{\mu}x^{\nu} x^{\rho}x^\sigma - x^2 x^\mu x^\nu \eta^{\rho\sigma} - x^2 x^{\rho} x^\sigma \eta^{\mu\nu} -x^2 x^\mu x^{\sigma} \eta^{\nu\rho}) \langle T_{\mu\nu}(x) T_{\rho\sigma}(0) \rangle|_{\mu|x| = 1} \ ,
\end{align}
which is nothing but the one defined by Zamolodchikov. The application of the Callan-Symanzik equation leads to the claim that $\mu$ dependence is only through the running coupling constants.

 We will come back to this point later when we discuss the renormalization scale dependence in the proof of the higher dimensional analogue of Zamolodchikov's $c$-theorem and its application to scale invariance and conformal invariance. Here we only emphasize that the crucial distinction between $d=2$ and $d>2$ in relation to the argument of this section is that the so-defined averaged $c$-function is not an intrinsic quantity of the fixed point, but it is a quantity of the flow. In particular, the equation (\ref{moroncccc}) by itself is consistent with the cyclic renormalization group flow with $\mathcal{B}^I \neq 0$ because there is no reason why $c^M_{(d)}$ should take a constant value when the theory is scale invariant. However, it is remarkable to mention that within a few orders in perturbation theory when the theory is classically conformal invariant, (\ref{moroncccc}) gives the same renormalization scale dependence as that for the higher dimensional analogue of Zamolodchikov's $c$-function we will discuss in the next section, which only depends on the running coupling constants at the scale $\mu$. 
 
\newpage

\section{Conjecture in $d>2$}\label{section6}

\subsection{Scale invariance vs Conformal invariance}
Given a proof in $d=2$ dimension reviewed in section \ref{proof2}, and various examples studied in section \ref{examples}, we conjecture that any scale invariant quantum field theory (in $d >2)$ is conformal invariant under the following assumptions
\begin{itemize}
\item unitarity
\item Poincar\'e invariance (causality)
\item discrete spectrum in scaling dimension
\item existence of scale current
\item unbroken scale invariance
\end{itemize}

The necessity of these assumptions may be found in examples listed in section \ref{examples}. Our focus in the following is $d=4$ dimension, but we will add some remarks for the other dimensions in section \ref{section9}.

In terms of the property of the energy-momentum tensor, the claim is that under the above assumptions, whenever the trace of the energy-momentum tensor is a divergence of the virial current
\begin{align}
T^\mu_{\ \mu} = \partial^\mu J_\mu \ , 
\end{align}
the virial current can be removed by the improvement. Or equivalently, it is a derivative of a certain local scalar operator
\begin{align}
T^\mu_{\ \mu} = \partial^\mu \partial_\mu L \ .
\end{align}

\subsection{Cardy's conjecture (a.k.a ``$a$-theorem")}\label{Cardyc}
In $d=2$ dimension, the proof of the enhancement from scale invariance to conformal invariance was almost identical to that of Zamolodchikov's $c$-theorem. It may be natural to look for a hint for the possible enhancement from scale invariance to conformal invariance in higher dimensions by considering the natural generalization of Zamolodchikov's $c$-theorem in higher dimension.

Cardy conjectured \cite{Cardy:1988cwa} that in $d=4$ dimension, the higher dimensional analogue of Zamolodchikov's $c$-function  is given by
the Weyl anomaly coefficient $a$. As discussed in section \ref{Weyla}, the Weyl anomaly in $d=4$ dimension is given by
\begin{align}
\langle T^{\mu}_{\ \mu} \rangle =c (\mathrm{Weyl})^2 -a \mathrm{Euler} +  \tilde{b} \Box R + d \epsilon^{\mu\nu\rho\sigma} R_{\mu\nu}^{\ \ \alpha\beta}R_{\alpha\beta \rho \sigma} \ .
\end{align}
We know $\tilde{b}$ is not universal because it can be changed by a local counterterm, and $c$ does not show monotonicity along the renormalization group flow (see the examples below) so they cannot be the candidate. The remaining possibility is $a$. We recall that in $d=2$ dimension, the scalar curvature $R$ can be also regarded as the Euler density, and it shares the common feature with the $a$ anomaly term in $d=4$ dimension. In general even dimensions, the Euler density (A-type Weyl anomaly in the classification of \cite{Deser:1993yx}) seems to be a good candidate for the analogue of Zamolodchikov's $c$-function. Cardy formulated it as $a = -\frac{1}{2 \cdot 32\pi^2}\int_{\mathbb{S}^4} d^4x \sqrt{|g|} \langle T^{\mu}_{\ \mu} \rangle$ because we can see that Weyl tensor vanishes on $\mathbb{S}^4$ since it is conformally flat and $\Box R$ term is zero after integration (and we assumed $d = 0$). Note that the Euler characteristic of $\mathbb{S}^{2n}$ is given by $\chi(\mathbb{S}^{2n}) = \int_{\mathbb{S}^{2n}} d^{2n} x \sqrt{|g|} \frac{1}{(8\pi^2)\Gamma(n+1)}\epsilon_{\mu_1\cdots \mu_{2n}}\epsilon_{\nu_1\cdots \nu_{2n}}R^{\mu_1\mu_2\nu_1\nu_2}\cdots R^{\mu_{2n-1}\mu_{2n}\nu_{2n-1}\nu_{2n}} = 2$.

The conjecture can be stated in different versions.
\begin{itemize}
\item Weak version: $a_{\mathrm{IR}} \ge a_{\mathrm{UV}}$ between the flow of two conformal field theories, which is proved in \cite{Komargodski:2011vj}\cite{Komargodski:2011xv} recently.
\item Strong version: $\frac{da(g(\mu))}{d\log\mu}\ge 0$ along the renormalization group flow.
\item Gradient formula: $\mathcal{B}^I  = \chi^{IJ} \partial_J a$ (we will make more precise about the statement).\footnote{Even in $d=2$ dimension, the proof of the gradient formula is much more non-trivial than that of the $c$-theorem discussed in section \ref{cfunc}. Within the power-counting renormalization scheme it was discussed in \cite{Osborn:1991gm}. With relevant perturbations, we find more recent discussions in \cite{Friedan:2009ik}.}
\end{itemize}
The gradient formula suggests $\frac{da(g(\mu))}{d\log\mu} = \mathcal{B}^I \chi_{IJ} \mathcal{B}^J$, so it is 
also called the strongest version (with a tacit assumption $\chi_{IJ}$ is positive definite). The gradient formula for the renormalization group flow in $d=4$ dimension was first discussed in \cite{Wallace:1974dx}\cite{Wallace:1974dy}.

So far, we have introduced the $a$-function that can be a candidate for the higher dimensional analogue of Zamolodchikov's $c$-function from the Weyl anomaly. Since Weyl anomaly appears once we put our conformal field theories on a curved background, we have a natural question if we can read the Weyl anomaly coefficients $a$ and $c$ from correlation functions in the flat space-time. We know that in $d=2$ dimension, the Weyl anomaly $c$ is related to the two-point function of the energy-momentum tensor. The point is that the Weyl anomaly $c$ appears in the contact terms of two-point functions of the trace of the energy-momentum tensor in $d=2$ dimension, and the conservation condition relates it to the two-point functions of the energy-momentum tensor $T_{zz}$. Therefore, we can compute the Weyl anomaly $c$ by studying the two-point functions of $T_{zz}$.

In $d=4$ dimension, the situation is more complicated. We can show that the two-point function of the energy-momentum tensor for conformal field theories is completely fixed by the Weyl anomaly $c$. The three-point function of the energy-momentum tensor depends on three independent numbers, which are determined by $a$ and $c$ (and one additional parameter). The way to read $a$ from the three-point functions of the energy-momentum tensor has been developed in \cite{Osborn:1993cr}\cite{Erdmenger:1996yc} (see also \ref{EMtensor}).

What else do we know about $a$ and $c$? In \cite{Hofman:2008ar}, the general bound on the value of $a$ and $c$ has been derived. The bound on $c\ge 0$ is easily obtained from the positivity of the energy-momentum tensor two-point function. The bound on $a$ is more non-trivial: they studied the energy flux operator
\begin{align}
E (\theta) = \int dt r^2 n^i T_{\  i}^{t} (t, r\vec{n})|_{r \to \infty}
\end{align}
and assumed its positivity from the averaged null energy condition.
Then by imposing the condition of the positivity of the energy flux operator for  the state constructed out of the energy-momentum operator $T_{\mu\nu}$, they schematically required
\begin{align}
\langle T_{\mu\nu} | E(\theta) |T_{\mu\nu} \rangle \ge 0 \ ,
\end{align}
which is related to $a$ and $c$ from the three-point functions of $T_{\mu\nu}$. The resulting condition is 
\begin{align}
 \frac{31}{18}c \ge a \ge \frac{1}{3}c \ . 
\end{align}
Since $c$ is bounded from below $ c \ge 0$, $a$ is also bounded from below $a\ge \frac{1}{3}c \ge 0$. The bound of $a$ will be important in our later discussions. The bound $a\ge 0$ was also discussed in the paper \cite{Latorre:1997ea}
by assuming the ``quantum modified null energy-condition".

Let us come back to the comparison between $a$ and $c$ under renormalization group flow.
To motivate $a$ rather than any other linear combinations of the Weyl anomaly coefficients, we will now show examples of renormalization group flow in which $c$ increases.
We consider $SU(N_c)$ SQCD with $N_f$ fundamental flavors within conformal window $ \frac{3}{2} N_c \le N_f \le 3N_c$. Supersymmetry
allows the exact computation of $a$ and $c$ both in UV and IR (see section \ref{amax}).
\begin{align}
a_{\mathrm{UV}} &= \frac{1}{48}(9N_c^2 - 9 + 2N_f N_c) \cr
c_{\mathrm{UV}} &= \frac{1}{16}(3N_c^2 -3 + 2N_f N_c)
\end{align}
and
\begin{align}
a_{\mathrm{IR}} &= \frac{3}{16}\left(2N_c^2 - 1 -3\frac{N_c^4}{N_f^2}\right) \cr
c_{\mathrm{IR}} &= \frac{1}{16}\left(7N_c^2 -2 - 9\frac{N_c^4}{N_f^2}\right) \ .
\end{align}
One can see that near $N_f \sim 3N_c$, $c_{\mathrm{UV}}-c_{\mathrm{IR}} \le 0$ so $c$ cannot be monotonically decreasing
along the renormalization group flow \cite{Anselmi:1997am}. On the other hand, one can check that $a_{\mathrm{UV}}-a_{\mathrm{IR}} \ge 0$ always hold in the above SQCD example. By  taking $N_f \to 3N_c$ limit, we can further conclude that 
any other combinations of $a + kc$ are not monotonically decreasing for $k>0$.

%\subsection{Toward a proof in $d=4$ dimension}
%We give a review of some earlier attempts to prove Cardy's conjecture in $d=4$.  They will give some insights about the renormalization group flow even though the argument is not complete.

\subsection{Scale invariant energy-momentum tensor}\label{scale2}
Let us try to generalize the argument given in section \ref{sad} in higher dimensions. While we will not be able to derive the conformal invariance from the following argument alone, we learn some interesting structures of the dilatation anomaly in two-point functions of the energy-momentum tensor.

For simplicity, we assume that the energy-momentum tensor has a canonical scaling dimension $d$ under the assumed dilatation symmetry. This may not necessarily hold in general when the theory contains dimension $d-2$ scalar operators as mentioned in section \ref{techni}. When this happens the two-point function may contain $\log$ in the position space. 

In the momentum space, with the scale invariance, the two-point function of the energy-momentum tensor must take the form \cite{Dorigoni:2009ra}
\begin{align}
\langle T_{\mu\nu}(k) T_{\lambda \sigma}(p) \rangle &= \delta(k+p) f_d(k^2) \left[A_1 k^2 (\eta_{\mu\lambda} \eta_{\nu\sigma} + \eta_{\nu\lambda}\eta_{\mu\sigma}) + A_2 k^2 \eta_{\mu\nu} \eta_{\lambda\sigma} \right. \cr
&\left. +A_{3}(\eta_{\mu\lambda}k_\nu k_\sigma +(\text{3 perms})) + A_{4}(\eta_{\mu\nu} k_{\lambda}k_{\sigma} + \eta_{\lambda \sigma} k_\mu k_\nu) + A_5 k_{\mu}k_\nu k_{\lambda} k_{\sigma}/k^2 \right] \ ,
\end{align}
where the scale invariance demands $f_d \propto (k^2)^{d/2}$ when $d$ is odd, and $f_d \propto (k^2)^{d/2} \log k^2$ when $d$ is even (up to contact terms). 
The fact that $f_{d}$ with even $d$ contains $\log$ is due to the conformal (or more precisely dilatation) anomaly manifested in the momentum space. 

The conservation $\partial^\mu T_{\mu\nu} = 0$ demands
\begin{align}
A_1 + A_3 = A_2 + A_4 = 2A_3+ A_4 + A_5 = 0 \ .
\end{align}
On the other hand, the trace component of the two-point function is given by
\begin{align}
\langle T^{\mu}_{\ \mu} (k) T^{\mu}_{\ \mu}(p) \rangle = \delta(k+p) (d-1) (2A_1 + (d-1) A_2)f_{d}(k^2) \ ,
\end{align}
which does not vanish automatically. The unitarity and the Reeh-Schlieder theorem demand that the trace of the energy-momentum tensor vanishes (and hence conformal invariant) if and only if $2A_1 + (d-1) A_2 = 0$. There is an additional unitarity condition $A_1 \ge 0$ by studying the traceless component of the energy-momentum tensor two-point function.

In $d=4$ dimension, one can relate the two-point functions of the energy-momentum tensor to the Weyl anomaly coefficients. The traceless component of the two-point function is related to $c \mathrm{Weyl}^2$ term in the trace anomaly (see \ref{EMtensor}) while the trace component of the two-point function is related to $b R^2$ term in the trace anomaly. The above discussion concludes that unitary scale invariant field theories is conformal invariant if and only if $b R^2$ anomaly vanishes when the energy-momentum tensor has the canonical scaling dimension (see also \cite{Farnsworth:2013osa}\cite{Skenderis}\cite{Dymarsky:2013pqa} for more recent discussions).
%\footnote{If we had not assumed the canonical scaling dimensions, the $\log k^2$ in the two-point function could have been explained by the Jordan structure of the dilatation matrix in addition to the dilatation anomaly. In such a case, the relation between $b$ coefficient in trace anomaly and the two-point function of the trace component of the energy-momentum tensor would have been shifted. The claim that  unitary scale invariant field theories is conformal invariant if and only if $b R^2$ anomaly vanishes would be only true in a particular choice of the improvement in which the shift does not affect the two-point function. Other than this aspect, the discussion in this subsection still holds.}

\newpage

\section{Local renormalization group and perturbative proof in $d=4$ dimension}\label{localr}
\subsection{Local renormalization group}\label{localrg}
Local renormalization group analysis gives a very strong constraint on the renormalization group flow even in the flat space-time limit.
Indeed, the analysis gives a perturbative proof of the strong version of Cardy's conjecture  \cite{Jack:1990eb}\cite{Osborn:1991gm} as well as the gradient formula. 
Since the argument is based on the generic consistency conditions of the effective action, the positivity of the target space metric appearing in the effective action, for instance, was
not derived (because their argument works also in non-unitary field theories). Nevertheless, it was shown that perturbatively, $a$-theorem is true, and scale invariance implies conformal invariance from the subsequent result.

The idea to use the local renormalization group in this problem is to generalize the Wess-Zumino consistency condition for the Weyl anomaly mentioned in section \ref{Weyla} not only in the non-trivial metric background but with the space-time dependent coupling constants (a.k.a Schwinger's source theory). This is conceptually very natural because if we consider the non-trivial renormalization group flow, the Weyl transformation acts on coupling constants non-trivially, so the coupling constants must be treated in a space-time dependent way after the Weyl transformation even if we started with a constant background.  In addition to the space-time dependent coupling constants, for each  (conserved or non-conserved) current operators, we will introduce the background gauge field $a_{\mu}$. If the currents are non-conserved, we further transform the coupling constants under the background gauge transformation so that the theory is spuriously invariant. We also note that the space-time dependent source is natural in AdS/CFT correspondence as we will discuss in section \ref{section10}.

As a consequence of the space-time dependent sources, in order to properly reguralized and renormalize the theory, we have to introduce various additional counterterms that are not present in the flat space-time limit, and the consistency of the renormalization group flow will give more non-trivial constraints, whose consequence will be the main subject of our analysis. Unfortunately, the entire analysis is slightly complicated partly because there are many terms, which is not essential but technical, so we will focus on the points relevant for our discussions, and leave the other aspects to the original literature \cite{Osborn:1991gm} (see also \ref{localWZC}).

Let us first revisit the operator Weyl anomaly. In addition to the scalar beta functions corresponding to background coupling constants $g^I$, we have to introduce the term given by the beta function for the background currents $a_{\mu}^a$. Within power-counting renormalization scheme,  we have the field dependent part of the trace of the energy-momentum tensor as
\begin{align}
T^{\mu}_{\ \mu;\mathrm{field}} = \beta^I {O}_I + \rho^{a}_I(g) (D_\mu g)^I J^\mu_a + D_{\mu} ( v^a(g) J^{\mu}_a)  \  \label{opano}
\end{align}
up to the terms that vanish upon using the equations of motion.
The second term is particular to the space-time dependent coupling constants, and the third term is related to the divergence part of the vector beta function term discussed in section \ref{Intera}. 

As discussed in section \ref{local}, due to the operator identities, or in this case, due to the background gauge independence, the last term (\ref{opano}) can be removed 
\begin{align}
T^{\mu}_{\ \mu;\mathrm{field}} = \mathcal{B}^I {O}_I + \hat{\rho}^{a}_I(g) (D_\mu g)^I J^\mu_a \ \label{opannn}
\end{align}
up to the term that vanishes by equations of motion, where $\mathcal{B}^I = \beta^I -(vg)^I$ and $\hat{\rho}^{a}_I (D_\mu g)^I = \rho^{a}_I (D_\mu g)^I + (D_{\mu} v^a)$. This is nothing but the broken current conservation, and $\mathcal{B}^I$ are the same $\mathcal{B}$ functions introduced in section \ref{red}. 

Correspondingly, the local renormalization group generator is given by
\begin{align}
\delta_{\sigma(x)} = - \int d^4x \sqrt{|g|}  \sigma(x) \left( 2g^{\mu\nu} \frac{\delta}{\delta g^{\mu\nu}} - \mathcal{B}^I \frac{\delta}{\delta g^I} - \hat{\rho}^{a}_I D_{\mu} g^I \frac{\delta}{\delta a_{\mu}^a} \right) \ . \label{weylg}
\end{align}
If there were no anomaly, the Schwinger functional would satisfy 
\begin{align}
\delta_{\sigma(x)} W[g_{\mu\nu}, g^I, a_\mu]  =-\int d^4x \sqrt{|g|}  \sigma\left(T^{\mu}_{\ \mu} - \mathcal{B}^I {O}_I - \hat{\rho}^{a}_I(g) (D_\mu g)^I J^\mu_a \right)= 0 \ , 
\end{align}
and this would be nothing but the trace identity. Physically, it means that the change of the space-time dependent renormalization scale can be cancelled by the change of the scalar as well as vector sources with the amount given by the beta functions.
 In the following, we study consistency condition on the anomalous terms in the Weyl variation.

To go further,  we demand the Wess-Zumino consistency condition as in section \ref{Weyla}, 
\begin{align}
[\delta_{\sigma(x)}, \delta_{\tilde{\sigma}(x')}]  W[g_{\mu\nu}, g^I, a_\mu^a] = 0 \ ,
\end{align} 
but now the Schwinger functional $W[g_{\mu\nu}, g^I, a_\mu^a]$ depends not only on the background metric but also background space-time dependent coupling constants as well as background gauge fields.

In section \ref{Weyla}, we wrote down all the possible first order variation of $W$ from the background metric alone as a candidate for the Weyl anomaly. Similarly, we should consider all the possible  invariant terms (within power-counting renormalization scheme) from $g_{\mu\nu}$, $g^I$ and $a_{\mu}^a$, and study the consistency equations. We only focus on three terms that are relevant for our discussions:
\begin{align}
-\delta_{\sigma(x)} W[g_{\mu\nu}, g^I, a_\mu^a] = - \int d^4x \sqrt{|g|}  \left( a \sigma \mathrm{Euler} + \frac{1}{2}\sigma G^{\mu\nu} \chi^{g}_{IJ}  D_\mu g^I D_{\nu} g^J + \partial_\mu \sigma  G^{\mu\nu} (w_I D_\nu g^I) + \cdots \right) \ , \label{threet}
\end{align}
where we recall $G_{\mu\nu}$ is the Einstein tensor.
The right hand side is regarded as the Weyl anomaly on the curved background with space-time dependent coupling constants because (\ref{weylg}) gives the trace of the energy-momentum tensor from the left hand side of (\ref{threet}).
A particular class of the Wess-Zumino consistency condition demands
(we refer \cite{Osborn:1991gm}  and \ref{localWZC} for the full details)
\begin{align}
8 \partial_I a &= \chi_{IJ}^{g} \mathcal{B}^J - \partial_J w_I \mathcal{B}^J - \partial_I \mathcal{B}^J w_J + (\hat{\rho}_I g)^J w_J \cr
\mathcal{B}^I \hat{\rho}^a_I  &= 0 \ . \label{consistwww}
\end{align}
Here $(\hat{\rho}_Ig)^J = h_{ab}\hat{\rho}^a_I T^{bJ}_{\ \ \ K} g^K$ with some representation matrix $T^a$ of the ``flavor symmetry". The former equation in (\ref{consistwww}) comes from the term proportional to $G^{\mu\nu} D_\mu g^I \sigma \partial_\nu \tilde{\sigma}$, and the latter comes from the consistency of 
\begin{align}
0 &= \left[\int d^4x \sqrt{|g|}  \sigma(x) \left(\mathcal{B}^I \frac{\delta}{\delta g^I} + \hat{\rho}^{a}_I D_{\mu} g^I \frac{\delta}{\delta a_{\mu}^a}\right), \int d^4y \sqrt{|g|}  \tilde{\sigma}(x') \left(\mathcal{B}^I \frac{\delta}{\delta g^I} + \hat{\rho}^{a}_I D_{\mu} g^I \frac{\delta}{\delta a_{\mu}^a}\right)\right] \cr
 &= \int d^4x \sqrt{|g|} (\sigma \partial_\mu \tilde{\sigma} - \tilde{\sigma} \partial_\mu \sigma)\mathcal{B}^I\hat{\rho}_I^a \frac{\delta}{\delta a^a_{\mu}} \ .
\end{align}
It is instructive to see that these conditions are indeed satisfied in the conformal perturbation theory results in section  \ref{reddd} (while we did not specify the space-time dimension there).

Now we proceed to the physical interpretation of the Wess-Zumino consistency condition.
If we define $\tilde{a} = a + \frac{1}{8}w_I \mathcal{B}^I$, the first line of (\ref{consistwww}) gives the flow equation or ``gradient formula"
\begin{align}
8 \partial_I \tilde{a} &= (\chi_{IJ}^{g} + w_{[IJ]}) \mathcal{B}^J + (\hat{\rho}_I g)^J w_J \cr
& = (\chi_{IJ}^{g} + w_{[IJ]} + \hat{\rho}_{[I} Q_{J]}) \mathcal{B}^J 
\end{align}
where $w_{[IJ]} = \partial_I w_J - \partial_J w_I$, and in the second line, we have used the formula in \ref{wzconstfull} in Appendix. Explicit checks of the consistency conditions in perturbation theory can be found in \cite{Antipin:2013pya}\cite{Antipin:2013sga}\cite{Jack:2013sha}.

One important consequence of the ``gradient formula" is 
\begin{align}
\frac{d \tilde{a}}{d\log\mu} \equiv \mathcal{B}^I \partial_I \tilde{a} = \mathcal{B}^I \chi^{g}_{IJ} \mathcal{B}^J \ \label{ggggaaa}
\end{align}
from $\mathcal{B}^I\hat{\rho}_I^a=0$. This means that $\tilde{a}$-function would be decreasing monotonically along the renormalization group flow (defined by $\frac{dg^I}{d\log\mu} = \mathcal{B}^I$) {\it if} the metric $\chi^{g}_{IJ}$ is positive definite. Since we have not assume any physical requirement such as unitarity, the argument here cannot say the positivity of the metric, but in perturbation theory, we can check that this metric is positive definite in all known unitary renormalizable quantum field theories.\footnote{Indeed, we can show $\chi_{IJ}^g$ is always positive definite at the unitary conformal fixed point when $\mathcal{B}^I =0$ (with a suitable choice of counterterms). Thus, the deviation is small as long as $\mathcal{B}^I$ are small in perturbation theory.}
This gives the perturbative proof of strong $a$-theorem by identifying $\tilde{a}$ as the interpolating $a$-function. Note that $\tilde{a}$ coincides with the trace anomaly coefficients $a$ proposed by Cardy at the conformal fixed point.

The $\tilde{a}$-function of the local renormalization group flow is not unique.
 The flow equation itself is invariant under the dressing transformation
\begin{align}
\delta \chi^{g}_{IJ} &= \mathcal{L}_{\mathcal{B}} C_{IJ} = \mathcal{B}^K \partial_K C_{IJ} + C_{KJ}(\partial_I \mathcal{B}^K - (\hat{\rho}_Ig)^K ) + C_{IK}(\partial_J \mathcal{B}^K - (\hat{\rho}_Jg)^K) \cr
  \ \delta w_I &= -8\partial_I A + C_{IJ}\mathcal{B}^J \ , \ \ \delta \tilde{a} =  \mathcal{B}^I C_{IJ} \mathcal{B}^J \ , \label{ambigu}
\end{align}
where  $C_{IJ}$ and $A$ are curved space-time counterterms that can be chosen as an arbitrary tensor of coupling constants. Note that $(\hat{\rho}_Jg)^K \partial_K A =0$ due to the gauge invariance of $A$.

The reason why we have this ambiguity is that we can add coupling constant dependent local counterterms 
\begin{align}
\mathcal{S}_{\mathrm{ct}} = -\int d^4x\sqrt{|g|} \left( -\frac{1}{2} G^{\mu\nu} C_{IJ}(g^I) D_{\mu} g^I D_\nu g^J - A(g^I) \mathrm{Euler}\right) \ , \label{ambct}
\end{align}
which generates the additional terms in the trace of the energy-momentum tensor  so that we have the dressing transformation as in (\ref{ambigu}). These are related to the ambiguities in the contact terms in various correlation functions among $T^{\mu}_{\ \mu}$ and $O^I$ (see \cite{Osborn:1991gm} for details). There are more terms we could add than (\ref{ambct}) but they do not contribute to our discussions on our $a$-theorem.\footnote{By using such ambiguities, one can show that the $R^2$ Weyl anomaly is given by $\mathcal{B}^I \chi^{a}_{IJ} \mathcal{B}^J$, which is expected because when a theory is conformal invariant (i.e. $\mathcal{B}^I = 0$), $R^2$ anomaly must vanish.  This is in accord with our discussions at the end of section \ref{scale2}. We note that $2\chi_{IJ}^{a}$ agrees with $\chi_{IJ}^{g}$ in a certain order of perturbation theory (indeed $2\chi_{IJ}^a = \chi_{IJ}^g$ when $\mathcal{B}^I = 0 $ and if we set $S_{IJ}=0$ by using the further ambiguity), but they can deviate at the higher order.}

Let us point out one important consequence of the formula (\ref{ggggaaa}). As pointed out in \cite{Nakayama:2012sn}\cite{Nakayama:2012nd}\cite{Fortin:2012hn}, the scale invariance demands that Osborn's $\tilde{a}$-function must take a constant value. By assuming the positivity of $\chi_{IJ}^g$, it means that $\mathcal{B}^I=0$ with the scale invariance, forbidding the cyclic behavior \cite{Fortin:2012hn}. The trace identity (\ref{opannn}) tells that the energy-momentum tensor is traceless in the flat space-time limit, and the theory must be conformal invariant.

The similar ambiguity existed in $d=2$ dimension \cite{Osborn:1991gm}, in which we can introduce the scheme dependent $\tilde{c}$-function from the local renormalization group analysis with the ambiguity as in (\ref{ambigu}). 
From this viewpoint, the main claim of Zamolodchikov is that one can choose a good counterterm $C_{IJ}$ so that $\chi^g_{IJ}$ agrees with Zamolodchikov metric and positive definite. Or more precisely what Zamolodchikov did is he first read the counterterms from the two-point functions and then defined the monotonically decreasing $c$-function  by considering a particular combination to cancel the ambiguity.
We may not be able to remove the antisymmetric part $w_{IJ}$ (as well as $\hat{\rho}_{[I} Q_{J]}$ term) unless $w_{I}$ is exact, but this is unimportant for the strong version of Zamolodchikov's $c$-theorem which we derived in section \ref{cfunc}. Also note the above ambiguity does not affect the value of the $a$-function at the conformal fixed point because $\mathcal{B}^I = 0$ there. It is consistent with the fact that at the conformal fixed point the trace anomaly does not have a local counterterm (except for $\Box R$ term in $d=4$ dimension).

\subsection{Mixing of energy-momentum tensor with scalars and improvement}\label{mixingsi}
As one technical side remark, we would like to discuss the improvement of the energy-momentum tensor and a possible choice of ``canonical scaling" mentioned in section \ref{techni} from the viewpoint of the local renormalization group approach. In section \ref{localrg}, we have only considered the dimensionless coupling constants, and the energy-momentum tensor is not renormalized. However, as discussed in section \ref{techni} the presence of the dimension $d-2$ operators (e.g. scalar mass operator) will introduce the renormalization of the energy-momentum tensor with its associated ambiguity.

In order to understand the mixing, we introduce the source for the dimension $d-2$ operator $\delta S = \int d^dx \sqrt{|g|} M^i O_i$. 
The mixing is generated by the additional contribution to the local renormalization group operator as
\begin{align}
\Delta_{\sigma,M} &= -\int d^dx \sqrt{|g|}\left(\sigma(d-2-\gamma^i_{j(M)}) M^j + \frac{1}{2(d-1)} \sigma R \eta^i + \sigma \delta^i_I (D^2 g^I) + \sigma \epsilon^i_{IJ} (D^\mu g^I D_\mu g^J) \right. \ \cr
 & \left. \left. + 2\partial_\mu \sigma (\theta^i_I D^\mu g^I) + (D^2 \sigma) \tau^i  \right) \frac{\delta}{\delta M^i} \right) \ . \label{mderiv}
\end{align}
By studying the local Callan-Symanzik equation, we find that there is a further renormalization of the operators by
\begin{align}
\frac{d}{d\log\mu} O_i^{(M)} &= \gamma^{^j}_{i(M)} O_j^{(M)} \cr 
\frac{d}{d\log\mu} O_I &= \gamma^{J}_I O_J + \delta_{I}^i D^2 O_{i}^{(M)}  \cr
\frac{d}{d\log\mu} J_\mu^a &= \gamma^{a}_{b} J_\mu^a + 2\theta_I^i (T^{aI}_{\ \ J} g^J)D_\mu O_{i}^{(M)}  \cr
\frac{d}{d\log \mu}T^{\mu}_{\ \mu} & = \eta^i \Box O_{i}^{(M)} \  \label{opmixing}
\end{align}
in the flat space-time limit with no space-time dependence on coupling constants.
However, one may use the ambiguity in the renormalization group so that we may make some of the mixing vanish. Such ambiguity was called Class 2 ambiguity (scheme ambiguity) of the local renormalization group in \cite{Nakayama:2013wda}. In addition, for these equations to be consistent with the trace identity, the coefficients satisfy the integrability conditions which may be found in \cite{Osborn:1991gm}\cite{JO}\cite{Nakayama:2013wda}.

For instance, we can always set $\tau^i = 0$ by using the local counter-term of the form $\int d^dx \sqrt{|g|} h^iRO_i^{(M)}$. This convention is know as the Callan-Coleman-Jackiw improved energy-momentum tensor \cite{Callan:1970ze}. One advantage of the choice is that when $\mathcal{B}^I = 0$ at the fixed point, the theory is manifestly conformal invariant in the flat space-time and we keep the same property during the renormalization group flow by adjusting $h^i$ at each energy scale.

However, away from the conformal fixed point, this improved energy-momentum tensor may be renormalized according to \eqref{opmixing} due to the operator mixing  from $\eta^i \neq 0$ in \eqref{opmixing}. 
For this reason, it may be sometimes more useful to define the non-renormalized energy-momentum tensor by demanding $\eta^i = 0$ rather than $\tau^i = 0$. This is known as Zamolodchikov's canonically scaling energy-momentum tensor \cite{Zamolodchikov:1986gt}\cite{Polchinski:1987dy} (see also \cite{Yonekura:2012kb}). As argued by Polchinski,\footnote{There is a typo in eq (18) of \cite{Polchinski:1987dy}. We would like to thank Z.~Komargodski for the related discussion.} this is always possible by adjusting $h^i$ when $d-2-\gamma_{(M)}$ does not contain any zero eigenvalues, being invertible. Otherwise, due to a potential obstruction to choose $\eta^i=0$, it is logically possible that the theory is scale invariant, but the energy-momentum tensor is still logarithmically renormalized. 
 When the theory is conformal invariant (i.e. $\mathcal{B}^I = 0$) then such a possibility is unavailable from the consistency conditions in agreement with the fact that in unitary conformal field theory, the energy-momentum tensor cannot mix with any other operators. In any case, away from the fixed point, it is important to understand that the Callan-Coleman-Jackiw improved energy-momentum tensor and Zamolodchikov's non-renormalized energy-momentum tensor (if any) may differ. Note however that this ambiguity does not affect the main part of the argument in section \ref{localrg} within the power-counting renormalization scheme.

\newpage

\section{Nonperturbative aspects, $a$-theorem, and dilaton scattering amplitudes}\label{section8}
In section \ref{localr}, we have discussed the perturbative proof of the $a$-theorem as well as the enhancement from scale invariance to conformal invariance.
 In this section we will further discuss the non-perturbative aspects of $a$-theorem and its implication in the question about the enhancement of conformal invariance from scale invariance in relativistic quantum field theories.
\subsection{$a$-maximization}\label{amax}
In supersymmetric field theories, the conformal anomaly is directly related to the anomaly of the superconformal $\mathcal{R}$-current. From the structure of the superconformal supermultiplet \cite{Anselmi:1997am}, we can show that $a$ and $c$ coefficients in Weyl anomaly in $d=4$ dimension are related to the $\mathcal{R}$-current anomaly as
\begin{align}
a &= \frac{9}{16 \cdot (8\pi)^2} \left(3 \mathrm{Tr}\mathcal{R}^3  - \mathrm{Tr} \mathcal{R} \right)\cr
c &=\frac{9}{16 \cdot (8\pi)^2} \left(3\mathrm{Tr} \mathcal{R}^3 - \frac{5}{3} \mathrm{Tr} \mathcal{R} \right)\ . \label{aandc}
\end{align}
The $\mathrm{Tr}$ here means a schematic notation to compute the triangle anomaly of the $\mathcal{R}$-currents. 
 By using the formula, once we can determine the superconformal $\mathcal{R}$-current, we may compute $a$ and $c$. In particular, if the conformal $\mathcal{R}$-symmetry is a well-defined  symmetry in the UV theory, we can use the 't Hooft anomaly matching argument \cite{'tHooft:1979bh} and evaluate (\ref{aandc}) by using the free field theory computation.
 
 To compute $a$ (and $c$), all we need is to determine the superconformal $\mathcal{R}$-current, and there is a principle so called ``$a$-maximization" \cite{Intriligator:2003jj}. The principle gives a three-line proof of the weak version of   Cardy's conjecture in supersymmetric field theories under some technical assumptions. The idea of $a$-maximization is that under all possible candidates of $U(1)_{\mathcal{R}}$ symmetry, the superconformal one is the one that maximizes the ``trial $a$-function": $a_{\mathrm{trial}} = 3 \mathrm{Tr}\mathcal{R}^3  - \mathrm{Tr} \mathcal{R}$. Since relevant deformations generically break the flavor symmetries, the set for candidates of the $\mathcal{R}$-symmetry in IR is a subset of that in UV. Thus, the maximized $a$ must satisfy $a_{\mathrm{UV}} \ge a_{\mathrm{IR}}$.

While $a$-maximization is very powerful in practice, we have a limitation. One limitation is that  we might have accidental symmetries in IR that could spoil the  above argument because the set for the candidate $\mathcal{R}$-symmetry is no longer a subset  of that for UV. However, since we have a more general proof of the $a$-theorem, this means that the mixing of accidental symmetry at the superconformal fixed point is somehow bounded (see \cite{Buican:2011ty} for discussions). We also note that the $a$-maximization argument relies on the existence of the superconformal fixed point, so we cannot exclude scale invariant but non-conformal supersymmetric field theories from this argument (see e.g. \cite{Nakayama:2011tk}\cite{Nakayama:2012nd}). Furthermore, in principle, supersymmetric field theories can be scale invariant without any $\mathcal{R}$-symmetry \cite{Nakayama:2012nd}. For instance, the non-renormalizable theory with the supersymmetric action
\begin{align}
\mathcal{S} = \int d^4x d^4\theta |\Phi|^2 (\Phi^2 + (\Phi^\dagger)^2) + \int d^4x d^2\theta \Phi^6 + c.c  \ ,
\end{align}
where $\Phi$ is a chiral superfield,
is classically scale invariant, but not $\mathcal{R}$-symmetric, and therefore it is not superconformal invariant \cite{Nakayama:2012nd}.

\subsection{Proof of weak $a$-theorem}\label{weak}
There have been various attempts to prove the $a$-theorem in $d=4$ dimension. Finally, Komargodski and Schwimmer gave a reasonable and ingenious physical argument for the  weak version of the theorem \cite{Komargodski:2011vj}.
Consider the renormalization group flow from $\mathrm{CFT}_{\mathrm{UV}}$ to $\mathrm{CFT}_{\mathrm{IR}}$ in $d=4$ dimension. For technical simplicity, we assume that both
are Weyl  invariant for a while. We assume that the flow is induced by adding a relevant 
deformation ${O}$ with the conformal dimension $\Delta$ to the UV CFT so that under the Weyl transformation $g_{\mu\nu} \to e^{2\sigma}g_{\mu\nu}$ it transforms as ${O} \to e^{-\Delta \sigma} {O}$.\footnote{Depending on the regularization scheme we choose, we implicitly add all the relevant counterterms to arrive at the fixed point $\mathrm{CFT}_{\mathrm{IR}}$ we desired.}
From unitarity, the deformation must be a  conformal primary operator. 

The deformed theory is no-longer Weyl invariant, but we may introduce the ``dilaton"  $\tau$ to compensate the violation of the Weyl invariance due to the deformation.
 We can always do it  by dressing the deformation with $e^{-(4-\Delta)\tau} {O}$. 
Under the Weyl transformation, we assume that the dilaton transforms as $\tau \to \tau + \sigma$ to make the deformation spuriously Weyl invariant:
$ \sqrt{|g|} e^{-(4-\Delta)\tau} {O} \to \sqrt{|g|} e^{-(4-\Delta)\tau} {O}$.
The dilaton compensated UV action is schematically given by
\begin{align}
\mathcal{S}_{\mathrm{UV}} = \mathcal{S}_{\mathrm{CFT}_{\mathrm{UV}}} + \int d^4x \sqrt{|g|} e^{-(4-\Delta)\tau}{O} + \mathcal{S}_{\mathrm{ct}} +\tilde{f}^2 \int d^4 x \sqrt{|g|} e^{-2\tau} \left((\partial_\mu \tau)^2 +\frac{1}{6}R\right) + \tilde{\mathcal{S}}_\mathrm{{nu}} \ ,
\end{align}
where $\mathcal{S}_{\mathrm{ct}}$ is the dilaton compensated (relevant) counterterms that contain various relevant operators in the UV CFT (including cosmological constant) that  will be fine-tuned during the renormalization group flow so that it will end up with the desired IR fixed point. The kinetic term   for the dilaton (added by hand) is Weyl invariant by itself. Here $\tilde{f}$ is arbitrarily large dimensionful decay constant of the dilaton, which we could add if we wish. $\tilde{\mathcal{S}}_{\mathrm{{nu}}}$ is Weyl invariant non-universal dilaton counterterms that can again be introduced by hand.

The dilaton is very weakly coupled as long as we take large $\tilde{f}$ (or we can even regard it as an external source in the extreme limit) so it will not affect the dynamics
or the properties of the IR CFT. This is equivalent to the claim that the dilaton will decouple 
in the IR physics so that the IR effective action has the decoupled form
\begin{align}
\mathcal{S}_{\mathrm{eff}} = \mathcal{S}_{\mathrm{CFT}_{\mathrm{IR}}} + f^2 \int d^4 x \sqrt{|g|} e^{-2\tau} \left((\partial_\mu \tau)^2+\frac{1}{6}R\right) + \mathcal{S}_{\mathrm{WZ}} + \mathcal{S}_{\mathrm{nu}} \ .
\end{align}
The dilaton decay constant $f$ and the non-universal term $\mathcal{S}_{\mathrm{nu}}$ can be different from those of UV, but this does not affect the following discussions. 
In addition, we may want to introduce counterterms that are associated with the position dependent coupling constants in relation to the local renormalization group flow discussed in section \ref{localr}. As will be discussed in section \ref{spacet}, these do not affect the analysis in this section basically because we assume UV and IR theories are Weyl invariant and $\mathcal{B}$ function vanishes.
 
The 't Hooft anomaly matching condition \cite{'tHooft:1979bh} (for Weyl invariance \cite{Schwimmer:2010za}) fixes the form of the Wess-Zumino term $\mathcal{S}_{\mathrm{WZ}}$ 
\begin{align}
\mathcal{S}_{\mathrm{WZ}} =& \int d^4x \sqrt{|g|}\left( (a_{\mathrm{UV}}-a_{\mathrm{IR}}) \left(\tau \mathrm{Euler} + 4 G_{\mu\nu}\partial^\mu \tau \partial^\nu \tau -4(\partial_\mu \tau)^2\Box \tau + 2(\partial_\mu \tau)^4\right)  \right. \cr
  &-(c_{\mathrm{UV}}-c_{\mathrm{IR}})\tau (\mathrm{Weyl})^2 \cr
   &\left. -(\tilde{b}_{\mathrm{UV}}-\tilde{b}_{\mathrm{IR}}) \frac{R^2}{12} \right) \ .
\end{align}
If the theory breaks CP invariance, there is a potential addition of Pontryagin term 
\begin{align}
(d_{UV} - d_{IR}) \tau (\epsilon^{\mu\nu\rho\sigma} R_{\mu\nu}^{\ \ \alpha \beta}R_{\alpha\beta\rho\sigma}) 
\end{align}
but it will play no role in the following (see footnote \ref{footz}). 
The necessity of the anomaly matching is as follows. Suppose we would like to hypothetically gauge the Weyl symmetry. We had to cancel the Weyl anomaly of the UV theory. We do it by adding (non-unitary) spectator Weyl invariant theory with  the opposite Weyl anomaly. The consistency of the gauging suggests that the IR theory must show the same anomaly ('t Hooft matching condition) to cancel the contribution from the added spectators. Because the anomaly of the IR theory with $\mathcal{S}_{\mathrm{CFT}_{\mathrm{IR}}}$ is different than that of the original theory, it must be somehow compensated. This is precisely what Wess-Zumino terms do. Under the classical Weyl variation $g_{\mu\nu} \to e^{2\sigma} g_{\mu\nu}$ and $\tau \to \tau + \sigma$, the Wess-Zumino terms give 
\begin{align}
\delta_\sigma \mathcal{S}_{\mathrm{WZ}} =\int d^4x \sqrt{|g|} \left( (a_{\mathrm{UV}} - a_{\mathrm{IR}}) \sigma \mathrm{Euler} - (c_{\mathrm{UV}} - c_{\mathrm{IR}}) \sigma (\mathrm{Weyl})^2 + (\tilde{b}_{\mathrm{UV}} - \tilde{b}_{\mathrm{IR}}) \sigma \Box R \right) \ .
\end{align}
The Wess-Zumino term can be obtained by trial and error or by using the Wess-Zumino trick \cite{Schwimmer:2010za} (see also \cite{Elvang:2012yc} for higher dimensional computations)\footnote{Up to the non-universal term, the Wess-Zumino action itself was known in \cite{Riegert:1984kt}. See also \cite{Antoniadis:1991fa}\cite{Antoniadis:1992xu}\cite{Deser:1996na}.}
\begin{align}
\mathcal{S}_{\mathrm{WZ}} &= (a_{\mathrm{UV}}-a_{\mathrm{IR}}) \int_0^1 dt \int d^4x  \tau \sqrt{|g|}\mathrm{Euler}(g \to g e^{-2t\tau}) \cr
&- (c_{\mathrm{UV}}-c_{\mathrm{IR}}) \int_0^1 dt \int d^4x  \tau \sqrt{|g|}\mathrm{Weyl}^2(g \to g e^{-2t\tau}) \ \label{trickf}
\end{align}
and it is constructed so that it cancels the Weyl anomaly at the conformal fixed point. The relation of this extra integration over $t$ in \eqref{trickf} and the holographic direction was discussed in \cite{Schwimmer:2013jma}. 

The non-universal terms contain all possible Weyl invariant counterterms
\begin{align}
\mathcal{S}_{\mathrm{nu}} = \int d^4x \sqrt{|\hat{g}|} \left( \hat{a}_c\mathrm{Euler}(\hat{g}) + \hat{c}_c(\mathrm{Weyl}^2(\hat{g})) + \hat{b}_c\hat{R}^2 \right) \ .
\end{align}
We define $\hat{R} = \hat{g}^{\mu\nu} R_{\mu\nu}[\hat{g}]$ with $\hat{g}_{\mu\nu} = e^{-2\tau}g_{\mu\nu}$, which is the Weyl compensated metric so that $\hat{g}_{\mu\nu} \to \hat{g}_{\mu\nu}$ under the Weyl transformation.
We cannot determine these terms from the symmetry alone. In UV, we can just add them by hand, and in IR, they can be generated from the renormalization group flow. This ambiguity will turn out to be irrelevant for our argument in this section.

Indeed, one may use the counterterms to rewrite the above Wess-Zumino term in the Riegert form 
\begin{align}
\int d^4x \sqrt{|\hat{g}|}a\left(\tau (\hat{\mathrm{Euler}} -\frac{2}{3}\hat{\Box} \hat{R}) + 2\tau \hat{\Delta}_4 \tau \right) \ ,
\end{align}
where $\Delta_4$ is the Riegert operator \cite{Riegert:1984kt}
\begin{align}
\Delta_4 = \Box^2 +  2R^{\mu\nu}D_\mu D_\nu - \frac{2}{3} R \Box + \frac{1}{3}(D^\mu R)D_\mu \ .
\end{align}
 See \cite{Antoniadis:1992xu}\cite{Schwimmer:2010za} for more details. 
This dilaton effective action with this Riegert form does look like that for a (higher derivative) free field for $\tau$ in the flat-space limit, but the following dilaton scattering argument is not affected after imposing the second order on-shell condition.

Following the strategy of Komargodski and Schwimmer \cite{Komargodski:2011vj}, we will study the scattering amplitudes of the dilaton in the flat Minkowski
background $g_{\mu\nu} = \eta_{\mu\nu}$. \\\\\\\\\\\\
We focus on the two-two dilaton scattering amplitudes. 
In particular, we are interested in the $s$-channel forward scattering amplitude ($t = 0$) in the leading order of $s$ (see \ref{smatrix} for a brief review of the scattering theory). For this purpose, we can assume the  on-shell dilaton condition $(\partial_\mu \tau)^2 = \Box  \tau$. In other words, we introduce the canonically normalized dilaton field $\varphi$ defined by $e^{-\tau} = 1 + \frac{\varphi}{f}$ with $\Box \varphi = 0$ as the on-shell condition. The error of using the on-shell condition in the interaction term is suppressed by $\frac{s}{f^2}$ compared with the leading term we are interested in.
With the on-shell condition, the flat space-time IR effective action is given by\footnote{It is instructive to see what happened to the other trace anomaly terms. The Wess-Zumino terms for $c(\mathrm{Weyl})^2$ and $\tilde{b}R^2$ vanish by $g_{\mu\nu}=\eta_{\mu\nu}$. Similarly, that for Pontryagin term (if any) does not affect the dilaton scattering amplitudes. The non-universal term $\mathcal{S}_{\mathrm{nu}}$ does not contribute either due to the on-shell condition $(\partial_\mu  \tau)^2 = \Box \tau$.\label{footz}}
\begin{align}
\mathcal{S}_{\mathrm{eff}} = \mathcal{S}_{\mathrm{CFT}_{\mathrm{IR}}} + \int d^4x \left( f^2e^{-2\tau}(\partial_\mu \tau)^2 - 2(a_{\mathrm{UV}}-a_{\mathrm{IR}}) (\partial_\mu \tau)^4 \right) \ . \label{ireffd}
\end{align}

In the following, we would like to argue that the coefficient of $(\partial_\mu \tau)^4$ must be negative definite\footnote{Our metric convention is opposite to that used in \cite{Komargodski:2011vj}, and we have a negative sign here.}
from causality and unitarity so that the weak version of the $a$-theorem $a_{\mathrm{UV}} \ge a_{\mathrm{IR}}$ must hold for its consistency.
A quick heuristic way to show the necessity of the requirement is the following argument: if we consider a particular non-trivial background $\varphi = c_\mu x^\mu$ for the dilaton effective action (\ref{ireffd}) with small $c_\mu/f$,
then the propagation of the dilaton $\varphi$ around the background is superluminal unless $a_{\mathrm{UV}}-a_{\mathrm{IR}}\ge0$, suggesting the violation of causality \cite{Adams:2006sv}.

A more rigorous argument can be made by using the dispersion relation \cite{Komargodski:2011vj}\cite{Komargodski:2011xv} (see also \cite{Luty:2012ww}).
We study the forward scattering ($t=0$: see again \ref{smatrix} for a brief review of the scattering theory) of the two-two dilaton scattering in the $s\to 0$ limit. 
The behavior of the forward scattering amplitude $A_4(s)=A(s,t=0)$ in the $s\to 0$ limit is governed by the IR effective action (\ref{ireffd}) and 
\begin{align}
A_4(s) = 8(a_{\mathrm{UV}} -a_{\mathrm{IR}})\frac{s^2}{f^4} + \mathcal{O}(s^{\Delta_{\mathrm{IR}}-2}) \ ,
\end{align}
where $\Delta_{\mathrm{IR}}>4$ is the lowest dimension of the irrelevant deformations at the IR fixed point. 
Note that relevant deformations are fine-tuned to be absent (otherwise it does not flow to the fixed point
we are focusing on). 

\begin{figure}[tbh]
\begin{center}
\includegraphics[width= 0.8 \linewidth]{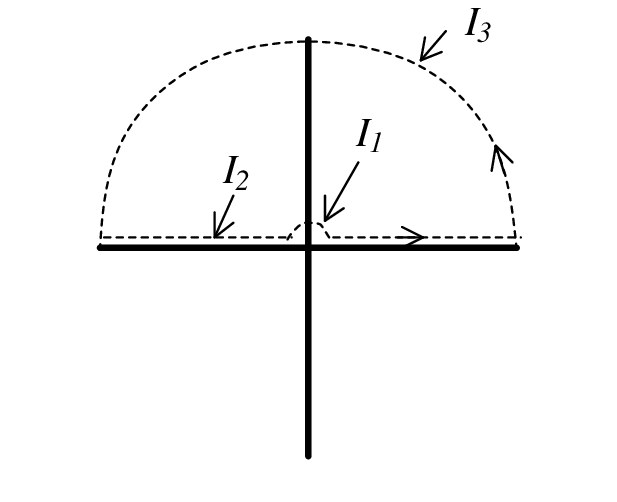}
\end{center}
\caption{The $s$ channel scattering amplitude shows positivity of $a_{\mathrm{UV}}-a_{\mathrm{IR}}$.}
\label{cont}
\end{figure}

From Cauchy's theorem and analyticity of S-matrix in the upper half plane, we have (see fig \ref{cont} for the contour)
\begin{align}
0 = \frac{1}{2\pi i} \oint ds \frac{A_4(s)}{s^3} \ . 
\end{align}
Around the $s=0$ pole, we have $I_1 = -\frac{8(a_{\mathrm{UV}}-a_{\mathrm{IR}})}{2f^4}$.
Just above the cut on the real axis, by noting that $A_4(s) = A_4(-s)$ from crossing symmetry (see \ref{smatrix} for more details), we obtain
\begin{align}
I_2 = \frac{1}{\pi} \int_{\epsilon}^{\infty} ds \frac{\mathrm{Im}A_4(s)}{s^3} = \frac{1}{\pi} \int_{\epsilon}^{\infty} ds \frac{\sigma(s)}{s^2} \label{crossingf}
\end{align}
The integral is convergent both in UV and IR. Here, $\sigma(s)$ is the total cross section of $\varphi \varphi \to \mathrm{CFT}$ from the
optical theorem (\ref{opticalt}),\footnote{Although in perturbative examples, we can directly check it from Feynmann diagrams, the validity of the usage of optical theorem may cause some suspicion because the dilaton is not physical, and the use of the unitarity may be invalid (in particular due to non-renormalizability). It would be more desirable if we had a better understanding.} so it must be manifestly positive from unitarity.
Finally, the large semi circle contribution is zero by noting in UV there is no  irrelevant deformation from renormalizability. Thus $a_{\mathrm{UV}} \ge a_{\mathrm{IR}}$. 

The above discussion applies when $\mathcal{B}$ function near the fixed point has a first  order zero both in UV and IR, but we can study the case with
higher order zero (which corresponds to marginally relevant/irrelevant couplings like UV gauge coupling constants), and we can still prove the convergence,
so the proof is also valid \cite{Luty:2012ww} in more generality.

We may wonder whether the argument here suggests a possibility to define the
 $a$-function not only at the fixed point but also along the renormalization group flow to derive the strong version of the $a$-theorem: $\frac{da(g)}{d\log\mu} \ge 0$. One candidate \cite{Komargodski:2011vj} is  $a^{\mathrm{KS}} (\mu)= \int_{\mu}^{\infty} ds \frac{\sigma(s)}{s^2}$. As we will see in the next subsection, this behaves very similarly to Osborn's $\tilde{a}$-function at least  within perturbation theory, and since $\sigma(s) \ge 0$, it is manifestly monotonically decreasing. However, we still have to show that this is a function of the running coupling constants at the energy scale $\mu$ alone, and does not depend on the path of the renormalization group flow to get the precise equivalence. Otherwise, the monotonicity along the renormalization group flow itself is not physically relevant (see also our discussions in section \ref{moron} on averaged $c$-theorem).\footnote{For instance, we can always define $2a(\mu) = (a_{\mathrm{UV}}-a_{\mathrm{IR}}) \tanh(\log\mu) + (a_{\mathrm{IR}} + a_{\mathrm{UV}})$. as ``$c$-function" in any renormalization group flow (without unitarity, Poincar\'e invariance and so on), which is monotonically decreasing by definition (one can even choose whatever number for $a_{\mathrm{UV}} \ge a_{\mathrm{IV}}$ here). This does not reflect the intrinsic properties of the flow, and it is completely useless. Needless to say, we cannot conclude anything about scale invariance and conformal invariance from this function.}

The discussion here is made sharper in \cite{Luty:2012ww}: They introduced the averaged amplitude over the semi-circle $C_{(\mu)}$ of radius $\mu$ as
\begin{align}
\bar{\alpha} (\mu) = -\frac{2f^4}{\pi} \int_{C_{(\mu)}} \frac{ds}{s^3} A_4(s) \ \label{avea}
\end{align}
with the differential relation 
\begin{align}
\frac{d\bar{\alpha}(\mu)}{d\log\mu} = \frac{2f^4}{\pi \mu^2} \mathrm{Im}{A_4(\mu)} \ .
\end{align}
The optical theorem implies that $\bar{\alpha}$ is a monotonically decreasing function of $\mu$. By Cauchy's theorem, $\bar{\alpha}(\mu)$ is same as $a^{\mathrm{KS}}(\mu)$ up to a constant. The relevance of this quantity in relation to the enhancement of conformal invariance from scale invariance will be discussed below.

\subsection{Scale vs Conformal in $d=4$ dimension}\label{svc}
Given the non-perturbative proof of the weak version of the  $a$-theorem, can we say anything about scale invariance
vs conformal invariance? Recall that in $d=2$ dimension, the argument was essentially based on the strong version of the 
$c$-theorem as discussed in section \ref{cfunc}, so we can imagine we have to be a little bit more creative here since the strong version of the $a$-theorem is not proved yet.

Let us suppose the IR limit of the deformed theory is not Weyl invariant but only scale invariant.\footnote{Here we still assume that the UV theory is conformal invariant, but one may relax the condition. In this case, it may be more convenient to  introduce the external gauge field $C_\mu$ corresponding to the virial current to compensate the Weyl transformation \cite{Luty:2012ww}. Since $C_{\mu}$ and $\partial_\mu \tau$ transform in the same way under the Weyl transformation, we have some freedom to choose.}
Can  we infer any inconsistency to exclude such a possibility? Note this is a slightly more subtle problem because we allow non-Weyl invariant but conformally equivalent IR fixed point such as Nambu-Goldstone bosons, which must be, of course, consistent.

The first thing we have to notice is that the dilaton does not decouple from the IR effective action of the matter when the IR theory is not Weyl invariant \cite{Nakayama:2011wq}.
The reason is that  we are enforcing the Weyl invariance, so the non-zero trace  of the energy-momentum tensor
of the IR limit of the deformed theory with non-trivial virial current (recall $T^{\mu}_{\ \mu} = \partial^\mu J_\mu$) must be 
cancelled by the coupling between the dilaton and the matter:
\begin{align}
\mathcal{S}_{\mathrm{eff}} = \mathcal{S}_{\mathrm{SFT}_{\mathrm{IR}}} + \mathcal{S}_{\mathrm{dilaton}}+ \int d^4x \left( \tau \partial^\mu J_\mu  + \mathcal{O}(\tau^2) {O}_{\mathrm{SFT}_{\mathrm{IR}}} \right) \ , \label{SFTdil}
\end{align}
where $\mathcal{S}_{\mathrm{SFT}_{\mathrm{IR}}}$ is the effective action for the scale invariant field theory (SFT).
The first order term $\tau \partial^\mu J_\mu$ is uniquely specified by the Noether procedure, but the $\mathcal{O}(\tau^2)$ terms are non-universal
and model dependent. It is related to the Weyl transformation property of $J_\mu$ and it is not specified by the scale invariance,
so we have a certain degree of freedom here. In addition, we recall that  $J_\mu$ is determined up to the equations of motion 
of $\mathcal{S}_{\mathrm{SFT}_{\mathrm{IR}}}$ with $\tau =0$. After the inclusion of $\tau$ coupling, the ``up to  EOM" term contributes to the higher order contact terms 
 so we cannot  say much about what  $\mathcal{O}(\tau^2)$ are from the generic argument alone. In principle, in a given theory, one can determine these terms order by order (while they may not be unique).

For instance, within the powercounting renormalization, the renormalization of the trace of the energy-momentum tensor can be specified by
\begin{align}
\delta_\sigma T^{\mu}_{\ \mu} = \sigma \eta^i \Box O^{(M)}_i + \mathcal{O}(\partial^\mu \sigma)  \ 
\end{align}
with operator $O^{(M)}_i$ whose tree-level scale dimension is $2$ (e.g. scalar mass term), so at least we have to add $\tau^2 \eta^i \Box O^{(M)}_i$ to cancel the scale variation. We should stress that this effect did not include the possibility that under the non-constant Weyl transformation, $T^{\mu}_{\ \mu}$ could contain the additional terms proportional to $\partial_\mu \sigma$. One important remark here is that if we choose the renormalization scheme so that $T^{\mu}_{\ \mu} = \mathcal{B}^I O_I$, then the consistency requires $\eta^i$ is always proportional to $\mathcal{B}^I$ multiplied by an extra loop factor \cite{Osborn:1991gm} so that this contribution is always small in perturbation theory. Furthermore, we have discussed in section \ref{localr} that $\eta^i$ depends on the renormalization scheme.

The second thing we have to notice is that  the structure of the Weyl anomaly is modified. For instance, the scale invariant 
but non-conformal invariant field theory introduces additional terms such  as $R^2$ (see non-conformal scalar example).
This may raise a puzzle in deriving the dilaton effective action: $R^2$ does not satisfy the  Wess-Zumino consistency condition,
so is there any $\mathcal{S}_{\mathrm{WZ}} [\tau]$ that reproduces the $R^2$ Weyl anomaly?
This seems impossible without introducing any other background fields as is clear from the fact that Wess-Zumino consistency condition is not satisfied. The naive try and error immediately reveals that there is no completion that depends only on metric and dilaton by starting from $\tau \hat{R}^2$.

The solutions to these problems are  related. Now the IR theory is modified by the external source coupling  to  the dilaton, 
 we have to reconsider the  Weyl anomaly of the  dilaton coupled scale invariant field theories.
 
 There are two approaches we can take:

\begin{itemize}
 \item The first approach is to assume that the dilaton is fully dynamical. Then the total theory is Weyl invariant by construction \cite{Nakayama:2011wq}. 
 The Wess-Zumino consistency condition dictates that there is no $R^2$ anomaly. The on-shell Wess-Zumino action is given by $\mathcal{S}_{\mathrm{WZ}} = -2 \int d^4 x (a_{\mathrm{UV}} - \hat{a}_{\mathrm{IR}}) (\partial_\mu \tau)^4$. Here $\hat{a}_{\mathrm{IR}}$ may be different from $a_{\mathrm{SFT}_{\mathrm{IR}}}$. 
 \item The second approach is to assume that the dilaton is still an external source. The possible Weyl anomaly is larger than without dilaton
 because we can construct more non-trivial solutions to  the Wess-Zumino consistency condition. Note that $\hat{g}$ is always Weyl invariant, so for instance 
 \begin{align}
 T_{\mathrm{new}} =  b e^{-2\tau}\hat{R}^2 = b(R + 6(\Box \tau) - 6(\partial_\mu\tau)^2)^2 
 \end{align}
are completely allowed trace anomaly term.
 The above mentioned $R^2$ anomaly is completed to be $\hat{R}^2$ by adding extra dilaton depending terms so that 
 it satisfies the Wess-Zumino consistency condition. If we assume that these have local Wess-Zumino action, there are $11$ terms that could appear in the dilaton effective action (see \ref{localWZC} and the discussions in section \ref{spacet}: for instance the dressing transformation by $C_{IJ}^g$ induces $(\Box \tau)^2$ term in the dilaton effective action and the corresponding Weyl anomaly term). Alternatively, we can compensate the anomaly from the  of the virial current $C_\mu$ by replacing $\partial_\mu \tau$ by $C_\mu$. 

If we impose the on-shell condition, the Wess-Zumino terms that contribute to the two-two dilaton scattering all reduce to
 \begin{align}
 \mathcal{S}_{\mathrm{WZ}}  = -2 \int d^4 x (a_{\mathrm{UV}} - \hat{a}_{\mathrm{IR}}) (\partial_\mu \tau)^4 \ .
 \end{align}
 Here $\hat{a}_{\mathrm{IR}}$ may be different from $a_{\mathrm{IR}}$ which gives the pure Euler term in the Weyl anomaly.  We realize that  $\hat{R}^2$ in the Weyl anomaly did not matter, but there are many other terms that could contribute to the dilaton scattering amplitudes. We will discuss relevant terms coming from the space-time dependent coupling constant counterterms later in section \ref{spacet}. The complete dilaton effective action in this case was presented in \cite{Jack:2013sha}\cite{Baume:2014rla} within the power-counting renormalization scheme.
 \end{itemize}
 Presumably, this modification of the four-dilaton scattering amplitude explains the ``modification" of $\tilde{a}$ from $a$ away from 
 the fixed point to see the monotonicity in Osborn's argument (see \cite{Fortin:2012hn} for the similar argument).\footnote{We should note that the amplitude is now scheme dependent because of the existence of trivial anomaly terms involving $\tau$. This is rather consistent with local renormalization group analysis, in which Osborn's $\tilde{a}$-function away from the conformal fixed point is not unique. We will revisit the effect in section \ref{spacet} from the space-time dependent coupling constant counterterms.} In the following argument, we do not need 
 the precise relation, but the only crucial thing is that the dilaton scattering amplitudes must be bounded. As emphasized in \cite{Fortin:2012hn}, actually we do not need the explicit form of the Wess-Zumino terms, either, for the following argument. We will identify $\hat{a}$ with the averaged dilaton scattering amplitude $\bar{\alpha} (\mu)$ in (\ref{avea}).

In either approaches we take, the dilaton does  not decouple and we have to regard the action (\ref{SFTdil}) as the effective action renormalized at a particular energy scale $\mu$. We now argue that generically $\hat{a}_{\mathrm{IR}}$ is logarithmically renormalized \cite{Nakayama:2011wq} suggesting its inconsistency with the strict scale invariance \cite{Luty:2012ww}.
To see this, we again compute the two-two forward dilaton scattering amplitude.
From the effective action (\ref{SFTdil}), we know that the computation of the S-matrix schematically follows from the matrix element of
\begin{align}
A_4 = \int \langle \varphi \varphi |TT| \varphi\varphi \rangle + \int \langle \varphi \varphi |TT T  T| \varphi\varphi \rangle + \int \langle \varphi \varphi |{O} {O}| \varphi\varphi \rangle \  + \cdots . \label{smatrixp}
\end{align}
The first two terms come from universal coupling to the virial current, and the last term comes from the non-universal terms in the dilaton
effective action. We again emphasize that $T^{\mu}_{\ \mu}$ is determined up to the equations of motion, so there is an intrinsic ambiguities in defining $T^{\mu}_{\ \mu}$ correlation functions. In particular, away from the fixed point, the contact terms contain various ambiguities from the renormalization group prescriptions. These explain the presence of the last term in (\ref{smatrixp}). We note that the on-shell condition makes this ambiguity milder as we will further discuss in the following. 

Whenever the two-two dilaton scattering amplitude $A_4$ has a logarithmic divergence in IR, it is inconsistent with the assumption of scale invariance.\footnote{We refer to \cite{Luty:2012ww} for the absence of scale anomaly for the on-shell dilaton scattering amplitudes.} The optical theorem tells that logarithmic divergence comes from the on-shell $\varphi\varphi \to \mathrm{SFT}$ amplitudes and  they must vanish. 
The cancellation is very unlikely unless $T = 0$ (and ${O}=0$) \cite{Luty:2012ww} up to improvement, and this is how the scale invariance without conformal invariance  may be excluded. Alternatively speaking, generically, $\hat{a}$ is monotonically decreasing along the renormalization group flow for scale invariant but non-conformal field theory, and as long as we demand that $\hat{a}$ is determined by the local running coupling constants, it must be finite and the logarithmic divergence here is inconsistent. If we can modify the IR behavior of the scale invariant but not conformal theory by adding further relevant deformations so that it will flow to a conformal invariant fixed point (which could be a trivial one), the inconsistency is more transparent: at the conformal fixed point, $\hat{a}$ or $\bar{\alpha}$ is identical to $a$ and is bounded from below (recall $a \ge 0$ as discussed in section \ref{Cardyc}) so $\hat{a}$ cannot decrease forever.

To make the  above discussion more precise, we assume that $T^{\mu}_{\ \mu}$ is small and ${O}$ is negligible. 
Recall that  $T^{\mu}_{\ \mu} = \mathcal{B}^I {O}_I$, so in perturbation theory, the first assumption is valid. The second assumption is 
more tricky, but in dimensional regularization with the most generic renormalizable action (as discussed in section \ref{red}),
it is true as long as we use the suitably renormalized new-improved energy-momentum tensor.

Under these assumptions, the two-two dilaton scattering amplitude is dominated by $\langle \varphi \varphi |TT| \varphi\varphi \rangle $.
Then the logarithmic divergence of the scattering amplitude is equivalent to
\begin{align}
\frac{d\hat{a}_{\mathrm{IR}}}{d\log\mu} = \frac{d\bar{\alpha}}{d\log\mu} =  \mathcal{B}^I \chi_{IJ}\mathcal{B}^J + \mathcal{O}(\mathcal{B}^4) \ , 
\end{align}
where $T=\mathcal{B}^I {O}_I$, and $\chi_{IJ} \sim \langle O_I(x) O_J(0) \rangle x^8|_{|x|=\mu^{-1}}$. As discussed, we cannot absorb the logarithmic divergence in running coupling constants, so for the scale invariance,
it must vanish, which means $\mathcal{B}^I = 0$ and the theory is conformal invariant.

Therefore under some technical assumptions, scale invariant fixed points must be conformal invariant, and the  strong $a$-theorem
(or more precisely strong $\tilde{a}$-theorem) must hold in  perturbation theory.
Beyond the perturbation theory, the proof is not complete because although implausible $TTTT$ and ${O} {O}$ term
might cancel the divergence (the latter may be more dangerous because it can be only proportional to $\mathcal{B}^I$). This is related to the open question of the positivity of the metric in the strong $a$-theorem from the local renormalization group flow analysis.

Clearly, there is a parallel between the dilaton scattering amplitude and $\tilde{a}$-function discussed in the local renormalization group flow. There is a slight difference, however, beyond the leading order in $\mathcal{B}^I$ discussed above. One advantage of the dilaton scattering amplitude is that the derivative with respect to the energy scale is always positive (if we assume the optical theorem) unlike $\tilde{a}$-function of Osborn. One disadvantage, on the other hand, is that it is not obvious whether the averaged amplitude is a function of the running coupling constants. We will give a further comment on this point below.

In relation to the running coupling constant dependence on the $\tilde{a}$-function as well as $\bar{\alpha}$ from the dilaton scattering amplitude, it is crucial that the dependence is not a multi-valued function without any monodromy. Clearly, if we allow the monodromy structure, the monotonically decreasing function along the renormalization group flow is consistent with the cyclic renormalization group flow \cite{Cappelli:1991ke}\cite{Morozov:2003ik}\cite{Curtright:2011qg}. We have implicitly assumed this in the local renormalization group flow analysis but indeed we can check that $\tilde{a}$-function in perturbation theory is  a single-valued function on $g^I$. In any perturbation theory, we have an explicit power series expansion of $\tilde{a}$ with respect to coupling constants.
We expect the multivaluedness does not happen in the dilaton scattering amplitude either because it is a physical observable at a given energy. Furthermore when there is a monodromy in the $a$-function, it is inconsistent with the further possibility to deform the theory so that it flows to a well-defined conformal fixed point (in particular the trivial fixed point by mass deformation) with a fixed value of $a$. We will continue to assume that $a$-function is a single-valued function of the running coupling constants.

\subsubsection{Some technical comments on possible cancellation}
We recall that in section \ref{weak}, we assumed that the IR theory under study is Weyl invariant. We might  ask a pedantic question if all the IR fixed points of unitary quantum field theories are Weyl invariant (say in $d=4$ dimension). The answer to this question is no. Take massless QCD for example. In the IR limit,  the chiral symmetry is  spontaneously broken, and the IR theory is described by Nambu-Goldstone  bosons. They possess the shift symmetry, so the natural energy-momentum tensor is not traceless (such a question was raised in \cite{Cappelli:1990yc}\cite{Cappelli:2001pz} in the discussion of the $a$-theorem). After all, the scale transformation as well as Weyl transformation do not commute with the shift symmetry because the improved curvature coupling $R\phi^2$ is incompatible with the  shift symmetry. Note however, all the flat space local correlation functions of the Nambu-Goldstone bosons may be embedded in those of the conformal field theories (in the strict IR limit), so we do not call it as a counterexample.

We wanted to use the consequence of the $a$-theorem in  the above examples. For instance, the weak $a$-theorem (if applied) gives a constraint on the possibility of  symmetry breaking:  too many spontaneous symmetry breaking could be inconsistent with the constraint from the $a$-theorem \cite{Ball:2001wr} etc. 
Note that we can compare the Weyl anomaly of Nambu-Goldstone bosons with that of conformally coupled scalar to see the difference is only in $R^2$ term. In particular $a$ is same. From the viewpoint of counting degrees of freedom, it seems natural to count the Nambu-Goldstone boson and conformal scalar with the same unit. 

We may see a possible complication in general, but we would like to focus on the simplest Nambu-Goldstone boson case to  see what we would expect. The action for the Nambu-Goldstone boson has a ``trivial" virial current, so the coupling with the dilaton is non-zero $\int d^4 x \tau \partial^\mu J_\mu$ with $J_\mu = \phi \partial_\mu \phi$. Since we have used the equations of motion to derive $J_\mu$, we have to augment the contact terms. Also the virial current does not  transform covariantly under the Weyl transformation, so  we have to add higher terms. In the end, we obtain the complete coupling between the dilaton and Nambu-Goldstone boson as
\begin{align}
\mathcal{S} = \tilde{f}^2 \int d^4x e^{-2\tau} (\partial^\mu \phi \partial_\mu \phi) \ , \label{golda}
\end{align}
which is Weyl invariant and shift-symmetric. Note that the Weyl weight of $\phi$ becomes zero by higher dilaton couplings, which should be contrasted with the usual free scalar whose Weyl weight is one in $d=4$ dimension.

The two-two dilaton scattering may obtain logarithmic divergence due to the intermediate Nambu-Goldstone boson channel. However, we can easily see that various terms cancel against the virial current  exchange so that the  total cross-section  is zero \cite{Nakayama:2011wq}, and there is no logarithmic divergence, which would lead to the same inconsistency that we encountered in the perturbative study of scale but non-conformal IR theories due to $\varphi\varphi \to \mathrm{SFT}$ channels. A more direct way to see the cancellation is to realize  that we can perform a simple field redefinition of dilaton and Nambu-Goldstone boson so that they are completely decoupled (by going to ``Cartesian coordinate"  in  (\ref{golda})).

This is an example in which the cancellation of various terms can occur in (\ref{smatrixp}) when the contribution of each term is of same order (even in unitary examples). As emphasized in \cite{Luty:2012ww}, this kind of cancellation is generically very unlikely with no good physical reason (in the example here because of the Nambu-Goldstone symmetry). In non-unitary scale but non-conformal field theory such as Riva-Cardy model, we also encounter a similar cancellation.

As we have discussed, in  perturbation theory,  the cancellation does not occur  as long as we use the new-improved energy momentum tensor. See also general arguments in \cite{Luty:2012ww} how the improvement terms such as $\int d^4 x \sqrt{|g|}R O$ term in the effective action do not affect the argument for the weak $a$-theorem. We see that the terms that can be improved away and the terms that cannot be improved away give qualitatively different contributions to the dilaton scattering amplitude. Thus we conclude that the non-zero dilaton scattering is really an obstruction not only for the Weyl invariance but also for the conformal invariance up to the improvement since it is insensitive to the improvement terms. The converse, however, is slightly subtle because it is harder to tell if vanishing of the dilaton scattering directly means that the energy-momentum tensor is always improved to be zero as we will further discuss it in section \ref{nn}.

This types of coupling between dilaton and Nambu-Goldstone boson naturally appear in supersymmetric extension of our construction because the dilaton must accompany the $\mathcal{R}$-axion from superconformal symmetry, and the supersymmetric extension of the dilaton effective action always contains such couplings (see e.g. \cite{Schwimmer:2010za} and more recent discussions in \cite{Bobev:2013vta}). 

\subsubsection{Space-time dependent coupling constant counterterms}\label{spacet}
One of the key ingredients in local renormalization group flow analysis is the introduction of the space-time dependent coupling constants. In particular, we recall that the local counterterms associated with the space-time dependent coupling constants introduce the ambiguity in $\tilde{a}$-function. We would like to argue that the similar ambiguity can be introduced in the dilaton effective action as well as in the dilaton scattering amplitudes away from the conformal fixed point.

We recall that the effective action of the original theory (before the Weyl compensation) may contain additional space-time dependent coupling constant counterterms (in $d=4$ dimension) \cite{Osborn:1991gm}
\begin{align}
\mathcal{S}_{\mathrm{ct}} = &\int d^4x \sqrt{|g|} \left( \frac{1}{2} C^g_{IJ}(g) D_{\mu} g^I D^\mu g^J G^{\mu\nu} + A(g) \mathrm{Euler} \right. \cr
 &\left. -C^a_{IJ}(g) D^2 g^I D^2 g^J -\frac{1}{4} C^c_{IJKL}(g) D_\mu g^I D^\mu g^J D_\nu g^K D^\nu g^L + \cdots \right) \ . \label{psc}
\end{align}
These counterterms are related to various contact terms of operator $O^I$ and $T_{\mu\nu}$ in the flat space-time limit (see \cite{Osborn:1991gm} for details).
Once they are at the fixed point, the counterterms are Weyl invariant since $\mathcal{B}^I = \frac{dg^I}{d\log\mu} = 0$, so the introduction of these counterterms does not change the dilaton effective action. 
 However, away from the conformal fixed point, the counterterms are not Weyl invariant, and we need the dilaton compensation so that they will give extra contributions to the dilaton scattering amplitudes. Alternatively speaking, the counterterms (\ref{psc}) parametrize the ambiguities of the dilaton coupling beyond the leading order, which we have mentioned in (\ref{SFTdil}).

Let us focus on the first two terms $C^g_{IJ}$ and $A$, which will be the most important ones for our discussions.\footnote{Most of the other terms will not contribute to the dilaton scattering amplitudes in flat space-time in the leading order $\mathcal{O}(\mathcal{B}^2)$ once we impose the on-shell condition. See the argument below. The complete form of the integrated dilaton effective action away from the conformal fixed point can be found in \cite{Jack:2013sha}\cite{Baume:2014rla}.} For the $C^g_{IJ}$ term, within the order of perturbation we are interested in, one can replace $D_\mu g^I$ with $D_\mu g^I + \mathcal{B}^I \partial_\mu \tau$ and replace $G^{\mu\nu}$ with $\hat{G}^{\mu\nu}$ to make it Weyl invariant. Once we impose the on-shell condition and restrict ourselves to the flat space-time (i.e. $g_{\mu\nu} = \eta_{\mu\nu}$ and $D^{\mu}g^I = 0$), we obtain the additional four-dilaton terms
\begin{align}
\mathcal{S}_{\mathrm{ct}} = \int d^4x 2( C^g_{IJ} \mathcal{B}^I \mathcal{B}^J) (\partial _\mu \tau)^4 \ .
\end{align}
This is nothing but the ambiguity in defining $\tilde{a}$-function in local renormalization group flow (see section \ref{localr}) because $\tilde{a}$ appearing in the Wess-Zumino action is modified by the extra contribution $ \mathcal{B}^I C^g_{IJ} \mathcal{B}^J$. Correspondingly, the energy scale dependence of the dilaton scattering amplitude must be modified by various terms as in (\ref{ambigu}). Similarly, $A$ term (after dilaton compensation by replacing $A \to A - \tau \mathcal{B}^I \partial_I A$ in the leading order) does not contribute to the dilaton scattering amplitude  to the order we are interested in, which corresponds to the fact that $A$ term gives no correction to $\tilde{a}$ (but not in $a$ !) in local renormalization group flow. However $A$ does change the form of the gradient formula. It will give the gauge transformation to $w_I$.

The origin of these terms are the contact terms in higher order interaction terms in defining the dilaton scattering amplitudes. Although we may implicitly fix the scheme as in \cite{Luty:2012ww} or \cite{Yonekura:2012kb}, we need not. A different renormalization group prescription gives different dilaton scattering amplitudes, and in $d=2$ dimension, we have used this ambiguity to demonstrate that the metric $\chi_{IJ}$ is manifestly positive definite. 

We should realize that by choosing the counterterms, it is possible to make the $(\partial_\mu \tau)^4$ term positive in the dilaton effective action at a certain energy scale if we wish. Apparently, such a choice would be inconsistent with the unitarity and the optical theorem that follows. The contact terms must be chosen so that the unitarity of the dilaton scattering amplitudes must be intact  at any energy scale (and we believe there is such a choice). At this point, one may rephrase one of the implicit assumptions of \cite{Luty:2012ww}. There are good choices of counterterms so that the dilaton scattering amplitudes are unitary. Essentially, this is a different way to impose the positivity of the metric in the strong ${a}$-theorem. Obviously the renormalization by itself may not know the unitarity, and we have to respect it by choosing the good counterterms.

We emphasize that at the conformal fixed point, there is no such ambiguity, so our discussion for the weak $a$-theorem in section \ref{weak} is not affected. At the UV conformal fixed point, the space-time dependent counterterms for irrelevant perturbations are fine-tuned away because they are non-renormalizable, and those for relevant perturbations are zero because they have vanishing beta functions at the UV fixed point. Similarly, at the IR conformal fixed point, the space-time dependent counterterms for relevant perturbations are fine-tuned away because they must sit at the fixed point, and those for irrelevant perturbations are zero because they have vanishing beta functions at the IR fixed point.

As for the other terms (e.g. second line of (\ref{psc})), Osborn \cite{Osborn:1991gm}  used the freedom  to cancel various  local Weyl anomaly terms with space-time dependent coupling constant. In spirit, it is close to removing $\Box R$ term in Weyl anomaly.
In this prescription, the remaining ambiguity in the dilaton scattering amplitudes from the counterterms (\ref{psc}) is the above mentioned ambiguity in $\tilde{a}$-function from $C^g_{IJ}$. If we did not use this particular  prescription, then the relation between $\tilde{a}$ and dilaton scattering amplitude is less clear. They seem to deviate at order $\mathcal{O}(\mathcal{B}^4)$.

There are some other subtleties in identifying the averaged dilaton scattering amplitude $\bar{\alpha}$ and $\tilde{a}$-function. As mentioned at the end of section \ref{localr}, $\tilde{a}$-function is a function of the running coupling constants alone and does not depend on the energy scale $\mu$ explicitly, while it is not immediately obvious whether the averaged dilaton scattering amplitude  $\bar{\alpha}$ can be written as a function of the running coupling constants alone and does not depend on the energy scale in addition. 

Indeed, if we truncate the computation at order $\mathcal{O}(\mathcal{B}^2)$ as was done in \cite{Luty:2012ww} and use the leading order Zamolodchikov metric, the averaged dilaton scattering amplitude  $\bar{\alpha}$ is nothing but the averaged $c$-function discussed in section \ref{moron} (by assuming that the energy-momentum tensor used there was suitably improved). 
Thus it may seem more natural to identify the dilaton scattering amplitudes with the averaged $c$-function.
As discussed in section \ref{moron}, however, the averaged $c$-function is known to depend on the trajectory of the renormalization group flow, which is equivalent to our concern that the averaged dilaton scattering amplitude  $\bar{\alpha}$ is not a function of the running coupling constants alone, but it may depend on the energy scale $\mu$ separately or on the renormalization group trajectory.

The reason why this kind of complication happens is that the renormalization group flow $\frac{d c^M_d}{d\log\mu}$ of the  averaged $c$-function (assuming the improved energy-momentum tensor) and  that of Osborn's $\tilde{a}$-function $\frac{d \tilde{a} }{d\log\mu}$ as well as that of the averaged dilaton scattering amplitude $\frac{d\bar{\alpha}}{d \log \mu}$ all coincide with one another up to the order we have computed. In particular, the ``Zamolodchikov metric" $\chi_{IJ}$ appearing in the averaged $c$-function (as well as in the averaged dilaton scattering amplitude $\bar{\alpha}$) is essentially identical to the metric $\chi_{IJ}^g$ appearing in the $\tilde{a}$-flow equation in many usual renormalization schemes such as dimensional regularization (at least up to two-loops).
In \cite{Fortin:2012hn}, however, they pointed out that the ``metric" $\chi_{IJ}$ for the averaged dilaton scattering amplitude at the leading order (which is identical to that for the averaged $c$-function in section \ref{moron}) is related to $\chi_{IJ}^a$,  and it can be different from that of the $\tilde{a}$-function, namely $\chi^g_{IJ}$, at the third order or higher.\footnote{We recall $\chi_{IJ}^a$ is derived from $D^2 g^I D^2 g^J$ term in the Weyl anomaly, so it governs the Weyl non-invariance of the contact terms in $\langle O_I(x) O_J(0) \rangle$. Thus, it is related to the Weyl non-invariance of $\langle \Theta (x) \Theta (0) \rangle$. On the other hand $\chi_{IJ}^g$ appears at higher point functions of $T_{\mu\nu}$ and $O_I$. See \cite{Osborn:1991gm} for more details.}

Of course, we have neglected the higher order corrections in the dilaton scattering amplitudes, too. The ``metric" that appears in the dilaton scattering amplitude can be different from $\chi_{IJ}$ in the averaged $c$-theorem beyond the leading order.
 In addition, the ambiguities discussed here are not treated carefully, so it is premature to say whether or not there is a good ``scheme" in which the metric in the dilaton scattering amplitude and $\chi_{IJ}^a$  really coincide to give a clue for the non-perturbative result as in $d=2$ dimension.

Let us summarize the non-perturbative situation. Osborn's $\tilde{a}$-function is a function of the running coupling constants at the given energy scale alone, but it is not obvious if it shows monotonicity. The averaged $c$-function $c^M_d$ is by definition monotonically decreasing, but it is not (a priori) defined as a function of the running coupling constants alone and presumably it is not. Finally, the averaged dilaton scattering amplitude $\bar{\alpha}$ is monotonically decreasing once we assume the optical theorem, but it is again not (a priori) obvious if it is a function of the running coupling constants alone. In the perturbative approach, they differ at $\mathcal{O}(\mathcal{B}^3)$ or higher but we could not see this distinction at $\mathcal{O}(\mathcal{B}^2)$.\footnote{The difference in the integrated form was discussed in \cite{Anselmi:2001yp} as a sum rule. The positivity of the difference is tricky due to the contact terms.}

In principle, the counterterms discussed here might modify the argument for the enhancement from scale invariance to conformal invariance, but the effect is always higher order $\sim \mathcal{O}(\mathcal{B}^4)$, so it does not change the perturbative conclusion in section \ref{svc}. The perturbative proof --- both from the analysis of $\tilde{a}$-function \cite{Fortin:2012hn} based on the local renormalization group analysis \cite{Osborn:1991gm} and from the dilaton scattering amplitude \cite{Luty:2012ww} --- that scale invariance implies conformal invariance is therefore robust.\footnote{See also appendix of \cite{Nakayama:2012sn} for the comment that Osborn's argument essentially implies that scale invariance must be enhanced to conformal invariance in perturbation theory. The argument there is not modified either.}

\subsection{$n-n$ dilaton scattering amplitude}\label{nn}

We have seen that the on-shell two-two dilaton scattering amplitude must vanish for the scale invariant field theories due to the absence of the scale invariant local counterterms. Generalizations to $n-n$ dilaton scattering amplitudes are possible \cite{Dymarsky:2013pqa} and they give a further severe constraint on the scale invariant but non-conformal field theories.

In perturbation theories, these higher generalizations can be intuitively understood as follows. We recall that the on-shell dilaton $\varphi$ couples to the beta function as $\int d^4x \log(1+\varphi/f) \mathcal{B}^I {O}_I$ in the leading order. We can clearly see that the leading contribution to the $n-n$ dilaton scattering amplitude comes from $\mathcal{B}^I \mathcal{B}^J \langle {O}_I {O}_J \rangle$ as in two-two dilaton scattering amplitude whenever beta functions are small in the perturbative regime. We will see that the same logic for the two-two dilaton scattering amplitude applies here and all of them must vanish.
 The vanishing of $n-n$ dilaton scattering amplitude for any $n$ seems unlikely unless the on-shell contribution from the energy-momentum tensor is trivial, which is nothing but the condition for the conformal invariance upon improvement.

A more precise discussion on the generalization to the $n-n$ scattering amplitude goes as follows \cite{Dymarsky:2013pqa}.
We first note that once we impose the on-shell condition $\Box \varphi = 0$, there is no scale invariant local counterterms for the dilaton scattering amplitude. In particular, the forward scattering amplitude, where $p_1 = -p_{n+1}$, $p_2 = -p_{n+2} \cdots$ must have a finite dispersion relation. The dimensional analysis tells that the $n-n$ dilaton forward scattering amplitude ${A}_{2n}$ must satisfy the canonical scaling 
\begin{align}
\mathrm{Im} A_{2n}(\lambda p) = \lambda^2 \mathrm{Im} A_{2n}(p) \ ,
\end{align}
but this can be consistent with the finiteness of the dilaton scattering amplitude from the dispersion relation as discussed in section \ref{weak} in the case of $n=1$ if and only if $\mathrm{Im} A_{2n} (p) = 0$. Otherwise, the dispersion integral is logarithmically divergent. This could only mean that the S-matrix is not well defined. However, the following heuristic argument applies. Suppose we reguralize our scale invariant field theory both in IR and UV. The low energy $n-n$ on-shell dilaton scattering amplitude are completely specified by the IR Wess-Zumino term in the dilaton effective action with finite $a_{\mathrm{UV}} -a_{\mathrm{IR}}$ coefficients, but the difference must be finite and cannot be arbitrarily large. If the unregulated amplitude is logarithmically divergent, we could make it arbitrarily large and it is inconsistent.

Now we can use the unitarity and optical theorem to argue that on-shell $n$ dilaton does not couple to any physical states. Apparently, the infinitely many conditions like this is very hard to achieve unless the on-shell dilaton is essentially a free field, and the theory is improved to be conformal with $\mathcal{B}^I = 0$. However, note again that this still does not exclude the possibility that the higher order contributions (with possible semi-local contact term contributions from the operators other than the energy-momentum tensor) do cancel against the leading contributions from the beta functions with effectively degenerate metric $\chi_{IJ}$ for all $n-n$ dilaton scattering amplitudes simultaneously.\footnote{A similar analysis was done in the momentum space correlation functions with careful treatment of the semi-local terms in \cite{Skenderis}, where they have found that the semi-local term not encoded in the dilaton scattering amplitudes may explain the dilatation anomaly so that the relation between vanishing of the on-shell dilaton scattering and vanishing of the trace of the energy-momentum tensor is not direct from the operator analysis alone. Their analysis revealed that the dimension 2 scalar operator, which is related to the second variation of the energy-momentum tensor under Weyl transformation that we have mentioned in section \ref{svc}, is needed to explain the semi-local terms.}
 In non-unitary scale invariant but non-conformal field theories, this simultaneous cancellation indeed does happen.

The folklore in quantum field theories says that the triviality of the on-shell scattering amplitude implies that the coupling can be completely removed by the field redefinition. In our case, this field redefinition is nothing but the improvement transformation. Within the class of theories for which this argument applies, the scale invariance is equivalent to conformal invariance in $d=4$ dimension.

We emphasize the role of the on-shell condition $\Box \varphi = 0$. First of all, we do not expect to be able to show $T_{\mu}^{\mu} = 0$ as an operator identity directly in higher dimensions because of the improvement ambiguities. In $d=2$ dimension, it was possible because the unitarity does not allow non-trivial improvement at the fixed point, but we cannot expect the similar argument in higher dimensions.
The on-shell condition is convenient for this purpose because it precisely encodes the possibility that $T_{\mu}^{\mu}$ can be non-zero but can be improved to be traceless (at the leading order). At least, on-shell dilaton scattering amplitudes are not affected by the improvement. 
On the other hand, one drawback is we cannot use the powerful field theory theorem like Reeh-Schlieder theorem for the on-shell restricted correlation functions, so we had to resort to a more intuitive argument of triviality of physical S-matrix. In relation, the triviality of the S-matrix may be explained by assuming the trace of the energy momentum tensor $T^\mu_{\ \mu}$ is a generalized free field (see e.g. section \ref{without}) rather than being improved to be traceless because the S-matrix has a contribution from the ``connected" correlation functions, and the generalized free field gives only disconnected correlation functions although the operator itself is non-zero \cite{Dymarsky:2013pqa}. 
If there are no dimension 2 scalar operators, however, such a possibility can be ruled out by studying the dilatation anomaly in three-point functions as pointed out in \cite{Dymarsky:2014zja} (see also \cite{Skenderis}).

 Finally, as a technical remark, we note that the use of S-matrix may have an intrinsic problem if the theory cannot be IR regularized because  due to the possible IR divergence it may not be well-defined. Most of the known scale invariant quantum field theories can be deformed in the IR so that the IR theory is gapped, but this is not always the case like chiral gauge theories. Without the assumption of the existence of the IR regularization, the claim that the dilaton scattering amplitude must be bounded is not obvious.\footnote{Actually, the same subtlety appeared in Zamolodchikov's argument in $d=2$ dimension. His $c$-function is not manifestly positive definite in its definition away from the conformal fixed point, so it is logically possible that it decreases forever because there is no way to make the theory gapped by deformations.}

\subsection{Physical reason why scale invariance implies conformal invariance in perturbative fixed point}\label{reason}

In the last couple of sections, we have discussed how scale invariant but non-conformal field theories are inconsistent with the perturbative $a$-theorem. We have showed that the suitably defined $a$-function (whose particular realization is the dilaton scattering amplitudes) decreases at the rate dominantly proportional to the bilinear of $\mathcal{B}$-functions in perturbation theory. In order to achieve the finite dilaton scattering amplitude, it must be accompanied with vanishing $\mathcal{B}$ functions near the asymptotic IR region, and therefore it must show conformal invariance (up on possible improvement of the energy-momentum tensor). 

If we perform an explicit computation, however, we may realize something a little bit more about the structure of the renormalization group flow. We realize that $\mathcal{B}$ functions in certain directions are identically zero irrespective of if the coupling constants are at the fixed point or not. In addition, within a few orders of perturbation theory, the zero-directions directly correspond to the ``would-be" virial current direction that may induce scale invariance without conformal invariance by using the equations of motion. We have seen this phenomena explicitly in section \ref{reddd} when we discussed the conformal perturbation theory.

This may sound unexpected because we are solving the weaker equation $T^{\mu}_{\ \mu} = \partial^\mu J_\mu$ rather than $T^{\mu}_{\ \mu} =0$ by introducing extra parameters in $J_\mu$ (as many as the number of non-conserved current). Why can't we expect more solutions generically?
The argument based on the $a$-theorem gives the constraint near the fixed point, but what is the origin of these zero directions during the entire renormalization group flow?

To see it in a simple example, we consider Yukawa interaction $y\phi \psi^2$ in $d=4$ dimension. The Yukawa interaction has the one-loop $\mathcal{B}$ function
\begin{align}
 T^\mu_{\ \mu} = \mathcal{B}^I O_I = \frac{|y|^2}{16\pi^2} y (\psi \psi \phi) + c.c. +\mathcal{O}(\phi^4) \ .
\end{align}
We realize that the $\mathcal{B}$ function is orthogonal to the direction $iy (\psi \psi \phi) + c.c.$, which can be rewritten as the divergence of the non-conserved current (which is given by $i\partial^\mu (\bar{\psi} \gamma_\mu \psi)$). In other words, the phase of the Yukawa coupling constants are not renormalized (as observed in \cite{Dorigoni:2009ra}\cite{Fortin:2011ks}), and this is the reason why the ``would-be" virial current does not appear in this one-loop computation. We can easily see that the phase of the Yukawa coupling is unphysical from the beginning because we can remove it by a field redefinition of $\psi$. Since  the phase is not a parameter of the theory nor does it affect any observables of the theory, it had better not show any physical consequences in the renormalization group flow. 
Of course,  the strong $a$-theorem  $\frac{d\tilde{a}}{d\log\mu} \sim \frac{1}{16\pi}|y|^6$ does tell that the $\tilde{a}$-function must be decreasing along the renormalization group flow, but again note that the phase of the Yukawa coupling does not appear in the $\tilde{a}$-function, either.

As discussed in section  \ref{reddd}, the $\mathcal{B}$ function that can be completely rewritten as a divergence of a virial current is related to redundant directions in the renormalization group flow. The physical reason why the scale invariant but non-conformal field theory is difficult to achieve in perturbation theory can be understood as the claim that redundant directions do not acquire $\mathcal{B}$ functions as can be checked directly within a first few orders of perturbation theory. At higher orders, $\mathcal{B}$ functions in the virial current direction can be non-zero, but they still possess the zero directions in the $\mathcal{B}$ function flow, which essentially excludes the cyclic renormalization group flow.

The above observation can be made more precise when we can find the counterterm in which $w_I = 0$ in local renormalization group flow discussed in section \ref{localr}.\footnote{This is shown to be possible within the first few orders in perturbation theory. The superpotential flows in holography also suggest this is the case in explicit holographic models. In general this may not be true unless $w_I$ is exact, but the author does not know any explicit examples that show $w_I$ is not exact.} 
In this situation, the $\mathcal{B}$ functions must have the same number of zero directions from the consistency condition $\mathcal{B}^I = g^{IJ} \partial_J \tilde{a}$ together with the fact that $\tilde{a}$ is a singlet under the ``flavor" symmetries generated by the candidates of the virial current.

Let us label $i,j$ as the direction corresponding to the ``would-be" virial current direction, and $M,N$ as the direction that cannot be transformed to currents. The Wess-Zumino consistency condition of the local renormalization group and the ``flavor" invariance of $\tilde{a}$ requires 
\begin{align}
0 = \partial_i \tilde{a} = g_{iM} \beta^M + g_{ij} \mathcal{B}^j \ . \label{zeromode}
\end{align}
Suppose $\beta^M = 0$ but $\mathcal{B}^i \neq 0$ so that we obtain scale invariant but non-conformal fixed point. This cannot occur because it is inconsistent with the expected zero directions in $\mathcal{B}$ function flow (\ref{zeromode}) as long as $g_{ij}$ is non-degenerate (as expected because $g_{ij}$ must be isometric under the ``flavor" symmetry group). 

Even when $w_I$ is not zero, we know that $\mathcal{B}^I$ functions are not independent because $\partial_i \tilde{a} = 0$ gives the constraint along the renormalization group flow, which reflects the fact that there are as many redundant perturbations in renormalization group flow as the number of candidates for the virial current. The condition (\ref{zeromode}) is replaced by
\begin{align}
 0 = \partial_i \tilde{a} = (g_{iM} +w_{iM}) \beta^M + (g_{ij} + w_{ij}) \mathcal{B}^j + (\hat{\rho}_ig)^j w_j  
\end{align}
By demanding $\beta^M=0$ and contracting it with $\mathcal{B}^i$, we obtain the same conclusion that $\mathcal{B}^i = 0$ (and hence it must be conformal invariant) as long as $g_{ij}$ is non-degenerate.

In summary, we have two physical intuitions for non-existence of (perturbative) scale invariant but non-conformal field theories in power-counting renormalization scheme. The first one is the $a$-theorem: the dilaton scattering amplitude (or Osborn's $\tilde{a}$-function) must be bounded and it cannot decrease forever from non-trivial $\mathcal{B}$ functions in scale invariant but non-conformal field theories. The second one is that $\mathcal{B}$ functions in the directions that might be used to  construct a non-trivial virial current are actually redundant directions in perturbation theory. Even at higher orders, we  expect to keep the same number of zero directions in the $\mathcal{B}$ function flow as the number of redundant directions. 
 The both intuitions are beautifully realized by the generalized ``gradient formula" if true.
In holographic computations, we will precisely see  both obstructions if we try to construct the gravitational dual of scale invariant but non-conformal field theories.

In the above argument, we have not talked about the importance of dimension 2 operators in $d=4$ dimension, and the argument is very similar to the one in $d=2$ dimension. Within power-counting renormalization scheme, the dimension 2 operators do not show any essential contributions to the structure of the local renormalization group in relation to the $a$-theorem, but the mixing with the energy-momentum tensor may give additional subtlety in our discussions on the relation between scale invariance and conformal invariance beyond perturbation theory as we have mentioned a few times in this section. Clearly, the possibility of the renormalization of the energy-momentum tensor is one novel complexity in higher dimensions that did not appear in $d=2$ dimension, and it has given a small gap between the establishment of $a$-theorem and the enhancement of conformal invariance in higher dimensions compared with the situations in $d=2$.
It would be very interesting to obtain more intuitive physical understanding of the role of these dimension 2 operators.

\newpage

\section{Other dimensions or less symmetry}\label{section9}

We have discussed the problem of the possible enhancement of conformal invariance from scale invariance mainly in $d=2$ and $d=4$ dimensions. In this section, we would like to review the situations in the other dimensions, which is less understood. Also we would like to review the extension of our program in less symmetric situations e.g. without assuming the full Poincar\'e invariance. 

\subsection{Summary of the situations in other dimensions}

Let us first summarize the situations of scale invariance vs conformal invariance in various dimensions other than $d=2$ and $d=4$ dimensions.

\begin{itemize}

\item  In $d=1$, due to the lack of Poincar\'e invariance, we cannot use the Reeh-Schlieder theorem. This is a major drawback. 
If we assume its validity then scale invariance
implies conformal invariance \cite{Nakayama:2012ed}. Similarly, the  boundary $g$-theorem \cite{Affleck:1991tk}, which claims that boundary entropy of the two-dimensional system is monotonically decreasing along the renormalization group flow, can be proved \cite{Friedan:2003yc}\cite{Friedan:2005dj}. On the other hand, a quantum field theory in $d=1$ is equivalent to a simple quantum mechanical system, and there are examples of cyclic renormalization group flow \cite{Efimov:1970zz} realized in non-relativistic field theories \cite{Nishida:2007de} as well as the system with scale invariance without conformal invariance \cite{Nakayama:2012ed}. In these cases, the Reeh-Schlieder theorem does not hold so the formal argument does not apply.

\item In $d=3$ dimension, a candidate of Zamolodchikov's $c$-function is the finite part of the $\mathbb{S}^3$  partition function $F=-\log Z_{\mathbb{S}^3}|_{\mathrm{reg}}$ as we will elaborate a little more in section \ref{ftheorem}. This is equivalent to the finite part of  the entanglement entropy of the half $\mathbb{S}^3$ when the theory is at the conformal fixed point. It is an interesting open question if there is a strong version of the $F$-theorem that would imply enhancement from scale invariance to  conformal invariance 
in $d=3$ dimension. We have more to say in section \ref{ftheorem}.

\item In even dimensions, Cardy's conjecture (or $a$-theorem) has a natural generalization: the coefficient in front of the Euler density in the Weyl anomaly must be monotonically decreasing along the renormalization group flow.
In $d=6$ dimension, so far we have not been successful in using the dilaton-scattering argument to show the weak version of the $a$-theorem. A reported problem  \cite{Elvang:2012st} is that
it is hard to  show the positivity of the dilaton scattering amplitudes in $d=6$ dimension. On the other hand, there is no counterexample of $a$-theorem reported and there is no known scale invariant but non-conformal invariant field theories (with gauge invariant scale current). Within perturbation theory, the argument similar to the one presented in section \ref{localr} can be found in \cite{Grinstein:2013cka} and will be reviewed in the following. In $d=6$ dimension, a possible non-trivial (super)conformal fixed point may be related to the so-called little string theory\footnote{We believe that the little string theory with $(2,0)$ supersymmetry becomes superconformal in the IR limit. On the other hand, the little string theory with $(1,1)$ supersymmetry  becomes free super Yang-Mills theory in the IR limit, which is scale invariant but not conformal invariant. We refer \cite{Aharony:1999ks} for a review of the little string theory.}

\item In higher dimension $d\ge 7$, it is likely that there is no interacting unitary conformal field theory, but there is no proof of it. Certainly, there is no classical scale invariant Lagrangian with two-derivative kinetic terms other than free field theories.
The reason why higher dimensional free Maxwell theory cannot be conformal invariant is consistent with the fact that there is no superconformal algebra in $d\ge7$, but we know supersymmetric Maxwell theories exist in $d\le 10$. If it were conformal, it would be inconsistent with the non-existence of the superconformal algebra \cite{Nahm:1977tg} (unless it breaks supersymmetry). We note that Nahm's classification is based on the assumption of the existence of the S-matrix, so it does not exclude the possibility of superconformal membrane field theories in higher dimension than $6$.

\end{itemize}

\subsection{Possible directions}
\subsubsection{F-theorem in $d=3$ dimension}\label{ftheorem}
In section \ref{amax}, we briefly discussed the relation between the problem of finding the superconformal $\mathcal{R}$-symmetry in superconformal field theories and the $a$-theorem via the $a$-maximization principle. A similar question arises in the other dimensions. In $d=2$ dimension, we can apply the completely similar argument: in order to determine the superconformal $\mathcal{R}$-symmetry, we can maximize the trial $c$-function \cite{Benini:2012cz}, and it gives a new perspective of the $c$-theorem in supersymmetric field theories in $d=2$ dimension. 
What about the situation in $d=3$ dimension, where we have no Weyl anomaly constructed out of the background metric?

The question was answered in the paper \cite{Jafferis:2010un} by Jafferis. He proposed that in order to determine the superconformal $\mathcal{R}$-symmetry of $\mathcal{N}=2$ superconformal field theories in $d=3$ dimension, we can minimize the (real part of the) supersymmetric free energy on $\mathbb{S}^3$ (hence it is known as $F$-theorem, $F$ referring to the free energy). 
Due to a subtle anomaly in the contact terms in $\mathcal{N}=2$ superconformal field theories in $d=3$ dimension, the partition function acquires a non-trivial phase \cite{Closset:2012vp}, but the phase is irrelevant for the discussion here. The derivation is based on the supersymmetric localization technique to compute the partition function and the holomorphic dependence on the real mass parameters of the $\mathcal{N}=2$ supersymmetric field theories. We refer to \cite{Jafferis:2010un} for the discussions.

Given the analogy with Cardy's conjecture in even dimensions, it is natural to conjecture that the free-energy on $\mathbb{S}^3$ should give a natural candidate for Zamolodchikov's $c$-function in $d=3$ dimension \cite{Jafferis:2011zi}\cite{Klebanov:2011gs}. Indeed, there have been various non-trivial checks  if this is indeed the case. Perturbative arguments have been presented both with slightly relevant deformations \cite{Klebanov:2011gs} as well as with marginal deformations in \cite{Yonekura:2012kb}. Unlike the supersymmetric situation, where the localization computation is available, the regularization can be very tricky and there are some issues \cite{Klebanov:2012va}.\footnote{It is important to note that the topological Chern-Simons gauge theories give non-zero contributions to the $\mathbb{S}^3$ partition function although they do not carry any dynamical degrees of freedom. Thus, the argument based on the local properties of the theory may not capture some physics here.} However, the formal argument that the $\mathbb{S}^3$ partition function is related to the entanglement entropy at the conformal fixed point suggests that it seems a promising direction to pursue (see the following sections on the entanglement entropy). The establishment of the strong $F$-theorem is related to the question of enhancement from scale invariance to conformal invariance in $d=3$ dimension (see e.g. \cite{Nakayama:2012ed}\cite{Yonekura:2012kb} for perturbative argument), and the non-perturbative understanding is desirable. 

The heuristic argument goes as follows. We first note that the conformal invariance and the absence of the conformal anomaly in $d=3$ dimension dictate that if we evaluate the one-point function of a conformal primary operator on $\mathbb{S}^3$, it must vanish. This means that the sphere partition function does not depend on the exactly marginal deformation (by assuming the deformation is given by conformal primary operators from unitarity):
\begin{align}
\frac{\partial}{\partial g^I} Z^{\mathrm{CFT}}_{\mathbb{S}^3} = 0 \ .
\end{align} 
Near the conformal fixed point,\footnote{Once we are away from the conformal fixed point, there is a certain ambiguity in defining the partition function. The improvement terms in the action do change the partition function itself, but it will not change the expectation value of the trace of the energy-momentum tensor. In addition, there are counterterm ambiguities that may introduce ambiguities  in $\hat{g}_{IJ}$.}
 the one-point function does not vanish, but the deviation must be proportional to $\mathcal{B}^I$ function:
\begin{align}
\frac{\partial}{\partial g^I} Z_{\mathbb{S}^3} = \mathcal{B}^J \hat{g}_{IJ} \ .
 \label{gradientt}
\end{align} 
We can show (e.g. \cite{Nakayama:2012ed}\cite{Yonekura:2012kb}) that in perturbation theory $\hat{g}_{IJ}$ is equivalent to the Zamolodchikov metric in certain loop orders, but it can contain the antisymmetric part at  higher loop orders. 

We now compute the scale dependence of the sphere partition function.
Since $T^{\mu}_{\ \mu} = \mathcal{B}^I O_I$, we can derive
\begin{align}
\frac{\partial}{\partial\log \mu} Z_{\mathbb{S}^3} = \mathcal{B}^I \hat{g}_{IJ} \mathcal{B}^I \ , \label{secondderiv}
\end{align}
which gives the perturbative F-theorem with the renormalizability as long as $\hat{g}_{IJ}$ is positive definite. 

On the other hand, if the theory is scale invariant (but not necessarily conformal invariant), the energy-momentum tensor is divergence of a certain current: $T^{\mu}_{\ \mu} = D^\mu J_\mu$. Thus, the partition function cannot depend on the scale as long as the virial current is well-defined.\footnote{A counterexample is the three-dimensional free Maxwell field theory because the virial current in the dual frame is not gauge invariant as we showed in section \ref{examples}. The author would like to thank Z.~Komargodski for pointing out the fact.} Therefore we should obtain 
\begin{align}
\frac{\partial}{\partial\log \mu} Z^{\mathrm{SFT}}_{\mathbb{S}^3} = \mathcal{B}^I \hat{g}_{IJ} \mathcal{B}^I = 0 \ ,
\end{align}
which is possible only if $\mathcal{B}^I=0$ as long as the metric $\hat{g}_{IJ}$ is positive definite. Therefore, we can conclude that  scale invariance must imply conformal invariance within perturbation theory in $d=3$ dimension.

For a consistency check of the above argument, we would like to point out the effect of the operator identity $\mathcal{B}^I O_I = \beta^I O_I + v^a \partial_\mu J^\mu_a$. Since the total derivative does not contribute to the partition function after spatial integration, \eqref{secondderiv} can be also written as
\begin{align}
\frac{\partial}{\partial\log \mu} Z_{\mathbb{S}^3} = \beta^I \hat{g}_{IJ} \mathcal{B}^I \ . \label{secondderiv} 
\end{align}
However, the partition function is invariant under the ``flavor rotation" induced by $J^\mu_a$, so \eqref{gradientt} tells that $\beta^I \hat{g}_{IJ} \mathcal{B}^I=  \mathcal{B}^I \hat{g}_{IJ} \mathcal{B}^I$. At the scale invariant fixed point, $\mathcal{B}^I = 0$ and the conformal invariance is recovered irrespective  of our gauge freedom in the definition of the beta functions. The structure is  in complete parallel with that in $d=4$ dimension.

\subsubsection{In relation to entanglement entropy}
It was observed that the Weyl anomaly is intimately related to the entanglement entropy of the vacuum states in relativistic field theories (see e.g. \cite{Calabrese:2004eu} for a review of the entanglement entropy in quantum field theories). 
The entanglement entropy is defined as follows. We first divide the space into two domains $A$ and $\bar{A}$. We compute the partial density matrix $\rho_A = \mathrm{Tr}_{\bar{A}} \rho_{\mathrm{tot}}$, where $\rho_{\mathrm{tot}}$ is the density matrix of the total system, and in particular in our case it is the pure vacuum state $\rho_{\mathrm{tot}} = |0\rangle \langle 0|$. The entanglement entropy is given by 
\begin{align}
S_A = -\mathrm{Tr}_A \rho_A \log \rho_A \ .
\end{align}

It satisfies various interesting properties such as 
\begin{align}
S_A = S_{\bar{A}} \ .
\end{align}
In particular, the so-called strong subadditivity holds (see e.g. \cite{entangle} for a review. See \cite{subadditivity} for its original derivation):
\begin{align}
S_{A+B+C} + S_B \le S_{A+B} + S_{B+C} \ . \label{suba}
\end{align}

In relativistic quantum field theories on $\mathbb{R}^{d-1} \times \mathbb{R}^t$, one can argue that the (regularized) entanglement entropy has the following structure
\begin{align}
S = \cdots + (-1)^{\frac{d}{2}-1} a_d \log(R_{\mathrm{IR}}/\epsilon) + \cdots \cr
\end{align}
for even dimension $d$, and
\begin{align}
S = \cdots + (-1)^{\frac{d}{2}-1} a_d + \cdots \cr \ 
\end{align}
for odd dimension $d$,
where $R_{\mathrm{IR}}$ is a typical IR scale of the entangling surface $\partial A$ that divides $A$ and $\bar{A}$, and $\epsilon$ is the UV cut-off to regularize the trace over the field theory Hilbert space.
In these expressions, we have neglected both the UV diverging terms $a_2 \left(\frac{R_{\mathrm{IR}}}{\epsilon}\right)^{{d-2}} + a_3\left(\frac{R_{\mathrm{IR}}}{\epsilon}\right)^{d-3} + \cdots$ and the finite terms $a_{d+1} \left(\frac{R_{\mathrm{IR}}}{\epsilon}\right)^{-1} + a_{d+2} \left(\frac{R_{\mathrm{IR}}}{\epsilon}\right)^{-2} \cdots$. The leading diverging part is known as the ``area law" because it is proportional to the area of the entangling surface.

The universal contribution $a_d$ does depend on the shape of the entangling surfaces $\partial A$ as well as the background geometry, but if we take $\partial A $ to be $(d-2)$ dimensional sphere $\mathbb{S}^{d-2}$ inside $\mathbb{R}^{d-1}$, it can be argued that in even dimension $d$, the coefficient $a_d$ coincides with the coefficients of the Euler density in the Weyl anomaly if the quantum field theory under consideration is a conformal field theory \cite{Solodukhin:2008dh}\cite{Casini:2011kv}. Thus we have an alternative way to define the $a$-function in even dimension by using the universal part of the entanglement entropy of the vacuum state divided by $\mathbb{S}^{d-2}$.

Moreover, in $d=2$ dimension, one may even prove the inequality $a_{\mathrm{UV}} \ge a_{\mathrm{IR}}$ by using the strong subadditivity condition (\ref{suba}) together with the Lorentz invariance  in a clever way \cite{Casini:2004bw}. Here the strong subadditivity is replacing the unitarity condition in Zamolodchikov's argument: of course the unitarity (in the sense that there is no negative norm state) was crucially assumed in the derivation of the strong subadditivitiy.
It is an open question if the similar argument is possible in $d=4$ dimension to derive the weak $a$-theorem. It is also an interesting question if there is any pathology if we consider the scale invariant but non-conformal field theory and study the properties of the entanglement entropy. In a recent paper \cite{Solodukhin:2013yha}, the author computed the entanglement entropy of the dilaton compensated effective field theories in the flat Minkowski space-time in $d=4$ dimension, and argued that it is governed by the $G_{\mu\nu} \partial^\mu \tau \partial^\nu \tau$ term in the dilaton effective action. The  evaluation of the term led to the similar expression to the averaged $c$-function $c_d^M$.\footnote{Again, we should point out that the effective action term has the contact term ambiguity as discussed in section \ref{spacet}.}

In odd dimensions, there is no Weyl anomaly, but it is conjectured that the universal part of the entanglement entropy can be used as a candidate of the $a$-function. There are some supporting evidence from AdS/CFT correspondence \cite{Casini:2011kv}. In $d=3$ dimension, more detained analysis is needed, but there have been some attempts in this direction, suggesting that the (renormalized) entanglement entropy is monotonically decreasing along the holographic renormalization group flow \cite{Casini:2012ei}. Again, it is an interesting question to ask if the enhancement  from scale invariance to conformal invariance follows from such an argument.

One remark is that away from the conformal fixed point, the entanglement entropy and the $c$-function may differ. Of course, with the dimensionful parameters, the way to assign the ``diverging part" of the entanglement entropy is ambiguous. However, one important observation made in the literature is that the monotonically decreasing entanglement entropy in $d=2$ dimension is slightly different from Zamolodchikov's $c$-function defined away from the critical point by the two-point functions of the energy-momentum tensor \cite{Casini:2004bw}. 
By itself, it is not so crucial because after all, the interpolating $c$-function has some arbitrariness associated with the renormalization group ambiguities,  and the monotonically decreasing one is not unique. 
However, it was argued that the gradient properties do not seem to hold for the entanglement entropy \cite{Klebanov:2012va} and it means that there may be no good renormalization scheme that gives the direct relation to the $c$-function. 
Because of these subtleties, it is not obvious if the argument from the entanglement entropy (even if proved to be monotonically decreasing under the renormalization group flow) can be used to show the enhancement of conformal invariance from scale invariance. Even in $d=2$ dimension, the direct argument has not been established yet.

\subsubsection{Local renormalization group in other dimensions}
The local renormalization group analysis discussed in section \ref{localr} can be generalized to the other dimensions, possibly to argue for the enhancement of conformal invariance from scale invariance by providing a suitable generalization of Zamolodchikov's $c$-theorem. The generalization depends on the dimensionality of the space-time.

In any space-time dimension, the consistency of the local renormalization group demands that the local renormalization group operator $\Delta_{\sigma}$ must commute. This leads to many constraints on various renormalization group beta functions. For instance, if we consider the not-necessarily conserved vector operators in  any $d$-dimension, their beta functions $ \hat{\rho}^a_I D_\mu g^I$ must be orthogonal to the scalar beta functions.
\begin{align}
\hat{\rho}^a_I \mathcal{B}^I = 0 \ .
\end{align}

On the other hand, the consistency conditions coming from anomaly depends on the dimension of the space-time, in particular under the assumption of power-counting renormalization. The situations drastically differ in odd and even dimensions as we have already discussed from different viewpoints in this section.

In odd dimension, there is no CP even trace anomaly in massless local renormalization group.\footnote{There exists CP even trace anomaly once we include the dimensionful coupling constants.} However, there can be CP violating trace anomaly such as
\begin{align}
T^{\mu}_{ \ \mu} = -\epsilon^{\mu\nu\rho}  C_{IJK} D_\mu g^I D_\mu g^J D_\rho g^K - \epsilon^{\mu\nu\rho} f_{\mu\nu} C_I D_\rho g^I
\end{align}
in $d=3$ dimension \cite{Nakayama:2013wda}. Beta functions must satisfy various consistency conditions from the Wess-Zumino integrability condition. In the above case, integrability condition demands
\begin{align}
3\mathcal{B}^I C_{IJK} + \hat{\rho}_J C_K - \hat{\rho}_K C_J &= 0 \cr
\mathcal{B}^I C_I & = 0 \ .
\end{align}
These consistency conditions give non-trivial constraint on the local renormalization group flow, but it is not immediately obvious if we could derive the analogue of Zamolodchikov's $c$-theorem or enhancement from scale invariance to conformal invariance.

In even dimensions, the situation is more similar to $d=2$ and $d=4$ dimensions discussed in section \ref{localr}. The only main difference is that we have terms with more and more derivatives in the trace anomaly (even within the power-counting renormalization scheme). The most important term we should consider is the Euler term as a natural generalization of Cardy's conjecture:
\begin{align}
T^{\mu\nu} = a \mathrm{Euler} + \chi^H_{IJ} \partial_\mu g^I \partial_\nu g^J H^{\mu\nu} + D_\mu( w_I \partial_\nu g^I H^{\mu\nu}) \ ,
\end{align} 
where $H^{\mu\nu}$ is the divergence-free tensor that appear in the Weyl variation of the Euler density
\begin{align}
\delta_{\sigma}(\sqrt{|g|} \mathrm{Euler}) = \sqrt{|g|} H^{\mu\nu} D_\mu \partial_\nu \sigma \ .
\end{align}
In $d=4$ dimension, $H^{\mu\nu}$ is the Einstein tensor as can be seen from the formula in \ref{convention}.

Here, we only consider the massless renormalization group flow without any  non-conserved vector operators with engineering dimension $d-1$.
In \cite{Grinstein:2013cka}, it was shown that the $a$-function satisfies the gradient property
\begin{align}
\partial_I \tilde{a} = (\chi^H_{IJ} + \partial_{[I} w_{J]}) \beta^J \ 
\end{align}
with $\tilde{a} = a + w_I \beta^I$ similarly to the situation in $d=4$ dimension discussed in section \ref{localr}. If we assumed that $\chi^H_{IJ}$ is positive definite, it shows the enhancement from scale invariance to conformal invariance in the massless renormalization group flow. 
See also \cite{Yonekura:2012kb} for the approach from the dilaton scattering amplitude in the perturbative regime. The argument here is very formal because there are very limited number of examples (e.g. $\phi^3$ interaction in $d=3$) dimension for which we can compute the metric $\chi^{H}_{IJ}$ in perturbation theory.

\subsection{Reduced symmetry}\label{reduced}
\begin{itemize}

\item For chiral field theories in $d=2$ dimension, we actually have two central charges, $c$ and $\bar{c}$, which are the central charges for the (left-moving and right-moving) independent component of the energy-momentum tensor and can take different values (modulo gravitational anomaly). It is easy to generalize our argument in section \ref{cfunc} by taking CP transformation and see that both $c$ and $\bar{c}$ are monotonically decreasing with the same rate governed by $\langle \Theta(x) \Theta(0)\rangle$ \cite{Bastianelli:1996gh}. Thus the difference $c-\bar{c}$ is a constant along the renormalization group flow, which is consistent with the 't Hooft anomaly matching condition for the gravitational anomaly. In our discussions in $d=4$ dimension, we have not discussed the gravitational anomaly but it is possible to incorporate the gravitational anomaly in our discussions of the local renormalization group flow of section \ref{localr}. See also \cite{Luty:2012ww} for a related discussion on the gravitational anomaly in dilaton scattering amplitudes.

\item In finite temperature situations, thermodynamic $c$-theorem was proposed in \cite{CastroNeto:1992ie}. See also \cite{Danchev:1998nx} for applications in condensed matter physics. Note that it is known that the naive free energy does not decrease along the renormalization group flow.

\item It is an interesting question whether the scale invariant boundary conditions will lead to the conformal invariant boundary condition. Let us consider a $d$-dimensional bulk conformal field theory and put a $(d-1)$-dimensional Poincar\'e invariant boundary at $r=0$ in $(t,x_1,\cdots ,r)$ plane. The condition for the Poincar\'e invariant boundary is given by \cite{McAvity:1993ue}\cite{1995NuPhB.455..522M} 
\begin{align}
T_{ri} (r=0) = \partial^j \tau_{ji} \ , \label{pcons}
\end{align} 
where $T_{ri} (r=0)$ is the bulk energy-momentum tensor evaluated at the boundary and  $\tau_{ji}$ is the symmetric ``boundary energy-momentum tensor", where $i$ runs through $(0, \cdots d-1)$.

The boundary scale invariance further requires
\begin{align}
\tau^{i}_{ \ \ i} = \partial^i j_i \ ,
\end{align}
where we call $j_i$ the boundary virial current. Much like in the bulk situation, if we can improve the boundary energy-momentum tensor so that it is traceless, then the boundary is conformal invariant. 

With the assumption of the canonical scaling of the (boundary) energy-momentum tensor, we can argue that Cardy's condition \cite{Cardy:1984bb} $T_{ri}(r=0) =0$ is a necessary condition for the conformal boundary. This is because boundary conformal invariance demands $\tau_{ij}$ is a symmetric traceless tensor whose conformal dimension is $d-1$. The unitarity of the boundary conformal algebra then demands it must be conserved. The Poincar\'e invariance (\ref{pcons}) furthermore dictates that $T_{ri}(r=0) = 0$, and Cardy's condition follows.

The sufficiency of Cardy's condition is more non-trivial. We believe that the scale invariant boundary condition (with some extra assumptions) implies the conformal boundary condition as in the bulk case, but no rigorous derivation is available (in $d>3$). In $d=2$ dimension, the argument on the boundary $g$-theorem implies that the scale invariance without conformal invariance is inconsistent as in the bulk situation. We have already addressed that the boundary $g$-theorem can be proved as long as we assume the analogue of the Reeh-Schlieder theorem.
In $d=3$ dimension, we can show that Cardy's condition is a sufficient condition, but it has not been proved whether this can be derived from the scale invariance alone.

Within the boundary perturbation theory, one can show that the higher dimensional analogue of the $g$-theorem should apply in boundary deformations, and therefore, the scale invariance must imply vanishing of the boundary $\mathcal{B}$ function, resulting in the boundary conformal invariance \cite{Nakayama:2012ed}. It would be very interesting to see if this argument can hold beyond the leading order in perturbation theory. 

\item
The chiral scale invariance studied in \cite{Hofman:2011zj} (see also \cite{Nakayama:2011fe}) states that the theory is invariant under the translation
\begin{align}
 t \to t + \epsilon_t \ , \ \ x \to x + \epsilon_x \ ,
\end{align}
and the chiral dilatation
\begin{align}
 t \to \lambda t \ 
\end{align}
in $(1+1)$ dimensional local quantum field theories.
Correspondingly, the theory possesses three conserved charges $H$, $P$, and $D$ with the commutation relation
\begin{align}
i[D,H] = H \ , \ \ i[D,P] = 0\ , \ \ i[H,P] = 0 \ .
\end{align}
We assume that all the symmetries are linearly realized in a unitary manner.

From the Noether assumption, the translational invariance requires the existence of a conserved energy-momentum tensor
\begin{align}
\partial_x T_{tx} + \partial_t T_{xx} = 0 \ , \ \ \partial_x T_{tt} + \partial_t T_{xt} = 0 \ ,
\end{align}
which is not necessarily symmetric $T_{xt} \neq T_{tx}$ due to the lack of Lorentz invariance.
The chiral scale invariance implies that the ``trace" of the energy-momentum tensor must be  given by the ``divergence" of the ``virial current":
\begin{align}
T_{xt} = \partial_t J_x + \partial_x J_t \ .
\end{align}
Then the chiral dilatation current 
\begin{align}
D_t = t T_{tt} - J_t \ , \ \ D_x = t T_{xt} - J_x \ 
\end{align}
is conserved: $\partial_x D_t + \partial_t D_x = 0$.

As discussed in  \cite{Hofman:2011zj}, we can always remove $J_t$ by defining the new conserved energy-momentum tensor
\begin{align}
\tilde{T}_{tt} = T_{tt} + \partial_t J_t \ , \ \ \tilde{T}_{xt} = T_{xt} - \partial_x J_t \ . \label{impr}
\end{align}
When, in addition, $\partial_tJ_x$ vanishes, the theory possesses the chiral special conformal transformation induced by the conserved current
\begin{align}
K_t = t^2 \tilde{T}_{tt} \ , \ \ K_x = 0 \ 
\end{align}
together with the infinite tower of the chiral Virasoro symmetry ($L^n_t = t^{n} \tilde{T}_{tt}, \ L^n_x = 0$). The chiral special conformal transformation $K$ with the chiral dilatation will generate the $SL(2) \times U(1)$ subalgebra
\begin{align}
i[K,H] = D \ , \ \ i[D,K] = -K \ , \ \ i[K,P] = 0 \ .
\end{align}

The vanishing of $\partial_t J_x$ in unitary quantum field theories comes from the fact that the chiral scale invariance demands $\langle J_x(x,t) J_x(0) \rangle = f(x)$, indicating $\partial_t J_x (x,t) |0\rangle = 0$ from the unitarity and translational invariance \cite{Hofman:2011zj}. Furthermore, {\it if} the analogue of the Reeh-Schlieder theorem \cite{RS} is true, then $\partial_t J_x (x,t)|0 \rangle=0$ is equivalent to the vanishing of the local operator itself $\partial_t J_x(x,t) =0$ (in any correlation functions): in relativistic field theories, the proof requires the microscopic causality in addition to the unitarity (see section \ref{uniReeh}).  
This shows that the chiral scale invariant field theories in (1+1) dimension are automatically invariant under the full chiral conformal transformation (with various technical assumptions). However, we should stress that without assuming Lorentz invariance, the role of the Reeh-Schlieder theorem is not obvious.

\item Since our primary interest in this review article is relativistic field theories, we have little to say about the non-relativistic scale invariance and  conformal invariance. Some interesting classes of scale invariant algebra in non-relativistic systems include Schr\"odinger algebra \cite{Hagen}\cite{Niederer}, Lifshitz algebra \cite{Lifshitz}, and Galilean conformal algebra \cite{Barut}\cite{Havas} with rotation, time-translation, and space-translation in common. The Lifshitz algebra does not contain ``special conformal symmetry", so indeed we have an example of scale invariant but non-conformal field theories simply because there is no way to enhance the symmetry.

Let us consider the Galilean invariant field theories with scale invariance. We can ask the question if we automatically obtain the non-relativistic conformal transformation, leading to the Schr\"odinger algebra \cite{Nakayama:2009ww}. The question is very similar to the one we have been working on in this review article with Poincar\'e invariance.

We begin with a (possibly non-symmetric) conserved energy-momentum tensor (see e.g. \cite{Jackiw:1990mb}) from time-translation and space-translation
\begin{align}
\partial_t T^{0i} + \partial_j T^{ji} &= 0 \cr
\partial_t T^{00} + \partial_i T^{i0} &= 0 \ .
\end{align}
The $U(1)$ particle number conservation demands
\begin{align}
m \dot{\rho} = -\partial_i T^{0i} \ ,
\end{align}
where $m$ is the mass ``central charge" of the Galilean algebra.
Then, the Galilean boost density $G^i = tT^{0i} - mx^i \rho$ is conserved. Note in general $T^{0i} \neq T^{i0}$ in non-relativistic systems.

Suppose the energy-momentum tensor satisfies the condition
\begin{align}
2T^{00} - T^{ij} \delta_{ij} = \partial_t S  + \partial_j A^j \ , \label{tracenon}
\end{align}
then the dilatation density $D= tT^{00} - \frac{1}{2}x_i T^{0i} - \frac{S}{2}$
is conserved. We can always improve the energy-momentum tensor to remove $A_j$ by $2T^{00} \to 2T^{00} + \partial_j A^j$.
When $S$ is a total divergence $S = \partial_i \sigma^i$, one can further improve the energy-momentum tensor by $T^{00} \to T^{00} + \partial_j \partial_i \sigma^j$ so that the right hand side of (\ref{tracenon}) is zero \cite{Nakayama:2009ww}.

When we can improve the energy-momentum tensor in this way, we are able to construct the non-relativistic special conformal density
\begin{align}
K = t^2 T^{00} - tx_i T^{0i} + \frac{m}{2}x^2 \rho \ ,
\end{align}
which is conserved. We see that the structure of the symmetry enhancement is very close to the relativistic situation.

It would be interesting to prove if the non-relativistic special conformal invariance can be derived from the non-relativistic scale invariance possibly with additional assumptions. We have tried some perturbative searches in \cite{Nakayama:2009ww}, and as far as we are aware, there are no known counterexamples. Again, it seems that the absence of the Reeh-Schlieder theorem can be a major obstruction for the proof. Once normal ordered, the two-point functions of the energy-momentum tensor vanishes in vacuum in stark contrast with the relativistic cases.
Another difficulty would be the absence of the analogue of Zamolodchikov's $c$-theorem. In non-relativistic systems, it would be possible to have limit cycles in renormalization group flow, and it would make it more difficult to imagine the proof similar to the one for the relativistic field theories.

\end{itemize}

\

\

\begin{shadebox}
Renormalization group and irreversibility

\

The main subject of the previous sections has been the irreversibility of the renormalization group flow. Presumably, the concept of the irreversibility was not envisaged when the terminology of the renormalization group (by Stueckelberg and Petermann \cite{Stueckelberg:1951gg}) was first invented in the context of the quantum field theories with high energy physics in mind.

What do they mean by ``group" in the renormalization group? Can a group transformation be irreversible? These are interesting questions, and a formal answer is that if we throw away the irrelevant parameters (as we implicitly did in our studies when we interpret the renormalization group flow in Wilson's sense), then the transformation is only semi-group, and clearly there is a preferred direction. In the conventional field theory language, the renormalization group was just the $U(1)$ Abelian group of the scale transformation\footnote{At that time, because of the ``Gruppen Pest", The author imagines everyone wanted to use the terminology ``group".}, but 
 the gradient property makes it possible to introduce the notion of the preferred direction. In a sense, the Wess-Zumino consistency condition is the most advanced way to use the ``group properties". We should recall that we have to supplement the unitarity constraint to say anything about the irreversibility.

With this respect, it is very much similar to time. The time translation is Abelian, and from the group structure, there seems no preferred direction. However,  if we throw away information along the evolution, the entropy increases, and there is a preferred future direction. We have seen that the notion of ``time" (or unitarity and causality) actually played a hidden but crucial role in our discussions of the irreversibility of the renormalization group flow. For instance, our argument based on the correlation functions or dilaton scattering amplitudes relies on the assumption of unitarity and causality.

The absence of scale invariant field theories without conformal invariance may be tightly related to the irreversibility of the renormalization group flow. It will be interesting to see if we could understand the nature of time in this way. Does the hidden enhanced symmetry tell us the nature of the time? We will see the concept of emergent space-time in holography in the following section. There again, the holographic argument crucially depends on the notion of time.
While the discussion is yet to be scrutinised, it seems a promising direction to pursue.

\end{shadebox}

\newpage

\section{Holography}\label{section10}
In this section, we study the relation between scale invariance and conformal invariance from the holographic perspective. The holography is an alternative but  powerful way to understand the strongly coupled quantum field theories in $d$-dimension by using the gravitational system in $d+1$ (or higher) dimension. We discuss the holographic realization of the higher dimensional analogue of Zamolodchikov's $c$-theorem and the enhancement from scale invariance to conformal invariance based on the energy-condition in general relativity. We will mention the validity and a possible violation of the energy-condition and its consequence  in the holographic argument.

The purpose of the holographic study is two-fold. The first obvious one is to understand the field theory better to give more confidence in the enhancement of conformal invariance from scale invariance through our experience in the holographic dual. By examining the holographic argument, we may learn what would be the crucial point in the possible field theory argument.
The second one is more related to the structure of the quantum gravity itself. We would like to understand the fundamental properties of emergent space-time from the renormalization group flow of the dual field theories. Even without the definite answer from the field theory side (so far), this poses a novel question in quantum gravity, which could be answered by itself, and might lead to our deeper understandings of quantum gravity.

\subsection{AdS/CFT and holography}\label{Basics}

 Holography is one of the most powerful guiding principles to understand the fundamental aspects of quantum gravity \cite{hol1}\cite{hol2}. Roughly speaking, the holography dictates that $d+1$ dimensional physics of the quantum gravity (referred to as ``bulk" hereafter) is described by $d$ dimensional non-gravitational physics (referred to as ``boundary" hereafter), typically realized by non-gravitational quantum field theories. The (in)consistency of the dual field theory would yield strong constraints on the properties of quantum theories of gravity in the bulk. For example, the information paradox of black holes is supposed to be resolved in the dual field theory via holography because the time evolution is manifestly unitary in the dual quantum field theory side.
Conversely, we may be able to answer some unsolved problems in boundary field theories by using the holographic bulk argument.

One of the most established examples of the holography is the duality between a gravitational theory on $d+1$ dimensional AdS space-time and a non-gravitational conformal field theory on $d$ dimensional boundary of the AdS space-time (AdS/CFT correspondence \cite{Maldacena:1997re}: see e.g. \cite{Aharony:1999ti} for an earlier review).
AdS space-time is defined by the maximally symmetric space-time with constant negative curvature. 
In Poincar\'e coordinate, the $d+1$ dimensional metric of the AdS space-time is given by
\begin{align}
ds^2 = g_{MN} dx^M dx^N = L^2 \frac{dz^2 + \eta_{\mu\nu} dx^\mu dx^\nu}{z^2} \ . \label{Poincare}
\end{align}
It solves the Einstein equation\footnote{We suppress the $d$-dimensional Planck constant throughout the review article.} $R_{\mu\nu} - \frac{R}{2} g_{\mu\nu} +\Lambda g_{\mu\nu} = 0$ with negative cosmological constant $\Lambda = -\frac{d(d-1)}{2L^2}$.

As its name suggests, Poincar\'e coordinate has the manifest $d$ dimensional Poincar\'e invariance acting on $x^\mu$. We can easily see that the metric (\ref{Poincare}) is also invariant under the scaling transformation
\begin{align}
z \to \lambda z \ , \ \ x^{\mu} \to \lambda x^{\mu} \ . \label{AdSS}
\end{align}
Thus, the dual field theory is expected to be scale invariant. The corresponding killing vectors satisfy the commutation relation of the scaling algebra (\ref{dilatationc}) acting on the Poincar\'e symmetry.
It is less obvious but the metric (\ref{Poincare}) is invariant under the ``special conformal transformation"
\begin{align}
\delta x^\mu = 2(\rho^{\nu}x_\nu)x^\mu - (z^2 + x^\nu x_\nu)\rho^\mu \ , \ \ \delta z = 2(\rho^{\nu}x_\nu)z  \ . \label{cnft}
\end{align}
The full isometry group of the AdS space-time is $SO(2,d)$ and it agrees with the conformal group of $d$-dimensional quantum field theories. This leads to the first conjecture by Maldacena that the $(d+1)$-dimensional gravitational theory on AdS space-time describes a $d$-dimensional conformal field theory.

In AdS/CFT correspondence, we identify the $SO(2,d)$ isometry of the AdS space-time with the conformal group of the dual conformal field theories. From the scaling transformation (\ref{AdSS}), it is natural that we identify the holographic direction $z$ with the direction of the renormalization group flow of the dual quantum field theory. The boundary $z \to 0$ corresponds to the UV limit of the dual field theory, and $z \to \infty$ corresponds to the IR limit of the dual field theory. For a later purpose, we note that the simple coordinate transformation (i.e. $z = e^{-Ar} A^{-1}$ with $A = L^{-1}$) makes the Poincar\'e  metric (\ref{Poincare}) into 
the warped metric 
\begin{align}
ds^2 = dr^2 + e^{2Ar} \eta_{\mu\nu} dx^\mu dx^\nu \ . 
\end{align}
We will identify $Ar$ with the energy scale of the renormalization group flow $Ar \sim \log \mu$. We will assume $A$ and $L$ are both positive hereafter.

The most studied example of AdS/CFT correspondence is the duality between the type IIB string theory on $\mathrm{AdS}_5 \times \mathbb{S}^5$ background with the $N$ unit of Ramond-Ramond five-form flux, and the $\mathcal{N}=4$ supersymmetric Yang-Mills theory with $SU(N)$ gauge group in $d=4$ dimension. The underlying string theory construction allows us to identify various parameters in the both sides (e.g $N^2 = 4\pi^2 L^3$ for the above $\mathcal{N}=4$ Yang-Mills case).   In our following discussions, we do not specify the concrete realization of the AdS/CFT correspondence from the string theory construction. We rather take an effective field theory approach of (quantum) gravity, and we will study the required consistency of the properties of the (quantum) gravitational system from the existence of the consistent dual quantum field theory interpretation. Of course, the following argument should apply to the string theory construction. Indeed, the validity of various assumptions such as the energy-condition or unitarity can be directly checked within the allegedly consistent string theory background.

To be more concrete, let us consider how to compute correlation functions via AdS/CFT correspondence.
For this purpose, we recall that one of the most basic recipes in AdS/CFT correspondence is the Gubser-Klebanov-Polyakov-Witten (GKP-W) prescription \cite{Gubser:1998bc}\cite{Witten:1998qj} that connects the computation of the generating function of correlation functions in the dual conformal field theory and the computation of the partition function in the gravity side. Schematically, we postulate the relation
\begin{align}
\left\langle \exp\left(\int d^d x \phi^{(0)}(x) {O}(x)\right) \right\rangle_{\mathrm{CFT}_d} = e^{-\mathcal{S}_{\mathrm{grav}}[\phi(x,z)|_{z=0} = \phi^{(0)}(x)]} \ \label{GKP}
\end{align}
within the classical approximation of the gravity side. Here $O(x)$ are operators in the dual conformal field theory and $\phi^{(0)}(x)$ are corresponding source functions. 
Obviously we should regard it as a certain hypothetical classical limit (saddle point approximation) of the quantum gravity ``path integral" in the right hand side. 
 It is not obvious such ``path integral" exists, or we should do the path integral from the beginning, but we do believe that the right hand side should exist in a string theory context while
the details are not well-established because of the difficulty in quantizing strings in AdS background with the Ramond-Ramond background.

Within the classical approximation, we can use the prescription (\ref{GKP}) for the study of correlation functions of conformal field theories from the corresponding classical equations of motion of the bulk theories.
For instance, let us demonstrate it in a free massive scalar field $\phi(x_\mu,z)$ in $\mathrm{AdS}_{d+1}$ with the minimally coupled action
\begin{align}
\mathcal{S}_{\mathrm{bulk}} = \int d^{d+1} x \sqrt{|g|} \left(\partial_M \phi \partial^M \phi + m^2 \phi^2\right) \ .
\end{align}
The asymptotic solution ($z \to 0$) of the equations of motion gives
\begin{align}
\phi \sim \phi^{(0)} z^{\Delta_-} + \langle O \rangle z^{\Delta_+} \ 
\end{align}
with
\begin{align}
\Delta_{\pm} = \frac{d}{2} \pm \sqrt{\frac{d^2}{4} + m^2L^2} \ .
\end{align}
The scaling shows that the dual operator $O$ has the scaling dimension $\Delta_+$
and the coupling constant $\phi^{(0)}$ has the scaling dimension $\Delta_-$. We are able to check that the GKP-W prescription gives the scalar two-point function
\begin{align}
\langle O(x) O(y)\rangle  = \frac{c_d^{\Delta}}{(x-y)^{2\Delta_+}} \ ,
\end{align}
by functionally differentiating the GKP-W partition function (\ref{GKP}) twice with respect to the source $\phi^{(0)}(x)$. Here $c_d^{\Delta}$ is a calculable number that depends on $\Delta$ and $d$, but after all it can be changed by the overall normalization of the scalar action, so we will not care at this point.

We have seen that scalar fields such as $\phi(x^\mu,z)$ in the bulk correspond to scalar operators in the dual conformal field theories. 
The other important classes of fields that we typically encounter in the bulk is the graviton ($d+1$ dimensional metric)  and gauge fields. Since the graviton couples with the conserved energy-momentum tensor and the gauge field couples with the conserved current, it is natural to postulate the duality between the $d+1$ dimensional metric fluctuation and $d$ dimensional energy-momentum tensor, and the duality between $d+1$ dimensional bulk gauge field with the $d$ dimensional conserved current. Indeed, the dimensional analysis similar to the above scalar example suggests that massless graviton must have $\Delta^+ = d$, and massless gauge field must have $\Delta^+ = d-1$, which are precisely the values for the scaling dimension of the energy-momentum tensor and the conserved current of the dual $d$-dimensional conformal field theories.

It is important to realize that the GKP-W formalism of computing the generating function for the correlation functions by adding space-time dependent source terms (near $z=0$) is very similar to the introduction of space-time dependent coupling constants discussed in the local renormalization group flow introduced in section \ref{emr} and developed in section \ref{localr}. The GKP-W partition function is nothing but the (renormalized) Schwinger functional.
In the following part of this section, we will see how the consistency condition for the local renormalization group flow is related to the dynamics of the $d+1$ dimensional space-time through gravity.

In the actual computation  of the GKP-W partition function, we encounter various divergence in the on-shell action due to the integration near $z\to 0$ limit (typically when the conformal dimension $\Delta$ takes an integer value). The resolution is obtained by first adding the finite cut-off at $z= {\epsilon}$ and  add local counterterms at the boundary of the AdS space-time. They are given by the boundary fields such as boundary metric or the value of the scalar fields at the boundary and their derivatives. Note that they are local functional of the boundary fields, and they only changes the GKP-W prescription in contact terms.
The structure of the holographic counterterms are very similar to the one we discussed in section \ref{localr} in relation to the local counterterms for the quantum effective actions with space-time dependent coupling constants. 
The procedure is called holographic renormalization and systematically developed in \cite{de Haro:2000xn}\cite{Bianchi:2001kw}.

Generically, we have to impose boundary conditions to solve the second order equations of motion in gravitational theory. In the following, we will be mostly interested in the so-called domain wall solution that interpolates two AdS space-time (with different cosmological constants). Suppose the gravitational theory under consideration admits multiple AdS vacua. We can consider the domain wall connecting the  two different vacua in the radial (i.e. $r$ or $z$) direction. By assuming that the domain wall preserves the $d$-dimensional Poincar\'e invariance, the metric must take the form
\begin{align}
ds^2 = dr^2 + e^{2A(r)} \eta_{\mu\nu} dx^\mu dx^\nu \ ,
\end{align}
where $A(r)$ approaches $A_{\mathrm{UV}} r$ in $r\to \infty$ and $A_{\mathrm{IR}} r$ in $r \to -\infty$ limit. It will interpolate the two different AdS vacua with different cosmological constants. As we will see, the holographic interpretation of the domain wall solution is the renormalization group flow between a UV conformal field theory described by one particular AdS vacuum (with $e^{2A_{\mathrm{UV}}r}$ as the warp factor) and an IR conformal field theory described by another  particular AdS vacuum (with $e^{2A_{\mathrm{IR}}r}$ as the warp factor).

In this flow, the boundary condition is fixed both at $r\to \pm \infty$, and the solution is uniquely specified by the choice of the vacua. The details of the flow depends on the potential of the theory that determines vacua, and it may not be simple to solve the equations of motion with the fixed boundary conditions both at UV and IR. However, there is a beautiful simple realization of such a flow by using the (fake) superpotential as we will review. Such a simple flow is motivated by the Hamilton-Jacobi formalism of the flow \cite{de Boer:1999xf} (see e.g. \cite{deBoer:2000cz}\cite{Fukuma:2002sb} for reviews) as well as the stability of the vacua in AdS space-time \cite{Townsend:1984iu}, and of course supersymmetry when available.\footnote{However, we emphasize the (fake) superpotential flow is not limited to supergravity.} In  section \ref{holct}, we will argue that it has a holographic interpretation as the gradient renormalization group flow of the corresponding dual quantum field theories.

\subsection{Holographic $c$-theorem}\label{holct}
As we mentioned, in the AdS/CFT correspondence, we can regard the radial direction $r$ as renormalization scale $\log\mu$ of the dual quantum field theory. The dynamics of the $r$ direction is governed by the gravitational equations of motion, and our task is to relate the gravitational dynamics with the renormalization group equation.
In the holographic renormalization group flow \cite{Akhmedov:1998vf}\cite{Alvarez:1998wr}\cite{Girardello:1998pd}\cite{Freedman:1999gp}\cite{Sahakian:1999bd}, we consider the  metric
\begin{align}
ds^2 = dr^2 + e^{2A(r)} \eta_{\mu\nu} dx^\mu dx^\nu \ , \label{flowh}
\end{align}
special case of which corresponds to the domain wall solution discussed at the end of section \ref{Basics}.
When $A(r) = A_* r$, the metric describes the Poincar\'e patch of the AdS space-time and the 
dual field theory is conformal invariant because of the AdS isometry. 
Here we consider the generalization with a non-trivial warp factor, which describes the renormalization group flow of the dual quantum field theory.

The first observation is that without any matter, in diffeomorphism invariant gravitational theories,
the requirement of the scale invariance is equivalent to the requirement of the AdS isometry. Indeed, under the scale transformation $ x^{\mu} \to \lambda x^{\mu}$, the Minkowski metric $\eta_{\mu\nu} dx^{\mu} dx^{\nu}$ acquires $\lambda^2$. On the other hand, we can always assume that the scaling transformation acts on $r$ as the shift transformation $r \to r + c \log \lambda$ with a certain constant $c$. We see that the condition of the isometry under the scale transformation fixes the metric to be AdS space-time
\begin{align}
ds^2 = dr^2 + e^{2A_* r} \eta_{\mu\nu} dx^\mu dx^\nu \ 
\end{align}
so that the isometry vector fields give the desired scaling transformations of the Poincar\'e algebra (\ref{dilatationc}).
We realize that the geometry is nothing but the AdS space-time, and we have obtained the enhanced AdS isometry $SO(2,d)$ even though we did not require the invariance under the special conformal transformation.

Once we have understood the scale invariant vacuum solutions, we would like to take a further look at the non-trivial flow solution with the metric ansatz (\ref{dilatationc}). Here, we will see the holographic analogue of Zamolodchikov's $c$-theorem.
To be concrete, our  starting point is a gravitational theory
described by the  classical Einstein Hilbert action (with negative cosmological constant) minimally coupled with a classical matter sector (we will relax some of the conditions in later sections by introducing higher derivative terms or quantum corrections).
\begin{align}
\mathcal{S} = -\frac{1}{2}\int d^{d+1}x \sqrt{|g|}\left(R + \frac{d(d-1)}{L^2} \right) + \mathcal{S}_{\mathrm{matter}} \ .
\end{align}

In this setup, we define the holographic $c$-function $a_d(r)$
by (hereafter prime denotes the derivative with respect to the radial direction: $A'(r) \equiv \frac{dA(r)}{dr}$) 
\begin{align}
a_d(r) = \frac{\pi^{d/2}}{\Gamma(d/2)((A)'(r))^{d-1}} \ . \label{holographicc}
\end{align}
The overall normalization factor is fixed from the holographic Weyl anomaly argument we will explain a few paragraphs below.
Then by using the Einstein equation, we obtain
\begin{align}
\frac{da_d(r)}{dr} = -\frac{(d-1)\pi^{d/2}}{\Gamma(d/2)(A'(r))^d}A''(r) =  -\frac{\pi^{d/2}}{\Gamma(d/2)(A'(r))^d} [T^t_{\ t} - T^{r}_{\ r} ] \ge 0 \ . \label{einsteinttt}
\end{align}
In the last equality we have assumed the null energy condition (NEC).\footnote{When $d$ is even, the positivity of the factor in front is obvious. When $d$ is odd, we can still argue \cite{Myers:2010tj} that the sign of $A'(r)$ cannot change along the renormalization group flow. We will also see it from (\ref{potflow}).} Therefore, the holographic $c$-function $a_d(r)$ is monotonically decreasing along the renormalization group flow.

The null energy condition is the condition for the energy-momentum tensor: for all null vectors $k^M$ such that
$g_{MN} k^M k^N =0$, the energy-momentum tensor satisfies 
$k^M k^N T_{MN} \ge 0$. In the fluid frame, it requires that sum of the energy and pressure is semi-positive: $\epsilon + p \ge 0$.
By using the Einstein equation, it constrains possible geometries through  relating it with the constraints on the energy-momentum tensor: $k^{M}k^{N} R_{MN} \ge 0$. 

Apart from the holographic $c$-theorem, the requirement of the null energy-condition  leads to a deep consequence in general relativity. Einstein equation itself allows many seemingly pathological solutions like worm-hole, superluminal propagation of information, classically decreasing black hole horizon, time-machine and so on because it is always possible to declare that such geometries are supported by the corresponding energy-momentum tensor if we have no restrictions on the energy-momentum tensor. However, the null energy condition forbids them (see e.g. \cite{Friedman:1993ty}\cite{Tipler:1976bi}\cite{Hawking:1991nk}\cite{Olum:1998mu}).

At the conformal fixed point, we can relate the above defined holographic $c$-function $a_d(r)$ to the Weyl anomaly when  space-time dimension of the dual field theory $d$ is even. The idea is to replace  the boundary Minkowski metric $\eta_{\mu\nu}$ with a general weakly curved metric $g_{\mu\nu}$ and study the expectation value of $T_{\mu\nu}$ through the GKP-W prescription. We will not go into the detailed computation (see \cite{Henningson:1998gx} for the original computation), but we only quote the result in $d=4$ dimension. For the Einstein gravity, the holographic Weyl anomaly is given by 
\begin{align}
\langle T^{\mu}_{\ \mu} \rangle = \frac{L^3}{16} (\mathrm{Weyl}^2-\mathrm{Euler}) \ .
\end{align}
We see that the Einstein gravity predicts $a=c = \frac{L^3}{16} = \frac{a_4}{16\pi^2}$ in holography. This holographic computation indeed suggests that the central charge $a$ has a natural interpolation function, which is monotonically decreasing along the holographic renormalization group flow. However, at this stage, we cannot distinguish two different Weyl anomaly coefficients $a$ and $c$ in the Einstein gravity.

Let us discuss more details on the structure of the holographic $c$-theorem.
Suppose the matter action is given by a generic non-linear sigma model (possibly with a potential):
\begin{align}
\mathcal{S}_{\mathrm{matter}} = \int d^{d+1} x \sqrt{|g|} \left(\mathcal{G}_{IJ}(\Phi) \partial^M \Phi^I \partial_M \Phi^J + V(\Phi) \right) \ .
\end{align}
Then the explicit computation of the energy-momentum tensor gives\footnote{Here, we assume the canonical 
(non-improved) energy-momentum tensor that corresponds to the  Einstein frame  of gravity. This is reasonable because
we use the gravity equations in the Einstein frame.} 
\begin{align}
T^{t}_{\ t} - T^{r}_{\ r} =  -g^{rr} \mathcal{G}_{IJ} \partial_r \Phi^I \partial_r \Phi^J \ .
\end{align}
We may regard $\partial_r \Phi^I$ as the beta function $\beta^I$ for the coupling constant corresponding 
to $\Phi^I$ under AdS/CFT correspondence (up to a choice of the renormalization scheme). We may also regard the target space metric $\mathcal{G}_{IJ}$
as the Zamolodchikov metric $\chi_{IJ}$ (up to a choice of the renormalization scheme and the associated constant multiplicative factor $a_d(r)$ we will discuss below). Indeed, at the conformal fixed point, $\mathcal{G}_{IJ}$ determines the two-point function via the GKP-W prescription.

Under the holographic renormalization group flow, the Einstein equation demands
\begin{align}
\frac{da_d}{dr} = \frac{\pi^{d/2}}{\Gamma(d/2)(A'(r))^d} g^{rr} \mathcal{G}_{IJ} \partial_r \Phi^I \partial_r \Phi^J \ ,
\end{align}
which is interpreted as
\begin{align}
\frac{d\tilde{a}}{d\log\mu} = \beta^I\chi_{IJ}  \beta^J \ .
\end{align}
This is nothing but the strong $c$-theorem (or what we called the strong $\tilde{a}$-theorem in $d=4$ dimension) we have discussed in section \ref{section6}.

To be more precise, we have to make the relation between the renormalization group scale $\log\mu$ and the radial direction $r$ because away from the scale invariant fixed point, there is some arbitrariness. The standard choice would be the  so-called holographic scheme \cite{Anselmi:2000fu}: 
\begin{align}
\log \mu \equiv A(r) \ , \ \ \beta^I = \frac{d\Phi^I}{dA(r)} \ , \label{holgraphics}
\end{align}
in which ``Zamolodchikov metric" $\chi_{IJ}$ is given by $a_d \mathcal{G}_{IJ}$. The proportional factor by $a_4 \sim N^2$ (in $d=4$ dimension) can be well-understood in the large $N$ CFT (e.g. $\mathcal{N}=4$ supersymmetric Yang-Mills theory), in which supergravity light operators are described by single trace operators like $\mathrm{Tr} (F_{\mu\nu}^2)$.

We also note that the coupling constant dependent Weyl anomaly we have discussed in section \ref{localr} can be computed by assuming (nearly) massless fields $\Phi^I$ are space-time dependent at boundaries \cite{Liu:1998bu}\cite{Balasubramanian:2000pq}\cite{Erdmenger:2001ja}.  In particular, it turns out that the Weyl anomaly with $\beta^I = 0$ in $d=4$ dimension is governed by the (boundary) Riegert operator $\Delta_4$ in (\ref{Riegert}) as $\mathcal{G}_{IJ} \Phi^I \Delta_4 \Phi^J$ (up to boundary counterterms), which is reasonable because the Wess-Zumino consistency condition is trivially solved due to the Weyl invariance of $\Delta_4$ then. Thus, by comparing it with the result in section \ref{localr}, we see that the AdS/CFT at the conformal fixed point predicts $\chi_{IJ}^{a} = 2a_4 \mathcal{G}_{IJ}$ and $\chi_{IJ}^{g} = 4 a_4 \mathcal{G}_{IJ}$.\footnote{The proportional relation $\chi^{a}_{IJ} = \frac{1}{2}\chi^g_{IJ}$ at the conformal fixed point may be explained by the Wess-Zumino consistency condition of the local renormalization group.} The computation can be generalized in which $\beta^I$ are small but non-zero and it agrees with our discussion in section \ref{localr}.

In addition, the holographic renormalization group flow shows the gradient property whenever the potential takes a ``holographically renormalizable form".
Suppose the potential admits the ``superpotential" $W(\Phi)$ so that
\begin{align}
V(\Phi) = \mathcal{G}^{IJ} \partial_I W(\Phi) \partial_J W(\Phi) - \frac{d}{4(d-1)}W(\Phi)^2 \ , \label{superpotential}
\end{align}
 then the flow of the scalar field $\Phi^I$ turns out to be a gradient flow: 
\begin{align}
\partial_r \Phi^I &= \mathcal{G}^{IJ} \partial_J |W| \cr
A'(r) &= \frac{1}{d-1} |W| . \label{potflow}
\end{align}
The first formula corresponds to the gradient  formula of the renormalization group. Indeed, the second formula suggests that the holographic $c$-function (\ref{holographicc}) is proportional to $|W|^{-d+1}$ and with the holographic scheme (\ref{holgraphics}), we can precisely reproduce the gradient formula of the field theory.

Note that for a given $V(\Phi)$ there could be many different superpotentials $W$ that satisfy (\ref{superpotential}) and each gives a different flow. On the other hand, not every $V(\Phi)$ possess the corresponding fake superpotential $W(\Phi)$. 
The potential flow (\ref{potflow}) has a natural interpretation as the gradient formula of the beta function. 
We recall that the gradient formula in the renormalization group could contain the antisymmetric part. 
If there were ``B-field", then it would have an antisymmetric part in $\mathcal{G}^{IJ}$ (which is related to $w_{IJ} = \partial_I w_J - \partial_J w_I$ in the local renormalization group discussed in section \ref{localr}), but the significance is not so clear at this point. It seems that the antisymmetric part vanishes in holographic computation.\footnote{Another interesting feature of the holographic computation is $\chi^{a}_{IJ} = \frac{1}{2}\chi^{g}_{IJ}$ in a natural holographic scheme as we mentioned above. Of course, this relation can be modified by adding local counterterms, so it is not a robust prediction.}

From the Hamilton-Jacobi formulation of the holographic renormalization group flow \cite{de Boer:1999xf}, the condition (\ref{superpotential}) is regarded as a holographic renormalizability from adding the boundary counterterms of scalar fields. In the literature it was suggested that the condition (\ref{superpotential}) is probably a necessary condition to guarantee a consistent stable renormalization group flow \cite{de Boer:1999xf}. For instance, the form of the potential (\ref{superpotential}) automatically guarantees  the Breitenlohner-Freedman bound \cite{Breitenlohner:1982jf}: $m^2 \ge - \frac{d^2}{4L^2}$ that assures the unitarity bound of the scaling dimensions of operators at the conformal fixed point.
It seems remarkable that the consistency of the renormalization group interpretation gives a constraint on the possible potential in the bulk gravity.

Before going on, let us say a few words about the situation in the case with odd $d$. When $d$ is odd, we do not have (holographic) Weyl anomaly, so the question arises what is the physical interpretation of the monotonically decreasing holographic $c$-function (\ref{holographicc}). A priori, there are many physical quantities related to the number: two-point functions (or higher-point functions) of energy-momentum tensor, thermal free-energy, entanglement entropy\footnote{The holographic discussions on the entanglement entropy first appeared in seminal papers by Ryu and Takayanagi \cite{Ryu:2006bv}\cite{Ryu:2006ef}. See e.g. \cite{Nishioka:2009un} for a review.} and so on. At the level of the Einstein gravity, however, they are indistinguishable. The study of the higher order derivative corrections revealed that it is related to the Euclidean $\mathbb{S}^d$ partition function and the entanglement entropy at the conformal fixed point. We refer to \cite{Casini:2011kv} for more details.

So far, we have not considered the beta function for the vector operators. In our applications of scale invariance
 vs conformal invariance, it  is important to realize the operator identity such as $\beta^I {O}_I = -v^a\partial_\mu J_a^\mu$ as discussed in section \ref{local}.
 The redundant operators in the conformal field theory are realized by gauge symmetries in the holographic renormalization group flow.
 Suppose we gauge the non-linear sigma-model by requiring it is invariant under the gauge transformation $\Phi \to e^{i\Lambda} \Phi$ and $A \to A + d\Lambda$ (we can easily generalize the situations with non-Abelian symmetry). 
 Then the gauged non-linear sigma model is described by the action 
\begin{align}
\mathcal{S}_{\mathrm{matter}} = \int d^d x\sqrt{|g|} \left(\mathcal{G}_{IJ}(\Phi) D_M \Phi^I D^M \bar{\Phi}^J + V(\Phi) \right) \ , 
\end{align}
where $D_{M} = \partial_M -A_M$ contains the gauge connection, and the kinetic term $\mathcal{G}_{IJ}$ and the potential $V(\Phi)$ must be gauge invariant.

Now, the energy-momentum tensor appearing in the holographic renormalization group flow is replaced by the gauged one
\begin{align}
T^{t}_{\ t} - T^{r}_{\ r} = -g^{rr} \mathcal{G}_{IJ} D_r \Phi^I D_r \bar{\Phi}^J \ .  \label{gaugedtt}
\end{align}
We can regard $D_r \Phi^I$ as the gauge invariant $\mathcal{B}^I$ function rather than the beta function $\beta^I \sim \partial_r \Phi^I$. 
Indeed, as we will discuss, the arbitrary separation of $\mathcal{B}^I O_I = \beta^I O_I + v^a \partial_\mu J^\mu_a$ is the corresponding gauge transformation of the Scwhinger functional that makes the beta functions ambiguous.
By substituting the energy-momentum tensor (\ref{gaugedtt}) into the holographic renormalization group flow (\ref{einsteinttt}),
 we interpret the holographic renormalization group flow in the gauged non-linear sigma-model 
\begin{align}
\frac{da_d}{dr} = \frac{\pi^{d/2}}{\Gamma(d/2)(A'(r))^d} g^{rr} \mathcal{G}_{IJ} D_r \Phi^I D_r \Phi^J \ ,
\end{align}
as the holographic realization of the strong $c$-theorem
with respect to the $\mathcal{B}$ function flow
\begin{align}
\frac{d\tilde{a}}{d\log\mu} =   \mathcal{B}^I \chi_{IJ} \mathcal{B}^J 
\end{align}
as expected from field theory discussions of the strong $c$-theorem in the previous sections.

It is interesting to observe that the gauge invariance of the action imposes some interesting restrictions of the 
holographic renormalization group flow with the operator identity $\mathcal{B}^I O_I = \beta^I O_I + v^a \partial_\mu J^\mu_a$.
First of all, the holographic $c$-function does not depend on the coupling constant that can be removed 
from the gauge transformation. This is due to the gauge invariance, and the holographic $c$-function has 
 flat directions corresponding to the redundant perturbations. 
 If the gradient formula holds, then this further suggests that the $\mathcal{B}^I$ functions have as many zero directions
 as the gauged directions. These directions must be in contrast with exactly marginal directions that are not gauged: physics changes along the exactly marginal but non-redundant directions. One example of the exactly marginal direction is the dilaton in type IIB string theory on $\mathrm{AdS}_5 \times \mathbb{S}^5$ which corresponds to the coupling constant of the $\mathcal{N}=4$ super Yang-Mills theory. On the other hand, the redundant directions appear in $\mathcal{N}=8$ gauged supergravity in which various scalar fields are gauged under the $\mathcal{R}$-symmetry \cite{Freedman:1999gp}. 
  These gauge directions in holographic renormalization group flow precisely correspond to redundant perturbations, and the flow in that direction (if any) should be regarded as physically equivalent. Indeed, the argument here is in complete parallel with the one in section \ref{reason}. In particular, $w_I$ is exact in holographic computation and can be gauged away, so the gradient formula does not contain the inhomogeneous terms in the holographic scheme.

\subsection{Scale vs Conformal from holography}\label{holproof}
As we have mentioned, the scale invariance dictates that the metric  must take the form
\begin{align}
ds^2 =  L^2\frac{dz^2+\eta_{\mu\nu} dx^\mu dx^\nu}{z^2}
\end{align}
up to diffeomorphism invariance. 
It is accidentally invariant under the full AdS isometry. The full AdS isometry means that without any other fields, the holographic dual would be invariant under the full conformal symmetry.
The holographic interpretation is obvious: only with the gravity, there is no other operators than the energy-momentum tensor
in the dual field theory, and in particular there is no candidate for the virial current.
Without the candidate for the virial current, the only possible way to realize scale invariance is $T^{\mu}_{\ \mu} = \partial^\mu J_\mu = 0$, and the theory is conformal invariant.

In order to break the conformal invariance, therefore, we need the matter sector that represents a non-trivial existence of the virial current. 
As discussed in \cite{Nakayama:2009qu}\cite{Nakayama:2009fe}, a certain non-trivial configuration of the matter may break the isometry corresponding to the special conformal transformation while preserving the isometry corresponding to the scale transformation.
Some examples are vector condensation
\begin{align}
A = A_M dx^M = \alpha \frac{dz}{z} \label{veccon}
\end{align}
and the scalar condensation (with gauge invariance)
\begin{align}
\Phi = \gamma z^{i\alpha} \ \label{scacon}
\end{align}
in the AdS background. 
The gauge invariance is needed because under the scale transformation $z \to \lambda z$, $\Phi$ acquires the phase: $\Phi \to \lambda^{i\alpha} \Phi $, and we need to cancel it by the gauge transformation on $\Phi$. Since the physical observations do not depend on the gauge, this recovers the invariance under the scale transformation.
Without the gauge invariance, it is only invariant under the discrete scale transformation.

Clearly these configurations are invariant under the scaling transformation, but it is not invariant under the AdS isometry that corresponds to the special conformal transformation (\ref{cnft}).
As we will discuss, such configurations are directly related to the non-zero contribution to the trace of the energy-momentum tensor of the dual field theory and gives a non-trivial existence of the virial current realized in the holographic renormalization group.

In the previous section, we have introduced the gauged non-linear sigma-model to realize the operator identity of the dual field theory, which is crucial to understand the emergence of the virial current as discussed in section \ref{local}. From this  perspective, the above two configurations  (\ref{veccon}) and (\ref{scacon})  are mutually
related by the gauge transformation. 
Indeed, suppose $\Phi$ and $A_M$ are related by the gauge transformation:$\Phi \to e^{i\Lambda} \Phi$ and $A_M \to A_M + \partial_M \Lambda$, then the configuration
\begin{align}
\Phi &= \gamma z^{i\alpha} \cr
A_M dx^M &=0 
\end{align}
which can be interpreted as $\beta^g = i\alpha g $ with $v^a\partial_\mu J^\mu_a = 0$ in the dual field theory, is gauge equivalent to
\begin{align}
\Phi &= \gamma \cr
A_M dx^M &=\frac{\alpha dz}{z}  
\end{align}
which can be interpreted as $\beta^g = 0$ with $v^a\partial_\mu J^\mu_a \neq 0$ in the dual field theory. 

The former shows a cyclic renormalization group flow because the phase of the field $\Phi$ (phase of the dual coupling constant) is rotating along the evolution in $z$ direction  in the holographic renormalization group flow from our identification $\beta^I \sim \partial_r \Phi^I$ discussed in section \ref{holct}. On the other hand the latter gives the non-zero background gauge field renormalization with the identification $\partial^\mu j_\mu$ with $A_z$, which is eventually related to the non-zero virial current.
Of course, the gauge invariant quantity $\mathcal{B}^I$ that appears in the trace of the energy-momentum tensor of the dual field theory is non-zero whichever gauge one uses because of the gauge invariant identification $\mathcal{B}^I \sim D_r \Phi^I$.

Our central question is whether such a flow is possible in a reasonable theory of holography. We argue that there are two main obstructions. The first one is that the potential of the gauge direction is always zero from the gauge invariance, 
so it is unlikely that such a flow is generated from the beginning. In particular in the superpotential flow or gradient flow, the radial evolution of the field $\Phi^I$ is uniquely specified by the gradient of the gauge invariant superpotential, and the field theory discussions in section \ref{reason} directly applies. The second one is the inconsistency with the 
holographic $c$-theorem. Suppose that the metric of the non-linear sigma-model in the holographic renormalization group flow is positive definite.
Then such a scale invariant but non-conformal  background gives a non-trivial flow for the warp-factor: we derived the holographic $c$-theorem
\begin{align}
\frac{da_d}{dr} = \frac{\pi^{d/2}}{\Gamma(d/2)(A'(r))^d} g^{rr} \mathcal{G}_{IJ} D_r \Phi^I D_r \Phi^J \ ,
\end{align}
or its field theoretic interpretation
\begin{align}
\frac{da_d}{d \log\mu} = a_d  \mathcal{B}^I \mathcal{G}_{IJ} \mathcal{B}^J
\end{align}
but whenever $D_r \Phi^I \neq 0$ these are inconsistent with the scale invariance because the warp-factor does not take the scale invariant form $e^{2A_*r}$ as long as the metric is non-degenerate.
Note that the requirement of the positivity of the metric is natural because of the unitarity of the bulk theory although strictly speaking it is not guaranteed by the null energy-condition alone. We will introduce the notion of  the strict null energy-condition to give a sufficient condition for the unitarity as well as the enhancement from scale invariance to conformal invariance.

In retrospect, the ``counterexample" of the scale invariant but non-conformal field theory in beta function flow (see section \ref{Intera}) can be understood in holography as follows. We start with the manifestly conformal invariant background
\begin{align}
\Phi &= \gamma \cr 
A_M dx^M &=0  \label{gaugeo}
\end{align}
and perform the gauge transformation
\begin{align}
\Phi &= \gamma z^{i\alpha } \cr
A_M dx^M &= -\frac{\alpha dz}{z}  \label{gaugev}
\end{align}
so now we interpret that beta function is non-zero $\beta_g = i\alpha g $ and the renormalization group flow appears to be cyclic from $\frac{dg^I}{d\log\mu} = \beta^I$ because the eigenvalues under the scale transformation is pure imaginary. However, there is an extra contribution from
the beta function for the background vector fields and the total trace of the energy-momentum tensor is zero
and the theory is conformal invariant as it must be. In spirit, this is close to the artificial separation between the scalar beta functions and vector beta functions we did in section \ref{reddd}.

The gauge transformation in effective $d+1$ dimensional gravity may be seen as the coordinate transformation in the higher dimensional gravity. Take $\mathrm{AdS}_5 \times \mathbb{S}_5$ solution in type IIB string theory with the metric
\begin{align}
ds^2 = L^2 \left(\frac{dz^2 + \eta_{\mu\nu} dx^\mu dx^\nu}{z^2} + d\gamma^2 + \cos^2\gamma d\varphi^2 + \sin^2\gamma (d\psi^2 + \cos^2\psi d\theta_1^2 + \sin^2\psi d\theta_2^2) \right)
\end{align}
and perform the coordinate transformation $\varphi \to \varphi + \alpha \log z$:
\begin{align}
ds^2 = L^2 \left(\frac{dz^2 + \eta_{\mu\nu} dx^\mu dx^\nu}{z^2} + d\gamma^2 + \cos^2\gamma (d\varphi + \frac{\alpha dz}{z})^2 + \sin^2\gamma (d\psi^2 + \cos^2\psi d\theta_1^2 + \sin^2\psi d\theta_2^2) \right) \ .
\end{align}
The geometry is still manifestly scale invariant, but the isometry corresponding to the special conformal transformation is obscured. The $\varphi$ direction was isometry so we are mixing the dilatation current with the conserved current. If we shifted the non-isometric direction, say $\gamma$ with $\log z$, we would have non-zero artificial virial current that is not conserved and it would induce the apparent cyclic renormalization group flow in holography. 
 The scaling transformation must accompany the shift of $\gamma$ as is consistent with (\ref{gaugev}). Note that in this example, the current mixed here (e.g. above $\gamma$ direction) is not necessarily the eigenstate under the scaling transformation. This is not inconsistent with the fact that the energy-momentum tensor has the definite scaling dimension as we see below.

Note that from the holographic perspective, there is nothing wrong with the second background (\ref{gaugev}), and it may be actually more mysterious 
why the perturbative computation of the beta functions (say in the minimal subtraction scheme) prefers the seemingly zero renormalization for 
the background vector fields. There seems to be a certain perturbative mechanism to choose a gauge.

In previous sections, we have discussed how the use of the equations of motion in $\mathcal{B}^I O_I = \beta^I O_I + v^a \partial_\mu J^\mu_a$ introduces the additional anti-symmetric wavefunction renormalization resulting in the anti-symmetric contributions in the anomalous dimensions. As long as we are interested in the physical spectrum of the AdS space-time, we cannot see it: the only gauge invariant object is the mass of the physical excitations that correspond to the dimensions of operators after the diagonalization. In this way, the anti-symmetric wavefunction renormalization does not play a significant role in the holographic approach, suggesting that they are simply an artifact of perturbative computations (with reference to the trivial fixed point) and are not intrinsic to the theory.

We have a small comment on the anomalous dimension of the current operator. As discussed above, we are led to the conformal invariant fixed point once we assume the scale invariance. We note that there is an operator identity $\beta^I O_I = -v^a \partial_\mu J^\mu_a$ so that $\mathcal{B}^I=0$. This means that $J^{\mu}_a$ are not conserved and must acquire anomalous dimensions: otherwise they must be conserved from unitarity. The holographic realization of the anomalous dimension is the Higgs mechanism. We note that for the operator identity to be realized in our holographic setup, the charged scalar must obtain a vacuum expectation value (as in (\ref{gaugeo})),  and then the gauge field becomes massive. Following the discussion in section \ref{Basics}, such a massive vector field corresponds to a non-conserved vector operator with the scaling dimension $\Delta > d-1$. In contrast, the combination $\beta^I O_I + v^a \partial_\mu J^\mu_a$ does not acquire the anomalous dimension because it is the gauge direction.

The holographic interpretation of the non-conserved current operators and their vector beta functions are further studied in \cite{Nakayama:2013fha}\cite{Nakayama:2013ssa}, where interested readers can find more information.
It may be worthwhile to mention that the local Callan-Symanzik equation can be regarded as the Hamiltonian constraint of the gravity, and the consistency of the local renormalization group should follow from the consistency of the bulk equations of motion. For instance the orthogonality relation between vector beta functions and scalar beta functions $\mathcal{B}^I \hat{\rho}_I = 0$ in holography can be understood in this way.

So far, we have assumed a particular classical action for the matter sector.
More generically, we argue that if the matter field satisfies the strict null energy condition, scale invariance
implies conformal invariance in holography. 
The strict null energy condition states that the equality of the null energy condition is saturated if and only if the field contributing to the energy momentum tensor takes the trivial configuration \cite{No}. More precisely, we demand that if there exists any null-vector that makes $T_{MN} k^M k^N =0$, then the field configuration must be invariant under all the isometry transformation of the space-time.  Note that the null energy-condition itself cannot exclude the degenerate metric for the (gauged) non-linear sigma-model but the strict null energy condition does. Supergravity analysis of the strict null energy-condition in string compactification can be found in \cite{Nakayama:2009fe}.

\subsubsection{Holographic counterexample 1: null energy violation}\label{countex1}
In section \ref{holproof}, we used a certain energy-condition to prove the holographic $c$-theorem as well as the enhancement from scale invariance to conformal invariance. 
Without imposing the strict null energy-condition, one can construct a counterexample of scale invariant but non-conformal dual field theories in holography \cite{Nakayama:2009qu}\cite{Nakayama:2009fe}. Let us consider the
vector field theory with the generic (gauge non-invariant) potential
\begin{align}
\mathcal{S}  = \int d^d x \sqrt{|g|}\left( -\frac{1}{2}R + \frac{1}{4}F_{MN}F^{MN} + V(A_MA^M) \right) \ . \label{massivev}
\end{align}
We assume that the potential $V(A_MA^M)$ for the vector field $A_M$ has a non-trivial extremum (e.g. Mexican hat potential $V(A_MA^M) = \Lambda_0 -m^2 (A_MA^M) + \lambda (A_MA^M)^2$).
We see that the theory may admit the non-trivial vector condensation solution
\begin{align}
ds^2 &= L^2 \frac{dz^2 + \eta_{\mu\nu}dx^\mu dx^{\nu}}{z^2} \cr
A &= \alpha \frac{dz}{z} \label{solutionvv}
\end{align}
with non-zero $\alpha$, depending on the shape of the potential. 
Note that the violation of the AdS isometry from the matter configuration due to the condensation of $A_M$ does not back-react to the metric, so it will keep the AdS metric.
Indeed, the contribution from the kinetic term vanishes because $F_{MN}=0$ in the above configuration. The contribution from the potential term is proportional to the metric, so it only changes the overall AdS radius.

As discussed in section \ref{holproof}, the configuration preserves the scale invariance, but does not preserve the special conformal invariance in the matter vector condensation. Therefore, we may regard it as a holographic realization of scale invariant but non-conformal field theories.

It is interesting to point out its relation to the ghost condensation \cite{ArkaniHamed:2003uy}.
The above vector condensation model can be made gauge invariant by introducing the Higgs field with  higher derivative kinetic terms:
\begin{align}
\mathcal{S} = \int d^{d+1} x \sqrt{|g|} \left( -\frac{1}{2} R + \frac{1}{4}F_{MN}F^{MN} + V(D_M\Phi D^M \Phi) \right) \ .
\end{align}
Here the gauge invariant higher derivative kinetic term is given by e.g. $V(D_M\Phi D^M \Phi) = \Lambda_0 -m^2 (D_M\Phi D^M \Phi) + \lambda (D_M\Phi D^M \Phi)^2$. 
If we fix the gauge in a unitary gauge $\Phi = \mathrm{const}$, then it is equivalent to the vector condensation model. Again this particular form of the kinetic term does not back-react to the AdS space-time even though the isometry is broken by the non-trivial field configuration.

On the other hand, if we ignore the gauge field, this is nothing but a model of the ghost condensation studied in relation to the alternative gravity with the action 
\begin{align}
\mathcal{S} = \int d^{d+1} x \sqrt{|g|} \left( -\frac{1}{2} R + V(D_M\Phi D^M \Phi) \right) \ .
\end{align}
Our discussion suggests that the holographic dual of the (gauged) ghost condensation in AdS space-time would be inconsistent (at least when $d=2$) because it is incompatible with the enhancement from scale invariance to conformal invariance, which we know must be true from the field theory argument. 
Presumably, the unitarity is sacrificed in order to reconcile with the situation, which is suggested by the violation of the (strict) null energy condition.

As emphasized in \cite{Nakayama:2009qu}, the situation can be different in dS space-time, where we allow time-like condensation $A = \frac{dt}{t}$ with the de-Sitter metric $ds^2 =L^2 \frac{-dt^2 + \delta_{\mu\nu} dx^{\mu} dx^{\nu}}{t^2}$. Although we can still postulate holography like dS/CFT \cite{Strominger:2001pn} to discuss the properties of this hypothetical dual of  the scale invariant but not conformal invariant Euclidean field theory realized in holography, they may not disagree with our  field theory arguments. The point is that the dS/CFT does not assure the unitarity of the dual field theory, and without unitarity (more precisely reflection positivity in the Euclidean signature), it is possible to construct  scale invariant but non-conformal field theories without unitarity. See some examples in section \ref{examples}. The cyclic behavior in the time-like condensation or dS/CFT is something like ``time crystal" studied in \cite{Shapere:2012nq}. There, again, the existence of ``time crystal" does not seem to have an immediate inconsistency with the quantum mechanics.

\subsubsection{Holographic counterexample 2: foliation preserving diffeomorphic theory of gravity}\label{folia}

Another interesting possibility to realize the holographic dual of scale invariant but not conformal field theories is to abandon the full space-time diffeomorphism \cite{Nakayama:2012sn}. 
We have discussed that the scale invariance and Poincar\'e invariance naturally leads to the AdS metric
\begin{align}
ds^2 = g_{MN} dx^M dx^N = L^2\frac{dz^2 + \eta_{\mu\nu} dx^\mu dx^\nu}{z^2} \ . \label{Poincccc}
\end{align}
The metric has a natural foliation with respect to the $d$-dimensional Minkowski space-time. In order to preserve the scale invariance, we do not have to assume the full $d+1$ dimensional diffeomorphism.

In \cite{Horava:2008ih}\cite{Horava:2009uw} (see e.g. \cite{Mukohyama:2010xz} for a review), they discussed a foliation preserving diffeomorphic theory of gravity. Their motivation is to improve the power-counting renormalizability of the quantum  gravity by adding higher spatial curvature terms without introducing higher time derivative terms to avoid ghost, or negative norm states. We are not particularly interested in the renormalizability problem, but we can borrow their idea, and consider the foliation preserving diffeomorphic theory of gravity in the holographic radial direction rather than time directions as in the original proposal.

We write the $d+1$ dimensional space-time metric in the form similar to the  Arnowitt-Deser-Misner (ADM) decomposition \cite{Arnowitt:1962hi}:
\begin{align}
ds^2 = N^2 dz^2 + g_{\mu\nu} (dx^\mu + N^\mu dz)(dx^{\nu} + N^{\nu}dz) \ ,
\end{align}
where $g_{\mu\nu}$ has the Lorentzian signature, and $N$ is an analogue of the lapse function and $N^\mu$ is that of the shift vector.
We demand the theory be invariant under the foliation preserving diffeomorphism\begin{align}
\delta x^{\mu} = \xi^{\mu}(z,x^{\mu}) \ , \ \ \delta z = f(z) \ . \label{foliap}
\end{align}
The simplest action up to the  second order derivatives would be
\begin{align}
\mathcal{S} = \int N dz \sqrt{|g|} d^d x (K^{\mu\nu} K_{\mu\nu} - \lambda K^2 + R_d + \Lambda) \ ,
\end{align}
where $K_{\mu\nu} = \frac{1}{2N}(\partial_z g_{\mu\nu} - D_{\mu} N_{\nu} - D_{\nu} N_\mu)$, and $K = g^{\mu\nu}K_{\mu\nu}$ with $D_\mu$ being the covariant derivative with respect to $d$-dimensional metric $g_{\mu\nu}$. $R_d$ is the $d$-dimensional Ricci scalar out of $g_{\mu\nu}$. The parameter $\lambda$ describes the deviation from the Einstein-Hilbert action, and $\lambda =1$ formally corresponds to the Einstein-Hilbert action up to  surface terms.

We can easily show that the theory has a solution of the Poincar\'e AdS-metric (\ref{Poincccc}), but crucial point is that although the scaling isometry is a foliation preserving diffeomorphism, the special conformal isometry is not: the coordinate transformation
\begin{align}
\delta x^\mu = 2(\rho^{\nu}x_\nu)x^\mu - (z^2 + x^\nu x_\nu)\rho^\mu \ , \ \ \delta z = 2(\rho^{\nu}x_\nu)z  \  \label{scnft}
\end{align}
is not the foliation preserving form (\ref{foliap}).
Therefore, if the foliation diffeomorphic theory of gravity with the Poincar\'e AdS-metric solution has a dual field theory interpretation, it cannot possess the full conformal invariance as the space-time symmetry.
It still possesses the scale invariance and the $d$-dimensional Poincar\'e invariance, so it should be dual to a scale invariant but non-conformal field theory.

Furthermore, if we perform the holographic Weyl anomaly computation in $d=4$ dimension, which is a generalization of the one we have reported in Einstein gravity in section \ref{holct},
we can derive \cite{Nakayama:2012sn}
\begin{align}
T^{\mu}_{\ \mu} = \frac{L^3}{16}\left(\mathrm{Weyl}^2 - \mathrm{Euler}  + \frac{2}{3}\frac{\lambda-1}{4\lambda-1} R^2 \right) \ ,
\end{align}
and the explicit appearance of $R^2$ term dictates that the dual field theory cannot be conformal invariant due to the Wess-Zumino consistency condition (see section \ref{Weyla}). We learned that the $R^2$ term cannot appear unless the theory violates the conformal invariance. In addition, the term is related to bilinear of $\mathcal{B}$ functions, so effectively, the deviation from the Einstein-Hilbert limit introduces non-trivial virial current.\footnote{The converse is not necessarily true. In non-unitary examples, vanishing $R^2$ anomaly does not immediately mean  conformal invariance. Indeed, if we computed the holographic Weyl anomaly for the model studied in section \ref{countex1}, we would obtain vanishing $R^2$ anomaly from the contribution of the gravity sector alone. The natural interpretation is that the metric $\chi_{IJ}$ is degenerate as can be suggested from the effective kinetic term of the ghost condensate.} 

At the same time, given our understanding of the importance of unitarity to show the enhancement from scale invariance to conformal invariance, it is very likely that the theory effectively violates the (strict) null energy-condition and the unitarity is lost. Again, the situation can be different in the original Horava time-like setup, in which we foliate the space-time with space-like Cauchy surface. Their setup may be consistent with the holography because as in the ghost condensation, the lack of the unitarity of the dual field theory may not be directly related to the inconsistency of the gravity dual (if any) in the Euclidean signature. It would be interesting to understand a possible (non reflection positive) scale invariant but non-conformal field theory as the dual of Horava gravity in the de-Sitter solution.

\subsection{Beyond classical Einstein gravity}
The holographic arguments so far have assumed the classical Einstein gravity coupled with a classical matter sector (except in section \ref{folia}). Some of the predictions such as $a=c$ are particular to the classical approximation to the gravity dual with Einstein-Hilbert action and it does not cover the entire space of the conformal field theories. In this subsection, we will discuss various attempts to introduce  corrections to the Einstein gravity in holography such as higher derivative corrections and quantum anomalous corrections. We discuss these aspects within the effective $d+1$ dimensional gravity. Ultimately, these must be embedded in the string theory to fully understand the quantum gravity, which we will leave for the future study.

\subsubsection{Higher derivative corrections}\label{hdc}
In quantum gravity (like string theory) the Einstein equation is modified in two different ways. The first one is higher derivative corrections that can be derived from the local action principle (e.g. $\alpha'$ corrections in string theory). The second one is possibly non-local corrections from quantum loop effects (e.g. $g_s$ corrections in string theory) including anomalous terms.

The effects of local higher derivative corrections in holographic renormalization group flow have been studied \cite{Myers:2010tj}\cite{Myers:2010xs} within the assumption that the gravity part of the action and the matter part of the action are separated in a minimal way. This restriction is ultimately related to the usage of the ``null energy condition" in higher derivative gravity: without the separation, the notion of the energy-condition becomes very much obscured.

Let us consider the higher derivative gravity with $\mathcal{O}(R^2)$ correction. We start with the action
\begin{align}
\mathcal{S} = -\frac{1}{2}\int d^{d+1} x \sqrt{|g|} \left(\frac{d(d-1)}{L^2} + R + b_1 L^2 R_{MNLK}^2 + b_2 L^2 R_{MN}^2 + b_3 L^2 R^2 \right) + \mathcal{S}_{\mathrm{matter}} \ .
\end{align}
The assumption of the minimal separation is that the matter action $\mathcal{S}_{\mathrm{matter}}$ does not contain curvature couplings.\footnote{Or more generally, we study the case in which any corrections can be thought as the corrections to the energy-momentum tensor, and the left hand side of the gravity equation is purely geometric.}
We do not expect that every gravitational theory satisfies the holographic $c$-theorem because the unitarity may be violated in a higher derivative gravity. 
Our strategy is to compute $T^{t}_{\ t} - T^{r}_{\ r}$ with the use of the higher derivative corrected equations of motion, and demand the null energy-condition. Can we still find the holographic $c$-function that monotonically decreases along the holographic renormalization group flow in the radial direction?

For this purpose, it is sufficient to require that $T^{t}_{\ t} - T^{r}_{\ r}$ can be written as the second order derivatives of the warp factor $A(r)$. Intuitively speaking, this eliminates the ghost mode in the fluctuations in the radial direction. If the equations of motion contain higher order in derivatives, it would suggest the ghost mode.\footnote{Strictly speaking, this explanation is rather superficial because the derivative here is the radial derivative and what is actually important is the time derivative. As discussed in \cite{Myers:2010tj}\cite{Myers:2010xs}, the condition is anyway necessary to avoid the ghost, so a posteriori, the argument here is justifiable.}
 This gives a constraint on the parameters in higher derivative gravity
\begin{align}
b_1 + \frac{d+1}{4} b_2 + d b_3 = 0 \ . \label{condition}
\end{align}
Then one can define the monotonically decreasing holographic $c$-function
\begin{align}
a_d(r) \equiv \frac{\pi^{d/2}}{\Gamma(d/2) A'(r)^{d-1}}( 1 - \hat{\lambda} L^2 A'(r)^2) \ ,  \label{defha}
\end{align}
where $\hat{\lambda} = 2(2b_1 + db_2 + d(d+1) b_3)$ with the constraint (\ref{condition}).
By using the higher derivative corrected Einstein equation with the condition (\ref{condition}) we obtain the higher derivative corrected holographic c-theorem:  
\begin{align}
a'_{d}(r) =  -\frac{\pi^{d/2}}{\Gamma(d/2) A'(r)^d} (T^{t}_{ \ t} - T^{r}_{\ r} ) \ge 0 \ ,
\end{align}
where we have assumed the null energy condition for $T_{\mu\nu}$ that is obtained from the matter action. 
Note that the requirement of vanishing higher derivative terms in $T^{t}_{\ t}- T^{r}_{\ r}$ in terms of the warp factor by using the modified Einstein equation gives the integrability condition on the holographic flow. Without the condition (\ref{condition}) we cannot find the good (or at least simple) monotonically decreasing function along the holographic renormalization group flow.

What was the physical origin of the constraint (\ref{condition})? We have seen that there are two  independent parameters that allow holographic $c$-theorem in $\mathcal{O}(R^2)$ gravity. We can check that these are the sum of the  Gauss-Bonnet term (which is Euler density in $d=4$ dimension) and Weyl$^2$ term with no independent $R^2$ term. The appearance of the Weyl$^2$ term is accidental because the geometry for the holographic renormalization group flow is conformally flat (irrespective of the shape of $A(r)$), so the  Weyl$^2$ term cannot affect the holographic renormalization group flow equation that we study at this point. 
The origin of the Gauss-Bonnet term is deeper. It does affect the equations of motion, but it does in such a way that there is no ghost mode along the renormalization group flow, and in addition, it assures the existence of the monotonically decreasing holographic $c$-function in any space-time dimension.

Actually, if we demand that there is no ghost mode not only along the radial direction  but along the other directions in the geometry of the holographic renormalization group flow, the only allowed $\mathcal{O}(R^2)$ corrections to the Einstein gravity is the Gauss-Bonnet term (if we did not include further higher derivative terms). Although the Weyl$^2$ term does not affect the holographic renormalization group flow, the fluctuation in the other directions contain a ghost.
 From the field theory viewpoint, unitarity of the theory is guaranteed by the absence of the ghost mode of the gravity and the null energy condition of the matter, so it seems reasonable that we have to assume the absence of the ghost mode in gravitational fluctuations to obtain the holographic $c$-theorem.\footnote{We should emphasize that we do not mean that the higher derivative terms other than the Gauss-Bonnet do not appear in consistent (quantum) gravity. Our discussion only suggests that we do not have to add further corrections in the Gauss-Bonnet case.}

In $d=4$ dimension, we can compute the value of $A'(r)$ at the AdS fixed point, and read the Weyl anomaly from the holographic renormalization analysis with higher derivative corrections. Since it requires a certain amount of computational details, we only quote the result \cite{Nojiri:1999mh}\cite{Blau:1999vz}: $a_d(r)$ defined in (\ref{defha}) agrees with the holographic Weyl anomaly $a$ (that couples to Euler density) at the conformal fixed point, but not with $c$ (that couples to Weyl$^2$) in $d=4$ dimension. This is non-trivial because within the Einstein gravity, we always obtain $a=c$ and we cannot make a distinction. We also note that there is no known way to construct the monotonically decreasing function ``$c(r)$" (in contrast with the above holographic $c$-function ``$a_d(r)$") that naturally interpolates $c$ in the higher derivative corrected holography. This seems in complete agreement with the field theory result in which $c$ is not monotonically decreasing but $a$ is. In general even dimension, we can show that the monotonically decreasing holographic $c$-function $a_d(r)$ is related to the Weyl anomaly that couples with the Euler density in even dimensions. It gives a supporting evidence for Cardy's conjecture in higher dimensions. In particular, we should note that the strong version of the $a$-theorem is realized in holography. Furthermore, the dilaton degrees of freedom can be introduced in the holography discussion (see \cite{Hoyos:2012xc}\cite{Bhattacharyya:2012tc}) in connection with the proof of the weak $a$-theorem reviewed in section \ref{weak}.

In odd dimensions, in particular in $d=3$ dimension, higher derivative corrections again enable us to distinguish various proposals for the interpretation of the monotonically decreasing function $a_d(r)$ along the holographic renormalization group flow. It turned out \cite{Myers:2010xs} that the higher derivative corrected $a_d(r)$ for the AdS background corresponds to the  $\mathbb{S}_d$ partition function and the entanglement entropy with sphere entangling surface which are equivalent at the conformal fixed point.

In reference \cite{Myers:2010tj}\cite{Myers:2010xs}  they generalize the above discussions by further including $\mathcal{O}(R^3)$ corrections and obtain the same conclusion. There are some restricted parameter regions in which the holographic renormalization group flow allows the monotonically decreasing holographic $c$-function $a_d(r)$. The parameter regions are interpreted as the combination of Weyl terms that do not change the holographic renormalization group equations, and quasi topological terms that avoid the ghost modes in holographic renormalization group flows.
Some other classes of higher derivative gravities (e.g. $f(R)$ gravity) have been studied in \cite{Liu:2010xc}\cite{Liu:2011iia}. In most of these examples, the decoupling between matter and the gravity sector is assumed, and it would be very interesting to see if we can generalize the discussion when the matter and gravity couple with each other through various curvature corrected terms because the notion of the energy-condition is very much obscured. Presumably, the unitarity of the total system must become important in the holographic realization of the generalized $c$-theorem.

Let us move on to our interest in the holographic enhancement from scale invariance to conformal invariance \cite{Nakayama:2010wx}. Once the holographic $c$-theorem is established, the argument in the last section can be naturally generalized. At the scale invariant fixed point, the metric must take the form of the AdS space-time. This do argument did not require any knowledge of the gravity equations of motion, but only the full $d+1$ dimensional diffeomorphism was assumed.  As for the dynamics, the higher derivative corrections do not affect the conclusion that the energy-momentum tensor $T^{t}_{\ t} - T^{r}_{\ r}$ must vanish for scale invariance because the AdS space-time is maximally symmetric. We postulate that this occurs if and only if the matter shows a trivial field configuration (a.k.a strict null energy-condition with higher derivative corrections), then the conformal invariance follows.

Of course, as long as we use the same matter action, the requirement of the strict null energy-condition is no different than in the Einstein theory. As long as we postulate the separation of the matter and gravity, the leading order unitarity is  governed by the same strict null energy condition, and we cannot relax it. The importance of the no-ghost mode is slightly indirect. The ghost mode in the radial direction would  allow non-trivial (non-AdS) holographic  renormalization group solution even if the matter saturates the null energy condition, which seems pathological, meaning that the effective matter metric responsible to the radial flow is singular.

\subsubsection{Quantum violation of null energy condition}
In the holographic argument above, the assumption of the null energy condition played a crucial role. It is interesting to observe, however, that the null energy condition can be violated quantum mechanically. Various sources of violations  \cite{Visser:1994jb}\cite{Urban:2009yt}\cite{Urban:2010vr} include
\begin{itemize}
\item Casimir effect
\item general squeezed quantum states
\item Weyl anomaly induced energy-momentum tensor in curved background
\item Hawking radiation
\item Orientifold 
\end{itemize}
It is an important question if these reported violations will invalidate the holographic $c$-theorem argument \cite{Nakayama:2012jv}. 

Since we do not know the general mechanism for the violation of the null energy condition, we focus on the universal violation from the Weyl anomaly, which should appear in any quantum field theories in curved backgrounds. For this purpose, we study the $\mathrm{AdS}_4$ case, whose field theory dual is a three-dimensional conformal field theory,\footnote{A canonical example of AdS/CFT correspondence in $d=3$ dimension is the so-called ABJM model \cite{Aharony:2008ug}, which is given by a quiver Chern-Simons gauge theory with $\mathcal{N}=6$ supersymmmetry.} based on the discussion in \cite{Nakayama:2012jv}. We emphasize that the correction of the equations of motion from anomaly are of order $\mathcal{O}(R^2)$, and it  gives the same order effect as the one studied in  section \ref{hdc}. The effect is clearly distinguishable, however, because the anomaly term cannot be  absorbed by the local curvature correction terms discussed in section \ref{hdc}. 
The Weyl anomaly induces the anomalous transformation for the energy-momentum tensor in $d=4$ dimension under the Weyl rescaling $\bar{g}_{\mu\nu} = \Omega^2 g_{\mu\nu}$ \cite{Page:1982fm} (see also \cite{Brown:1977sj}\cite{Cappelli:1988vw}\cite{Herzog:2013ed} for a conformally flat case):
\begin{align}
\bar{T}^{M}_{\ N} &= \Omega^{-4} T^{M}_{\ N} - 8c\Omega^{-4}\left(D^B D_A(C^{AM}_{\ BN} \log \Omega) + \frac{1}{2}R^{B}_{A} C^{A M}_{B N} \log\Omega \right) \cr
&-a\left(4\bar{R}^{B}_{A} \bar{C}^{AM}_{\ BN} - 2\bar{H}^{M}_N - \Omega^{-4}(4R^{B}_{A}C^{AM}_{\ BN} - 2H^{M}_{N}) \right) \ 
\end{align}
with 
\begin{align}
H_{MN} = -R_{M}^{A} R_{AN} + \frac{2}{3}RR_{MN} + \left(\frac{1}{2}R_{AB}R^{AB} - \frac{1}{4} R^2 \right) g_{MN} \ .
\end{align}
We have assumed $b' = 0$ in the Weyl anomaly by adding suitable counterterms $b' R^2$ because they are indistinguishable from the local higher derivative corrections studied in the previous subsection.\footnote{As discussed in  section \ref{hdc}, we have to tune this parameter to avoid the ghost mode and derive the holographic $c$-theorem.}

Let us begin with the AdS vacuum in which $T_{\mu\nu} \propto g_{\mu\nu}$.
The anomalous Weyl transformation of the energy-momentum tensor generates the universal contribution to the energy-momentum tensor in the holographic renormalization group background
\begin{align}
(\bar{T}^{r}_{\ r} - \bar{T}^t_{\ t})|_{\mathrm{anom}} = 4a A''(r) (A'(r))^2 \le 0 \ ,
\end{align}
which by itself violates the null energy condition because $A''(r) \le 0$ and $a \ge 0$. Accordingly, the equation of motion for the warp factor is modified as
\begin{align}
2A''(r) = (T^{t}_{\ t} - T^{r}_{\ r})|_{\mathrm{class}} - 4a A''(r) (A'(r))^2 \ .
\end{align}

Does this invalidate the holographic $c$-theorem? We claim that as long as the classical part of the energy-momentum tensor satisfies the null energy-condition, the holographic $c$-theorem is still intact in a slightly modified sense. To see this, we introduce the modified holographic $c$-function as
\begin{align}
a_3(r) \equiv \frac{\pi^{3/2}}{\Gamma(3/2) (A'(r))^2} - 4a \frac{\pi^{3/2}}{\Gamma(3/2)} \log A'(r) \ ,
\end{align}
and we observe it is monotonically decreasing
\begin{align}
a_3'(r) = -\frac{\pi^{3/2}}{\Gamma(3/2) (A'(r))^3)} (T^{t}_{\ t} - T^{r}_{\ r})|_{\mathrm{class}} \ge 0 \  
\end{align}
once we assume the null energy condition for the classical energy-momentum tensor.

The field theory interpretation of the logarithmic corrections to the holographic $c$-function seems very interesting. In \cite{Fuji:2011km}\cite{Marino:2011eh}, the localization computation showed that the $\mathbb{S}^3$ free-energy of the ABJM model contains $\log N$ corrections, and the supergravity 1-loop computation also suggests its existence \cite{Bhattacharyya:2012ye}. Since it was argued that classically the $\mathbb{S}^3$ free-energy is identified with the $\mathbb{S}^3$ free energy, the  appearance of the logarithmic corrections in holographic $c$-function is very promising.

As for the question of scale invariance and conformal invariance, if we assume the strict null energy-condition for the classical energy-momentum tensor, the argument is still valid. Actually, for AdS geometry, which is required from the scale invariance alone, the anomalous part of the energy-momentum tensor gives only a trivial contribution to the null energy-condition, so the discussion is in complete parallel with the higher derivative case. 

\subsection{Reduced symmetry}
The holography in a broader sense is applicable not only to Poincar\'e invariant quantum field theories, but also other non-gravitational systems with different space-time symmetries.
In this subsection, we give a holographic dual approach to the enhancement of conformal invariance from scale invariance in the case with reduced symmetries mentioned in section \ref{reduced}.

\begin{itemize}
\item
First we would like to consider the holographic dual of the chiral version of the $c$-theorem. Such a left-right asymmetric CFT in $d=2$ dimension is obtained by adding the gravitational Chern-Simons term. In \cite{Hotta:2009zn}, they computed the holographic renormalization group flow and showed the validity of the holographic $c$-theorem in dual chiral conformal field theories (with $c -\bar{c}$ constant along the renormalization group flow).

\item
A generalization of the AdS/CFT correspondence with a boundary was proposed and discussed in \cite{Takayanagi:2011zk}\cite{Fujita:2011fp}\cite{Nozaki:2012qd}, where they introduced the new boundary in the bulk space-time with the Neumann boundary condition (in contrast to the Dirichlet boundary condition at $z\to 0$ limit of the AdS space-time).

In this setup, we consider the Poincar\'e-AdS metric
\begin{align}
ds^2 = L^2 \frac{dz^2 + d\xi^2 + \eta_{ij} dx^i dx^j}{z^2} 
\end{align}
with the boundary at $\xi = \xi(z)$. The holographic $g$-function is defined by
\begin{align}
\log g(z) = \mathrm{Arcsinh}\left(\frac{d\xi(z)}{dz} \right) \ .
\end{align}
When the boundary energy-momentum tensor satisfies the null energy-condition, one can prove the holographic $g$-theorem (indices $a$ run through the boundary coordinate)
\begin{align}
-\frac{d\log g(z)}{d\log z} \sim T_{ab}k^a k^b \ge 0 \ . \label{holog}
\end{align}
As in the bulk case, the further assumption of the strict null energy condition demands that when the boundary $g$-function takes a constant value due to the scale invariance, there is no non-trivial field configurations at the boundary, and it results in the boundary conformal invariance \cite{Nakayama:2012ed}. If we assume the gauged non-linear sigma model as our boundary matter fields, then the argument is in complete parallel with the field theory argument.
In particular, (\ref{holog}) is understood as 
\begin{align}
\frac{d\log g(\mu)}{d\log \mu} = \mathcal{B}^I \chi_{IJ} \mathcal{B}^J , 
\end{align}
which we expect to hold in the boundary quantum field theories in general. Here $\mathcal{B}^I$ is the $\mathcal{B}$ functions for the boundary coupling and should be related to the radial evolution of the boundary field $\partial_z \phi^I$.
The stronger gradient formula was derived in $1+1$ dimensional setup in \cite{Friedan:2003yc} from the field theory argument. Presumably, the introduction of the boundary ``superpotential" will give the holographic gradient formula.

\item
In section \ref{reduced}, we have discussed the enhancement of chiral scale invariance to chiral conformal invariance. A similar argument applies \cite{Nakayama:2011fe}. If we assume time and space translations and chiral scale invariance, the most generic metric is given by the warped $\mathrm{AdS}_3$ space-time:
\begin{align}
ds^2 = \frac{1}{c}\left(\frac{bdt}{z} - cdx\right)^2 + \frac{e(dz)^2 - (a+\frac{b^2}{c})dt^2}{z^2} \ 
\end{align}
up to coordinate transformation, which has the enhanced $SL(2) \times U(1)$ isometry. This corresponds to the enhanced chiral conformal invariance.\footnote{A similar argument was presented by D.~Honda and M.~Nakamura \cite{HN}.} 

The matter contributions may break the chiral conformal invariance. For instance is it possible to introduce a vector condensation
\begin{align}
A_Mdx^M = \alpha \frac{dz}{z} + \beta dx + \gamma \frac{dt}{z} 
\end{align}
but again the (strict) null energy condition will forbid such a non-trivial field configuration unless it is consistent with the $SL(2) \times U(1)$ isometry (up to the field redefinition discussed in \cite{Nakayama:2011fe}).\footnote{The strict null energy condition in this context is actually stronger than the requirement in  our non-chiral discussions in the other sections. Although we do not know any physical counterexamples, it is interesting to see why this must be satisfied. Presumably, it is related to an extra assumption of the validity of the Reeh-Schlieder theorem in the field theory argument.} The chiral conformal invariance in the warped $\mathrm{AdS}_3$ space-time has been studied in the literature (see e.g. \cite{Compere:2008cv}\cite{Song:2011sr}), in which the emergence of the chiral Virasoro symmetry is also discussed.

\item 

Non-relativistic systems show various interesting scaling or conformal symmetries such as Schr\"odinger symmetry \cite{Son:2008ye}\cite{Balasubramanian:2008dm}, Lifshitz symmetry \cite{Kachru:2008yh}, Galilean conformal symmetry \cite{Bagchi:2009my} and so on. Correspondingly, gravitational dual descriptions with these symmetries have been investigated.

For our interest in the distinction and the relation between scale invariance and conformal invariance, let us take a look at the Schr\"odinger holography. The $d$ dimensional Schr\"odinger algebra can be realized as an isometry of $d+2$ dimensional space-time with one null direction $\zeta$ compactified.
\begin{align}
ds^2 = -2\frac{dt^2}{z^4} + \frac{-2dtd\zeta + dx_i^2 + dz^2}{z^2} \ .
\end{align}
The non-relativistic special conformal transformation is realized by the isometry
\begin{align}
(t,\zeta,x_i,z) \to \left(\frac{t}{1+\eta t}, \zeta - \frac{\eta}{2}\frac{x_i^2 + z^2}{1+\eta t}, \frac{x_i}{1+\eta t}, \frac{z}{1+\eta t}\right) \ .
\end{align}
It can be shown that if we try to deform the geometry so that we preserve the scale invariance and Galilei invariance acting as $(\zeta, x_i) \to (\zeta -v_i x_i + \frac{1}{2}v^2 t, x_i -v_it)$, then there is no such a geometric deformation that breaks non-relativistic special conformal invariance.

In order to support the Schr\"odinger geometry, which is not Einstein, we have to introduce the matter. A typical example is a massive vector field. From scale invariance and Galilean invariance, the matter vector field takes the form
\begin{align}
A_M dx^M = \alpha \frac{dt}{z^2} + \gamma \frac{dz}{z} \ .
\end{align}
The first term is compatible with the non-relativistic conformal invariance and needed for the geometry. The second term, however, is not invariant under the non-relativistic conformal invariance, and may lead to the scale invariant but non conformal field configuration. It may be possible to forbid such a configuration by introducing a certain stronger notion of the strict null energy-condition, but the validity is yet to be addressed.

\end{itemize}

\subsection{Further thoughts}
We have seen various approaches to the possible enhancement of conformal invariance from scale invariance from the holographic viewpoint. We would like to give further thoughts or somewhat more philosophical perspectives on this issue to conclude the section.

\subsubsection{More on literature}
A precursor of the AdS/CFT correspondence \cite{Brown:1986nw} is a discovery of the fact that the AdS space-time has the so-called asymptotic symmetry group isomorphic to the conformal group. In particular, this asymptotic symmetry group is how the Virasoro symmetry is realized in $\mathrm{AdS}_3$ space-time, in which the geometry itself does not possess the infinite dimensional isometry. In this  review article, we did not discuss the realization of scale invariance or conformal invariance as an asymptotic symmetry group. We have only focused on the realization by the isometry. It is an interesting question if such a realization would give a new perspective on the subject.

With this regard, the Kerr-CFT \cite{Guica:2008mu} (see e.g. \cite{Compere:2012jk} for a review) is one concrete example of realizing the (chiral) conformal algebra not as an isometry of the system, but as an asymptotic symmetry group. Since the asymptotic symmetry group does not specify whether the theory is in the vacuum state, the Kerr-CFT has its own temperature. Most of our discussion in review article is done in the vacuum state, and it would be interesting to see if we can generalize the argument with non-zero temperature. 

In \cite{Douglas:2010rc}, they proposed an interesting attempt to derive (a special class of) AdS/CFT from a free field theory. They showed that the (singlet) correlation functions of the free scalar field theory can be computed from the restricted sector of the higher spin gauge theory. They further showed that the vacuum solution of the higher spin gauge theory is the AdS space-time. This begs the question what will happen if we consider the free Maxwell theory (rather than the scaler theory) which is not conformal invariant (unless in $d=4$ dimension). A similar construction was done in \cite{Nakayama:2012sn}, and it was shown that the theory seems only invariant under the foliation preserving diffeomorphism, but the situation in $d=4$ dimension must be clarified.

In a recent paper, a possible way to circumvent the holographic $c$-theorem and and a possibility to construct the holographic geometry that shows the cyclic behavior (with manifest Poincar\'e invariance) was proposed in \cite{Balasubramanian:2013ux}. It was argued that the higher dimensional null energy condition does not necessarily lead to an immediate inconsistency with the holographic $c$-theorem when the warped compactification does not allow the effective truncation to the lower $d+1$ dimension. It would be very interesting to understand the physics of the dual field theory, if any. We, however, believe that the presented metric has the hidden AdS isometry.

Finally let us point out that there are a couple of interesting but slightly different geometric constructions of the  holographic dual for scale invariant but non-conformal field theories. In \cite{Awad:2000ac}, they discussed the violation of the conformal invariance by putting the boundary theory on a curved background while preserving the scale invariance. In \cite{ReyNakayama}, they discussed the violation of the conformal invariance by putting the boundary theory on a non-anti-commutative background while preserving the scale invariance.

\subsubsection{Final Project}

Originally, this review article was prepared as a lecture note, in which we have presented various exercises. Most of the exercises left for the reader have been answered in this review article.
 The final project of the original lecture note was the following question. We have decided to keep it without providing the answer (if any).

\

\begin{shadebox}
(Project) Prove the $\mathcal{N}=4$ supersymmetric Yang-Mills theory is superconformal invariant.
\end{shadebox}

\

First, we try to answer the question from the field theory perspective. 

\begin{enumerate}
\item We begin with the simplest case: show that the $U(1)$ $\mathcal{N}=4$ supersymmetric Yang-Mills theory is superconformal invariant. The Abelian theory is free, so it must be easy. What is the role of the improvement terms?

\item Let us move on the non-Abelian case. Our first task is to show that beta function vanishes for the $\mathcal{N}=4$ supersymmetric Yang-Mills theory. 
The rough argument goes like this. Use the holomorphic scheme and show that the  holomorphic gauge coupling constant is not renormalized. Then consider the wavefunction renormalization of the adjoint ``matter" superfields. Use the $\mathcal{N}=4$ supersymmetry to connect it to the gauge coupling constant and show it is not renormalized either. Then the physical beta function must vanish. See e.g. \cite{Seiberg:1988ur} for a more complete argument.

\item But we know that vanishing of the beta function is not enough to declare the (super)conformal invariance. Use the superfield structure and show that there is no vector beta functions. We can find the discussions in \cite{Nakayama:2012nd}\cite{Fortin:2012hc}.

\item Combining them together, we learn that the $\mathcal{N}=4$ supersymmetric Yang-Mills theory has vanishing $\mathcal{B}$ function, and it is superconformal.
\end{enumerate}

Alternatively, we could begin with the holographic dual.

\begin{enumerate}
\item Show that the type IIB supergravity on $\mathrm{AdS}_5 \times \mathbb{S}^5$ has the symmetry that corresponds to the $\mathcal{N}=4$ superconformal algebra. In this lecture, we discussed the bosonic part of the statement. What is the role of the improvement terms?

\item Make an argument that the symmetry is not anomalous in the perturbative string theory on $\mathrm{AdS}_5 \times \mathbb{S}^5$. 
Maybe we had better use Berkovits formulation \cite{Berkovits:2004xu}. 

\item Prove the AdS/CFT correspondence.

\end{enumerate}

If you are a firm believer of the holography, you may skip the last part of the holographic argument. We will discuss  possible directions to pursue in section \ref{derive} for the derivation of holography from the local renormalization group that we have heavily used during the discussions of scale invariance vs conformal invariance.

\subsubsection{Derivation of holography from local renormalization group?}\label{derive}

In this section, we have addressed the deep connection between holography and the renormalization group flow. As a final comment, we would like to mention one ambitious approach to the quantum gravity from the attempts to derive the holographic dual bulk theory constructively by utilizing the (local) renormalization group flow. While this approach is yet to be throughly scrutinized, we would like to give a brief comment on this approach with some emphasis on the possibility of scale invariant but non-conformal geometry in holography.

The first observation is that the local Callan-Symanzik equation in $d$ dimension may be interpreted as the Hamiltonian constraint, or the constraint coming from the variation of the Lapse function in the holographic dual bulk gravitational  system in $d+1$ dimension. This is reasonable because the invariance of the Schwinger functional under the change of the local scale transformation is the physical content of the local Callan-Symanzik equation and the invariance of the GKP-W partition functional under the change of the Lapse function (due to the $d+1$ dimensional diffeomorphism) is the physical content of the Hamiltonian constraint.\footnote{The momentum constraint can be also derived from the fact that the local renormalization group preserves the diffeomorphism at each local energy scale.}

Alternatively, we may say that the origin of the $d+1$ dimensional diffeomorphism in holography is the local renormalizability of the Schwinger functional. 
The form of the local renormalization group operator 
\begin{align}
\Delta_{\sigma} &= \int d^dx \sqrt{|g|} \left( 2\sigma g_{\mu\nu} \frac{\delta}{\delta g_{\mu\nu}} + \sigma \beta^I \frac{\delta}{\delta g^I}  \right. + \left. \left( \sigma \rho_I^a D_{\mu} g^I - (\partial_\mu \sigma) v^a \right) \frac{\delta}{\delta a^a_{\mu}} \right) \  
\end{align}
has a suggestive form of $H = \sum \dot{Q} \cdot P$ if we regard the momentum as the derivative operators acting on the ``wave functional of the universe", and $\dot{Q}$ is the radial change of the coupling constant (general coordinate) through the beta functions.

However, we need a little bit more information to address the more precise relation between the local renormalization group flow and the holographic bulk equations of motion. The point is that the renormalization group equation is always deterministic (and first order) for any quantum field theories, but in holography, we need a particular semi-classical limit to obtain the classical description. Otherwise, the bulk system is quantum mechanical and not deterministic.

A further  crucial observation was made by S.~S.~Lee \cite{Lee:2010ub}\cite{Lee:2012xba}\cite{Lee:2013dln} to circumvent this point: in the large $N$ limit, we may effectively project the renormalization group flow in the multi-trace deformations down to that of the single trace deformations provided we introduce averaging over all the possible single trace couplings. This leads to the so-called quantum local renormalization group flow, in which the sources (for the single trace operators) are all dynamical.
It was demonstrated that this quantum renormalization group was the origin of the quantum mechanical properties of the bulk system. Schematically, the dual Hamiltonian is obtained by
\begin{align}
H(P_I, Q^J) = (\beta^I(Q) + \partial_z Q^I) P_I  + G^{IJ}(Q) P_I P_J + V(Q^I) \ ,
\end{align}
where $\beta^I (Q)$ is the single trace beta function while $G_{IJ} (Q)$ is the double trace beta function. The potential term $V(Q^I)$ comes from the vacuum contribution to the local renormalization group (e.g. anomaly term). The theory is quantum mechanical in the sense that we do path integral over $P_I$ and $Q^I$.
Assuming that the single trace beta function satisfies the gradient flow property, we obtain the usual second order kinetic action for $Q^I$ after integrating out $P_I$ to reach the Lagrangian formulation.
A further check of the construction and its consistency with the holographic Weyl anomaly can be found in \cite{Nakayama:2014cca}. In particular, under certain assumptions, one may derive the Einstein-Hilbert action from the hypothetical conformal field theory which has the single trace energy-momentum tensor alone.

The Hamiltonian constraint and its consistency relations remain valid in the quantum local renormalization group, and as we have mentioned they give the origin of $d+1$ dimensional diffeomorphism rather than the $d$-dimensional diffeomorphism naturally equipped in the Schwinger functional. Similarly, the ambiguity of the beta functions we have discussed is the source of the $d+1$ dimensional gauge invariance rather than the mere $d$-dimensional gauge invariance. In this quantum local renormalization group approach, we can manifestly construct the $d+1$ dimensional gauge invariant bulk action from the $d$-dimensional dual field theory in large $N$ limit that satisfies the local renormalization group equations (with some assumptions on the derivative expansions), but the crucial point behind the construction is that the local renormalization group already knows the $d+1$ dimensional gauge symmetry.

In section \ref{folia} we argued that scale invariance without conformal invariance may be realized in foliation preserving diffeomorphic theory of gravity  rather than generally covariant gravitational theory. Our discussion may suggest that the validity of the general covariance in our universe (if it were holographically emergent) can be explained  by the fact that there is no scale invariant but non-conformal field theory in $d=3$ dimension.  Our journeys to find the reason for the possible enhancement from conformal invariance from scale invariance may shed a light on the space-time structure of our universe and the deep properties of the quantum gravity.

\newpage

\section{Conclusions}
So this review article based on the lectures given in the 5th Taiwan School on Strings and Fields is almost finished, 
but our journey still continues. 
In this review article, we have shown as many examples of scale invariant but possibly non-conformal field theories as possible. We have tried to argue such examples are extremely rare and most probably inconsistent with some important assumptions in quantum field theories. We have discussed various approaches to the question, and as we have provided various applications in the introduction, this may be just the beginning. 

We hope that in the near future, the enhancement from scale invariance to conformal invariance is proved (or disproved) in higher dimensions, and the necessary condition for the claim is stated in a clear manner. On the other hand, the implication of the enhancement of symmetry in holography would be very helpful to understand the detailed consistency conditions of the quantum gravity, and may even lead to the derivation of the holographic principle.  

The author always feels that there is a deep space-time structure behind the enhancement of conformal invariance from scale invariance. Let us consider Zamolodchikov's $c$-theorem and its higher dimensional analogues, which would play a significant role in understanding the enhancement.
Even though our understanding of the $c$-theorem is based on the intuition of ``coarse graining" in renormalization group flow, whenever we try to make the statement concrete, we had to assume the notion of ``time" such as unitarity, causality, energy-condition and so on, all of which are not always available in the Euclidean statistical systems. Probably there is a magic in Wick rotation and the renormalization group flow. With this regard, the author always finds the first chapter of the textbook by Polyakov \cite{polyakov} mesmerizing.

As Einstein once said, ``Subtle is the Lord, but malicious He is not". His own explanation of the meaning is ``Nature hides her secret because of her essential loftiness, but not by means of ruse." Indeed, a beautiful symmetry may be secretly hidden unless we try hard to understand it as our conformal invariance. We need to choose a good probe (e.g. energy-momentum-tensor) and respect it very carefully (in the renormalization prescription).

\begin{figure}[tbh]
\begin{center}
\includegraphics[width=0.8\linewidth]{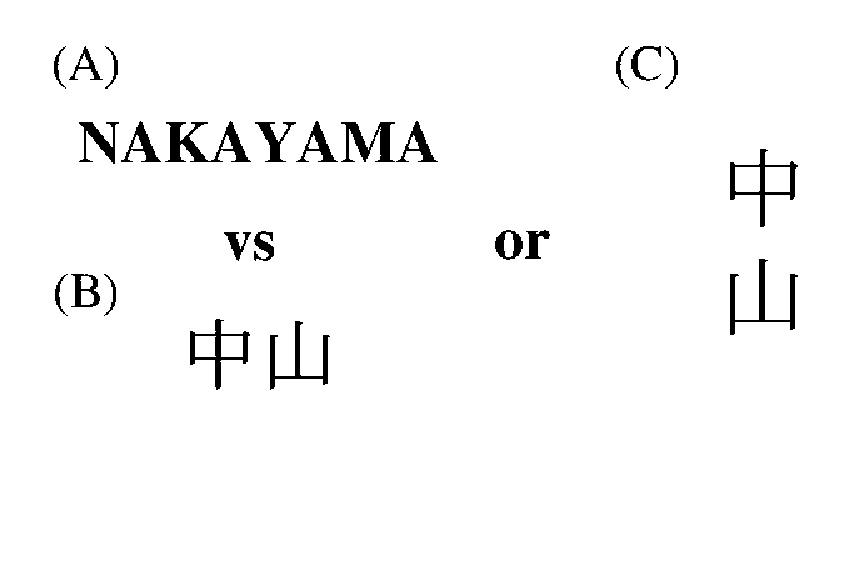}
\end{center}
\caption{(A) The author's family name is written alphabet. (B) Same but in Japanese (or Chinese) characters. (C) This is the traditional form which Sun Yat-sen must have encountered in Tokyo.}
\label{fig3}
\end{figure}

Let the author offer one analogy to finish this review article. The author's family name ``Nakayama" has a hidden symmetry. See fig \ref{fig3}. You cannot see it in alphabet. To uncover the hidden symmetry, you need to pay respect, and use a proper probe. In this case, you have to look at their Japanese or Chinese characters (which are the same in this case). Gradually you will see the symmetry pattern, but you need one more step. Those who use alphabet may be accustomed to writing the characters from left to right (and then top to bottom), but in traditional Japanese or Chinese, you write them from top to bottom (and right to left). Now you understand that there is a hidden axisymmetry in the author's family name. It literally means the middle mountain.

There is a further story to it.\footnote{The author would like to thank H.~Nakajima for informing me of the history after his lecture.}  A great leader of Taiwan, Sun Yat-sen (1866-1925) once visited Japan. He lived close to Hibiya-Park in Tokyo. Near the park, there was a mansion whose family name tag was ``Nakayama". 
He immediately liked the symmetry of the name very much (of course Japanese name tag is written from top to bottom), and he decided to call himself ``Zhongshan", which is the Chinese way to read the Japanese characters for ``Nakayama". Thanks to this great leader, the author's family name has become very popular in Taiwan. Unfortunately, since Japanese and Chinese read the same characters in a very different way, without writing down in characters, they do not recognize they are the same. That is why  the author wrote his name down in Chinese characters when he gave the lectures in Taiwan.
The symmetry is only shared after using the proper communication tool. Anyway, it was the author's greatest pleasure to give these lectures in Taiwan. The author wishes the participants (and the readers of this review!) had learned something from them. 

\section*{Acknowledgements}
This lecture note is prepared for the 5th Taiwan School on Strings and Fields.
The author would like to thank the organizers, in particular C.~M.~Chen for the host, and all the  participants for kind invitation and stimulating discussions.

I also thank all the people I talked with for stimulating discussions on the subject. I, in particular, thank  S.~El-Showk, C.~Ho, S.~Rey and S.~Rychkov for collaborations, and P.~Argyres, D.~Bak, K.~Balasubramanian, J.~Cardy,  S.~Deser, P.~Di Vecchia, S.~Dubovsky, M.~Duff, A.~Edery, H.~Elvang, J.~Fortin, P.~Horava, Y.~Iwasaki,  R.~Jackiw, C.~Keeler, A.~Konechny, S.~Kuzenko,  Z.~Komargodski, S.~S.~Lee, R.~Myers, A.~Migdal, E.~Mottola, S.~Mukohyama, C.~Nunez, H.~Osborn, J.~Polchinski, A.~Shapere, K.~Skenderis,  S.~Solodukhin, A.~Stergiou and Y.~Tachikawa for discussions and correspondence.

\newpage

\appendix 

\section{Useful formulae and miscellaneous topics}
\subsection{Weyl transformation}\label{convention}
In this appendix, we review the Weyl transformation properties  of various tensors. The sign convention in the review article is the same as that of Wald \cite{Wald:1984rg}, which is $s_1 = s_2 = s_3= +$ in the  Misner-Thorne-Wheeler convention \cite{Misner:1974qy}. Note, however, that  the action density used in this review article is minus of the Lagrangian density in the Lorentzian signature.
We start with the definition of curvature tensors
\begin{align}
\Gamma^\lambda_{\mu\nu} &= \frac{1}{2} g^{\lambda \sigma} (\partial_\nu g_{\sigma \mu} + \partial_{\mu} g_{\sigma \nu} - \partial_\sigma g_{\mu\nu} )  \cr
R^{\ \ \ \ \lambda}_{\mu\nu\sigma} &= \partial_\nu \Gamma^{\lambda}_{\mu\sigma}  + \Gamma^{\lambda}_{\tau\nu} \Gamma^{\tau}_{\mu\sigma} - \partial_\mu \Gamma^{\lambda}_{\nu\sigma} - \Gamma^{\lambda}_{\tau\mu} \Gamma^{\tau}_{\nu\sigma} \ \cr
R_{\mu\nu} &= R^{\ \ \ \ \lambda}_{\mu\lambda\nu} \cr
R &= g^{\mu\nu} R_{\mu\nu} \ \cr
G_{\mu\nu} &= R_{\mu\nu} -\frac{1}{2}R g_{\mu\nu} \ .
\end{align}

Under the finite Weyl transformation\footnote{Note that our Weyl transformation is minus that of \cite{Osborn:1991gm}.}
\begin{align}
g_{\mu\nu} \to e^{2\sigma(x)} g_{\mu\nu}
\end{align}
they transform as (see e.g. appendix of \cite{Wald:1984rg})
\begin{align}
R^{\ \ \ \ \lambda}_{\mu\nu\sigma} &\to R^{\ \ \ \ \lambda}_{\mu \nu \sigma} + \delta^{\lambda}_{\sigma}( D_\mu D_\nu\sigma - \partial_\mu \sigma \partial_\nu \sigma+ g_{\mu\nu} \partial_\rho \sigma \partial^\rho \sigma) + g_{\mu\nu}(D^\lambda \partial_\sigma \sigma - \partial^\lambda \sigma \partial_\sigma \sigma) - (\nu \leftrightarrow \sigma) \cr
R_{\mu\nu} &\to R_{\mu\nu} - g_{\mu\nu} \Box \sigma - (d-2) (D_\mu \partial_\nu\sigma-\partial_\mu \sigma \partial_\nu \sigma + g_{\mu\nu} \partial_\lambda \sigma \partial^\lambda \sigma) \cr
R &\to e^{-2\sigma} (R - 2(d-1) \Box \sigma - (d-1)(d-2) \partial_\mu \sigma \partial^\mu \sigma) \ .
\end{align}
The traceless part of the Riemann-tensor is known as Weyl tensor
\begin{align}
C^{\ \ \ \ \lambda}_{\mu\nu\sigma} &= R^{\ \ \ \ \lambda}_{\mu\nu\sigma} - \frac{1}{d-2} (\delta^{\lambda}_\nu R_{\mu\sigma} + g_{\mu\sigma}R^{\lambda}_{\ \nu} - (\nu \leftrightarrow \sigma)) - \frac{1}{(d-2)(d-1)}(\delta^{\lambda}_{\sigma}g_{\mu\nu} - \delta^{\lambda}_{\nu}g_{\mu\sigma} )R \ , 
\end{align}
and it is invariant under the Weyl transformation
\begin{align}
C^{\ \ \ \ \lambda}_{\mu\nu\sigma} \to C^{\ \ \ \ \lambda}_{\mu\nu\sigma} \ .
\end{align}
In $d=4$ dimension, the Weyl transformation of the Euler term is given by
\begin{align}
&\sqrt{|g|} \mathrm{Euler} \to \sqrt{|g|} \mathrm{Euler} \cr
&+ 4\sqrt{|g|} D^\mu \left(-R \partial_\mu \sigma + 2R_{\mu}^{ \ \nu} \partial_\nu \sigma - D_{\mu}(\partial_\nu \sigma\partial^\nu\sigma) + 2 \partial_\mu \sigma \Box \sigma + 2 \partial_\nu \sigma \partial^\nu \sigma \partial_\mu \sigma) \right) \ .
\end{align}
Note that the inhomogeneous term is a total derivative.

\subsection{Energy-momentum tensor correlation functions}\label{EMtensor}
In this appendix, we will show the correlation functions of the energy-momentum  tensor in conformal filed theories to read various Weyl anomaly coefficients, in particular $a$ and $c$ anomaly in $d=4$ dimension. We implicitly assume that the theory does not violate CP, so the following formulae contain no CP-violating term with Levi-Civita tensor. We also ignore the contact terms.

In two-dimensional conformal field theories, the two-point function and three-point function of the energy-momentum tensor are governed by the conformal invariance up to one-number, which is the central charge $c$:
\begin{align}
\langle T(z) T(w) \rangle &= \frac{1}{(2\pi)^2}\frac{c}{2(z-w)^4} \cr
\langle T(z_1) T(z_2) T(z_3) \rangle &= \frac{-1}{(2\pi)^3} \frac{c}{(z_1-z_2)^2(z_2-z_3)^2 (z_3-z_1)^2} \ ,
\end{align}
where we use the holomorphic coordinate.\footnote{Our normalization is different from the string theory literature.}

In higher dimensions, the two-point function is again uniquely specified up to an overall number
\begin{align}
\langle T_{\mu\nu} (x) T_{\sigma \rho}(0) \rangle = \frac{c_T}{x^{2d}} \mathcal{I}^T_{\mu\nu,\sigma \rho}(x) \ ,
\end{align}
where
\begin{align}
\mathcal{I}^T_{\mu\nu,\sigma \rho} (x) = \frac{1}{2} \left(I_{\mu\sigma}(x) I_{\nu\rho}(x) + I_{\mu\rho}(x)I_{\nu\sigma}(x) \right) -\frac{1}{d}\delta_{\mu\nu} \delta_{\sigma \rho} \ 
\end{align}
with
\begin{align}
I_{\mu\nu}(x) = \delta_{\mu\nu} - 2\frac{x_\mu x_\nu}{x^2} \ .
\end{align}
In $d=4$ dimension, the coefficient of the two-point function $c_T$ is given by the Weyl anomaly $c$ as 
\begin{align}
c_T = \frac{640}{\pi^2} c \ .
\end{align}

The three-point function is sufficiently complicated. It was shown in \cite{Osborn:1993cr} (see also \cite{Erdmenger:1996yc}) that it is given by
\begin{align}
\langle T_{\mu\nu}(x_1) T_{\sigma\rho}(x_2)T_{\alpha\beta}(x_3) \rangle = \frac{1}{x_{12}^d x_{13}^d x_{23}^d} \Gamma_{\mu\nu,\sigma\rho,\alpha\beta}(x_1,x_2,x_3)
\end{align}
with
\begin{align}
\Gamma_{\mu\nu,\sigma\rho,\alpha\beta}(x_1,x_2,x_3)  
=& \mathcal{E}^T_{\mu\nu,\mu'\nu'} \mathcal{E}^T_{\sigma \rho, \sigma' \rho'} \mathcal{E}^{T}_{\alpha \beta, \alpha' \beta'} \cr 
&\left[A I_{\nu'\sigma'}(x_{12})I_{\rho'\alpha'}(x_{23})I_{\beta'\mu'}(x_{31})  + BI_{\mu'\sigma'}(x_{12})I_{\nu'\alpha'}(x_{23})X^2_{\rho'}X^3_{\beta'}(x_2-x_3)^2 + \mathrm{perm}  \right] \cr
&+C\mathcal{I}^T_{\mu\nu,\sigma\rho}(x_{12})\left(\frac{X^3_{\alpha}X^3_{\beta}}{(X^3)^2} -\frac{1}{d}\delta_{\alpha\beta} \right) + \mathrm{perm} \cr
&+ D\mathcal{E}^T_{\mu\nu,\mu'\nu'} \mathcal{E}^{T}_{\sigma\rho,\sigma'\rho'} X^1_{\mu'}X^2_{\sigma'}(x_1-x_2)^2 I_{\nu'\rho'}(x_{12})\left(\frac{X^3_{\alpha}X^3_{\beta}}{(X^3)^2} -\frac{1}{d}\delta_{\alpha\beta} \right) + \mathrm{perm} \cr
&+ E \left(\frac{X^1_{\mu}X^1_{\nu}}{(X^1)^2} -\frac{1}{d}\delta_{\mu\nu} \right) \left(\frac{X^2_{\sigma}X^2_{\rho}}{(X^2)^2} -\frac{1}{d}\delta_{\sigma\rho} \right) \left(\frac{X^3_{\alpha}X^3_{\beta}}{(X^3)^2} -\frac{1}{d}\delta_{\alpha\beta} \right)
\end{align}
where 
\begin{align}
\mathcal{E}^{T}_{\mu\nu,\sigma\rho} = \frac{1}{2}(\delta_{\mu\sigma}\delta_{\nu\rho} + \delta_{\mu\rho} \delta_{\nu\sigma}) - \frac{1}{d}\delta_{\mu\nu} \delta_{\sigma \rho} 
\end{align}
is the projection operator onto symmetric traceless tensors.
We have also introduced ($i=1,2,3$ mod $3$)
\begin{align}
X^i_\mu = \frac{(x_{i+1}-x_{i})_\mu}{(x_{i+1}-x_{i})^2} - \frac{(x_{i+2}-x_{i})_\mu}{(x_{i+2}-x_{i})^2} \ .
\end{align}

The conservation of the energy-momentum tensor demands
\begin{align}
(d^2-4) A + (d+2) B -4d C -2D &= 0 \cr
(d-2)(d+4)B -2d(d+2)C + 8D - 4E &= 0 \ .
\end{align}
There are three free parameters left, two of which are related to $a$ and $c$ in $d=4$ dimension. The relation has been worked out in \cite{Osborn:1993cr}\cite{Erdmenger:1996yc} as
\begin{align}
c &= \frac{\pi^4}{640\times 12} (9A - B - 10C) \cr
a &= \frac{\pi^4}{512\times 90}(13A -2B -40C) \ .
\end{align}
In terms of free fields (in $d=4$ dimension), we have
\begin{align}
A &= \frac{1}{\pi^6} \left(\frac{8}{27}N_0 - 16N_1 \right) \cr
B &= -\frac{1}{\pi^6} \left(\frac{16}{27}N_0 + 4N_{1/2} + 32 N_1 \right) \cr
C &= -\frac{1}{\pi^6} \left(\frac{2}{27}N_0 + 2N_{1/2} + 16 N_1 \right) \ .
\end{align}

\subsection{Local Wess-Zumino consistency condition}\label{localWZC}
In  this appendix, we will summarize the Wess-Zumino consistency condition of the local renormalization group. We follow the convention of \cite{Erdmenger:2001ja} rather than that in \cite{Osborn:1991gm}. There are a couple of sign difference there.

We begin with the most generic candidates for the Weyl anomaly that depends
on the metric and space-time dependent coupling constants in $d=4$ dimension:
\begin{align}
-\delta_{\sigma} W  = \int d^4x \sqrt{|g|} \left( \sigma T + \partial^\mu \sigma Z_{\mu} \right)
\end{align}
where
\begin{align}
T &= c\mathrm{Weyl}^2 - a \mathrm{Euler} + \frac{1}{9}\tilde{b} R^2  \cr
&+ \frac{1}{3}\chi_I^e D_{\mu} g^{I}\partial^\mu R + \frac{1}{6}\chi_{IJ}^f D_{\mu} g^I D^{\mu} g^J R - \frac{1}{2}\chi_{IJ}^g D_{\mu}g^I D_{\nu}g^J G^{\mu\nu} + \frac{1}{2}\chi_{IJ}^a D^2 g^I D^2 g^J \cr
&+ \frac{1}{2}\chi_{IJK}^{b} D_{\mu} g^I D^{\mu} g^J D^2 g^K + \frac{1}{4}\chi_{IJKL}^c D_\mu g^I D^{\mu} g^J D_{\nu} g^K D^{\nu} g^L \cr
&+ \frac{1}{4}F_{\mu\nu} \kappa F^{\mu\nu} + \frac{1}{2} F^{\mu\nu} \zeta_{IJ} D_\mu g^I D_{\nu} g^J \ . \label{traceanomaly}
\end{align}
and
\begin{align}
Z_{\mu} &= -G_{\mu\nu} w_I D^{\nu}g^I + \frac{1}{3} \partial_\mu (qR) + \frac{1}{3} R Y_I D_{\mu} g^I + F_{\mu\nu} \eta_I D^\nu g^I \cr
& + \partial_\mu(U_I D^2 g^I + \frac{1}{2}V_{IJ} D_{\nu} g^I D^\nu g^J) + S_{IJ} D_{\mu}g^I D^2 g^J + \frac{1}{2}T_{IJK} D_{\nu}g^I D^{\nu} g^J D_{\mu} g^K \ .
\end{align}
$c,a,\tilde{b},\chi_I^e , \chi_{IJ}^f, \chi_{IJ}^g , \chi_{IJ}^a, \chi_{IJK}^b , \chi_{IJKL}^c, w_I, q, Y_I , U_I, V_{IJ}, S_{IJ}$ and $T_{IJK}$ are gauge invariant tensors on the coupling constant space $g^I$. $\kappa, \zeta_{IJ}$ and $\eta_{I}$ are tensors that take values on the Lie algebra of the ``flavor" symmetries. As discussed in the main text, our ``flavor" symmetries act on $g^I$, so the covariant derivative $D_{\mu} = \partial_\mu - a_\mu$ contains their connection, too. $F_{\mu\nu}$ here is the curvature constructed out of $a_\mu$.
For simplicity, as in \cite{Osborn:1991gm}, we do not consider the CP violating terms as well as anomaly for the ``flavor" symmetries.
If the ``flavor symmetries" are anomalous with each other, we have extra compensation needed in the gauge transformation. The discussions must be straightforward, but it has not been scrutinized in a complete manner. 

The Wess-Zumino consistency condition 
\begin{align}
[\delta_{\sigma(x)}, \delta_{\tilde{\sigma}(x')} ] W[g_{\mu\nu}, g^I, A_\mu^a] = 0 \ 
\end{align} 
with
\begin{align}
\delta_{\sigma(x)} = -\int d^4x \sqrt{|g|}  \sigma(x) \left( 2g^{\mu\nu} \frac{\delta}{\delta g^{\mu\nu}} - \mathcal{B}^I \frac{\delta}{\delta g^I} - \hat{\rho}^{a}_I D_{\mu} g^I \frac{\delta}{\delta A_{\mu}^a} \right) \ 
\end{align}
gives various integrability conditions on the Weyl anomaly. 
\begin{align}
8 \partial_I a - \chi^g_{IJ} \mathcal{B}^J &= -\mathcal{L}_{\mathcal{B}} w_I \cr
2\chi_I^e + \chi_{IJ}^a \mathcal{B}^J &= -\mathcal{L}_{\mathcal{B}} U_I \cr
8\tilde{b} - \chi_{IJ}^a \mathcal{B}^I\mathcal{B}^J &= -\mathcal{L}_{\mathcal{B}} (2q+ U_I\mathcal{B}^I) \cr
-\chi_{IJ}^g + 2\chi_{IJ}^a + \Lambda_{IJ} &= \mathcal{L}_{\mathcal{B}} S_{IJ} \cr
2(\chi_{IJ}^f + \chi_{IJ}^a) + \Lambda_{IJ} + \mathcal{B}^K(2\bar{\chi}^a_{K(IJ)} -\bar{\chi}^a_{IJK}) &= \mathcal{L}_{\mathcal{B}}(S_{IJ} - \chi_{IJ}^a -2 U_{(I,J)} + V_{IJ}) \cr
\chi_{IJK}^b - \chi_{K(I,J)}^g + \frac{1}{2}\chi^g_{IJ,K} + D_K \mathcal{B}^L \chi^b_{IJL} + \chi^c_{IJKL}\mathcal{B}^L &= \frac{1}{2}\mathcal{L}_{\mathcal{B}} T_{IJK} + D_I D_J {\mathcal{B}}^L S_{KL} \cr
\hat{\rho}_I \mathcal{B}^I &= 0 \cr
\eta_I \mathcal{B}^I &= g^I w_I \cr
\kappa \hat{\rho}_I + \zeta_{JI} \mathcal{B}^J &= \mathcal{L}_{\mathcal{B}} \eta_I + g^J \hat{\rho}_J \eta_I \ \ , \label{wzconstfull}
\end{align}
where
\begin{align}
\Lambda_{IJ} &= 2D_I \mathcal{B}^K \chi_{KJ}^a + \mathcal{B}^K \chi_{KIJ}^b \cr
\bar{\chi}^a_{IJK} &= \chi_{IJ,K}^a - \chi_{K(IJ)}^b \ ,
\end{align}
and $U_{(I,J)} = \frac{1}{2}(\partial_I U_J + \partial_J U_I)$ and so on.
The modified Lie derivative is defined as
\begin{align}
\mathcal{L}_{\mathcal{B}} t_I = \mathcal{B}^J \partial_J t_I + t_J(\partial_I \mathcal{B}^J - (\hat{\rho}_I g)^J)
\end{align}
for a 1-form and similarly for other tensors. 

As discussed in the main text, anomaly is defined up to the addition of the local counterterms. In this case, we can introduce various local counterterms given by 11 terms as in $T$ of (\ref{traceanomaly}). 
\begin{align}
\mathcal{S}_{\mathrm{ct}} &= -\int d^4x \sqrt{|g|} \left( C\mathrm{Weyl}^2 - A \mathrm{Euler} + \frac{1}{9}\tilde{B} R^2 \right. \cr 
&+ \frac{1}{3}C_I^e D_{\mu} g^{I}\partial^\mu R + \frac{1}{6}C_{IJ}^f D_{\mu} g^I D^{\mu} g^J R - \frac{1}{2}C_{IJ}^g D_{\mu}g^I D_{\nu}g^J G^{\mu\nu} + \frac{1}{2}C_{IJ}^a D^2 g^I D^2 g^J \cr
&+ \frac{1}{2}C_{IJK}^{b} D_{\mu} g^I D^{\mu} g^J D^2 g^K + \frac{1}{4}C_{IJKL}^c D_\mu g^I D^{\mu} g^J D_{\nu} g^K D^{\nu} g^L \cr
&\left. + \frac{1}{4}F_{\mu\nu} K F^{\mu\nu} + \frac{1}{2} F^{\mu\nu} Z_{IJ} D_\mu g^I D_{\nu} g^J \right) \ .
\end{align}
Here $C,A,\tilde{B},C_I^e , C_{IJ}^f, C_{IJ}^g , C_{IJ}^a, C_{IJK}^b$ and $C_{IJKL}^c$ are gauge invariant tensors of coupling constant spaces $g^I$ and $K$ and $Z_{IJ}$ are tensors that take values on the Lie algebra of the ``flavor" symmetries.
The induced local contributions to the Weyl anomaly are
\begin{align}
\delta(c,a,\tilde{b},\chi_I^e, \chi_{IJ}^f,\chi_{IJ}^g,\chi_{IJ}^a) &= \mathcal{L}_{\mathcal{B}}(C,A,\tilde{B},C_I^e,C_{IJ}^f, C_{IJ}^g, C_{IJ}^a) \cr
\delta \chi_{IJK}^b &= \mathcal{L}_{\mathcal{B}} C^b_{IJK} + 2D_ID_J \mathcal{B}^L C^a_{LK} \cr \delta \chi_{IJKL}^c &= \mathcal{L}_{\mathcal{B}} C^c_{IJKL} + D_ID_J \mathcal{B}^M C^b_{KLM} + D_K D_L \mathcal{B}^M C^b_{IJM} \cr
\delta w_I &= -8\partial_I A + C^g_{IJ} \mathcal{B}^J \cr
\delta q &= 4 \tilde{B} + C^e_I \mathcal{B}^I \cr
\delta U_I &= -2 C^e_I - C^a_{IJ} \mathcal{B}^J \cr
\delta Y_I &= -2 C^e_I - D_I (C_J^e \mathcal{B}^J) + C^f_{IJ} \mathcal{B}^J \cr
\delta V_{IJ} &= -4C^e_{(I,J)} + 2C^f_{IJ} + C^g_{IJ} - C^b_{IJK} \mathcal{B}^K \cr
\delta S_{IJ} &= C^g_{IJ} + 2C^a_{IJ} + 2D_I \mathcal{B}^K C^a_{KJ} + C^b_{KIJ} \mathcal{B}^K \cr
\delta T_{IJK} &= 2C^g_{K(I,J)} + C^g_{IJ,K} + 2C^b_{IJK} + 2D_K \mathcal{B}^L C^b_{IJL} + 2C^c_{IJKL} \mathcal{B}^L \ .
\end{align}
The above expression is directly taken from Osborn's paper \cite{Osborn:1991gm}. Since he did not discuss the $K $ and $Z_{IJ}$ term, it does not contain the effect of $K $ and $Z_{IJ}$ counterterms. If we introduced these counterterms we would schematically obtain
\begin{align}
\delta \zeta_{IJ} &\sim (g^K\hat{\rho}_K) Z_{IJ} + \mathcal{L}_{\mathcal{B}} Z_{IJ} \cr
\delta \eta_I &\sim - \hat{\rho}_I K + {\mathcal{B}}^J Z_{IJ} \cr
\delta \kappa &\sim \mathcal{L}_{\mathcal{B}} K 
\end{align}
and so on. The other terms like $T_{IJK}$ and $\chi^c_{IJKL}$ are also modified. An interested reader may complete the transformation rule or consult the more recent paper \cite{Jack:2013sha}.

As mentioned in \cite{Osborn:1991gm}, one can use the freedom to make $q$, $Y_I$ or $U_I$, $V_{IJ}$, $S_{(IJ)}$, and $T_{IJK}$ vanish. The remaining ambiguity for $\delta a$ and $\delta w_I$ are used in the dressing transformation of the gradient formula in section \ref{spacet}.

\subsection{Analytic properties  of  S-matrix}\label{smatrix}
We need some elementary facts about analytic properties  of S-matrix when we use  the dilaton scattering amplitudes to derive constraints on the renormalization group flow. We briefly summarize them here. One cautious  remark is that a formal textbook derivation of the following formulae on the S-matrix assume a mass gap in the spectrum, and strictly speaking, we need a careful treatment for massless theories like the deformed conformal field theories coupled with a dilaton.\footnote{To some extent, the problem is alleviated since we do not consider the internal loop of massless dilatons. The loop from the conformal field theory may be regularized by assuming adding relevant perturbation so that the IR theory is gapped, which, however, may not always be possible.} 

We are interested in two-two scattering of the identical massless particles
with the initial momenta $(p_1^\mu, p_2^\mu)$ to the final ones $(p_3^\mu, p_4^\mu)$.
Let us introduce the conventional Mandelstam variables
\begin{align}
s &= -(p_1 + p_2)^2 \cr
t &= -(p_1 - p_3)^2 \cr
u &= -(p_1 - p_4)^2 \ .
\end{align}
They are not independent because $s + t + u = 0$ from the conservation of the energy-momentum. The scattering amplitude is a function of the two of them, e.g. $A(s,t)$. We see that the forward scattering corresponds to $t=0$ (or $u=0$).
The scattering amplitude $A(s,t)$ is related to the S-matrix as
\begin{align}
\langle f|S|i \rangle = \delta_{fi} + i(2\pi)^4 \delta^{(4)}(p_i - p_f) \langle f|T|i \rangle
 \ ,
\end{align}
and we identify $A(s,t)$ with $\langle f|T|i \rangle$ for two-two scattering.

We recall that the S-matrix is unitary: $SS^\dagger = S^\dagger S =1$, therefore the T-matrix satisfies
\begin{align}
2 \mathrm{Im} \langle i|T |i \rangle = \sum_f (2\pi)^4 \delta^{(4)}(p_i-p_f) | \langle f|T| i \rangle|^2 \ . \label{tmatrix}
\end{align}
Now, in our two-two scattering,
 Fermi's golden rule tells that the right hand side of (\ref{tmatrix}) is proportional to the total cross section of the two-dilaton initial states (because we  summed over the final states), while the left hand side of (\ref{tmatrix}) is the imaginary part of the forward scattering amplitude of two-two dilatons (because initial state and final state are identical). Thus we obtain the special case of the optical theorem:
\begin{align}
\mathrm{Im} A(s, t=0) = s \sigma (s) \ . \label{opticalt}
\end{align}
$A(s,t=0)$ for the dilaton two-two scattering was denoted by $A_4(s)$ in the main text.

Scattering amplitudes have some important analytic structures. In particular, the diagrammatic computations do not distinguish the exchange of the initial state and final state (up on replacing particles with anti-particles). The $u$-channel exchange, therefore, gives
\begin{align}
A_{a+\bar{d} \to \bar{b}+c}(u,t,s) = A_{a+b\to c+d}(s,t,u) \ .
\end{align}
In the forward two-two dilaton scattering we are interested in, this leads to the crossing symmetry relation $A_4(s) = A_4(u) = A_4(-s)$ since $t=0$ and $u=-s$.

The two-two dilaton scattering amplitudes have a cut along the real $s$ axis. This is due to the massless multi-particle intermediate states. If the intermediate channels were massive, $A_4(s)$ would be real near $s=0$ on the real $s$ axis, and the analytic continuation of $s$ in complex plane should satisfy the Schwarz reflection  principle
\begin{align}
A_4(s^*) = A_4(s)^* \ . \label{schwarz}
\end{align}
Although our dilaton scattering may have a bad IR behavior, we postulate (\ref{schwarz}) holds. Then, for real $s$ we obtain
\begin{align}
A_4(-s+i\epsilon) = A_4(s-i\epsilon) = [A_4(s+i\epsilon)]^* \ ,
\end{align}
which is the basis of the first equality (\ref{crossingf}) in the main text.

\clearpage

%\section{Literature guides}

%\subsection{Literature guides} \label{guide1}
%At the end of each lecture, we put the literature guide section.
%This is intended to supplement materials that are not covered in the main part o%f the lecture note. 

%\end{itemize}

%\subsection{Literature guides}


\clearpage

\begin{thebibliography}{99}

\bibitem{Migdal} 
A.~Migdal, ``Ancient History of CFT", The 12th Claude Itzykson Meeting (2007).
%\cite{Wilson:1973jj}


\bibitem{Israel}
W.~Israel, Phys. Rev. 164, 1776 (1967).


\bibitem{Wilson:1973jj} 
  K.~G.~Wilson and J.~B.~Kogut,
  %``The Renormalization group and the epsilon expansion,''
  Phys.\ Rept.\  {\bf 12}, 75 (1974).
  %%CITATION = PRPLC,12,75;%%


%\cite{Wilson:1970ag}
\bibitem{Wilson:1970ag} 
  K.~G.~Wilson,
  %``The Renormalization Group and Strong Interactions,''
  Phys.\ Rev.\ D {\bf 3}, 1818 (1971).
  %%CITATION = PHRVA,D3,1818;%%


\bibitem{Bogo} N.~N.~Bogoliubov, D.~V.~Shirkov, Introduction to the Theory of Quantized Field. John Wiley and Sons Inc; 3rd edition (1980).


%\cite{Zamolodchikov:1986gt}
\bibitem{Zamolodchikov:1986gt}
  A.~B.~Zamolodchikov,
  %``Irreversibility of the Flux of the Renormalization Group in a 2D Field
  %Theory,''
  JETP Lett.\  {\bf 43} (1986) 730
  [Pisma Zh.\ Eksp.\ Teor.\ Fiz.\  {\bf 43} (1986) 565].
  %%CITATION = ZFPRA,43,565;%%
%\cite{Fortin:2011sz}
\bibitem{Fortin:2011sz} 
  J.~-F.~Fortin, B.~Grinstein and A.~Stergiou,
  %``Scale without Conformal Invariance: Theoretical Foundations,''
  JHEP {\bf 1207}, 025 (2012)
  [arXiv:1107.3840 [hep-th]].
  %%CITATION = ARXIV:1107.3840;%%



%\cite{Nakayama:2010zz}
\bibitem{Nakayama:2010zz} 
  Y.~Nakayama,
  %``Scale invariance vs conformal invariance from holography,''
  Int.\ J.\ Mod.\ Phys.\ A {\bf 25}, 4849 (2010).
  %%CITATION = IMPAE,A25,4849;%%

%\cite{Nakayama:2012nd}
\bibitem{Nakayama:2012nd} 
  Y.~Nakayama,
  %``Supercurrent, Supervirial and Superimprovement,''
  arXiv:1208.4726 [hep-th].
  %%CITATION = ARXIV:1208.4726;%%
%\cite{Luty:2012ww}
\bibitem{Luty:2012ww} 
  M.~A.~Luty, J.~Polchinski and R.~Rattazzi,
  %``The $a$-theorem and the Asymptotics of 4D Quantum Field Theory,''
  arXiv:1204.5221 [hep-th].
  %%CITATION = ARXIV:1204.5221;%%

%\cite{Fortin:2012hn}
\bibitem{Fortin:2012hn} 
  J.~-F.~Fortin, B.~Grinstein and A.~Stergiou,
  %``A generalized c-theorem and the consistency of scale without conformal invariance,''
  arXiv:1208.3674 [hep-th].
  %%CITATION = ARXIV:1208.3674;%%


\bibitem{Stanley}
H.~E.~Stanley, ``Introduction to Phase Transitions and Critical Phenomena".

\bibitem{Widom}
B.~Widom, J. Chem. Phys. 43, 3898 (1965).

%\cite{Gross:1973id}
\bibitem{Gross:1973id} 
  D.~J.~Gross and F.~Wilczek,
  %``Ultraviolet Behavior of Nonabelian Gauge Theories,''
  Phys.\ Rev.\ Lett.\  {\bf 30}, 1343 (1973).
  %%CITATION = PRLTA,30,1343;%%
  %3269 citations counted in INSPIRE as of 14 Feb 2014

%\cite{Politzer:1973fx}
\bibitem{Politzer:1973fx} 
  H.~D.~Politzer,
  %``Reliable Perturbative Results for Strong Interactions?,''
  Phys.\ Rev.\ Lett.\  {\bf 30}, 1346 (1973).
  %%CITATION = PRLTA,30,1346;%%
  %3233 citations counted in INSPIRE as of 14 Feb 2014


%\cite{Bjorken:1968dy}
\bibitem{Bjorken:1968dy} 
  J.~D.~Bjorken,
  %``Asymptotic Sum Rules at Infinite Momentum,''
  Phys.\ Rev.\  {\bf 179}, 1547 (1969).
  %%CITATION = PHRVA,179,1547;%%
  %1300 citations counted in INSPIRE as of 14 Feb 2014

%\cite{Luty:2004ye}
\bibitem{Luty:2004ye} 
  M.~A.~Luty and T.~Okui,
  %``Conformal technicolor,''
  JHEP {\bf 0609}, 070 (2006)
  [hep-ph/0409274].
  %%CITATION = HEP-PH/0409274;%%
  %122 citations counted in INSPIRE as of 14 Feb 2014

%\cite{Rattazzi:2008pe}
\bibitem{Rattazzi:2008pe} 
  R.~Rattazzi, V.~S.~Rychkov, E.~Tonni and A.~Vichi,
  %``Bounding scalar operator dimensions in 4D CFT,''
  JHEP {\bf 0812}, 031 (2008)
  [arXiv:0807.0004 [hep-th]].
  %%CITATION = ARXIV:0807.0004;%%
  %107 citations counted in INSPIRE as of 14 Feb 2014


%\cite{Shaposhnikov:2009pv}
\bibitem{Shaposhnikov:2009pv} 
  M.~Shaposhnikov and C.~Wetterich,
  %``Asymptotic safety of gravity and the Higgs boson mass,''
  Phys.\ Lett.\ B {\bf 683}, 196 (2010)
  [arXiv:0912.0208 [hep-th]].
  %%CITATION = ARXIV:0912.0208;%%
  %75 citations counted in INSPIRE as of 14 Feb 2014



%\cite{Tavares:2013dga}
\bibitem{Tavares:2013dga} 
  G.~Marques Tavares, M.~Schmaltz and W.~Skiba,
  %``Higgs mass naturalness and scale invariance in the UV,''
  Phys.\ Rev.\ D {\bf 89}, 015009 (2014)
  [arXiv:1308.0025 [hep-ph]].
  %%CITATION = ARXIV:1308.0025;%%
  %8 citations counted in INSPIRE as of 25 Feb 2014







%\cite{Weinberg:1980gg}
\bibitem{Weinberg:1980gg} 
  S.~Weinberg,
  %``Ultraviolet Divergences In Quantum Theories Of Gravitation,''
  790-831
  %7 citations counted in INSPIRE as of 14 Feb 2014


%\cite{Nagy:2012ef}
\bibitem{Nagy:2012ef} 
  S.~Nagy,
  %``Lectures on renormalization and asymptotic safety,''
  arXiv:1211.4151 [hep-th].
  %%CITATION = ARXIV:1211.4151;%%
  %8 citations counted in INSPIRE as of 14 Feb 2014




%\cite{Komatsu:2010fb}
\bibitem{Komatsu:2010fb} 
  E.~Komatsu {\it et al.}  [WMAP Collaboration],
  %``Seven-Year Wilkinson Microwave Anisotropy Probe (WMAP) Observations: Cosmological Interpretation,''
  Astrophys.\ J.\ Suppl.\  {\bf 192}, 18 (2011)
  [arXiv:1001.4538 [astro-ph.CO]].
  %%CITATION = ARXIV:1001.4538;%%
  %4315 citations counted in INSPIRE as of 14 Feb 2014


%\cite{Maldacena:2002vr}
\bibitem{Maldacena:2002vr} 
  J.~M.~Maldacena,
  %``Non-Gaussian features of primordial fluctuations in single field inflationary models,''
  JHEP {\bf 0305}, 013 (2003)
  [astro-ph/0210603].
  %%CITATION = ASTRO-PH/0210603;%%
  %1102 citations counted in INSPIRE as of 14 Feb 2014



\bibitem{Zipf}
G.~K.~Zipf (1935) The Psychobiology of Language. Houghton-Mifflin.

\bibitem{Dash}
J.~Dash, gQuantitative finance and risk management : a physicistfs approachh World Scientific Pub., 2004.

%\cite{Wald:1984rg}
\bibitem{Wald:1984rg} 
  R.~M.~Wald,
  ``General Relativity,''
  Chicago, Usa: Univ. Pr. ( 1984) 491p

%\cite{Collins:1984xc}
\bibitem{Collins:1984xc} 
  J.~C.~Collins,
  ``Renormalization. An Introduction To Renormalization, The Renormalization Group, And The Operator Product Expansion,''
  Cambridge, Uk: Univ. Pr. ( 1984) 380p
%\cite{ZinnJustin:2002ru}
\bibitem{ZinnJustin:2002ru} 
  J.~Zinn-Justin,
  ``Quantum field theory and critical phenomena,''
  Int.\ Ser.\ Monogr.\ Phys.\  {\bf 113}, 1 (2002).
  %%CITATION = IMPHA,113,1;%%
%\cite{Weinberg:1995mt}
\bibitem{Weinberg:1995mt} 
  S.~Weinberg,
  ``The Quantum theory of fields. Vol. 1: Foundations,''
  Cambridge, UK: Univ. Pr. (1995) 609 p
%\cite{Weinberg:1996kr}
\bibitem{Weinberg:1996kr} 
  S.~Weinberg,
  ``The quantum theory of fields. Vol. 2: Modern applications,''
  Cambridge, UK: Univ. Pr. (1996) 489 p


%\cite{Ginsparg:1988ui}
\bibitem{Ginsparg:1988ui} 
  P.~H.~Ginsparg,
  %``Applied Conformal Field Theory,''
  hep-th/9108028.
  %%CITATION = HEP-TH/9108028;%%

%\cite{DiFrancesco:1997nk}
\bibitem{DiFrancesco:1997nk} 
  P.~Di Francesco, P.~Mathieu and D.~Senechal,
  ``Conformal field theory,''
  New York, USA: Springer (1997) 890 p
\bibitem{Slava}
S.~Rychkov,
``EPFL Lectures on Conformal Field Theory in $D>=3$ Dimensions."
https://sites.google.com/site/slavarychkov/


%\cite{Wess:1992cp}
\bibitem{Wess:1992cp} 
  J.~Wess and J.~Bagger,
  ``Supersymmetry and supergravity,''
  Princeton, USA: Univ. Pr. (1992) 259 p
%\cite{Weinberg:2000cr}
\bibitem{Weinberg:2000cr} 
  S.~Weinberg,
  ``The quantum theory of fields. Vol. 3: Supersymmetry,''
  Cambridge, UK: Univ. Pr. (2000) 419 p




\bibitem{Argyres}

P.~Argyres. Lectures on Supersymmetry
http://www.physics.uc.edu/~argyres/661/index.html


%\cite{Polchinski:1987dy}
\bibitem{Polchinski:1987dy}
  J.~Polchinski,
  %``SCALE AND CONFORMAL INVARIANCE IN QUANTUM FIELD THEORY,''
  Nucl.\ Phys.\  B {\bf 303}, 226 (1988).
  %%CITATION = NUPHA,B303,226;%%%\cite{Callan:1970ze}


%\cite{Georgi:2007ek}
\bibitem{Georgi:2007ek} 
  H.~Georgi,
  %``Unparticle physics,''
  Phys.\ Rev.\ Lett.\  {\bf 98}, 221601 (2007)
  [hep-ph/0703260].
  %%CITATION = HEP-PH/0703260;%%

%\cite{Georgi:2007si}
\bibitem{Georgi:2007si} 
  H.~Georgi,
  %``Another odd thing about unparticle physics,''
  Phys.\ Lett.\ B {\bf 650}, 275 (2007)
  [arXiv:0704.2457 [hep-ph]].
  %%CITATION = ARXIV:0704.2457;%%

%\cite{Nakayama:2007qu}
\bibitem{Nakayama:2007qu} 
  Y.~Nakayama,
  %``SUSY unparticle and conformal sequestering,''
  Phys.\ Rev.\ D {\bf 76}, 105009 (2007)
  [arXiv:0707.2451 [hep-ph]].
  %%CITATION = ARXIV:0707.2451;%%

%\cite{Coleman:1967ad}
\bibitem{Coleman:1967ad} 
  S.~R.~Coleman and J.~Mandula,
  %``All Possible Symmetries Of The S Matrix,''
  Phys.\ Rev.\  {\bf 159}, 1251 (1967).
  %%CITATION = PHRVA,159,1251;%%
%\cite{Haag:1974qh}
\bibitem{Haag:1974qh}\
  R.~Haag, J.~T.~Lopuszanski and M.~Sohnius,
  %``All Possible Generators Of Supersymmetries Of The S Matrix,''
  Nucl.\ Phys.\  B {\bf 88}, 257 (1975).
  %%CITATION = NUPHA,B88,257;%%

%\cite{Maldacena:2011jn}
\bibitem{Maldacena:2011jn} 
  J.~Maldacena and A.~Zhiboedov,
  %``Constraining Conformal Field Theories with A Higher Spin Symmetry,''
  J.\ Phys.\ A {\bf 46}, 214011 (2013)
  [arXiv:1112.1016 [hep-th]].
  %%CITATION = ARXIV:1112.1016;%%
  %88 citations counted in INSPIRE as of 05 Feb 2014


%\cite{Coleman:1970je}
\bibitem{Coleman:1970je} 
  S.~R.~Coleman and R.~Jackiw,
  %``Why dilatation generators do not generate dilatations?,''
  Annals Phys.\  {\bf 67}, 552 (1971).
  %%CITATION = APNYA,67,552;%%

%\cite{Hortacsu:2001bp}
\bibitem{Hortacsu:2001bp} 
  M.~Hortacsu,
  %``Explicit examples on conformal invariance,''
  Int.\ J.\ Theor.\ Phys.\  {\bf 42}, 49 (2003)
  [hep-th/0106080].
  %%CITATION = HEP-TH/0106080;%%

%\cite{Weinberg:2010fx}
\bibitem{Weinberg:2010fx} 
  S.~Weinberg,
  %``Six-dimensional Methods for Four-dimensional Conformal Field Theories,''
  Phys.\ Rev.\ D {\bf 82}, 045031 (2010)
  [arXiv:1006.3480 [hep-th]].
  %%CITATION = ARXIV:1006.3480;%%

%\cite{Fubini:1972mf}
\bibitem{Fubini:1972mf} 
  S.~Fubini, A.~J.~Hanson and R.~Jackiw,
  %``New approach to field theory,''
  Phys.\ Rev.\ D {\bf 7}, 1732 (1973).
  %%CITATION = PHRVA,D7,1732;%%
\bibitem{Belinfante}
F.~J.~Belinfante, g
%On the Current and the Density of the Electric Charge, the Energy,
%the Linear Momentum and the Angular Momentum of Arbitrary Fields,h 
Physica 7, 449 (1940)

%\cite{Dumitrescu:2011iu}
\bibitem{Dumitrescu:2011iu} 
  T.~T.~Dumitrescu and N.~Seiberg,
  %``Supercurrents and Brane Currents in Diverse Dimensions,''
  JHEP {\bf 1107}, 095 (2011)
  [arXiv:1106.0031 [hep-th]].
  %%CITATION = ARXIV:1106.0031;%%

\bibitem{Callan:1970ze} 
  C.~G.~Callan, Jr., S.~R.~Coleman and R.~Jackiw,
  %``A New improved energy - momentum tensor,''
  Annals Phys.\  {\bf 59}, 42 (1970).
  %%CITATION = APNYA,59,42;%%





\bibitem{Zumino:1970}
B. Zumino, Lectures on elementary particles and quantum field theory, Brandeis University Summer
Institute 2 (1970).






%\cite{Osborn:1993cr}
\bibitem{Osborn:1993cr} 
  H.~Osborn and A.~C.~Petkou,
  %``Implications of conformal invariance in field theories for general dimensions,''
  Annals Phys.\  {\bf 231}, 311 (1994)
  [hep-th/9307010].
  %%CITATION = HEP-TH/9307010;%%




%\cite{Bzowski:2013sza}
\bibitem{Bzowski:2013sza} 
  A.~Bzowski, P.~McFadden and K.~Skenderis,
  %``Implications of conformal invariance in momentum space,''
  arXiv:1304.7760 [hep-th].
  %%CITATION = ARXIV:1304.7760;%%
  %4 citations counted in INSPIRE as of 13 Feb 2014



%\cite{Dymarsky:2014zja}
\bibitem{Dymarsky:2014zja} 
  A.~Dymarsky, K.~Farnsworth, Z.~Komargodski, M.~A.~Luty and V.~Prilepina,
  %``Scale Invariance, Conformality, and Generalized Free Fields,''
  arXiv:1402.6322 [hep-th].
  %%CITATION = ARXIV:1402.6322;%%

%\cite{Polyakov:1970xd}
\bibitem{Polyakov:1970xd} 
  A.~M.~Polyakov,
  %``Conformal symmetry of critical fluctuations,''
  JETP Lett.\  {\bf 12}, 381 (1970)
  [Pisma Zh.\ Eksp.\ Teor.\ Fiz.\  {\bf 12}, 538 (1970)].
  %%CITATION = JTPLA,12,381;%%
%\cite{Migdal:1971xh}
\bibitem{Migdal:1971xh} 
  A.~A.~Migdal,
  %``On hadronic interactions at small distances,''
  Phys.\ Lett.\ B {\bf 37}, 98 (1971).
  %%CITATION = PHLTA,B37,98;%%
%\cite{Migdal:1972tk}
\bibitem{Migdal:1972tk} 
  A.~A.~Migdal,
  %``Conformal invariance and bootstrap,''
  Phys.\ Lett.\ B {\bf 37}, 386 (1971).
  %%CITATION = PHLTA,B37,386;%% 


%\cite{Mack:1975je}
\bibitem{Mack:1975je} 
  G.~Mack,
  %``All Unitary Ray Representations of the Conformal Group SU(2,2) with Positive Energy,''
  Commun.\ Math.\ Phys.\  {\bf 55}, 1 (1977).
  %%CITATION = CMPHA,55,1;%%

%\cite{Minwalla:1997ka}
\bibitem{Minwalla:1997ka} 
  S.~Minwalla,
  %``Restrictions imposed by superconformal invariance on quantum field theories,''
  Adv.\ Theor.\ Math.\ Phys.\  {\bf 2}, 781 (1998)
  [hep-th/9712074].
  %%CITATION = HEP-TH/9712074;%%

\bibitem{Wilson}
K.~Wilson (1964), unpublished Cornell Report and Phys. Rev. 179 (1969) 1499.


%\cite{Dolan:2000ut}
\bibitem{Dolan:2000ut} 
  F.~A.~Dolan and H.~Osborn,
  %``Conformal four point functions and the operator product expansion,''
  Nucl.\ Phys.\ B {\bf 599}, 459 (2001)
  [hep-th/0011040].
  %%CITATION = HEP-TH/0011040;%%
%\cite{Dolan:2003hv}
\bibitem{Dolan:2003hv} 
  F.~A.~Dolan and H.~Osborn,
  %``Conformal partial waves and the operator product expansion,''
  Nucl.\ Phys.\ B {\bf 678}, 491 (2004)
  [hep-th/0309180].
  %%CITATION = HEP-TH/0309180;%%

%\cite{Pappadopulo:2012jk}
\bibitem{Pappadopulo:2012jk} 
  D.~Pappadopulo, S.~Rychkov, J.~Espin and R.~Rattazzi,
  %``OPE Convergence in Conformal Field Theory,''
  Phys.\ Rev.\ D {\bf 86}, 105043 (2012)
  [arXiv:1208.6449 [hep-th]].
  %%CITATION = ARXIV:1208.6449;%%
  %10 citations counted in INSPIRE as of 24 Feb 2014

%\cite{Belavin:1984vu}
\bibitem{Belavin:1984vu} 
  A.~A.~Belavin, A.~M.~Polyakov and A.~B.~Zamolodchikov,
  %``Infinite Conformal Symmetry in Two-Dimensional Quantum Field Theory,''
  Nucl.\ Phys.\ B {\bf 241}, 333 (1984).
  %%CITATION = NUPHA,B241,333;%%









%\cite{Grinstein:2008qk}
\bibitem{Grinstein:2008qk} 
  B.~Grinstein, K.~A.~Intriligator and I.~Z.~Rothstein,
  %``Comments on Unparticles,''
  Phys.\ Lett.\ B {\bf 662}, 367 (2008)
  [arXiv:0801.1140 [hep-ph]].
  %%CITATION = ARXIV:0801.1140;%%



%\cite{Schwinger:1951xk}
\bibitem{Schwinger:1951xk} 
  J.~S.~Schwinger,
  %``The Theory of quantized fields. 1.,''
  Phys.\ Rev.\  {\bf 82}, 914 (1951).
  %%CITATION = PHRVA,82,914;%%
  %365 citations counted in INSPIRE as of 18 Jul 2013


%\cite{Capper:1974ic}
\bibitem{Capper:1974ic} 
  D.~M.~Capper and M.~J.~Duff,
  %``Trace anomalies in dimensional regularization,''
  Nuovo Cim.\ A {\bf 23}, 173 (1974).
  %%CITATION = NUCIA,A23,173;%%


%\cite{Deser:1976yx}
\bibitem{Deser:1976yx} 
  S.~Deser, M.~J.~Duff and C.~J.~Isham,
  %``Nonlocal Conformal Anomalies,''
  Nucl.\ Phys.\ B {\bf 111}, 45 (1976).
  %%CITATION = NUPHA,B111,45;%%


%\cite{Duff:1993wm}
\bibitem{Duff:1993wm} 
  M.~J.~Duff,
  %``Twenty years of the Weyl anomaly,''
  Class.\ Quant.\ Grav.\  {\bf 11}, 1387 (1994)
  [hep-th/9308075].
  %%CITATION = HEP-TH/9308075;%%
%\cite{Deser:1996na}
\bibitem{Deser:1996na} 
  S.~Deser,
  %``Conformal anomalies: Recent progress,''
  Helv.\ Phys.\ Acta {\bf 69}, 570 (1996)
  [hep-th/9609138].
  %%CITATION = HEP-TH/9609138;%%

%\cite{Nakayama:2012gu}
\bibitem{Nakayama:2012gu} 
  Y.~Nakayama,
  %``CP-violating CFT and trace anomaly,''
  Nucl.\ Phys.\ B {\bf 859}, 288 (2012)
  [arXiv:1201.3428 [hep-th]].
  %%CITATION = ARXIV:1201.3428;%%


%\cite{Duff:1980qv}
\bibitem{Duff:1980qv} 
  M.~J.~Duff and P.~van Nieuwenhuizen,
  %``Quantum Inequivalence of Different Field Representations,''
  Phys.\ Lett.\ B {\bf 94}, 179 (1980).
  %%CITATION = PHLTA,B94,179;%%

%\cite{Bonora:2013rta}
\bibitem{Bonora:2013rta} 
  L.~Bonora and S.~Giaccari,
  %``Weyl transformations and trace anomalies in N=1, D=4 supergravities,''
  JHEP {\bf 1308}, 116 (2013)
  [arXiv:1305.7116 [hep-th]].
  %%CITATION = ARXIV:1305.7116;%%
  %1 citations counted in INSPIRE as of 24 Feb 2014


\bibitem{Bardeen:1984pm} 
  W.~A.~Bardeen and B.~Zumino,
  %``Consistent and Covariant Anomalies in Gauge and Gravitational Theories,''
  Nucl.\ Phys.\ B {\bf 244}, 421 (1984).
  %%CITATION = NUPHA,B244,421;%% 


%\cite{Bonora:1985cq}
\bibitem{Bonora:1985cq} 
  L.~Bonora, P.~Pasti and M.~Bregola,
  %``Weyl Cocycles,''
  Class.\ Quant.\ Grav.\  {\bf 3}, 635 (1986).
  %%CITATION = CQGRD,3,635;%%

%\cite{Boulanger:2007st}
\bibitem{Boulanger:2007st} 
  N.~Boulanger,
  %``General solutions of the Wess-Zumino consistency condition for the Weyl anomalies,''
  JHEP {\bf 0707}, 069 (2007)
  [arXiv:0704.2472 [hep-th]].
  %%CITATION = ARXIV:0704.2472;%%
%\cite{Bonora:1983ff}
\bibitem{Bonora:1983ff} 
  L.~Bonora, P.~Cotta-Ramusino and C.~Reina,
  %``Conformal Anomaly And Cohomology,''
  Phys.\ Lett.\ B {\bf 126}, 305 (1983).
  %%CITATION = PHLTA,B126,305;%%

%\cite{Cappelli:1988vw}
\bibitem{Cappelli:1988vw} 
  A.~Cappelli and A.~Coste,
  %``On The Stress Tensor Of Conformal Field Theories In Higher Dimensions,''
  Nucl.\ Phys.\ B {\bf 314}, 707 (1989).
  %%CITATION = NUPHA,B314,707;%%

%\cite{Osborn:1991gm}
\bibitem{Osborn:1991gm} 
  H.~Osborn,
  %``Weyl consistency conditions and a local renormalization group equation for general renormalizable field theories,''
  Nucl.\ Phys.\ B {\bf 363}, 486 (1991).
  %%CITATION = NUPHA,B363,486;%%


%\cie{Vassilevich:2003xt}
\bibitem{Vassilevich:2003xt} 
  D.~V.~Vassilevich,
  %``Heat kernel expansion: User's manual,''
  Phys.\ Rept.\  {\bf 388}, 279 (2003)
  [hep-th/0306138].
  %%CITATION = HEP-TH/0306138;%%
%\cite{Asorey:2003uf}
\bibitem{Asorey:2003uf} 
  M.~Asorey, E.~V.~Gorbar and I.~L.~Shapiro,
  %``Universality and ambiguities of the conformal anomaly,''
  Class.\ Quant.\ Grav.\  {\bf 21}, 163 (2003)
  [hep-th/0307187].
  %%CITATION = HEP-TH/0307187;%%
  %19 citations counted in INSPIRE as of 14 Feb 2013

%\cite{Fujikawa:2004cx}
\bibitem{Fujikawa:2004cx} 
  K.~Fujikawa and H.~Suzuki,
  %``Path integrals and quantum anomalies,''
  Oxford, UK: Clarendon (2004) 284 p

%\cite{Deser:1993yx}
\bibitem{Deser:1993yx} 
  S.~Deser and A.~Schwimmer,
  %``Geometric classification of conformal anomalies in arbitrary dimensions,''
  Phys.\ Lett.\ B {\bf 309}, 279 (1993)
  [hep-th/9302047].
  %%CITATION = HEP-TH/9302047;%%



%\cite{Jack:2013sha}
\bibitem{JO} 
  I.~Jack and H.~Osborn,
  %``Constraints on RG Flow for Four Dimensional Quantum Field Theories,''
  arXiv:1312.0428 [hep-th].
  %%CITATION = ARXIV:1312.0428;%%

%\cite{Antipin:2013pya}
\bibitem{Antipin:2013pya} 
  O.~Antipin, M.~Gillioz, E.~Molgaard and F.~Sannino,
  %``The a theorem for Gauge-Yukawa theories beyond Banks-Zaks,''
  Phys.\ Rev.\ D {\bf 87}, 125017 (2013)
  [arXiv:1303.1525 [hep-th]].
  %%CITATION = ARXIV:1303.1525;%%
  %8 citations counted in INSPIRE as of 25 Feb 2014

%\cite{Antipin:2013sga}
\bibitem{Antipin:2013sga} 
  O.~Antipin, M.~Gillioz, J.~Krog, E.~Molgaard and F.~Sannino,
  %``Standard Model Vacuum Stability and Weyl Consistency Conditions,''
  JHEP {\bf 1308}, 034 (2013)
  [arXiv:1306.3234, arXiv:1306.3234 [hep-ph]].
  %%CITATION = ARXIV:1306.3234,;%%
  %7 citations counted in INSPIRE as of 25 Feb 2014

%\cite{Closset:2012vp}
\bibitem{Closset:2012vp} 
  C.~Closset, T.~T.~Dumitrescu, G.~Festuccia, Z.~Komargodski and N.~Seiberg,
  %``Comments on Chern-Simons Contact Terms in Three Dimensions,''
  JHEP {\bf 1209}, 091 (2012)
  [arXiv:1206.5218 [hep-th]].
  %%CITATION = ARXIV:1206.5218;%%


%\cite{Jackiw:2011vz}
\bibitem{Jackiw:2011vz}
  R.~Jackiw, S.~-Y.~Pi,
  %``Tutorial on Scale and Conformal Symmetries in Diverse Dimensions,''
  J.\ Phys.\ A {\bf A44}, 223001 (2011).
  [arXiv:1101.4886 [math-ph]].

%\cite{ElShowk:2011gz}
\bibitem{ElShowk:2011gz}
  S.~El-Showk, Y.~Nakayama, S.~Rychkov,
  %``What Maxwell Theory in D<>4 teaches us about scale and conformal invariance,''
  Nucl.\ Phys.\  {\bf B848}, 578-593 (2011).
  [arXiv:1101.5385 [hep-th]].

%\cite{Riva:2005gd}
\bibitem{Riva:2005gd} 
  V.~Riva and J.~L.~Cardy,
  %``Scale and conformal invariance in field theory: A Physical counterexample,''
  Phys.\ Lett.\ B {\bf 622}, 339 (2005)
  [hep-th/0504197].
  %%CITATION = HEP-TH/0504197;%%

\bibitem{Elas}
L.~D.~Landau and E.~M.~Lifshitz, Theory of elasticity, Pergamon,
New York, 1970.


%\cite{Bialek:1986it}
\bibitem{Bialek:1986it} 
  W.~Bialek and A.~Zee,
  %``Statistical Mechanics And Invariant Perception,''
  Phys.\ Rev.\ Lett.\  {\bf 58}, 741 (1987).
  %%CITATION = PRLTA,58,741;%%
%\cite{Bialek:1987qc}
\bibitem{Bialek:1987qc} 
  W.~Bialek and A.~Zee,
  %``Understanding The Efficiency Of Human Perception,''
  Phys.\ Rev.\ Lett.\  {\bf 61}, 1512 (1988).
  %%CITATION = PRLTA,61,1512;%%
%\cite{Nakayama:2010ye}
\bibitem{Nakayama:2010ye} 
  Y.~Nakayama,
  %``Gravity Dual for a Model of Perception,''
  Annals Phys.\  {\bf 326}, 2 (2011)
  [arXiv:1003.5729 [hep-th]].
  %%CITATION = ARXIV:1003.5729;%%




%\cite{Brown:1980qq}
\bibitem{Brown:1980qq} 
  L.~S.~Brown and J.~C.~Collins,
  %``Dimensional Renormalization Of Scalar Field Theory In Curved Space-time,''
  Annals Phys.\  {\bf 130}, 215 (1980).
  %%CITATION = APNYA,130,215;%%
%\cite{Hathrell:1981zb}
\bibitem{Hathrell:1981zb} 
  S.~J.~Hathrell,
  %``TRACE ANOMALIES AND lambda phi**4 THEORY IN CURVED SPACE,''
  Annals Phys.\  {\bf 139}, 136 (1982).
  %%CITATION = APNYA,139,136;%%






%\cite{Caswell:1974gg}
\bibitem{Caswell:1974gg} 
  W.~E.~Caswell,
  %``Asymptotic Behavior of Nonabelian Gauge Theories to Two Loop Order,''
  Phys.\ Rev.\ Lett.\  {\bf 33}, 244 (1974).
  %%CITATION = PRLTA,33,244;%%


%\cite{Banks:1981nn}
\bibitem{Banks:1981nn} 
  T.~Banks and A.~Zaks,
  %``On the Phase Structure of Vector-Like Gauge Theories with Massless Fermions,''
  Nucl.\ Phys.\ B {\bf 196}, 189 (1982).
  %%CITATION = NUPHA,B196,189;%%


\bibitem{polyakov}
A.~Polyakov, ``Gauge Fields and Strings". 
CRC Press, (1987).

%\cite{Hortacsu:1972bw}
\bibitem{Hortacsu:1972bw} 
  M.~Hortacsu, R.~Seiler and B.~Schroer,
  %``Conformal symmetry and reverberations,''
  Phys.\ Rev.\ D {\bf 5}, 2519 (1972).
  %%CITATION = PHRVA,D5,2519;%%

\bibitem{Collins:1976yq} 
  J.~C.~Collins, A.~Duncan and S.~D.~Joglekar,
  %``Trace and Dilatation Anomalies in Gauge Theories,''
  Phys.\ Rev.\ D {\bf 16}, 438 (1977).
  %%CITATION = PHRVA,D16,438;%%

%\cite{Nielsen:1977sy}
\bibitem{Nielsen:1977sy} 
  N.~K.~Nielsen,
  %``The Energy Momentum Tensor in a Nonabelian Quark Gluon Theory,''
  Nucl.\ Phys.\ B {\bf 120}, 212 (1977).
  %%CITATION = NUPHA,B120,212;%%
%\cite{Hathrell:1981gz}
\bibitem{Hathrell:1981gz} 
  S.~J.~Hathrell,
  %``Trace Anomalies And Qed In Curved Space,''
  Annals Phys.\  {\bf 142}, 34 (1982).
  %%CITATION = APNYA,142,34;%% 




\bibitem{review}
J. Giedt, PoS(Lattice 2012) to be published;
E. Neil, PoS(Lattice 2011)009;
L. Del Debbio, PoS(Lattice 2010)004;
and references therein.

%\cite{DelDebbio:2013qta}
\bibitem{DelDebbio:2013qta} 
  L.~Del Debbio and R.~Zwicky,
  %``Conformal scaling and the size of $m$-hadrons,''
  Phys.\ Rev.\ D {\bf 89}, 014503 (2014)
  [arXiv:1306.4038 [hep-ph]].
  %%CITATION = ARXIV:1306.4038;%%
  %5 citations counted in INSPIRE as of 24 Feb 2014

 %\cite{Iwasaki:2012kv}
\bibitem{Iwasaki:2012kv} 
  Y.~Iwasaki,
  %``Conformal Window and Correlation Functions in Lattice Conformal QCD,''
  arXiv:1212.4343 [hep-lat].
  %%CITATION = ARXIV:1212.4343;%%
%\cite{Ishikawa:2013wf}
\bibitem{Ishikawa:2013wf} 
  K.~-I.~Ishikawa, Y.~Iwasaki, Y.~Nakayama and T.~Yoshie,
  %``Conformal Theories with IR cutoff,''
  arXiv:1301.4785 [hep-lat].
  %%CITATION = ARXIV:1301.4785;%%

%\cite{Ishikawa:2013tua}
\bibitem{Ishikawa:2013tua} 
  K.~-I.~Ishikawa, Y.~Iwasaki, Y.~Nakayama and T.~Yoshie,
  %``Global Structure of Conformal Theories in the SU(3) Gauge Theory,''
  arXiv:1310.5049 [hep-lat].
  %%CITATION = ARXIV:1310.5049;%%
  %6 citations counted in INSPIRE as of 24 Feb 2014


%\cite{Seiberg:1994pq}
\bibitem{Seiberg:1994pq} 
  N.~Seiberg,
  %``Electric - magnetic duality in supersymmetric nonAbelian gauge theories,''
  Nucl.\ Phys.\ B {\bf 435}, 129 (1995)
  [hep-th/9411149].
  %%CITATION = HEP-TH/9411149;%%

%\cite{Collins:1976vm}
\bibitem{Collins:1976vm} 
  J.~C.~Collins,
  %``The Energy-Momentum Tensor Revisited,''
  Phys.\ Rev.\ D {\bf 14}, 1965 (1976).
  %%CITATION = PHRVA,D14,1965;%%
%\cite{Dorigoni:2009ra}
\bibitem{Dorigoni:2009ra} 
  D.~Dorigoni and V.~S.~Rychkov,
  %``Scale Invariance + Unitarity => Conformal Invariance?,''
  arXiv:0910.1087 [hep-th].
  %%CITATION = ARXIV:0910.1087;%%

%\cite{Fortin:2011ks}
\bibitem{Fortin:2011ks} 
  J.~-F.~Fortin, B.~Grinstein and A.~Stergiou,
  %``Scale without Conformal Invariance: An Example,''
  Phys.\ Lett.\ B {\bf 704}, 74 (2011)
  [arXiv:1106.2540 [hep-th]].
  %%CITATION = ARXIV:1106.2540;%%
%\cite{Fortin:2012ic}
\bibitem{Fortin:2012ic} 
  J.~-F.~Fortin, B.~Grinstein and A.~Stergiou,
  %``Scale without Conformal Invariance at Three Loops,''
  JHEP {\bf 1208}, 085 (2012)
  [arXiv:1202.4757 [hep-th]].
  %%CITATION = ARXIV:1202.4757;%%




















%\cite{Hull:1985rc}
\bibitem{Hull:1985rc} 
  C.~M.~Hull and P.~K.~Townsend,
  %``Finiteness And Conformal Invariance In Nonlinear Sigma Models,''
  Nucl.\ Phys.\ B {\bf 274}, 349 (1986).
  %%CITATION = NUPHA,B274,349;%%
%\cite{Tseytlin:1986tt}
\bibitem{Tseytlin:1986tt} 
  A.~A.~Tseytlin,
  %``Conformal Anomaly in Two-Dimensional Sigma Model on Curved Background and Strings,''
  Phys.\ Lett.\ B {\bf 178}, 34 (1986).
  %%CITATION = PHLTA,B178,34;%%
%\cite{Shore:1986hk}
\bibitem{Shore:1986hk} 
  G.~M.~Shore,
  %``A Local Renormalization Group Equation, Diffeomorphisms, And Conformal Invariance In Sigma Models,''
  Nucl.\ Phys.\ B {\bf 286}, 349 (1987).
  %%CITATION = NUPHA,B286,349;%%
%\cite{Callan:1986jb}
\bibitem{Callan:1986jb} 
  C.~G.~Callan, Jr., I.~R.~Klebanov and M.~J.~Perry,
  %``String Theory Effective Actions,''
  Nucl.\ Phys.\ B {\bf 278}, 78 (1986).
  %%CITATION = NUPHA,B278,78;%%
%\cite{Friedan:1980jm}
\bibitem{Friedan:1980jm} 
  D.~H.~Friedan,
  %``Nonlinear Models in Two + Epsilon Dimensions,''
  Annals Phys.\  {\bf 163}, 318 (1985).
  %%CITATION = APNYA,163,318;%%




%\cite{Witten:1991yr}
\bibitem{Witten:1991yr} 
  E.~Witten,
  %``On string theory and black holes,''
  Phys.\ Rev.\ D {\bf 44}, 314 (1991).
  %%CITATION = PHRVA,D44,314;%%



%\cite{Bonneau:1986if}
\bibitem{Bonneau:1986if} 
  G.~Bonneau and F.~Delduc,
  %``Construction And Properties Of Quasi Ricci Flat Spaces,''
  Int.\ J.\ Mod.\ Phys.\ A {\bf 1}, 997 (1986).
  %%CITATION = IMPAE,A1,997;%%
  %1 citations counted in INSPIRE as of 15 Feb 2013
%\cite{Polchinski:1998rq}
\bibitem{Polchinski:1998rq} 
  J.~Polchinski,
  %``String theory. Vol. 1: An introduction to the bosonic string,''
  Cambridge, UK: Univ. Pr. (1998) 402 p

%\cite{Polchinski:1998rr}
\bibitem{Polchinski:1998rr} 
  J.~Polchinski,
  %``String theory. Vol. 2: Superstring theory and beyond,''
  Cambridge, UK: Univ. Pr. (1998) 531 p


%\cite{Wilson:1971dc}
\bibitem{Wilson:1971dc} 
  K.~G.~Wilson and M.~E.~Fisher,
  %``Critical exponents in 3.99 dimensions,''
  Phys.\ Rev.\ Lett.\  {\bf 28}, 240 (1972).
  %%CITATION = PRLTA,28,240;%%


%\cite{Smirnov:2007pm}
\bibitem{Smirnov:2007pm} 
  S.~Smirnov,
  %``Conformal invariance in random cluster models. I. Holomorphic fermions in the Ising model,''
  arXiv:0708.0039 [math-ph].
  %%CITATION = ARXIV:0708.0039;%%
  %16 citations counted in INSPIRE as of 14 Feb 2014

%\cite{Rychkov:2011et}
\bibitem{Rychkov:2011et} 
  S.~Rychkov,
  %``Conformal Bootstrap in Three Dimensions?,''
  arXiv:1111.2115 [hep-th].
  %%CITATION = ARXIV:1111.2115;%%

%\cite{ElShowk:2012ht}
\bibitem{ElShowk:2012ht} 
  S.~El-Showk, M.~F.~Paulos, D.~Poland, S.~Rychkov, D.~Simmons-Duffin and A.~Vichi,
  %``Solving the 3D Ising Model with the Conformal Bootstrap,''
  Phys.\ Rev.\ D {\bf 86}, 025022 (2012)
  [arXiv:1203.6064 [hep-th]].
  %%CITATION = ARXIV:1203.6064;%%


\bibitem{Gursoy}
 F.~G\"ursey, Nuovo Cimento 3, 988 (1956).



%\cite{Riegert:1984kt}
\bibitem{Riegert:1984kt} 
  R.~J.~Riegert,
  %``A Nonlocal Action for the Trace Anomaly,''
  Phys.\ Lett.\ B {\bf 134}, 56 (1984).
  %%CITATION = PHLTA,B134,56;%%


%\cite{Fradkin:1983tg}
\bibitem{Fradkin:1983tg} 
  E.~S.~Fradkin and A.~A.~Tseytlin,
  %``Conformal Anomaly in Weyl Theory and Anomaly Free Superconformal Theories,''
  Phys.\ Lett.\ B {\bf 134}, 187 (1984).
  %%CITATION = PHLTA,B134,187;%%

\bibitem{PA}
S.~Paneitz, 
``A Quartic Conformally Covariant Differential Operator for Arbitrary Pseudo-Riemannian Manifolds", MIT preprint, 1983. Published posthumously in SIGMA 4 (2008), 036. [arxiv:0803.4331]

%\cite{Fradkin:1982xc}
\bibitem{Fradkin:1982xc} 
  E.~S.~Fradkin, A.~A.~Tseytlin and ,
  %``Asymptotic Freedom In Extended Conformal Supergravities,''
  Phys.\ Lett.\ B {\bf 110}, 117 (1982).
  %%CITATION = PHLTA,B110,117;%%
  %23 citations counted in INSPIRE as of 01 Apr 2013
%\cite{Fradkin:1981jc}
\bibitem{Fradkin:1981jc} 
  E.~S.~Fradkin, A.~A.~Tseytlin and ,
  %``One Loop Beta Function In Conformal Supergravities,''
  Nucl.\ Phys.\ B {\bf 203}, 157 (1982).
  %%CITATION = NUPHA,B203,157;%%
  %28 citations counted in INSPIRE as of 01 Apr 2013


%\cite{Buchbinder:1988yu}
\bibitem{Buchbinder:1988yu} 
  I.~L.~Buchbinder and S.~M.~Kuzenko,
  %``Nonlocal Action For Supertrace Anomalies In Superspace Of N=1 Supergravity,''
  Phys.\ Lett.\ B {\bf 202}, 233 (1988).
  %%CITATION = PHLTA,B202,233;%%
  %18 citations counted in INSPIRE as of 13 Feb 2013
%\cite{Grosse:2007au}
\bibitem{Grosse:2007au} 
  J.~Grosse,
  %``Quantum Field Theories Coupled to Supergravity: AdS/CFT and Local Couplings,''
  Fortsch.\ Phys.\  {\bf 56}, 183 (2008)
  [arXiv:0711.0444 [hep-th]].
  %%CITATION = ARXIV:0711.0444;%%

%\cite{Butter:2013ura}
\bibitem{Butter:2013ura} 
  D.~Butter and S.~M.~Kuzenko,
  %``Nonlocal action for the super-Weyl anomalies: A new representation,''
  arXiv:1307.1290 [hep-th].
  %%CITATION = ARXIV:1307.1290;%%

\bibitem{Private}

Y.~Nakayama, unpublished. Private communication with J.~Erdmenger.

%\cite{Baumann:2011nm}
\bibitem{Baumann:2011nm} 
  D.~Baumann and D.~Green,
  %``Supergravity for Effective Theories,''
  JHEP {\bf 1203}, 001 (2012)
  [arXiv:1109.0293 [hep-th]].
  %%CITATION = ARXIV:1109.0293;%%
  %12 citations counted in INSPIRE as of 05 Jul 2013

%\cite{Manvelyan:1995hz}
\bibitem{Manvelyan:1995hz} 
  R.~P.~Manvelyan,
  %``SuperWeyl cocycle in d = 4 and superconformal invariant operator,''
  Phys.\ Lett.\ B {\bf 373}, 306 (1996)
  [hep-th/9512045].
  %%CITATION = HEP-TH/9512045;%%

\bibitem{FG}
C.~Feferman and C.~R.~Graham. 
%Conformal invariants. 
Asterisque. (1985), 95-116.
\bibitem{GG}
C. R. Graham, R. Jenne, L. Mason and G. Sparling. 
%Conformally invariant powers of the
%Laplacian, I: existence. 
Journal of London Mathematical Society. 46 (1992), 557-565.
\bibitem{G}
C.~R.~Graham. 
%Conformally invariant powers of the Laplacian, II: nonexistence. 
J. LondonvMath. Soc. (2). 46 (1992), 566-576.

\bibitem{Jackiw}
R.~Jackiw, Lectures on current algebra and its applications, by S.~B.~ Trieman, R.~Jackiw and D.~J.~ Gross (Princeton University Press, Princeton, 1972)
%\cite{Nakayama:2004vk}
\bibitem{Nakayama:2004vk} 
  Y.~Nakayama,
  %``Liouville field theory: A Decade after the revolution,''
  Int.\ J.\ Mod.\ Phys.\ A {\bf 19}, 2771 (2004)
  [hep-th/0402009].
  %%CITATION = HEP-TH/0402009;%%
%\cite{Iorio:1996ad}
\bibitem{Iorio:1996ad} 
  A.~Iorio, L.~O'Raifeartaigh, I.~Sachs and C.~Wiesendanger,
  %``Weyl gauging and conformal invariance,''
  Nucl.\ Phys.\ B {\bf 495}, 433 (1997)
  [hep-th/9607110].
  %%CITATION = HEP-TH/9607110;%%

%\cite{Ho:2008nr}
\bibitem{Ho:2008nr} 
  C.~M.~Ho and Y.~Nakayama,
  %``Dangerous Liouville Wave - Exactly marginal but non-conformal deformation,''
  JHEP {\bf 0807}, 109 (2008)
  [arXiv:0804.3635 [hep-th]].
  %%CITATION = ARXIV:0804.3635;%%




%\cite{Pons:2009nb}
\bibitem{Pons:2009nb} 
  J.~M.~Pons,
  %``Noether symmetries, energy-momentum tensors and conformal invariance in classical field theory,''
  J.\ Math.\ Phys.\  {\bf 52}, 012904 (2011)
  [arXiv:0902.4871 [hep-th]].
  %%CITATION = ARXIV:0902.4871;%%

%\cite{Siegel:1988gd}
\bibitem{Siegel:1988gd} 
  W.~Siegel,
  %``All Free Conformal Representations In All Dimensions,''
  Int.\ J.\ Mod.\ Phys.\ A {\bf 4}, 2015 (1989).
  %%CITATION = IMPAE,A4,2015;%%
%\cite{Mack:1969rr}
\bibitem{Mack:1969rr} 
  G.~Mack and A.~Salam,
  %``Finite component field representations of the conformal group,''
  Annals Phys.\  {\bf 53}, 174 (1969).
  %%CITATION = APNYA,53,174;%%
%\cite{Gross:1970tb}
\bibitem{Gross:1970tb} 
  D.~J.~Gross and J.~Wess,
  %``Scale invariance, conformal invariance, and the high-energy behavior of scattering amplitudes,''
  Phys.\ Rev.\ D {\bf 2}, 753 (1970).
  %%CITATION = PHRVA,D2,753;%%



%\cite{Bracken:1982ny}
\bibitem{Bracken:1982ny} 
  A.~J.~Bracken and B.~Jessup,
  %``Local Conformal Invariance Of The Wave Equation For Finite Component Fields. I. The Conditions For Invariance, And Fully Reducible Fields,''
  J.\ Math.\ Phys.\  {\bf 23}, 1925 (1982).
  %%CITATION = JMAPA,23,1925;%%



%\cite{Konishi:1983hf}
\bibitem{Konishi:1983hf} 
  K.~Konishi,
  %``Anomalous Supersymmetry Transformation of Some Composite Operators in SQCD,''
  Phys.\ Lett.\ B {\bf 135}, 439 (1984).
  %%CITATION = PHLTA,B135,439;%%


%\cite{Callan:1970yg}
\bibitem{Callan:1970yg} 
  C.~G.~Callan, Jr.,
  %``Broken scale invariance in scalar field theory,''
  Phys.\ Rev.\ D {\bf 2}, 1541 (1970).
  %%CITATION = PHRVA,D2,1541;%%


%\cite{Symanzik:1970rt}
\bibitem{Symanzik:1970rt} 
  K.~Symanzik,
  %``Small distance behavior in field theory and power counting,''
  Commun.\ Math.\ Phys.\  {\bf 18}, 227 (1970).
  %%CITATION = CMPHA,18,227;%%
%\cite{Symanzik:1971vw}
\bibitem{Symanzik:1971vw} 
  K.~Symanzik,
  %``Small distance behavior analysis and Wilson expansion,''
  Commun.\ Math.\ Phys.\  {\bf 23}, 49 (1971).
  %%CITATION = CMPHA,23,49;%%




\bibitem{GL}
M.~Gellman, F.~Low, Phys. Rev. 95, 1300 (1954).





%\cite{Yonekura:2012uk}
\bibitem{Yonekura:2012uk} 
  K.~Yonekura,
  %``On the Trace Anomaly and the Anomaly Puzzle in N=1 Pure Yang-Mills,''
  JHEP {\bf 1203}, 029 (2012)
  [arXiv:1202.1514 [hep-th]].
  %%CITATION = ARXIV:1202.1514;%%
%\cite{Jack:1990eb}
\bibitem{Jack:1990eb} 
  I.~Jack and H.~Osborn,
  %``Analogs For The C Theorem For Four-dimensional Renormalizable Field Theories,''
  Nucl.\ Phys.\ B {\bf 343}, 647 (1990).
  %%CITATION = NUPHA,B343,647;%%



\bibitem{Chis}
J.~S.~R.~Chisholm, 
%Change of variables in quantum field theories, 
Nucl. Phys. 26 (1961) 469;
\bibitem{Kame}
S.~Kamefuchi, L.~O'Raifeartaigh, and A.~Salam, 
%Change of variables and equivalence theorems in quantum field theories, 
Nucl. Phys. 28 (1961) 529;


\bibitem{Wegner}
F.~Wegner, g
%Some invariance properties of the renormalization grouph,
J.Phys. C7, 2098 (1974).



%\cite{Zamolodchikov:1987ti}
\bibitem{Zamolodchikov:1987ti} 
  A.~B.~Zamolodchikov,
  %``Renormalization Group and Perturbation Theory Near Fixed Points in Two-Dimensional Field Theory,''
  Sov.\ J.\ Nucl.\ Phys.\  {\bf 46}, 1090 (1987)
  [Yad.\ Fiz.\  {\bf 46}, 1819 (1987)].
  %%CITATION = SJNCA,46,1090;%%

%\cite{Cappelli:1989yu}
\bibitem{Cappelli:1989yu} 
  A.~Cappelli and J.~I.~Latorre,
  %``Perturbation Theory Of Higher Spin Conserved Currents Off Criticality,''
  Nucl.\ Phys.\ B {\bf 340}, 659 (1990).
  %%CITATION = NUPHA,B340,659;%%

%\cite{Friedan:2012hi}
\bibitem{Friedan:2012hi} 
  D.~Friedan and A.~Konechny,
  %``Curvature formula for the space of 2-d conformal field theories,''
  JHEP {\bf 1209}, 113 (2012)
  [arXiv:1206.1749 [hep-th]].
  %%CITATION = ARXIV:1206.1749;%%

\bibitem{Cun}
E.~Cunningham, 
%The Principle of Relativity in Electrodynamics and Extension Thereof , 
Proc. London.
Math. Soc. (ser. 2), 8, 77-98, (1910)
\bibitem{Bate}
 H.~Bateman, 
%The Transformation of the Electrodynamical Equations , 
Proc. London. Math. Soc. (ser. 2), 8, 223-264, (1910)




\bibitem{Weyl}
H.~Weyl,
%Gravitation and Electricity. Sitz. 
Berichte d. Preuss. Akad. d. Wissenschaften, 465 (1918).


%\cite{Kastrup:2008jn}
\bibitem{Kastrup:2008jn} 
  H.~A.~Kastrup,
  %``On the Advancements of Conformal Transformations and their Associated Symmetries in Geometry and Theoretical Physics,''
  Annalen Phys.\  {\bf 17}, 631 (2008)
  [arXiv:0808.2730 [physics.hist-ph]].
  %%CITATION = ARXIV:0808.2730;%%

%\cite{Seiberg:1997ax}
\bibitem{Seiberg:1997ax} 
  N.~Seiberg,
  %``Notes on theories with 16 supercharges,''
  Nucl.\ Phys.\ Proc.\ Suppl.\  {\bf 67}, 158 (1998)
  [hep-th/9705117].
  %%CITATION = HEP-TH/9705117;%%


%\cite{Schroer:1974ay}
\bibitem{Schroer:1974ay} 
  B.~Schroer and J.~A.~Swieca,
  %``Conformal Transformations for Quantized Fields,''
  Phys.\ Rev.\ D {\bf 10}, 480 (1974).
  %%CITATION = PHRVA,D10,480;%%
%\cite{Schroer:1974de}
\bibitem{Schroer:1974de} 
  B.~Schroer, J.~A.~Swieca and A.~H.~Volkel,
  %``Global Operator Expansions in Conformally Invariant Relativistic Quantum Field Theory,''
  Phys.\ Rev.\ D {\bf 11}, 1509 (1975).
  %%CITATION = PHRVA,D11,1509;%%

%\cite{Luscher:1974ez}
\bibitem{Luscher:1974ez} 
  M.~Luscher and G.~Mack,
  %``Global Conformal Invariance in Quantum Field Theory,''
  Commun.\ Math.\ Phys.\  {\bf 41}, 203 (1975).
  %%CITATION = CMPHA,41,203;%%

%\cite{Nikolov:2001iz}
\bibitem{Nikolov:2001iz} 
  N.~M.~Nikolov, Y.~S.~Stanev and I.~T.~Todorov,
  %``Four-dimensional CFT models with rational correlation functions,''
  J.\ Phys.\ A {\bf 35}, 2985 (2002)
  [hep-th/0110230].
  %%CITATION = HEP-TH/0110230;%%


%\cite{Strominger:2003fn}
\bibitem{Strominger:2003fn} 
  A.~Strominger and T.~Takayanagi,
  %``Correlators in time - like bulk Liouville theory,''
  Adv.\ Theor.\ Math.\ Phys.\  {\bf 7}, 369 (2003)
  [hep-th/0303221].
  %%CITATION = HEP-TH/0303221;%%
%\cite{Zamolodchikov:2005fy}
\bibitem{Zamolodchikov:2005fy} 
  A.~B.~Zamolodchikov,
  %``On the three-point function in minimal Liouville gravity,''
  hep-th/0505063.
  %%CITATION = HEP-TH/0505063;%%
%\cite{McElgin:2007ak}
\bibitem{McElgin:2007ak} 
  W.~McElgin,
  %``Notes on Liouville Theory at c <= 1,''
  Phys.\ Rev.\ D {\bf 77}, 066009 (2008)
  [arXiv:0706.0365 [hep-th]].
  %%CITATION = ARXIV:0706.0365;%%
%\cite{Harlow:2011ny}
\bibitem{Harlow:2011ny} 
  D.~Harlow, J.~Maltz and E.~Witten,
  %``Analytic Continuation of Liouville Theory,''
  JHEP {\bf 1112}, 071 (2011)
  [arXiv:1108.4417 [hep-th]].
  %%CITATION = ARXIV:1108.4417;%%


%\cite{LeClair:2003hj}
\bibitem{LeClair:2003hj} 
  A.~LeClair, J.~M.~Roman and G.~Sierra,
  %``Log periodic behavior of finite size effects in field theories with RG limit cycles,''
  Nucl.\ Phys.\ B {\bf 700}, 407 (2004)
  [hep-th/0312141].
  %%CITATION = HEP-TH/0312141;%%

%\cite{LeClair:2004ps}
\bibitem{LeClair:2004ps} 
  A.~LeClair and G.~Sierra,
  %``Renormalization group limit cycles and field theories for elliptic S matrices,''
  J.\ Stat.\ Mech.\  {\bf 0408}, P08004 (2004)
  [hep-th/0403178].
  %%CITATION = HEP-TH/0403178;%%

%\cite{Zamolodchikov:1979ba}
\bibitem{Zamolodchikov:1979ba} 
  A.~B.~Zamolodchikov,
  %``Z(4) Symmetric Factorized S Matrix In Two Space-time Dimensions,''
  Commun.\ Math.\ Phys.\  {\bf 69}, 165 (1979).
  %%CITATION = CMPHA,69,165;%%








\bibitem{Mack1}
M. L\"{u}scher and G. Mack,
%The energy momentum tensor of a critical quan-tum field theory,
 1976, unpublished;

G. Mack,
%Introduction to conformal invariant quantum field theory in two
%or more dimensions,in 
``NONPERTURBATIVE QUANTUM FIELD THEORY. PROCEEDINGS, NATO
ADVANCED STUDY INSTITUTE, CARGESE, FRANCE, JULY 16-30, 1987''.
%\cite{Nakayama:2009qu}

\bibitem{RS}
H.~Reeh, and S.~Schlieder, Nuovo Cimemento 22, 1051,
(1961) For a review see R.~Streater and A.~Wightman,
gPCT, Spin and Statistics, and All Thath.



%\cite{Cardy:1986ie}
\bibitem{Cardy:1986ie} 
  J.~L.~Cardy,
  %``Operator Content of Two-Dimensional Conformally Invariant Theories,''
  Nucl.\ Phys.\ B {\bf 270}, 186 (1986).
  %%CITATION = NUPHA,B270,186;%%


%\cite{Friedan:2009ik}
\bibitem{Friedan:2009ik} 
  D.~Friedan and A.~Konechny,
  %``Gradient formula for the beta-function of 2d quantum field theory,''
  J.\ Phys.\ A {\bf 43}, 215401 (2010)
  [arXiv:0910.3109 [hep-th]].
  %%CITATION = ARXIV:0910.3109;%%

%\cite{Behr:2013vta}
\bibitem{Behr:2013vta} 
  N.~Behr and A.~Konechny,
  %``Renormalization and redundancy in 2d quantum field theories,''
  JHEP {\bf 2014}, 1
  [arXiv:1310.4185 [hep-th]].
  %%CITATION = ARXIV:1310.4185;%%
  %1 citations counted in INSPIRE as of 25 Feb 2014


%\cite{Yonekura:2012kb}
\bibitem{Yonekura:2012kb} 
  K.~Yonekura,
  %``Perturbative c-theorem in d-dimensions,''
  arXiv:1212.3028 [hep-th].
  %%CITATION = ARXIV:1212.3028;%%
%\cite{Cappelli:1991ke}
\bibitem{Cappelli:1991ke} 
  A.~Cappelli, J.~I.~Latorre and X.~Vilasis-Cardona,
  %``Renormalization group patterns and C theorem in more than two-dimensions,''
  Nucl.\ Phys.\ B {\bf 376}, 510 (1992)
  [hep-th/9109041].
  %%CITATION = HEP-TH/9109041;%%
%\cite{Anselmi:1999xk}
\bibitem{Anselmi:1999xk} 
  D.~Anselmi,
  %``Anomalies, unitarity and quantum irreversibility,''
  Annals Phys.\  {\bf 276}, 361 (1999)
  [hep-th/9903059].
  %%CITATION = HEP-TH/9903059;%%
%\cite{Anselmi:2001yp}
\bibitem{Anselmi:2001yp} 
  D.~Anselmi,
  %``Kinematic sum rules for trace anomalies,''
  JHEP {\bf 0111}, 033 (2001)
  [hep-th/0107194].
  %%CITATION = HEP-TH/0107194;%%
%\cite{Anselmi:2002fk}
\bibitem{Anselmi:2002fk} 
  D.~Anselmi,
  %``Inequalities for trace anomalies, length of the RG flow, distance between the fixed points and irreversibility,''
  Class.\ Quant.\ Grav.\  {\bf 21}, 29 (2004)
  [hep-th/0210124].
  %%CITATION = HEP-TH/0210124;%%


%\cite{Cardy:1988cwa}
\bibitem{Cardy:1988cwa} 
  J.~L.~Cardy,
  %``Is There a c Theorem in Four-Dimensions?,''
  Phys.\ Lett.\ B {\bf 215}, 749 (1988).
  %%CITATION = PHLTA,B215,749;%%


%\cite{Komargodski:2011vj}
\bibitem{Komargodski:2011vj} 
  Z.~Komargodski and A.~Schwimmer,
  %``On Renormalization Group Flows in Four Dimensions,''
  JHEP {\bf 1112}, 099 (2011)
  [arXiv:1107.3987 [hep-th]].
  %%CITATION = ARXIV:1107.3987;%%


%\cite{Komargodski:2011xv}
\bibitem{Komargodski:2011xv} 
  Z.~Komargodski,
  %``The Constraints of Conformal Symmetry on RG Flows,''
  JHEP {\bf 1207}, 069 (2012)
  [arXiv:1112.4538 [hep-th]].
  %%CITATION = ARXIV:1112.4538;%%


%\cite{Wallace:1974dx}
\bibitem{Wallace:1974dx} 
  D.~J.~Wallace and R.~K.~P.~Zia,
  %``Gradient Flow and the Renormalization Group,''
  Phys.\ Lett.\ A {\bf 48}, 325 (1974).
  %%CITATION = PHLTA,A48,325;%%

%\cite{Wallace:1974dy}
\bibitem{Wallace:1974dy} 
  D.~J.~Wallace and R.~K.~P.~Zia,
  %``Gradient Properties of the Renormalization Group Equations in Multicomponent Systems,''
  Annals Phys.\  {\bf 92}, 142 (1975).
  %%CITATION = APNYA,92,142;%%













%\cite{Erdmenger:1996yc}
\bibitem{Erdmenger:1996yc} 
  J.~Erdmenger and H.~Osborn,
  %``Conserved currents and the energy momentum tensor in conformally invariant theories for general dimensions,''
  Nucl.\ Phys.\ B {\bf 483}, 431 (1997)
  [hep-th/9605009].
  %%CITATION = HEP-TH/9605009;%%






%\cite{Hofman:2008ar}
\bibitem{Hofman:2008ar} 
  D.~M.~Hofman and J.~Maldacena,
  %``Conformal collider physics: Energy and charge correlations,''
  JHEP {\bf 0805}, 012 (2008)
  [arXiv:0803.1467 [hep-th]].
  %%CITATION = ARXIV:0803.1467;%%

\bibitem{Latorre:1997ea} 
  J.~I.~Latorre and H.~Osborn,
  %``Positivity and the energy momentum tensor in quantum field theory,''
  Nucl.\ Phys.\ B {\bf 511}, 737 (1998)
  [hep-th/9703196].
  %%CITATION = HEP-TH/9703196;%%






%\cite{Anselmi:1997am}
\bibitem{Anselmi:1997am} 
  D.~Anselmi, D.~Z.~Freedman, M.~T.~Grisaru and A.~A.~Johansen,
  %``Nonperturbative formulas for central functions of supersymmetric gauge theories,''
  Nucl.\ Phys.\ B {\bf 526}, 543 (1998)
  [hep-th/9708042].
  %%CITATION = HEP-TH/9708042;%%



%\cite{Farnsworth:2013osa}
\bibitem{Farnsworth:2013osa} 
  K.~Farnsworth, M.~A.~Luty and V.~Prelipina,
  %``Scale Invariance plus Unitarity Implies Conformal Invariance in Four Dimensions,''
  arXiv:1309.4095 [hep-th].
  %%CITATION = ARXIV:1309.4095;%%
  %5 citations counted in INSPIRE as of 06 Feb 2014

%\cite{Dymarsky:2013pqa}
\bibitem{Dymarsky:2013pqa} 
  A.~Dymarsky, Z.~Komargodski, A.~Schwimmer and S.~Theisen,
  %``On Scale and Conformal Invariance in Four Dimensions,''
  arXiv:1309.2921 [hep-th].
  %%CITATION = ARXIV:1309.2921;%%
  %9 citations counted in INSPIRE as of 26 Feb 2014



%\cite{Bzowski:2014qja}
\bibitem{Skenderis} 
  A.~Bzowski and K.~Skenderis,
  %``Comments on scale and conformal invariance in four dimensions,''
  arXiv:1402.3208 [hep-th].
  %%CITATION = ARXIV:1402.3208;%%

%
%\cite{Nakayama:2013wda}
\bibitem{Nakayama:2013wda} 
  Y.~Nakayama,
  %``Consistency of local renormalization group in $d=3$,''
  Nucl.\ Phys.\ B {\bf 879}, 37 (2014)
  [arXiv:1307.8048 [hep-th]].
  %%CITATION = ARXIV:1307.8048;%%
  %5 citations counted in INSPIRE as of 06 Feb 2014






%\cite{Nakayama:2012sn}
\bibitem{Nakayama:2012sn} 
  Y.~Nakayama,
  %``Holographic Renormalization of Foliation Preserving Gravity and Trace Anomaly,''
  Gen.\ Rel.\ Grav.\  {\bf 44}, 2873 (2012)
  [arXiv:1203.1068 [hep-th]].
  %%CITATION = ARXIV:1203.1068;%%



%\cite{'tHooft:1979bh}
\bibitem{'tHooft:1979bh} 
  G.~'t Hooft,
  %``Naturalness, chiral symmetry, and spontaneous chiral symmetry breaking,''
  NATO Adv.\ Study Inst.\ Ser.\ B Phys.\  {\bf 59}, 135 (1980).
  %%CITATION = NASBD,59,135;%%


%\cite{Intriligator:2003jj}
\bibitem{Intriligator:2003jj} 
  K.~A.~Intriligator and B.~Wecht,
  %``The Exact superconformal R symmetry maximizes a,''
  Nucl.\ Phys.\ B {\bf 667}, 183 (2003)
  [hep-th/0304128].
  %%CITATION = HEP-TH/0304128;%%

%\cite{Buican:2011ty}
\bibitem{Buican:2011ty} 
  M.~Buican,
  %``A Conjectured Bound on Accidental Symmetries,''
  Phys.\ Rev.\ D {\bf 85}, 025020 (2012)
  [arXiv:1109.3279 [hep-th]].
  %%CITATION = ARXIV:1109.3279;%%

%\cite{Nakayama:2011tk}
\bibitem{Nakayama:2011tk} 
  Y.~Nakayama,
  %``Comments on scale invariant but non-conformal supersymmetric field theories,''
  Int.\ J.\ Mod.\ Phys.\ A {\bf 27}, 1250122 (2012)
  [arXiv:1109.5883 [hep-th]].
  %%CITATION = ARXIV:1109.5883;%%

%\cite{Schwimmer:2010za}
\bibitem{Schwimmer:2010za} 
  A.~Schwimmer and S.~Theisen,
  %``Spontaneous Breaking of Conformal Invariance and Trace Anomaly Matching,''
  Nucl.\ Phys.\ B {\bf 847}, 590 (2011)
  [arXiv:1011.0696 [hep-th]].
  %%CITATION = ARXIV:1011.0696;%%



%\cite{Armillis:2013wya}
\bibitem{Armillis:2013wya} 
  R.~Armillis, A.~Monin and M.~Shaposhnikov,
  %``Spontaneously Broken Conformal Symmetry: Dealing with the Trace Anomaly,''
  JHEP {\bf 1310}, 030 (2013)
  [arXiv:1302.5619 [hep-th]].
  %%CITATION = ARXIV:1302.5619;%%
  %8 citations counted in INSPIRE as of 25 Feb 2014


%\cite{Gretsch:2013ooa}
\bibitem{Gretsch:2013ooa} 
  F.~Gretsch and A.~Monin,
  %``Dilaton: Saving Conformal Symmetry,''
  arXiv:1308.3863 [hep-th].
  %%CITATION = ARXIV:1308.3863;%%
  %4 citations counted in INSPIRE as of 25 Feb 2014

%\cite{Elvang:2012yc}
\bibitem{Elvang:2012yc} 
  H.~Elvang and T.~M.~Olson,
  %``RG flows in d dimensions, the dilaton effective action, and the a-theorem,''
  arXiv:1209.3424 [hep-th].
  %%CITATION = ARXIV:1209.3424;%%

%\cite{Schwimmer:2013jma}
\bibitem{Schwimmer:2013jma} 
  A.~Schwimmer and S.~Theisen,
  %``Comments on the Algebraic Properties of Dilaton Actions,''
  arXiv:1311.4746 [hep-th].
  %%CITATION = ARXIV:1311.4746;%%



%\cite{Antoniadis:1991fa}
\bibitem{Antoniadis:1991fa} 
  I.~Antoniadis and E.~Mottola,
  %``4-D quantum gravity in the conformal sector,''
  Phys.\ Rev.\ D {\bf 45}, 2013 (1992).
  %%CITATION = PHRVA,D45,2013;%%
%\cite{Antoniadis:1992xu}
\bibitem{Antoniadis:1992xu} 
  I.~Antoniadis, P.~O.~Mazur and E.~Mottola,
  %``Conformal symmetry and central charges in four-dimensions,''
  Nucl.\ Phys.\ B {\bf 388}, 627 (1992)
  [hep-th/9205015].
  %%CITATION = HEP-TH/9205015;%%


%\cite{Adams:2006sv}
\bibitem{Adams:2006sv} 
  A.~Adams, N.~Arkani-Hamed, S.~Dubovsky, A.~Nicolis and R.~Rattazzi,
  %``Causality, analyticity and an IR obstruction to UV completion,''
  JHEP {\bf 0610}, 014 (2006)
  [hep-th/0602178].
  %%CITATION = HEP-TH/0602178;%%

%\cite{Nakayama:2011wq}
\bibitem{Nakayama:2011wq} 
  Y.~Nakayama,
  %``On \epsilon-conjecture in a-theorem,''
  Mod.\ Phys.\ Lett.\ A {\bf 27}, 1250029 (2012)
  [arXiv:1110.2586 [hep-th]].
  %%CITATION = ARXIV:1110.2586;%%
%\cite{Morozov:2003ik}
\bibitem{Morozov:2003ik} 
  A.~Morozov and A.~J.~Niemi,
  %``Can renormalization group flow end in a big mess?,''
  Nucl.\ Phys.\ B {\bf 666}, 311 (2003)
  [hep-th/0304178].
  %%CITATION = HEP-TH/0304178;%%
  %29 citations counted in INSPIRE as of 30 Dec 2013

%\cite{Curtright:2011qg}
\bibitem{Curtright:2011qg} 
  T.~L.~Curtright, X.~Jin and C.~K.~Zachos,
  %``RG flows, cycles, and c-theorem folklore,''
  Phys.\ Rev.\ Lett.\  {\bf 108}, 131601 (2012)
  [arXiv:1111.2649 [hep-th]].
  %%CITATION = ARXIV:1111.2649;%%



%\cite{Cappelli:1990yc}
\bibitem{Cappelli:1990yc} 
  A.~Cappelli, D.~Friedan and J.~I.~Latorre,
  %``C theorem and spectral representation,''
  Nucl.\ Phys.\ B {\bf 352}, 616 (1991).
  %%CITATION = NUPHA,B352,616;%%
%\cite{Cappelli:2001pz}
\bibitem{Cappelli:2001pz} 
  A.~Cappelli, R.~Guida and N.~Magnoli,
  %``Exact consequences of the trace anomaly in four-dimensions,''
  Nucl.\ Phys.\ B {\bf 618}, 371 (2001)
  [hep-th/0103237].
  %%CITATION = HEP-TH/0103237;%%


%\cite{Ball:2001wr}
\bibitem{Ball:2001wr} 
  R.~D.~Ball and P.~H.~Damgaard,
  %``The C theorem and chiral symmetry breaking in asymptotically free vector - like gauge theories,''
  Phys.\ Lett.\ B {\bf 510}, 341 (2001)
  [hep-th/0103249].
  %%CITATION = HEP-TH/0103249;%%

%\cite{Jack:2013sha}
\bibitem{Jack:2013sha} 
  I.~Jack and H.~Osborn,
  %``Constraints on RG Flow for Four Dimensional Quantum Field Theories,''
  arXiv:1312.0428 [hep-th].
  %%CITATION = ARXIV:1312.0428;%%
  %2 citations counted in INSPIRE as of 06 Feb 2014

%\cite{Baume:2014rla}
\bibitem{Baume:2014rla} 
  F.~Baume, B.~Keren-Zur, R.~Rattazzi and L.~Vitale,
  %``The local Callan-Symanzik equation: structure and applications,''
  arXiv:1401.5983 [hep-th].
  %%CITATION = ARXIV:1401.5983;%%


%\cite{Bobev:2013vta}
\bibitem{Bobev:2013vta} 
  N.~Bobev, H.~Elvang and T.~M.~Olson,
  %``Dilaton Effective Action with $\mathcal{N}=1$ Supersymmetry,''
  arXiv:1312.2925 [hep-th].
  %%CITATION = ARXIV:1312.2925;%%




%\cite{Nakayama:2012ed}
\bibitem{Nakayama:2012ed} 
  Y.~Nakayama,
  %``Is boundary conformal in CFT?,''
  arXiv:1210.6439 [hep-th].
  %%CITATION = ARXIV:1210.6439;%%




%\cite{Affleck:1991tk}
\bibitem{Affleck:1991tk} 
  I.~Affleck and A.~W.~W.~Ludwig,
  %``Universal noninteger 'ground state degeneracy' in critical quantum systems,''
  Phys.\ Rev.\ Lett.\  {\bf 67}, 161 (1991).
  %%CITATION = PRLTA,67,161;%%
%\cite{Friedan:2003yc}
\bibitem{Friedan:2003yc} 
  D.~Friedan and A.~Konechny,
  %``On the boundary entropy of one-dimensional quantum systems at low temperature,''
  Phys.\ Rev.\ Lett.\  {\bf 93}, 030402 (2004)
  [hep-th/0312197].
  %%CITATION = HEP-TH/0312197;%%
%\cite{Friedan:2005dj}
\bibitem{Friedan:2005dj} 
  D.~Friedan and A.~Konechny,
  %``Infrared properties of boundaries in 1-d quantum systems,''
  J.\ Statist.\ Phys.\  {\bf 0603}, P014 (2006)
  [hep-th/0512023].
  %%CITATION = HEP-TH/0512023;%%

%\cite{Efimov:1970zz}
\bibitem{Efimov:1970zz} 
  V.~Efimov,
  %``Energy levels arising form the resonant two-body forces in a three-body system,''
  Phys.\ Lett.\ B {\bf 33}, 563 (1970).
  %%CITATION = PHLTA,B33,563;%%
%\cite{Nishida:2007de}
\bibitem{Nishida:2007de} 
  Y.~Nishida,
  %``Renormalization group analysis of resonantly interacting anyons,''
  Phys.\ Rev.\ D {\bf 77}, 061703 (2008)
  [arXiv:0708.4056 [hep-th]].
  %%CITATION = ARXIV:0708.4056;%%




%\cite{Elvang:2012st}
\bibitem{Elvang:2012st} 
  H.~Elvang, D.~Z.~Freedman, L.~-Y.~Hung, M.~Kiermaier, R.~C.~Myers and S.~Theisen,
  %``On renormalization group flows and the a-theorem in 6d,''
  JHEP {\bf 1210}, 011 (2012)
  [arXiv:1205.3994 [hep-th]].
  %%CITATION = ARXIV:1205.3994;%%

%\cite{Aharony:1999ks}
\bibitem{Aharony:1999ks} 
  O.~Aharony,
  %``A Brief review of 'little string theories',''
  Class.\ Quant.\ Grav.\  {\bf 17}, 929 (2000)
  [hep-th/9911147].
  %%CITATION = HEP-TH/9911147;%%

%\cite{Nahm:1977tg}
\bibitem{Nahm:1977tg} 
  W.~Nahm,
  %``Supersymmetries and their Representations,''
  Nucl.\ Phys.\ B {\bf 135}, 149 (1978).
  %%CITATION = NUPHA,B135,149;%%


%\cite{Benini:2012cz}
\bibitem{Benini:2012cz} 
  F.~Benini and N.~Bobev,
  %``Exact two-dimensional superconformal R-symmetry and c-extremization,''
  arXiv:1211.4030 [hep-th].
  %%CITATION = ARXIV:1211.4030;%%
%\cite{Jafferis:2010un}
\bibitem{Jafferis:2010un} 
  D.~L.~Jafferis,
  %``The Exact Superconformal R-Symmetry Extremizes Z,''
  JHEP {\bf 1205}, 159 (2012)
  [arXiv:1012.3210 [hep-th]].
  %%CITATION = ARXIV:1012.3210;%%
%\cite{Jafferis:2011zi}
\bibitem{Jafferis:2011zi} 
  D.~L.~Jafferis, I.~R.~Klebanov, S.~S.~Pufu and B.~R.~Safdi,
  %``Towards the F-Theorem: N=2 Field Theories on the Three-Sphere,''
  JHEP {\bf 1106}, 102 (2011)
  [arXiv:1103.1181 [hep-th]].
  %%CITATION = ARXIV:1103.1181;%%
%\cite{Klebanov:2011gs}
\bibitem{Klebanov:2011gs} 
  I.~R.~Klebanov, S.~S.~Pufu and B.~R.~Safdi,
  %``F-Theorem without Supersymmetry,''
  JHEP {\bf 1110}, 038 (2011)
  [arXiv:1105.4598 [hep-th]].
  %%CITATION = ARXIV:1105.4598;%%

%\cite{Klebanov:2012va}
\bibitem{Klebanov:2012va} 
  I.~R.~Klebanov, T.~Nishioka, S.~S.~Pufu and B.~R.~Safdi,
  %``Is Renormalized Entanglement Entropy Stationary at RG Fixed Points?,''
  JHEP {\bf 1210}, 058 (2012)
  [arXiv:1207.3360 [hep-th]].
  %%CITATION = ARXIV:1207.3360;%%
%\cite{Calabrese:2004eu}
\bibitem{Calabrese:2004eu} 
  P.~Calabrese and J.~L.~Cardy,
  %``Entanglement entropy and quantum field theory,''
  J.\ Stat.\ Mech.\  {\bf 0406}, P06002 (2004)
  [hep-th/0405152].
  %%CITATION = HEP-TH/0405152;%%

\bibitem{entangle}
M.~Nielsen, and I.~L.~Chuang, gQuantum Computation and Quantum
Informationh, Cambridge university press, (2000).
\bibitem{subadditivity}
E.~H.~Lieb and M.~B.~Ruskai, 
%gA fundamental property of quantum-mechanical entropy,h
Phys. Rev. Lett. 30 (1973) 434-436; 
%g Proof of the strong subadditivity of quantummechanical entropy,h 
J. Math. Phys. 14 (1973) 1938?1941.


%\cite{Solodukhin:2008dh}
\bibitem{Solodukhin:2008dh} 
  S.~N.~Solodukhin,
  %``Entanglement entropy, conformal invariance and extrinsic geometry,''
  Phys.\ Lett.\ B {\bf 665}, 305 (2008)
  [arXiv:0802.3117 [hep-th]].
  %%CITATION = ARXIV:0802.3117;%%

%\cite{Casini:2011kv}
\bibitem{Casini:2011kv} 
  H.~Casini, M.~Huerta and R.~C.~Myers,
  %``Towards a derivation of holographic entanglement entropy,''
  JHEP {\bf 1105}, 036 (2011)
  [arXiv:1102.0440 [hep-th]].
  %%CITATION = ARXIV:1102.0440;%%


%\cite{Casini:2004bw}
\bibitem{Casini:2004bw} 
  H.~Casini and M.~Huerta,
  %``A Finite entanglement entropy and the c-theorem,''
  Phys.\ Lett.\ B {\bf 600}, 142 (2004)
  [hep-th/0405111].
  %%CITATION = HEP-TH/0405111;%%


%\cite{Solodukhin:2013yha}
\bibitem{Solodukhin:2013yha} 
  S.~N.~Solodukhin,
  %``The a-theorem and entanglement entropy,''
  arXiv:1304.4411 [hep-th].
  %%CITATION = ARXIV:1304.4411;%%


%\cite{Casini:2012ei}
\bibitem{Casini:2012ei} 
  H.~Casini and M.~Huerta,
  %``On the RG running of the entanglement entropy of a circle,''
  Phys.\ Rev.\ D {\bf 85}, 125016 (2012)
  [arXiv:1202.5650 [hep-th]].
  %%CITATION = ARXIV:1202.5650;%%


%\cite{Grinstein:2013cka}
\bibitem{Grinstein:2013cka} 
  B.~Grinstein, A.~Stergiou and D.~Stone,
  %``Consequences of Weyl Consistency Conditions,''
  JHEP {\bf 1311}, 195 (2013)
  [arXiv:1308.1096 [hep-th]].
  %%CITATION = ARXIV:1308.1096;%%
  %3 citations counted in INSPIRE as of 05 Jan 2014

%\cite{Bastianelli:1996gh}
\bibitem{Bastianelli:1996gh} 
  F.~Bastianelli and U.~Lindstrom,
  %``C theorem for two-dimensional chiral theories,''
  Phys.\ Lett.\ B {\bf 380}, 341 (1996)
  [hep-th/9604001].
  %%CITATION = HEP-TH/9604001;%%


%\cite{CastroNeto:1992ie}
\bibitem{CastroNeto:1992ie} 
  A.~H.~Castro Neto and E.~H.~Fradkin,
  %``The Thermodynamics of quantum systems and generalizations of Zamolodchikov's C theorem,''
  Nucl.\ Phys.\ B {\bf 400}, 525 (1993)
  [cond-mat/9301009].
  %%CITATION = COND-MAT/9301009;%%

%\cite{Danchev:1998nx}
\bibitem{Danchev:1998nx} 
  D.~M.~Danchev and N.~S.~Tonchev,
  %``On the finite temperature generalization of the C theorem and the interplay between classical and quantum fluctuations,''
  J.\ Phys.\ A {\bf 32}, 7057 (1999)
  [cond-mat/9806190].
  %%CITATION = COND-MAT/9806190;%%

%\cite{McAvity:1993ue}
\bibitem{McAvity:1993ue} 
  D.~M.~McAvity and H.~Osborn,
  %``Energy momentum tensor in conformal field theories near a boundary,''
  Nucl.\ Phys.\ B {\bf 406}, 655 (1993)
  [hep-th/9302068].
  %%CITATION = HEP-TH/9302068;%%

\bibitem{1995NuPhB.455..522M} 
McAvity, D.~M., \& Osborn, H.\ 1995, Nucl Phys B, 455, 522 

%\cite{Cardy:1984bb}
\bibitem{Cardy:1984bb} 
  J.~L.~Cardy,
  %``Conformal Invariance and Surface Critical Behavior,''
  Nucl.\ Phys.\ B {\bf 240}, 514 (1984).
  %%CITATION = NUPHA,B240,514;%%



%\cite{Hofman:2011zj}
\bibitem{Hofman:2011zj} 
  D.~M.~Hofman and A.~Strominger,
  %``Chiral Scale and Conformal Invariance in 2D Quantum Field Theory,''
  Phys.\ Rev.\ Lett.\  {\bf 107}, 161601 (2011)
  [arXiv:1107.2917 [hep-th]].
  %%CITATION = ARXIV:1107.2917;%%

%\cite{Nakayama:2011fe}
\bibitem{Nakayama:2011fe} 
  Y.~Nakayama,
  %``Gravity Dual for Hofman-Strominger Theorem,''
  Phys.\ Rev.\ D {\bf 85}, 085032 (2012)
  [arXiv:1112.0635 [hep-th]].
  %%CITATION = ARXIV:1112.0635;%%



\bibitem{Hagen}
C.~R.~Hagen, Phys. Rev. D 5, 377 (1972).

\bibitem{Niederer}
U.~Niederer, Helv. Phys. Acta 45 (1972) 802.

\bibitem{Lifshitz}
E.~M.~Lifshitz, Zh. Eksp. Teor. Fiz. 11, 255 (1941); 
Zh. Eksp. Teor. Fiz. 11, 269 (1941).

\bibitem{Barut}
A.~O.~Barut,
Helv. Phys. Acta 46 (1973) 496.

\bibitem{Havas}
P.~Havas, J.~Plebanski,
Schr\"odinger group, J. Math. Phys. 19 (1978) 482.

%\cite{Jackiw:1990mb}
\bibitem{Jackiw:1990mb} 
  R.~Jackiw and S.~-Y.~Pi,
  %``Classical and quantal nonrelativistic Chern-Simons theory,''
  Phys.\ Rev.\ D {\bf 42}, 3500 (1990)
  [Erratum-ibid.\ D {\bf 48}, 3929 (1993)].
  %%CITATION = PHRVA,D42,3500;%%


%\cite{Nakayama:2009ww}
\bibitem{Nakayama:2009ww} 
  Y.~Nakayama,
  %``Gravity Dual for Reggeon Field Theory and Non-linear Quantum Finance,''
  Int.\ J.\ Mod.\ Phys.\ A {\bf 24}, 6197 (2009)
  [arXiv:0906.4112 [hep-th]].
  %%CITATION = ARXIV:0906.4112;%%


%\cite{Stueckelberg:1951gg}
\bibitem{Stueckelberg:1951gg} 
  E.~C.~G.~Stueckelberg and A.~Petermann,
  %``The normalization group in quantum theory,''
  Helv.\ Phys.\ Acta {\bf 24}, 317 (1951).
  %%CITATION = HPACA,24,317;%%



%\cite{Gauntlett:2006vf}
\bibitem{Gauntlett:2006vf} 
  J.~P.~Gauntlett, D.~Martelli, J.~Sparks and S.~-T.~Yau,
  %``Obstructions to the existence of Sasaki-Einstein metrics,''
  Commun.\ Math.\ Phys.\  {\bf 273}, 803 (2007)
  [hep-th/0607080].
  %%CITATION = HEP-TH/0607080;%%


%\cite{Nakayama:2007sb}
\bibitem{Nakayama:2007sb} 
  Y.~Nakayama,
  %``Black Hole: String transition and rolling D-brane,''
  hep-th/0702221.
  %%CITATION = HEP-TH/0702221;%%

%\cite{Shapere:2008un}
\bibitem{Shapere:2008un} 
  A.~D.~Shapere and Y.~Tachikawa,
  %``A Counterexample to the 'a-theorem',''
  JHEP {\bf 0812}, 020 (2008)
  [arXiv:0809.3238 [hep-th]].
  %%CITATION = ARXIV:0809.3238;%%


%\cite{Cachazo:2001sg}
\bibitem{Cachazo:2001sg} 
  F.~Cachazo, B.~Fiol, K.~A.~Intriligator, S.~Katz and C.~Vafa,
  %``A Geometric unification of dualities,''
  Nucl.\ Phys.\ B {\bf 628}, 3 (2002)
  [hep-th/0110028].
  %%CITATION = HEP-TH/0110028;%%

%\cite{Bah:2011je}
\bibitem{Bah:2011je} 
  I.~Bah and B.~Wecht,
  %``New N=1 Superconformal Field Theories In Four Dimensions,''
  arXiv:1111.3402 [hep-th].
  %%CITATION = ARXIV:1111.3402;%%


%\cite{Gaiotto:2010jf}
\bibitem{Gaiotto:2010jf} 
  D.~Gaiotto, N.~Seiberg and Y.~Tachikawa,
  %``Comments on scaling limits of 4d N=2 theories,''
  JHEP {\bf 1101}, 078 (2011)
  [arXiv:1011.4568 [hep-th]].
  %%CITATION = ARXIV:1011.4568;%%



%\cite{Amariti:2012wc}
\bibitem{Amariti:2012wc} 
  A.~Amariti and K.~Intriligator,
  %``(Delta a) curiosities in some 4d susy RG flows,''
  JHEP {\bf 1211}, 108 (2012)
  [arXiv:1209.4311 [hep-th]].
  %%CITATION = ARXIV:1209.4311;%%
  %2 citations counted in INSPIRE as of 25 Feb 2014


%\cite{Buican:2013ica}
\bibitem{Buican:2013ica} 
  M.~Buican,
  %``Minimal Distances Between SCFTs,''
  arXiv:1311.1276 [hep-th].
  %%CITATION = ARXIV:1311.1276;%%
  %1 citations counted in INSPIRE as of 25 Feb 2014

 \bibitem{hol1}
G.~'t Hooft, 
%Dimensional Reduction in Quantum Gravity 
`Salamfest' ed A Ali, J Ellis and S Randjbar-Daemi 1993
(Singapore: World Scientific) pp 284-96 

  \bibitem{hol2}
     L.~Susskind,
  %``The World As A Hologram,''
  J.\ Math.\ Phys.\  {\bf 36}, 6377 (1995)
  [arXiv:hep-th/9409089].
  %%CITATION = JMAPA,36,6377;%%


%\cite{Maldacena:1997re}
\bibitem{Maldacena:1997re} 
  J.~M.~Maldacena,
  %``The Large N limit of superconformal field theories and supergravity,''
  Adv.\ Theor.\ Math.\ Phys.\  {\bf 2}, 231 (1998)
  [hep-th/9711200].
  %%CITATION = HEP-TH/9711200;%%

%\cite{Aharony:1999ti}
\bibitem{Aharony:1999ti} 
  O.~Aharony, S.~S.~Gubser, J.~M.~Maldacena, H.~Ooguri and Y.~Oz,
  %``Large N field theories, string theory and gravity,''
  Phys.\ Rept.\  {\bf 323}, 183 (2000)
  [hep-th/9905111].
  %%CITATION = HEP-TH/9905111;%%



%\cite{Gubser:1998bc}
\bibitem{Gubser:1998bc} 
  S.~S.~Gubser, I.~R.~Klebanov and A.~M.~Polyakov,
  %``Gauge theory correlators from noncritical string theory,''
  Phys.\ Lett.\ B {\bf 428}, 105 (1998)
  [hep-th/9802109].
  %%CITATION = HEP-TH/9802109;%%


%\cite{Witten:1998qj}
\bibitem{Witten:1998qj} 
  E.~Witten,
  %``Anti-de Sitter space and holography,''
  Adv.\ Theor.\ Math.\ Phys.\  {\bf 2}, 253 (1998)
  [hep-th/9802150].
  %%CITATION = HEP-TH/9802150;%%


%\cite{de Haro:2000xn}
\bibitem{de Haro:2000xn} 
  S.~de Haro, S.~N.~Solodukhin and K.~Skenderis,
  %``Holographic reconstruction of space-time and renormalization in the AdS / CFT correspondence,''
  Commun.\ Math.\ Phys.\  {\bf 217}, 595 (2001)
  [hep-th/0002230].
  %%CITATION = HEP-TH/0002230;%%
%\cite{Bianchi:2001kw}
\bibitem{Bianchi:2001kw} 
  M.~Bianchi, D.~Z.~Freedman and K.~Skenderis,
  %``Holographic renormalization,''
  Nucl.\ Phys.\ B {\bf 631}, 159 (2002)
  [hep-th/0112119].
  %%CITATION = HEP-TH/0112119;%%


%\cite{de Boer:1999xf}
\bibitem{de Boer:1999xf} 
  J.~de Boer, E.~P.~Verlinde and H.~L.~Verlinde,
  %``On the holographic renormalization group,''
  JHEP {\bf 0008}, 003 (2000)
  [hep-th/9912012].
  %%CITATION = HEP-TH/9912012;%%

%\cite{deBoer:2000cz}
\bibitem{deBoer:2000cz} 
  J.~de Boer,
  %``The Holographic renormalization group,''
  Fortsch.\ Phys.\  {\bf 49}, 339 (2001)
  [hep-th/0101026].
  %%CITATION = HEP-TH/0101026;%%

%\cite{Fukuma:2002sb}
\bibitem{Fukuma:2002sb} 
  M.~Fukuma, S.~Matsuura and T.~Sakai,
  %``Holographic renormalization group,''
  Prog.\ Theor.\ Phys.\  {\bf 109}, 489 (2003)
  [hep-th/0212314].
  %%CITATION = HEP-TH/0212314;%%

%\cite{Townsend:1984iu}
\bibitem{Townsend:1984iu} 
  P.~K.~Townsend,
  %``Positive Energy And The Scalar Potential In Higher Dimensional (super)gravity Theories,''
  Phys.\ Lett.\ B {\bf 148}, 55 (1984).
  %%CITATION = PHLTA,B148,55;%%



%\cite{Akhmedov:1998vf}
\bibitem{Akhmedov:1998vf} 
  E.~T.~Akhmedov,
  %``A Remark on the AdS / CFT correspondence and the renormalization group flow,''
  Phys.\ Lett.\ B {\bf 442}, 152 (1998)
  [hep-th/9806217].
  %%CITATION = HEP-TH/9806217;%%
%\cite{Alvarez:1998wr}
\bibitem{Alvarez:1998wr} 
  E.~Alvarez and C.~Gomez,
  %``Geometric holography, the renormalization group and the c theorem,''
  Nucl.\ Phys.\ B {\bf 541}, 441 (1999)
  [hep-th/9807226].
  %%CITATION = HEP-TH/9807226;%%

%\cite{Girardello:1998pd}
\bibitem{Girardello:1998pd} 
  L.~Girardello, M.~Petrini, M.~Porrati and A.~Zaffaroni,
  %``Novel local CFT and exact results on perturbations of N=4 superYang Mills from AdS dynamics,''
  JHEP {\bf 9812}, 022 (1998)
  [hep-th/9810126].
  %%CITATION = HEP-TH/9810126;%%

%\cite{Freedman:1999gp}
\bibitem{Freedman:1999gp} 
  D.~Z.~Freedman, S.~S.~Gubser, K.~Pilch and N.~P.~Warner,
  %``Renormalization group flows from holography supersymmetry and a c theorem,''
  Adv.\ Theor.\ Math.\ Phys.\  {\bf 3}, 363 (1999)
  [hep-th/9904017].
  %%CITATION = HEP-TH/9904017;%%

%\cite{Sahakian:1999bd}
\bibitem{Sahakian:1999bd} 
  V.~Sahakian,
  %``Holography, a covariant c function, and the geometry of the renormalization group,''
  Phys.\ Rev.\ D {\bf 62}, 126011 (2000)
  [hep-th/9910099].
  %%CITATION = HEP-TH/9910099;%%

%\cite{Myers:2010tj}
\bibitem{Myers:2010tj} 
  R.~C.~Myers and A.~Sinha,
  %``Holographic c-theorems in arbitrary dimensions,''
  JHEP {\bf 1101}, 125 (2011)
  [arXiv:1011.5819 [hep-th]].
  %%CITATION = ARXIV:1011.5819;%%




%\cite{Friedman:1993ty}
\bibitem{Friedman:1993ty} 
  J.~L.~Friedman, K.~Schleich and D.~M.~Witt,
  %``Topological censorship,''
  Phys.\ Rev.\ Lett.\  {\bf 71}, 1486 (1993)
  [Erratum-ibid.\  {\bf 75}, 1872 (1995)]
  [gr-qc/9305017].
  %%CITATION = GR-QC/9305017;%%

%\cite{Tipler:1976bi}
\bibitem{Tipler:1976bi} 
  F.~J.~Tipler,
  %``Causality violation in asymptotically flat space-times,''
  Phys.\ Rev.\ Lett.\  {\bf 37}, 879 (1976).
  %%CITATION = PRLTA,37,879;%%

%\cite{Hawking:1991nk}
\bibitem{Hawking:1991nk} 
  S.~W.~Hawking,
  %``The Chronology protection conjecture,''
  Phys.\ Rev.\ D {\bf 46}, 603 (1992).
  %%CITATION = PHRVA,D46,603;%%

%\cite{Olum:1998mu}
\bibitem{Olum:1998mu} 
  K.~D.~Olum,
  %``Superluminal travel requires negative energies,''
  Phys.\ Rev.\ Lett.\  {\bf 81}, 3567 (1998)
  [gr-qc/9805003].
  %%CITATION = GR-QC/9805003;%%








%\cite{Henningson:1998gx}
\bibitem{Henningson:1998gx} 
  M.~Henningson and K.~Skenderis,
  %``The Holographic Weyl anomaly,''
  JHEP {\bf 9807}, 023 (1998)
  [hep-th/9806087].
  %%CITATION = HEP-TH/9806087;%%


%\cite{Anselmi:2000fu}
\bibitem{Anselmi:2000fu} 
  D.~Anselmi, L.~Girardello, M.~Porrati and A.~Zaffaroni,
  %``A Note on the holographic beta and C functions,''
  Phys.\ Lett.\ B {\bf 481}, 346 (2000)
  [hep-th/0002066].
  %%CITATION = HEP-TH/0002066;%%

%\cite{Liu:1998bu}
\bibitem{Liu:1998bu} 
  H.~Liu and A.~A.~Tseytlin,
  %``D = 4 superYang-Mills, D = 5 gauged supergravity, and D = 4 conformal supergravity,''
  Nucl.\ Phys.\ B {\bf 533}, 88 (1998)
  [hep-th/9804083].
  %%CITATION = HEP-TH/9804083;%%
%\cite{Balasubramanian:2000pq}
\bibitem{Balasubramanian:2000pq}
  V.~Balasubramanian, E.~G.~Gimon, D.~Minic and J.~Rahmfeld,
  %``Four-dimensional conformal supergravity from AdS space,''
  Phys.\ Rev.\ D {\bf 63} (2001) 104009
  [hep-th/0007211].
  %%CITATION = HEP-TH/0007211;%%

%\cite{Erdmenger:2001ja}
\bibitem{Erdmenger:2001ja} 
  J.~Erdmenger,
  %``A Field theoretical interpretation of the holographic renormalization group,''
  Phys.\ Rev.\ D {\bf 64}, 085012 (2001)
  [hep-th/0103219].
  %%CITATION = HEP-TH/0103219;%%


%\cite{Breitenlohner:1982jf}
\bibitem{Breitenlohner:1982jf} 
  P.~Breitenlohner and D.~Z.~Freedman,
  %``Stability in Gauged Extended Supergravity,''
  Annals Phys.\  {\bf 144}, 249 (1982).
  %%CITATION = APNYA,144,249;%%

%\cite{Ryu:2006bv}
\bibitem{Ryu:2006bv} 
  S.~Ryu and T.~Takayanagi,
  %``Holographic derivation of entanglement entropy from AdS/CFT,''
  Phys.\ Rev.\ Lett.\  {\bf 96}, 181602 (2006)
  [hep-th/0603001].
  %%CITATION = HEP-TH/0603001;%%
%\cite{Ryu:2006ef}
\bibitem{Ryu:2006ef} 
  S.~Ryu and T.~Takayanagi,
  %``Aspects of Holographic Entanglement Entropy,''
  JHEP {\bf 0608}, 045 (2006)
  [hep-th/0605073].
  %%CITATION = HEP-TH/0605073;%%
%\cite{Nishioka:2009un}
\bibitem{Nishioka:2009un} 
  T.~Nishioka, S.~Ryu and T.~Takayanagi,
  %``Holographic Entanglement Entropy: An Overview,''
  J.\ Phys.\ A {\bf 42}, 504008 (2009)
  [arXiv:0905.0932 [hep-th]].
  %%CITATION = ARXIV:0905.0932;%%

%\cite{Nakayama:2013fha}
\bibitem{Nakayama:2013fha} 
  Y.~Nakayama,
  %``Holographic interpretation of renormalization group approach to singular perturbations in non-linear differential equations,''
  Phys.\ Rev.\ D {\bf 88}, 105006 (2013)
  [arXiv:1305.4117 [hep-th]].
  %%CITATION = ARXIV:1305.4117;%%
  %2 citations counted in INSPIRE as of 06 Feb 2014

%\cite{Nakayama:2013ssa}
\bibitem{Nakayama:2013ssa} 
  Y.~Nakayama,
  %``Vector Beta function,''
  Int.\ J.\ Mod.\ Phys.\ A {\bf 28}, 1350166 (2013)
  [arXiv:1310.0574 [hep-th]].
  %%CITATION = ARXIV:1310.0574;%%
  %3 citations counted in INSPIRE as of 06 Feb 2014





%\cite{Nakayama:2009qu}
\bibitem{Nakayama:2009qu} 
  Y.~Nakayama,
  %``Forbidden Landscape from Holography,''
  JHEP {\bf 0911}, 061 (2009)
  [arXiv:0907.0227 [hep-th]].
  %%CITATION = ARXIV:0907.0227;%%
%\cite{Nakayama:2009fe}
\bibitem{Nakayama:2009fe}
  Y.~Nakayama,
  %``No Forbidden Landscape in String/M-theory,''
  JHEP {\bf 1001}, 030 (2010)
  [arXiv:0909.4297 [hep-th]].
  %%CITATION = JHEPA,1001,030;%%
\bibitem{No}
G.~W.~Gibbons, in: F.~de Aguila, J.~A.~de Azcarraga, L.E.
Ibanez (Eds.), Supersymmetry, Supergravity and Related Topics,
World Scientific, Singapore, 1985;

%\cite{ArkaniHamed:2003uy}
\bibitem{ArkaniHamed:2003uy} 
  N.~Arkani-Hamed, H.~-C.~Cheng, M.~A.~Luty and S.~Mukohyama,
  %``Ghost condensation and a consistent infrared modification of gravity,''
  JHEP {\bf 0405}, 074 (2004)
  [hep-th/0312099].
  %%CITATION = HEP-TH/0312099;%%

%\cite{Strominger:2001pn}
\bibitem{Strominger:2001pn} 
  A.~Strominger,
  %``The dS / CFT correspondence,''
  JHEP {\bf 0110}, 034 (2001)
  [hep-th/0106113].
  %%CITATION = HEP-TH/0106113;%%

%\cite{Shapere:2012nq}
\bibitem{Shapere:2012nq} 
  A.~Shapere and F.~Wilczek,
  %``Classical Time Crystals,''
  Phys.\ Rev.\ Lett.\  {\bf 109}, 160402 (2012)
  [arXiv:1202.2537 [cond-mat.other]].
  %%CITATION = ARXIV:1202.2537;%%


%\cite{Horava:2008ih}
\bibitem{Horava:2008ih} 
  P.~Horava,
  %``Membranes at Quantum Criticality,''
  JHEP {\bf 0903}, 020 (2009)
  [arXiv:0812.4287 [hep-th]].
  %%CITATION = ARXIV:0812.4287;%%

%\cite{Horava:2009uw}
\bibitem{Horava:2009uw} 
  P.~Horava,
  %``Quantum Gravity at a Lifshitz Point,''
  Phys.\ Rev.\ D {\bf 79}, 084008 (2009)
  [arXiv:0901.3775 [hep-th]].
  %%CITATION = ARXIV:0901.3775;%%
%\cite{Mukohyama:2010xz}
\bibitem{Mukohyama:2010xz} 
  S.~Mukohyama,
  %``Horava-Lifshitz Cosmology: A Review,''
  Class.\ Quant.\ Grav.\  {\bf 27}, 223101 (2010)
  [arXiv:1007.5199 [hep-th]].
  %%CITATION = ARXIV:1007.5199;%%


%\cite{Arnowitt:1962hi}
\bibitem{Arnowitt:1962hi} 
  R.~L.~Arnowitt, S.~Deser and C.~W.~Misner,
  %``The Dynamics of general relativity,''
  gr-qc/0405109.
  %%CITATION = GR-QC/0405109;%%




%\cite{Myers:2010xs}
\bibitem{Myers:2010xs} 
  R.~C.~Myers and A.~Sinha,
  %``Seeing a c-theorem with holography,''
  Phys.\ Rev.\ D {\bf 82}, 046006 (2010)
  [arXiv:1006.1263 [hep-th]].
  %%CITATION = ARXIV:1006.1263;%%






%\cite{Nojiri:1999mh}
\bibitem{Nojiri:1999mh} 
  S.~'i.~Nojiri and S.~D.~Odintsov,
  %``On the conformal anomaly from higher derivative gravity in AdS / CFT correspondence,''
  Int.\ J.\ Mod.\ Phys.\ A {\bf 15}, 413 (2000)
  [hep-th/9903033].
  %%CITATION = HEP-TH/9903033;%%

%\cite{Blau:1999vz}
\bibitem{Blau:1999vz} 
  M.~Blau, K.~S.~Narain and E.~Gava,
  %``On subleading contributions to the AdS / CFT trace anomaly,''
  JHEP {\bf 9909}, 018 (1999)
  [hep-th/9904179].
  %%CITATION = HEP-TH/9904179;%%



%\cite{Hoyos:2012xc}
\bibitem{Hoyos:2012xc} 
  C.~Hoyos, U.~Kol, J.~Sonnenschein and S.~Yankielowicz,
  %``The a-theorem and conformal symmetry breaking in holographic RG flows,''
  arXiv:1207.0006 [hep-th].
  %%CITATION = ARXIV:1207.0006;%%

%\cite{Bhattacharyya:2012tc}
\bibitem{Bhattacharyya:2012tc} 
  A.~Bhattacharyya, L.~-Y.~Hung, K.~Sen and A.~Sinha,
  %``On c-theorems in arbitrary dimensions,''
  Phys.\ Rev.\ D {\bf 86}, 106006 (2012)
  [arXiv:1207.2333 [hep-th]].
  %%CITATION = ARXIV:1207.2333;%%
%\cite{Liu:2010xc}
\bibitem{Liu:2010xc} 
  J.~T.~Liu, W.~Sabra and Z.~Zhao,
  %``Holographic c-theorems and higher derivative gravity,''
  Phys.\ Rev.\ D {\bf 85}, 126004 (2012)
  [arXiv:1012.3382 [hep-th]].
  %%CITATION = ARXIV:1012.3382;%%

%\cite{Liu:2011iia}
\bibitem{Liu:2011iia} 
  J.~T.~Liu and Z.~Zhao,
  %``A holographic c-theorem for higher derivative gravity,''
  arXiv:1108.5179 [hep-th].
  %%CITATION = ARXIV:1108.5179;%%

%\cite{Nakayama:2010wx}
\bibitem{Nakayama:2010wx} 
  Y.~Nakayama,
  %``Higher derivative corrections in holographic Zamolodchikov-Polchinski theorem,''
  Eur.\ Phys.\ J.\ C {\bf 72}, 1870 (2012)
  [arXiv:1009.0491 [hep-th]].
  %%CITATION = ARXIV:1009.0491;%%


%\cite{Visser:1994jb}
\bibitem{Visser:1994jb} 
  M.~Visser,
  %``Scale anomalies imply violation of the averaged null energy condition,''
  Phys.\ Lett.\ B {\bf 349}, 443 (1995)
  [gr-qc/9409043].
  %%CITATION = GR-QC/9409043;%%

%\cite{Urban:2009yt}
\bibitem{Urban:2009yt} 
  D.~Urban and K.~D.~Olum,
  %``Averaged null energy condition violation in a conformally flat spacetime,''
  Phys.\ Rev.\ D {\bf 81}, 024039 (2010)
  [arXiv:0910.5925 [gr-qc]].
  %%CITATION = ARXIV:0910.5925;%%
%\cite{Urban:2010vr}
\bibitem{Urban:2010vr} 
  D.~Urban and K.~D.~Olum,
  %``Spacetime Averaged Null Energy Condition,''
  Phys.\ Rev.\ D {\bf 81}, 124004 (2010)
  [arXiv:1002.4689 [gr-qc]].
  %%CITATION = ARXIV:1002.4689;%%



%\cite{Nakayama:2012jv}
\bibitem{Nakayama:2012jv} 
  Y.~Nakayama,
  %``Does anomalous violation of null energy condition invalidate holographic c-theorem?,''
  arXiv:1211.4628 [hep-th].
  %%CITATION = ARXIV:1211.4628;%%


%\cite{Aharony:2008ug}
\bibitem{Aharony:2008ug} 
  O.~Aharony, O.~Bergman, D.~L.~Jafferis and J.~Maldacena,
  %``N=6 superconformal Chern-Simons-matter theories, M2-branes and their gravity duals,''
  JHEP {\bf 0810}, 091 (2008)
  [arXiv:0806.1218 [hep-th]].
  %%CITATION = ARXIV:0806.1218;%%




%\cite{Page:1982fm}
\bibitem{Page:1982fm} 
  D.~N.~Page,
  %``Thermal Stress Tensors in Static Einstein Spaces,''
  Phys.\ Rev.\ D {\bf 25}, 1499 (1982).
  %%CITATION = PHRVA,D25,1499;%%

%\cite{Brown:1977sj}
\bibitem{Brown:1977sj} 
  L.~S.~Brown and J.~P.~Cassidy,
  %``Stress Tensors and their Trace Anomalies in Conformally Flat Space-Times,''
  Phys.\ Rev.\ D {\bf 16}, 1712 (1977).
  %%CITATION = PHRVA,D16,1712;%%


%\cite{Herzog:2013ed}
\bibitem{Herzog:2013ed} 
  C.~P.~Herzog and K.~-W.~Huang,
  %``Stress Tensors from Trace Anomalies in Conformal Field Theories,''
  arXiv:1301.5002 [hep-th].
  %%CITATION = ARXIV:1301.5002;%%


%\cite{Fuji:2011km}
\bibitem{Fuji:2011km} 
  H.~Fuji, S.~Hirano and S.~Moriyama,
  %``Summing Up All Genus Free Energy of ABJM Matrix Model,''
  JHEP {\bf 1108}, 001 (2011)
  [arXiv:1106.4631 [hep-th]].
  %%CITATION = ARXIV:1106.4631;%%
%\cite{Marino:2011eh}
\bibitem{Marino:2011eh} 
  M.~Marino and P.~Putrov,
  %``ABJM theory as a Fermi gas,''
  J.\ Stat.\ Mech.\  {\bf 1203}, P03001 (2012)
  [arXiv:1110.4066 [hep-th]].
  %%CITATION = ARXIV:1110.4066;%%
%\cite{Bhattacharyya:2012ye}
\bibitem{Bhattacharyya:2012ye} 
  S.~Bhattacharyya, A.~Grassi, M.~Marino and A.~Sen,
  %``A One-Loop Test of Quantum Supergravity,''
  arXiv:1210.6057 [hep-th].
  %%CITATION = ARXIV:1210.6057;%%





%\cite{Hotta:2009zn}
\bibitem{Hotta:2009zn} 
  K.~Hotta, Y.~Hyakutake, T.~Kubota, T.~Nishinaka and H.~Tanida,
  %``Left-Right Asymmetric Holographic RG Flow with Gravitational Chern-Simons Term,''
  Phys.\ Lett.\ B {\bf 680}, 279 (2009)
  [arXiv:0906.1255 [hep-th]].
  %%CITATION = ARXIV:0906.1255;%%
%\cite{Takayanagi:2011zk}
\bibitem{Takayanagi:2011zk} 
  T.~Takayanagi,
  %``Holographic Dual of BCFT,''
  Phys.\ Rev.\ Lett.\  {\bf 107}, 101602 (2011)
  [arXiv:1105.5165 [hep-th]].
  %%CITATION = ARXIV:1105.5165;%%
%\cite{Fujita:2011fp}
\bibitem{Fujita:2011fp} 
  M.~Fujita, T.~Takayanagi and E.~Tonni,
  %``Aspects of AdS/BCFT,''
  JHEP {\bf 1111}, 043 (2011)
  [arXiv:1108.5152 [hep-th]].
  %%CITATION = ARXIV:1108.5152;%%
%\cite{Nozaki:2012qd}
\bibitem{Nozaki:2012qd} 
  M.~Nozaki, T.~Takayanagi and T.~Ugajin,
  %``Central Charges for BCFTs and Holography,''
  JHEP {\bf 1206}, 066 (2012)
  [arXiv:1205.1573 [hep-th]].
  %%CITATION = ARXIV:1205.1573;%%


\bibitem{HN}
D.~Honda and M.~Nakamura,
private communication.
%\cite{Compere:2008cv}
\bibitem{Compere:2008cv}
  G.~Compere and S.~Detournay,
  %``Semi-classical central charge in topologically massive gravity,''
  Class.\ Quant.\ Grav.\  {\bf 26} (2009) 012001
   [Erratum-ibid.\  {\bf 26} (2009) 139801]
  [arXiv:0808.1911 [hep-th]].
  %%CITATION = ARXIV:0808.1911;%%
%\cite{Song:2011sr}
\bibitem{Song:2011sr} 
  W.~Song and A.~Strominger,
  %``Warped AdS3/Dipole-CFT Duality,''
  JHEP {\bf 1205}, 120 (2012)
  [arXiv:1109.0544 [hep-th]].
  %%CITATION = ARXIV:1109.0544;%%



%\cite{Son:2008ye}
\bibitem{Son:2008ye} 
  D.~T.~Son,
  %``Toward an AdS/cold atoms correspondence: A Geometric realization of the Schrodinger symmetry,''
  Phys.\ Rev.\ D {\bf 78}, 046003 (2008)
  [arXiv:0804.3972 [hep-th]].
  %%CITATION = ARXIV:0804.3972;%%
%\cite{Balasubramanian:2008dm}
\bibitem{Balasubramanian:2008dm} 
  K.~Balasubramanian and J.~McGreevy,
  %``Gravity duals for non-relativistic CFTs,''
  Phys.\ Rev.\ Lett.\  {\bf 101}, 061601 (2008)
  [arXiv:0804.4053 [hep-th]].
  %%CITATION = ARXIV:0804.4053;%%


%\cite{Kachru:2008yh}
\bibitem{Kachru:2008yh} 
  S.~Kachru, X.~Liu and M.~Mulligan,
  %``Gravity Duals of Lifshitz-like Fixed Points,''
  Phys.\ Rev.\ D {\bf 78}, 106005 (2008)
  [arXiv:0808.1725 [hep-th]].
  %%CITATION = ARXIV:0808.1725;%%


%\cite{Bagchi:2009my}
\bibitem{Bagchi:2009my} 
  A.~Bagchi and R.~Gopakumar,
  %``Galilean Conformal Algebras and AdS/CFT,''
  JHEP {\bf 0907}, 037 (2009)
  [arXiv:0902.1385 [hep-th]].
  %%CITATION = ARXIV:0902.1385;%%

%\cite{Seiberg:1988ur}
\bibitem{Seiberg:1988ur} 
  N.~Seiberg,
  %``Supersymmetry and Nonperturbative beta Functions,''
  Phys.\ Lett.\ B {\bf 206}, 75 (1988).
  %%CITATION = PHLTA,B206,75;%%

%\cite{Fortin:2012hc}
\bibitem{Fortin:2012hc} 
  J.~-F.~Fortin, B.~Grinstein, C.~W.~Murphy and A.~Stergiou,
  %``On Limit Cycles in Supersymmetric Theories,''
  arXiv:1210.2718 [hep-th].
  %%CITATION = ARXIV:1210.2718;%%


%\cite{Berkovits:2004xu}
\bibitem{Berkovits:2004xu} 
  N.~Berkovits,
  %``Quantum consistency of the superstring in AdS(5) x S**5 background,''
  JHEP {\bf 0503}, 041 (2005)
  [hep-th/0411170].
  %%CITATION = HEP-TH/0411170;%%


%\cite{Brown:1986nw}
\bibitem{Brown:1986nw} 
  J.~D.~Brown and M.~Henneaux,
  %``Central Charges in the Canonical Realization of Asymptotic Symmetries: An Example from Three-Dimensional Gravity,''
  Commun.\ Math.\ Phys.\  {\bf 104}, 207 (1986).
  %%CITATION = CMPHA,104,207;%%


%\cite{Guica:2008mu}
\bibitem{Guica:2008mu} 
  M.~Guica, T.~Hartman, W.~Song and A.~Strominger,
  %``The Kerr/CFT Correspondence,''
  Phys.\ Rev.\ D {\bf 80}, 124008 (2009)
  [arXiv:0809.4266 [hep-th]].
  %%CITATION = ARXIV:0809.4266;%%

%\cite{Compere:2012jk}
\bibitem{Compere:2012jk} 
  G.~Compere,
  %``The Kerr/CFT correspondence and its extensions: a comprehensive review,''
  Living Rev.\ Rel.\  {\bf 15}, 11 (2012)
  [arXiv:1203.3561 [hep-th]].
  %%CITATION = ARXIV:1203.3561;%%
%\cite{Douglas:2010rc}
\bibitem{Douglas:2010rc} 
  M.~R.~Douglas, L.~Mazzucato and S.~S.~Razamat,
  %``Holographic dual of free field theory,''
  Phys.\ Rev.\ D {\bf 83}, 071701 (2011)
  [arXiv:1011.4926 [hep-th]].
  %%CITATION = ARXIV:1011.4926;%%

%\cite{Balasubramanian:2013ux}
\bibitem{Balasubramanian:2013ux} 
  K.~Balasubramanian,
  %``Gravity duals of cyclic RG flows, with strings attached,''
  arXiv:1301.6653 [hep-th].
  %%CITATION = ARXIV:1301.6653;%%

%\cite{Awad:2000ac}
\bibitem{Awad:2000ac} 
  A.~M.~Awad and C.~V.~Johnson,
  %``Scale versus conformal invariance in the AdS / CFT correspondence,''
  Phys.\ Rev.\ D {\bf 62}, 125010 (2000)
  [hep-th/0006037].
  %%CITATION = HEP-TH/0006037;%%


\bibitem{ReyNakayama}
Y.~Nakayama and S.~Rey.
Unpublished.



%\cite{Lee:2010ub}
\bibitem{Lee:2010ub} 
  S.~-S.~Lee,
  %``Holographic description of large N gauge theory,''
  Nucl.\ Phys.\ B {\bf 851}, 143 (2011)
  [arXiv:1011.1474 [hep-th]].
  %%CITATION = ARXIV:1011.1474;%%
  %17 citations counted in INSPIRE as of 07 Feb 2014

%\cite{Lee:2012xba}
\bibitem{Lee:2012xba} 
  S.~-S.~Lee,
  %``Background independent holographic description : From matrix field theory to quantum gravity,''
  JHEP {\bf 1210}, 160 (2012)
  [arXiv:1204.1780 [hep-th]].
  %%CITATION = ARXIV:1204.1780;%%
  %12 citations counted in INSPIRE as of 07 Feb 2014

%\cite{Lee:2013dln}
\bibitem{Lee:2013dln} 
  S.~-S.~Lee,
  %``Quantum Renormalization Group and Holography,''
  JHEP {\bf 1401}, 076 (2014)
  [arXiv:1305.3908 [hep-th]].
  %%CITATION = ARXIV:1305.3908;%%
  %11 citations counted in INSPIRE as of 07 Feb 2014


%\cite{Nakayama:2014cca}
\bibitem{Nakayama:2014cca} 
  Y.~Nakayama,
  %``$a-c$ test of holography vs quantum renormalization group,''
  arXiv:1401.5257 [hep-th].
  %%CITATION = ARXIV:1401.5257;%%
  %1 citations counted in INSPIRE as of 07 Feb 2014

%\cite{Misner:1974qy}
\bibitem{Misner:1974qy} 
  C.~W.~Misner, K.~S.~Thorne and J.~A.~Wheeler,
  ``Gravitation,''
  San Francisco 1973, 1279p

\end{thebibliography}
\end{document}